\documentclass[11pt,a4paper]{article}
\pdfoutput=1
\usepackage{amssymb,amsmath,amsfonts, mathtools, mathrsfs}
\usepackage[utf8]{inputenc} % per usare le lettere accentate come à,è etc...
\usepackage[dvipsnames]{xcolor}

\newif\ifnatbibsort\natbibsorttrue

\relax

\ifnatbibsort\RequirePackage[numbers,sort&compress]{natbib}\else\RequirePackage[numbers,compress]{natbib}\fi
\RequirePackage[colorlinks=true
,urlcolor=blue
,anchorcolor=blue
,citecolor=blue
,filecolor=blue
,linkcolor=blue
,menucolor=blue
,pagecolor=blue
,linktocpage=true
,pdfproducer=medialab
,pdfa=true
]{hyperref}

\usepackage{graphicx}
\usepackage{caption}
\usepackage{tensor}
\usepackage{subfigure}
\usepackage{enumerate}
\usepackage{dsfont}
\usepackage{verbatim}
\usepackage{amsfonts,euscript,amssymb,stmaryrd,braket}
\usepackage{tikz}
\usetikzlibrary{arrows,decorations.markings,patterns}
\usepackage{slashed}

\def\clock{{\count0=\time
		\divide\count0 60
		\ifnum\count0<10 0\fi\the\count0
		\multiply\count0 -60 \advance\count0 \time
		:\ifnum\count0<10 0\fi \the\count0
}}
\newcommand{\timestamp}{{\small\vbox{\hbox{\tt\jobname.tex}
			\hbox{\the\day/\the\month/\the\year, \clock}}}}

\newcommand{\CE}{\mathcal{E}}
\newcommand{\CH}{\mathcal{H}}

\newcommand{\CJ}{\mathcal{J}}
\newcommand{\CV}{\mathcal{V}}

\newcommand{\CT}{\mathcal{T}}
\newcommand{\A}{\mathcal{A}}
\newcommand{\Q}{\mathcal{M}}

\newcommand{\nn}{\nonumber}
\newcommand{\bea}{\begin{eqnarray}}
\newcommand{\eea}{\end{eqnarray}}

\newcommand{\be}{\begin{equation}}
\newcommand{\ee}{\end{equation}}

\makeatletter
\let\old@startsection=\@startsection
\let\oldl@section=\l@section
\renewcommand{\@startsection}[6]{\old@startsection{#1}{#2}{#3}{#4}{#5}{#6\mathversion{bold}}}
\renewcommand{\l@section}[2]{\oldl@section{\mathversion{bold}#1}{#2}}
\makeatother

\numberwithin{equation}{section}
\def \T {{ \cal T}}
\def \O {{\mathcal O}}

\usepackage{color}

\def\E{\mathcal{E}}

\def\O{\mathcal{O}}

\def \RR {{\mathbb R}}

\def\ri {{\rm i}}
\def\rd {{\rm d}}
\def\e {{\rm e}}

\begin{document}
	\renewcommand{\thefootnote}{\arabic{footnote}}

	\overfullrule=0pt
	\parskip=2pt
	\parindent=12pt
	\headheight=0in \headsep=0in \topmargin=0in \oddsidemargin=0in

	\vspace{ -3cm} \thispagestyle{empty} \vspace{-1cm}
	\begin{flushright} 
		\footnotesize
		\textcolor{red}{\phantom{print-report}}
	\end{flushright}

\begin{center}
	\vspace{.0cm}

	{\Large\bf \mathversion{bold}
	Modular transport in two-dimensional conformal field theory
	}
%	\\
%	\vspace{.25cm}
%	\noindent
%	{\Large\bf \mathversion{bold}
%	Or not?}

%	{\Large\bf \mathversion{bold}
%	Entanglement generated quantum transport
%	}
%	\\
%	\vspace{.25cm}
%	\noindent
%	{\Large\bf \mathversion{bold}
%	in two-dimensional conformal field theory}

	\vskip  0.8cm
	{
		Mihail Mintchev$^{\,a}$,
		Diego Pontello$^{\,b}$
		and Erik Tonni$^{\,b}$
	}
	\vskip  1.cm
	
	\small
	{\em
		$^{a}\,$Dipartimento di Fisica, Universit\'a di Pisa and INFN Sezione di Pisa, \\
		largo Bruno Pontecorvo 3, 56127 Pisa, Italy
		\vskip 0.05cm
		$^{b}\,$SISSA and INFN Sezione di Trieste, via Bonomea 265, 34136, Trieste, Italy 
	}
	\normalsize

\end{center}

\vspace{0.3cm}
\begin{abstract} 

We study the quantum transport generated by the bipartite entanglement in two-dimensional conformal field theory
at finite density with the $U(1) \times U(1)$ symmetry associated to the conservation of the electric charge and of the helicity.
The bipartition given by an interval is considered, either on the line or on the circle.
The continuity equations and the corresponding conserved quantities
for the modular flows of the currents and of the energy-momentum tensor are derived. 
We investigate the mean values of the associated currents and their quantum fluctuations in the finite density representation,
which describe the properties of the modular quantum transport.
The modular analogues of the Johnson-Nyquist law 
and of the fluctuation-dissipation relation are found,
which encode the thermal nature of the modular evolution. 
\end{abstract}

\vspace{1cm}

\begin{center}
{\it Dedicated to the memory of Ivan Todorov}
\end{center}

%\vspace{1cm}
%\noindent
%\textcolor{red}{\bf $\bullet$ TO DO:}
%\\
%\textcolor{red}{\bf $\Longrightarrow$ Understand/fix the additive constant in $K$ (see (\ref{fmh})). Longo?}
%\\
%\textcolor{red}{\bf $\Longrightarrow$ Understand/fix whether $|\,\mathsf{j}'(u)|$ or $\mathsf{j}'(u)$ must be used throughout the manuscript 
%(e.g. (\ref{J-on-phi-1}), (\ref{J-on-phi-2}), etc.). Hollands?}
%\\

\vspace{.5cm}

%\noindent
%\textcolor{orange}{\bf 
%$\bullet$ Modular evolution in $P$ (see Sec.\,\ref{sec-momentum-evo}) and corresponding KMS condition
%}
%\\
%\textcolor{orange}{
%$\longrightarrow$ The chiral part is trivial. It also satisfies KMS. Add a discussion at the end of Sec.\,\ref{sec-mod-corr-line}
%}
%\\
%\textcolor{orange}{
%$\longrightarrow$ Discuss the {\bf modular transport generated by the modular momentum}
%}
%\\
%\textcolor{red}{\bf 
%$\bullet$ Extend our results for {\it boosted interval} by employing $(a_\pm, b_\pm)$ (new appendix?)
%}
%\\
%\textcolor{red}{
%$\longrightarrow$ Half page to add at the end of Sec.\,\ref{subsec-mod-H-vacuum}
%}
%\\

%\noindent
%\textcolor{red}{\bf 
%$\bullet$  Revisit the finite volume case: the vector fields are not PERIODIC in $u$
%}
%\\
%\textcolor{red}{\bf 
%$\bullet$  Are $\mu_\pm$ quantised at finite volume? NO, but see red remark about (\ref{ftd5}).
%Maybe we have problems in (\ref{mod-cor-phi-circ})-(\ref{mod-cor-phi-circ-2}).
%%$\Longrightarrow$ All the FIGURES in Sec.\,\ref{sec-finite-volume} and Sec.\,\ref{sec-noise-circle} should be changed (like in [JMT])
%}
%\\

\newpage

%%%%%%%%%%%%%%%%%%%%%%%%%%%%%%%%%%%%%
\tableofcontents
%%%%%%%%%%%%%%%%%%%%%%%%%%%%%%%%%%%%%

%%%%%%%%%%%%%%%%%%%%%%%%%%%%%%%%%%%%%%%%%%%
%\newpage
\section{Introduction}
\label{sec_intro}
%%%%%%%%%%%%%%%%%%%%%%%%%%%%%%%%%%%%%%%%%%%

Entanglement is a fundamental property of quantum systems
which can be investigated e.g. by considering a spatial bipartition identified by 
a subsystem $A$ and its complement $B$.
Assuming that the Hilbert space is factorised accordingly
and denoting by $\rho$ the state of the entire system,
the reduced density matrix $\rho_A \equiv \textrm{Tr}_B \rho$
obtained by tracing out the degrees of freedom associated to $B$
determines an intrinsic internal dynamics known as modular evolution \cite{Haag:1992hx}.
This evolution is generated by the modular Hamiltonian $K_A$ of the subsystem,
which is defined as $\rho_A \sim \e^{-K_A}$ and depends on the representation of the theory. 
%
%\textcolor{red}{\bf [modular evolution is thermal (KMS) with $\tilde{\beta} = 1$ (necessario o complica solo il discorso?)]}
%
A basic feature of the evolution of any quantum system is the existence of conserved charges. 
Their propagation in spacetime is implemented by the corresponding currents,
which generate the quantum transport
\cite{Datta-book, Blanter00, Nazarov-book}. 
In this paper we study the conserved charges, 
the currents and transport properties associated to the modular evolution. 
%
%In the present paper we study this aspect along a specific 
%time flow generated by the so called modular (entanglement) Hamiltonian \cite{}. 
%Associating the modular time $\tau$ with $K_A$, one can investigate the 
%transport properties of the system along the $\tau$-evolution. 
%The study of this quantum transport, called in what follows modular, is the main goal of our study below. 
%\bigskip 

The modular Hamiltonian is known analytically in very few cases. 
The most important one corresponds to the theorem of Bisognano and Wichmann
\cite{Bisognano:1975ih, Bisognano:1976za}, which considers 
the bipartition associated to half space of a local relativistic quantum field theory 
in its fundamental representation. 
Another important class of examples is given by 
conformal field theories (CFT) in two spacetime dimensions,
where some modular Hamiltonians $K_A$ of an interval $A$
are explicitly known and take a local form
in various inequivalent representations
\cite{Hislop:1981uh, Casini:2011kv, Wong:2013gua, Cardy:2016fqc, Mintchev:2022xqh}.

In two-dimensional CFT,
besides the fundamental representation, where the quantum transport is rather trivial,
other representations at finite particle density and/or finite temperature have been explored
\cite{Araki:1977iz, Liguori:1999tw, Bernard:2013aru, Hollands:2016svy, Bernard:2016nci, Akhmedov:2022gzt}.
When the left and right moving excitations have different temperature, 
the underlying states are non-equilibrium steady states (NESS) \cite{Ruelle2000, Sasa-Tasaki-06, Kita-10} 
and display interesting transport properties \cite{Liguori:1999tw, Bernard:2013aru, Hollands:2016svy, Bernard:2016nci, Cappelli:2001mp}. 
Focussing on the zero temperature case for the sake of simplicity,
in this paper we study the modular evolution generated by  
the modular Hamiltonian of an interval $A$
for a CFT in the state characterised by non-vanishing chemical potentials
and investigate the corresponding quantum transport.

We consider a local CFT in $1\!+\!1$ spacetime dimensions with $U(1)\times U(1)$ symmetry, 
implementing the electric charge and the helicity conservation. 
Along the modular flow,
conformal invariance fixes the one-point and two-point functions.
In this way one determines the mean values of the currents and their quadratic quantum 
fluctuations (noise), which provide a physical picture the modular quantum transport in the system.

The paper is organised as follows. 
In Sec.\,\ref{sec-algebraic-1} we discuss the evolution of the basic CFT chiral fields 
generated by a Hamiltonian depending on a smooth 
inhomogeneous velocity for each chirality. 
In Sec.\,\ref{sec-algebraic-2} 
these evolutions are employed to construct currents,
continuity equations and conserved quantities in this inhomogeneous CFT.
In Sec.\,\ref{sec-line} we consider the modular evolution 
for a CFT on the line in the finite density representation 
and the bipartition determined by an interval is considered
and in Sec.\,\ref{sec-transport-line} 
the corresponding modular transport properties are investigated.
In Sec.\,\ref{sec-finite-volume} and Sec.\,\ref{sec-noise-circle} 
these analyses are performed for the finite density representation of a CFT at finite volume. 
Some conclusions are drawn in Sec.\,\ref{sec_conclusions}.
Further results and technical details supporting the analyses described in the main text
are discussed in the Appendices\;\ref{app-comm-checks}, 
\ref{app-current-phi}, \ref{app-reps}, \ref{app-evolution-B-region}, 
\ref{app-michele-modular} and \ref{app-noise integrals}.

%%%%%%%%%%%%%%%%%%%%%%%%%%%%%%%%%%%%%%%%%%%
%\newpage
\section {CFT with spacetime dependent velocities} 
\label{sec-algebraic-1}
%%%%%%%%%%%%%%%%%%%%%%%%%%%%%%%%%%%%%%%%%%%

In this section we focus on some universal algebraic features of CFT 
in the two-dimensional Minkowski spacetime which hold in any representation.

\subsection{Commutation relations} 
\label{subsec-commutators}

Consider the chiral field algebras $\A_\pm = \{T_\pm(u), \phi_\pm(u), j_\pm(u), \dots;\, u \in \RR\}$
generating the right $(+)$ and left $(-)$ sectors of a CFT with $U(1)$ symmetry.
The algebraic properties defining the theory are encoded in commutation relations
involving the following chiral field \cite{luschermacknotes, Belavin:1984vu, luschernotes,  di-francesco}:

\medskip 

(i) the chiral components of the energy-momentum tensor $T_\pm$, that satisfy 
\be 
\big[\,T_\pm(u)\, ,T_\pm (v)\,\big] 
= 
\mp  \,\ri \,\delta(u-v) \,\partial_v T_\pm (v) 
\pm 2 \ri \, \delta' (u-v) \,T_\pm (v) 
\mp  \ri \, \frac{c}{24 \pi}\, \delta''' (u-v) 
\label{cft1}
\ee
where $c$ is the central charge of the CFT model;

\medskip 

(ii) the complex primary fields $\phi_\pm$, with dimensions $h_\pm $, occurring into
\bea
\label{cft4a}
\big[\,T_\pm(u)\, , \phi_\pm (v)\,\big] 
&=&
\mp \,\ri \, \delta(u-v) \, \partial_v \phi_\pm (v) 
\pm \ri \, h_\pm \, \delta^\prime (u-v) \, \phi_\pm (v) 
\\
\label{cft4b}
\big[\, T_\pm(u)\, , \phi^*_\pm (v)\, \big] 
&=&
\mp \,\ri\, \delta(u-v) \ \partial_v \phi^*_\pm (v) 
\pm \ri \, h_\pm \delta^\prime (u-v) \, \phi^*_\pm (v) 
\eea
which are consistent\footnote{For instance, the Jacobi identity involving (\ref{cft1}) and (\ref{cft4a}) holds.} 
with (\ref{cft1}).

\medskip 

(iii) the chiral components of a conserved current $j_\pm$, with dimension $h_{j_\pm}=1$, which 
generate the $U(1)$ transformations and satisfy 
\be 
\label{cft6ab}
\big[\, j_\pm(u)\, , \phi_\pm (v)\, \big] = - \,\delta(u-v) \, \phi_\pm (v)  
\;\;\;\qquad\;\;\;
\big[\, j_\pm(u)\, ,\, \phi^*_\pm (v)\, \big] =  \delta(u-v) \, \phi^*_\pm (v)   \,.
\ee
The commutators (\ref{cft6ab}) imply that the charge of $\phi$ and $\phi^*$ are equal to $-1$ and $1$ respectively.
Moreover, 
%\textcolor{red}{\bf [notation: should we replace $\frac{\kappa}{2\pi} $ with $\kappa$ below (and in the following)?]}
%\\
%\textcolor{red}{[maybe not, otherwise, for the same reason, we should also replace $\frac{c}{24\pi} $ with $c$ in (\ref{cft1})]}
\be 
\big[ \, j_\pm(u)\, ,\, j_\pm (v) \, \big] = \mp \, \ri \, \frac{\kappa}{2\pi} \, \delta^\prime(u-v) 
\label{cft7}
\ee
where $\kappa $ is a real constant and the r.h.s. is known as Schwinger term.

%\medskip 

In the Appendix \ref{app-comm-checks} we report some consistency checks for the commutators (\ref{cft1}) 
and (\ref{cft7}) through the two-point functions in the fundamental representation 
which determines the normalization of the two-point functions for $j_\pm$ and $T_\pm$.

In order to obtain a conventional quantum field theory structure from the above algebraic setting, 
one should fix a Hilbert space representation $\CH_+$ and $\CH_-$ of the chiral algebras $\A_+$ and 
$\A_-$ respectively. After smearing with smooth test functions, the elements of $\A_\pm$ act as operators in 
$\CH_+\otimes \CH_-$, that represents the physical state space of the system. 
Before fixing $\CH_\pm$, in the following we employ (\ref{cft1})-(\ref{cft7}) 
to obtain some results which are independent of the representation.

%%%%%%%%%%%%%%%%%%%%%%%%%%%%%%%%%%%%%%%%%

%%%%%%%%%%%%%%%%%%%%%%%%%%%%%%%%%%%%%%%%%
%\newpage
\subsection{Evolution through spacetime dependent velocities} 
\label{sec-mod-evolution}
%%%%%%%%%%%%%%%%%%%%%%%%%%%%%%%%%%%%%%%%%

%In a 2D CFT, some modular Hamiltonians corresponding to a spatial bipartition made by a single interval
%involve a specific position dependent velocity (called in that context modular temperature)
%\textcolor{red}{[refs: Hislop-Longo, CHM11, Klich-Wong, CT16]}. 

In the following 
we consider a general setting where the temporal evolution of 
the right and the left sectors of the CFT is characterised by two independent velocities $V_\pm (u_\pm)$, 
which are smooth real functions depending on the light-cone coordinates $u_\pm \equiv x \pm t$. 
In particular, we focus on the Hamiltonian 
\be 
H  = 
\int_{-\infty}^\infty \! V_+(u_+)\, {\cal T}_+(u_+)\, \rd u_+
+ 
\int_{-\infty}^\infty \! V_-(u_-)\, {\cal T}_-(u_-)\, \rd u_-
\label{me3b}
\ee
where
\be
\label{cal-T-pm-def}
{\cal T}_\pm (u) \equiv T_\pm(u) - \mu_\pm\, j_\pm(u)
\ee
being $\mu_\pm$ defined as the chemical potentials associated to the two chiralities.
%which provide the ones associated to the non-vanishing electric charge and helicity densities. 
The temporal evolution generated by (\ref{me3b}) with $\mu_{+} = \mu_{-} = 0$ and $V_{+} = V_{-}$ independent of $t$
has been studied also in \cite{Moosavi:2019fas}.

The evolution of the primary fields $\phi_\pm(u)$ generated by (\ref{me3b}) is  
\be 
\phi_\pm (\tau ,u) \equiv \e^{\ri \tau H }\, \phi_\pm(u)\, \e^{-\ri \tau H}  
\label{nme1}
\ee
where $\tau$ will be called modular time to highlight its different role with respect to the physical time $t$.
Notice that the dimension of $V_\pm$ determines the dimension of $\tau$.
Furthermore, we remark that adding a real constant in the r.h.s. of (\ref{me3b}) does not influence the corresponding evolution (\ref{nme1}).

Taking the derivative of (\ref{nme1}) with respect to $\tau$ provides the corresponding equation of motion
\be
\label{mem2}
\partial_\tau \phi_\pm(\tau,u) 
\,=\, 
\ri \, \big[\, H  , \phi_\pm(\tau,u)\, \big] 
\,=\, 
\ri \, \big[\, H \, , \e^{\ri \tau H} \, \phi_\pm(u)\,  \e^{-\ri \tau H } \, \big] 
\,=\, 
\ri \,  \e^{\ri \tau H }\, \big[\, H \, ,  \phi_\pm(u) \, \big]\, \e^{- \ri \tau H }  \,.
\ee
By using (\ref{cft4a}), the first commutator in (\ref{cft6ab}) and (\ref{me3b}), for finite values of $u$ one finds 
\bea
\partial_\tau \phi_\pm(\tau  ,u) 
&=&
\e^{\ri \tau  H }\, 
\int_{-\infty}^\infty  
\! V_\pm (v) \,
\Big\{
\delta(v - u) \, (\pm \,\partial_{u} +\ri \mu_\pm )\phi_\pm(u) 
\mp h_\pm \,\partial_{v} \delta(v-u) \phi_\pm (u) 
\Big\}\, \rd v
\,\e^{-\ri \tau  H} 
\nonumber 
\\
& = & 
%\hspace{-2.5cm}
\e^{\ri \tau  H }\, 
\Big\{\!
\pm V_\pm (u) \big(\partial_{u} \pm \ri \mu_\pm \big) \phi_\pm(u) 
\pm h_\pm \big[\partial_u V_\pm (u)\big] \phi_\pm (u) 
\Big\}
\,\e^{-\ri \tau  H } 
\qquad \qquad \qquad  
\label{mem3}
\eea
which leads to 
\be
\partial_\tau \phi_\pm(\tau  ,u) 
= 
\pm \,V_\pm (u) \big(\partial_{u} \pm \ri \mu_\pm \big) \phi_\pm(\tau  ,u) 
\pm h_\pm \big[\partial_u V_\pm (u)\big] \phi_\pm (\tau  ,u)  \,.
\label{mem4}
\ee  
%\textcolor{red}{[in the above discussion we integrate by parts, we used that $V_\pm$ vanish at the boundary. 
%Should we restrict the integration to an interval where $w^{-1}$ is defined?]}

In order to solve (\ref{mem4}),
let us  consider the following linear first order equation 
\be 
\partial_\tau  \xi_\pm (\tau ,u) = \pm \, V_\pm (u) \, \partial_{u} \xi_\pm (\tau ,u) 
\label{me2a}
\ee 
with initial condition 
\be 
\xi_\pm (0,u) = u    \,.
\label{me2b}
\ee 
We assume that $V_\pm(u)$ are smooth functions with a finite number of zeros;
hence they are finite functions for any finite value of $u$.
We introduce $w_\pm(u)$ through the following condition
\be
\label{s1}
w'_\pm(u) = \frac{1}{V_\pm(u)}
\ee
and consider the interval $A \equiv [a,b]$  identified by two consecutive zeros of $V_\pm(u)$,  i.e. 
\be 
V_\pm(a) = V_\pm(b) = 0  \,.
%\quad \Longrightarrow \quad \partial_t Q_{[a,b]} = 0 
\label{cond}
\ee
When $u\in A$, the function $w_\pm(u)$ is monotonic and therefore its inverse function $w^{-1}_\pm(u)$ is well defined.
%The solution is well known and is expressed in terms of the inverse $w_\pm^{-1}(x)$ of the functions 
%\be 
%w_\pm(x) \equiv  \int_{x_0}^x \frac{\rd y }{v_\pm (y)} 
%\label{s1}
%\ee 
In this case, we define
\be 
\xi_\pm (\tau,u) 
\equiv
w_\pm^{-1} \big(w_\pm(u) \pm \tau \big)  \,.
\label{s2}
\ee 
The functions $\xi_\pm (\tau,u)$ satisfy the initial condition (\ref{me2b}) and 
\be 
\partial_\tau \xi_\pm (\tau,u) = 
\pm \, V_\pm\big(\xi_\pm (\tau,u)\big) 
\;\;\; \qquad\;\;\; 
\partial_u \xi_\pm (\tau,u) = 
\frac{V_\pm \big(\xi_\pm (\tau,u)\big)}{V_\pm(u)}  \;.
\label{s4}
\ee

The solution of the equation of motion (\ref{mem4}) is 
\be 
\phi_\pm (\tau,u) 
\,=\, 
\e^{\pm \ri \mu_\pm [\xi_\pm (\tau,u) - u]}\, \big[\partial_u \xi_\pm (\tau,u)\big]^{h_\pm} \, \phi _\pm\big(\xi_\pm (\tau,u)\big) 
\label{me9a}
\ee
and satisfies the initial condition 
\be 
\phi_\pm(0,u) = \phi_\pm (u)  \,.
\label{ic1}
\ee
Analogously, the evolution of the primary fields  $\phi_\pm^*$ is described by  
\be 
\phi^*_\pm (\tau,u) \,=\, \e^{\mp \ri \mu_\pm [\xi_\pm (\tau,u) - u]}\, \big[\partial_u \xi_\pm (\tau,u) \big]^{h_\pm} \, \phi^*_\pm(\xi_\pm (\tau,u))  \,.
\label{me9b}
\ee

The above considerations can be applied to explore also the evolution of the chiral currents 
generated by (\ref{me3b}), namely
$j_\pm (\tau ,u) \equiv  \e^{\ri \tau H}\, j_\pm(u)\, \e^{- \ri \tau H} $.
%\be 
%j_\pm (\tau ,u) \equiv  \e^{\ri \tau H}\, j_\pm(u)\, \e^{- \ri \tau H} 
%\label{me10}
%\ee
The corresponding equation of motion is affected by the Schwinger term in (\ref{cft7}) 
and has the form 
\bea
\label{mem4j}
\partial_\tau j_\pm(\tau,u) 
&=&
\pm \,V_\pm (u) \,\partial_u j_\pm(\tau, u) 
\pm  \big[ \partial_u V_\pm(u) \big]  j_\pm (\tau, u) 
\pm \frac{\kappa \mu_\pm}{2\pi} \,\partial_u V_\pm (u) 
\\
\label{mem4j-bis}
\rule{0pt}{.5cm}
&=&
%\pm \,V_\pm (u) \,\partial_u j_\pm(\tau, u) 
\pm\, \partial_u  \big[ V_\pm(u) \,  j_\pm (\tau, u)  \big] 
\pm \frac{\kappa \mu_\pm}{2\pi} \,\partial_u V_\pm (u) 
\eea
with initial condition $ j_\pm(0,u) = j_\pm(u)$;
whose solution is 
\be 
j_\pm (\tau,u) 
= 
\big[\partial_u \xi_\pm (\tau,u) \big] \, j _\pm\big(\xi_\pm (\tau,u)\big) 
- \frac{\kappa \mu_\pm}{2\pi} \,
\big[1- \partial_u \xi_\pm (\tau,u) \big]  \,.
\label{cp4}
\ee

The continuity equation (\ref{mem4j-bis}) can be written also as
\be
\label{cp4-mu-0}
\partial_\tau  j_\pm (\tau, u)  
=
\pm\, \partial_u  \!\left[ \,
V_\pm(u) 
\left( j_\pm (\tau, u)  + \frac{\kappa \mu_\pm}{2\pi} \right)
\right]
\ee
or, equivalently, as
\be
\label{cp4-mu}
\partial_\tau \!
\left( j_\pm (\tau, u)  + \frac{\kappa \mu_\pm}{2\pi} \right)
=
\pm\, \partial_u  \!\left[ \,
V_\pm(u) 
\left( j_\pm (\tau, u)  + \frac{\kappa \mu_\pm}{2\pi} \right)
\right] .
\ee
In the representations investigated in this manuscript,
the mean value of $j_\pm (\tau, u)  + \tfrac{\kappa \mu_\pm}{2\pi} $ vanishes
(see (\ref{fd1}) and (\ref{ftd1-L}), with the velocities given by (\ref{velocity_fund}) and (\ref{velocity_fund-circle}) respectively).
The previous observation leads us to realise that (\ref{mem4j-bis}) 
is equivalent to
\be
\label{cp4-mu-alpha}
\partial_\tau \!
\left( j_\pm (\tau, u)  + \tilde{\alpha}_\pm \frac{\kappa \mu_\pm}{2\pi} \right)
=
\pm\, \partial_u  \!\left[ \,
V_\pm(u) 
\left( j_\pm (\tau, u)  + \alpha_\pm \frac{\kappa \mu_\pm}{2\pi} \right)
\right] 
\pm \big(1-\alpha_\pm\big) \frac{\kappa \mu_\pm}{2\pi} \, \partial_u  V_\pm(u) 
\ee
where $\alpha_\pm$ and $\tilde{\alpha}_\pm$ are real constants.
However, in our analyses we mainly consider the case where 
$\alpha_\pm=\tilde{\alpha}_\pm=0$, 
as discussed in Sec.\,\ref{subsec-charge-continuity}.

The evolution of $T_\pm(u)$ generated by (\ref{me3b}), 
i.e. $T_\pm (\tau,u) \equiv  \e^{\ri \tau H}\, T_\pm(u)\, \e^{- \ri \tau H}$,
can be explored by employing (\ref{cft1}), (\ref{cft4a}) and (\ref{cft7}),
which lead to\footnote{The evolution in the physical time $t$ is generated by (\ref{me3b})
with constant velocities $V_\pm $ and $\mu_\pm = 0$.
In this case, from (\ref{cft1}), one finds 
\be
\partial_t T_\pm(t,u) = \pm V_\pm \,\partial_u T_\pm(t,u) 
\ee
whose solution is
\be
T_\pm(t,u) = T_\pm(u \pm V_\pm  t) 
\ee
We choose the convention where $V_\pm =1$.}
\be
\partial_\tau T_\pm(\tau,u) 
= 
\pm \,V_\pm (u) \, \partial_u T_\pm(\tau, u) 
\pm 2 \big[\partial_u V_\pm (u)\big] T_\pm (\tau, u) 
\mp \frac {c }{24\pi} \, \partial^3_u V_\pm (u) 
\mp \mu_\pm  V_\pm(u)\, \partial_u j_\pm(\tau,u)
\label{met3-0}
\ee
where $j_\pm(\tau,u)$ is given by (\ref{cp4}).
%\noindent
%\textcolor{blue}{
%{\bf $\bullet$ Constant velocities case [maybe in a footnote around (\ref{met3-0})?]:}
%\\
%For the physical time $t$ evolution, generated by (\ref{me3b})
%with constant velocities $V_\pm $ and $\mu_\pm = 0$, 
%from (\ref{cft1}) one finds 
%\be
%\partial_t T_\pm(t,u) = \pm V_\pm \,\partial_u T_\pm(t,u) 
%\ee
%whose solution is
%\be
%T_\pm(t,u) = T_\pm(u \pm V_\pm  t) 
%\ee
%We choose the convention where $V_\pm =1$.
%}
The occurrence of  $j_\pm(\tau,u)$ in (\ref{met3-0})
leads us to consider the evolution of $\T_\pm(u)$, 
namely $\T_\pm (t,x) \equiv  \e^{\ri t H}\, \T_\pm(x)\, \e^{- \ri t H}$.
%\be 
%\T_\pm (t,x) \equiv  \e^{\ri t H}\, \T_\pm(x)\, \e^{- \ri t H}
%\ee 
Indeed, the equation of motion for these fields reads
\be
\partial_\tau \T_\pm(\tau,u) 
= 
\pm \,V_\pm (u) \partial_u \T_\pm(\tau, u) 
\pm 2 \big[\partial_u V_\pm (u) \big] \T_\pm (\tau, u) 
\mp \frac{\kappa \mu_\pm^2}{2\pi} \, \partial_u V_\pm(u) 
\mp \frac {c }{24\pi} \, \partial^3_u V_\pm (u) 
\label{met3}
\ee 
with the initial condition $ \T_\pm(0,u) = \T_\pm(u)$,
where only the field $\T_\pm(\tau, u) $ occurs.
%and depends on both $\kappa$ and the central charge $c$.

Multiplying both sides of (\ref{met3})  by $V_\pm (u) $, we obtain the following equivalent form
\be
\partial_\tau \big[ V_\pm (u) \, \T_\pm(\tau,u) \big]
= 
\pm \,\partial_u \big[ V_\pm (u)^2 \,\T_\pm(\tau, u) \big]
%\pm 2 \big[\partial_u V_\pm (u) \big] \T_\pm (\tau, u) 
\mp \frac{\kappa \mu_\pm^2}{4\pi} \, \partial_u V_\pm(u)^2
\mp \frac {c }{24\pi} \, V_\pm (u) \,\partial^3_u V_\pm (u) \,.
\label{met3-bis}
\ee 
This differential equation can be arranged as 
\be
\partial_\tau \! 
\left[ \,
V_\pm (u)  \left( \T_\pm(\tau, u) - \frac{\kappa \mu_\pm^2}{4\pi}  \right)
\right]
= 
\pm \,\partial_u \! 
\left[ \,
V_\pm (u)^2  \left( \T_\pm(\tau, u) - \frac{\kappa \mu_\pm^2}{4\pi}  \right)
\right]
%\pm 2 \big[\partial_u V_\pm (u) \big] \T_\pm (\tau, u) 
%\mp \frac{\kappa \mu_\pm^2}{4\pi} \, \partial_u V_\pm(u)^2
\mp \frac {c }{24\pi} \, V_\pm (u) \,\partial^3_u V_\pm (u) \,.
\label{met3-ter}
\ee
As done in (\ref{cp4-mu-alpha}) for the differential equation for $ j_\pm (\tau, u) $,
(\ref{met3-ter}) can be written also in a general form 
involving the constants $\alpha_\pm$ and $\tilde{\alpha}_\pm$
that we do not find worth reporting here.

In the finite density representation on the line,
where the mean values (\ref{fd1}) and the velocity 
$V_\pm (u) = V(u)$ in (\ref{velocity_fund}) must be used,
the operators $\T_\pm(\tau, u) - \tfrac{\kappa \mu_\pm^2}{4\pi} $
have vanishing mean values and $\partial^3_u V (u) =0$.
Instead, in the finite density representation on the circle of length $L$,
where the mean values (\ref{ftd1-L}) and the velocity 
$V_\pm (u) = V_L(u)$ in (\ref{velocity_fund-circle}) must be employed,
it is more convenient to write (\ref{met3-ter}) as follows
\bea
\label{met3-ter-circle}
\partial_\tau \! 
\left[ \,
V_\pm (u)  \left( \T_\pm(\tau, u) - \frac{\kappa \mu_\pm^2}{4\pi}  + \frac{\pi c}{12 L^2} \right)
\right]
&=&
\pm \,\partial_u \! 
\left[ \,
V_\pm (u)^2  \left( \T_\pm(\tau, u) - \frac{\kappa \mu_\pm^2}{4\pi}  + \frac{\pi c}{12 L^2} \right)
\right]
\nn
\\
\rule{0pt}{.7cm}
& &
%\pm 2 \big[\partial_u V_\pm (u) \big] \T_\pm (\tau, u) 
%\mp \frac{\kappa \mu_\pm^2}{4\pi} \, \partial_u V_\pm(u)^2
\mp\, 
\frac {c }{24\pi} 
\left(
 \frac{2 \pi^2}{L^2}\, \partial_u V_\pm (u)^2 
+
V_\pm (u) \,\partial^3_u V_\pm (u) 
\right) .
\hspace{1.1cm}
\eea
Indeed, in this representation 
the operators within the round brackets in (\ref{met3-ter-circle})
have vanishing mean values and 
$ \tfrac{2 \pi^2}{L^2}\, \partial_u V_L (u)^2 +V_L (u) \,\partial^3_u V_L (u) =0$
for the velocity (\ref{velocity_fund-circle}).

The solution of (\ref{met3})  reads
\be 
\T_\pm (\tau,u) 
= 
\big[\partial_u \xi_\pm (\tau,u)\big]^2 \, \T _\pm\big(\xi_\pm (\tau,u)\big) 
+ \frac{\kappa \mu^2_\pm}{4\pi} \,
\Big\{1- \big[\partial_u \xi_\pm (\tau,u)\big]^2\Big\} 
- \frac {c}{24\pi} \, \mathcal{S}_u[\xi_\pm] (\tau,u) 
\label{met4}
\ee
where $\mathcal{S}_u[\xi ] (\tau,u)$ is the Schwarzian derivative of the functions $\xi_\pm (\tau,u)$, i.e. 
\be 
\mathcal{S}_u[\xi ] (\tau,u) 
= 
\frac{\partial^3_u \xi (\tau,u)}{\partial_u \xi (\tau,u)} 
- 
\frac{3}{2} \left [\frac{\partial^2_u \xi (\tau,u)}{\partial_u \xi (\tau,u)}\right ]^2  .
\label{cft3}
\ee
The solution (\ref{met4}) has been found by exploiting the following identity
\be 
\partial_\tau \mathcal{S}_u[\xi_\pm] (\tau,u) 
= 
\pm \,V_\pm (u) \, \partial_u \mathcal{S}_u[\xi_\pm] (\tau,u) 
\pm  2 \big[\partial_u V_\pm (u)\big] \mathcal{S}_u[\xi_\pm] (\tau,u) 
\pm \partial^3_u V_\pm (u) 
\label{met6}
\ee 
which is a consequence of (\ref{me2a}) and (\ref{cft3}). 
Multiplying (\ref{met6}) by $V_\pm(u)$, it can be written as
\be 
\partial_\tau \big\{ V_\pm(u) \, \mathcal{S}_u[\xi_\pm] (\tau,u) \big\}
=
\pm \, \partial_u \big\{ V_\pm (u)^2 \, \mathcal{S}_u[\xi_\pm] (\tau,u)  \big\}
\pm  V_\pm(u) \, \partial^3_u V_\pm (u)  \,.
\ee 
By employing the second expression in (\ref{s4}), one finds that (\ref{cft3}) can be expressed as follows 
\bea
\label{Schw-explicit-0}
\mathcal{S}_u[\xi_\pm] (\tau,u) 
&=&
%\frac{1}{v_\beta(x)^2} \,
%\left\{
%v''_\beta\big(\xi(\tau,x)\big) \, v_\beta\big(\xi(\tau,x)\big)  - \frac{v'_\beta\big(\xi(\tau,x)\big) ^2}{2}
%-
%\left[\,
%v''_\beta(x) \, v_\beta(x) - \frac{v'_\beta(x)^2}{2}
%\,\right]
%\right\}
%\nonumber
%\\
%\rule{0pt}{1.cm}
%&=&
\frac{ V_\pm \big(\xi(\tau,u)\big)^2 }{V_\pm(u)^2} 
\left[\,
\frac{ \partial^2_uV_\pm\big(\xi(\tau,u)\big) }{ V_\pm\big(\xi(\tau,u)\big) } 
- \frac{1}{2} \Bigg( \frac{\partial_uV_\pm\big(\xi(\tau,u)\big)}{ V_\pm\big(\xi(\tau,u)\big) } \Bigg)^2
\,\right]
\nonumber
\\
\rule{0pt}{.8cm}
& &
- 
\left[\,
\frac{ \partial^2_uV_\pm(u) }{ V_\pm(u) } - \frac{1}{2} \bigg( \frac{\partial_uV_\pm(u)}{V_\pm(u) } \bigg)^2
\,\right]  .
%\nonumber
%\\
%\rule{0pt}{1.cm}
%&=&
%\big( \partial_x \xi(\tau,x) \big)^2
%\left[\,
%\frac{ v''_\beta\big(\xi(\tau,x)\big) }{ v_\beta\big(\xi(\tau,x)\big) }    - \frac{1}{2} \left( \frac{v'_\beta\big(\xi(\tau,x)\big)}{ v_\beta\big(\xi(\tau,x)\big) } \right)^2
%\,\right]
%-
%\left[\,
%\frac{ v''_\beta(x) }{ v_\beta(x) } - \frac{1}{2} \left( \frac{v'_\beta(x)}{v_\beta(x) } \right)^2
%\,\right]
%\nonumber
%\\
%& &
\eea
This result will be applied in Sec.\,\ref{subsec-mod-H-vacuum} and Sec.\,\ref{sec-mod-ham-circle} for specific expressions of $V_\pm(u)$.

%%%%%%%%%%%%%%%%%%%%%%%%%%%%%%%%%%%%%%%%%%%
%\newpage
\section {Conservation laws, currents and charges} 
\label{sec-algebraic-2}
%%%%%%%%%%%%%%%%%%%%%%%%%%%%%%%%%%%%%%%%%%%

In this section we discuss the conservation of some charges in the spacetime diamond
\be
\label{diamond}
\mathcal{D}_A \equiv \big\{ (x,t)  : \, a \leqslant u_+ \leqslant  b\, ,\; a \leqslant u_- \leqslant  b \,\big\}
%\;\;\;\qquad\;\;\;
%u_\pm \equiv x \pm t
\ee
corresponding to the domain of dependence of the interval $A \equiv [a,b]$,
which can be identified by two consecutive zeros $a < b$ of $V_\pm (u)$. 
The spacetime coordinates $(x,t)$ of the vertices of $\mathcal{D}_A$ are 
$P_a = (a,0)$, $P_b = (b,0)$, 
$P_{+\infty} = (\tfrac{a+b}{2}, \tfrac{b-a}{2})$ and $P_{-\infty} = (\tfrac{a+b}{2}, -\tfrac{b-a}{2})$.
The function (\ref{s2}) provides $\xi_\pm(\tau, u_\pm) $, 
which allow to construct the trajectory in $\mathcal{D}_A$ 
parameterised by $\tau \in \RR$
whose generic point has spacetime coordinates
\be
\label{mod-traj-tau}
x(\tau) = \frac{\xi_+(\tau, u_+) + \xi_-(\tau, u_-) }{2}
\;\;\;\qquad\;\;\;
t(\tau) = \frac{\xi_+(\tau, u_+) - \xi_-(\tau, u_-) }{2}
\ee
where $(u_{+}, u_{-})$ are the light-cone coordinates of the initial point 
corresponding to $\tau=0$.

\subsection{Electric charge and helicity}
\label{subsec-charge-continuity}

%{\color{purple}

From the commutation relations (\ref{cft6ab}),
the electric charge density at $\tau=0$ is
\be 
\label{eden-0-tau}
\varrho (\tau=0;x,t) \equiv  j_+(u_+ ) + j_-(u_- )
%\;\;\;\qquad\;\;\;
%u_\pm \equiv x \pm t
\ee
where $u_\pm = x\pm t$ are the light-cone coordinates.
Hence, its modular evolution reads
\be 
\label{eden}
\varrho (\tau;x,t) \equiv  j_+(\tau, u_+ ) + j_-(\tau, u_- )
%\;\;\;\qquad\;\;\;
%u_\pm \equiv x \pm t
\ee
where $ j_\pm(\tau, u_\pm )$ are given by (\ref{cp4}).

We remark that the quantity (\ref{eden}) and all the subsequent ones defined in a similar way 
depend both on the physical time $t$ associated to the Hamiltonian of the CFT
and on the evolution parameter $\tau$ associated to the Hamiltonian (\ref{me3b}),
which will play the role of the modular Hamiltonian from the next section.

By using (\ref{mem4j}) and $\partial_x = \partial_{u_+} +\partial_{u_-}$, for $\mu_\pm =0$ 
we find the continuity relation 
\be 
\partial_\tau \varrho(\tau;x,t) = - \,\partial_x j_x(\tau;x,t) 
\label{ecr0}
\ee
where 
\be 
\label{ecurr}
j _x(\tau;x,t) 
\equiv 
-\,V_+(u_+)\,j_+(\tau, u_+ ) + V_-(u_-)\,j_-(\tau, u_- )
\ee
defines the space component of the electric current when spacetime dependent velocities $V_\pm (u_\pm )$ occur. 
%Notice that $j_x$ and $j_\pm$ do not have the same dimensions.
In the special case of $V_+(u_+) = V_{-}(u_{-})=1$ identically and for $\tau=0$, from (\ref{cp4}) we find that the current (\ref{ecurr}) becomes 
\be 
j _x(\tau=0;x,t)\big|_{V_+ = V_{-}=1} 
\,=\,
-\,j_+(u_+ ) + j_-(u_- )
\label{ecurr-0-1}
\ee
i.e. the standard CFT expressions employed e.g. in \cite{Liguori:1999tw, Hollands:2016svy, Bernard:2013aru, Bernard:2016nci},
which has vanishing expectation value in the fundamental representation.

When $\mu_\pm \not= 0$, 
the Schwinger term  in (\ref{cft7}) generates an additional contribution with respect to (\ref{ecr0}). 
In the case we are exploring, given the definition (\ref{ecurr}),
by employing (\ref{mem4j}) and $\partial_t = \partial_{u_+} - \partial_{u_-}$
we find the following continuity equation
\be 
\partial_\tau \varrho(\tau;x,t)  
= - \,\partial_x j_x(\tau;x,t) + 
\frac{\kappa}{2\pi}\,
\partial_t \big[ \,\mu_+ \, V_+(u_+) + \mu_- \, V_-(u_-) \,\big]
\label{ecr-time}
\ee
which naturally leads us to introduce the following  $t$-component for the charge current
\be
\label{j-t-def}
j_t(\tau;x,t)   \equiv   \frac{\kappa}{2\pi}\,\big[ \,\mu_+ \, V_+(u_+) + \mu_- \, V_-(u_-) \,\big]  
\ee
which is independent of $\tau$.

%\textcolor{red}{\bf [should we interpret the second term in the r.h.s. of (\ref{ecr-time}) as a classical source?]}

We emphasize that, from either (\ref{mem4j}) or (\ref{cp4-mu-0}) and $\partial_x = \partial_{u_+} + \partial_{u_-}$,
it is straightforward to observe that (\ref{ecr-time}) can be written also as 
\bea
\partial_\tau \varrho(\tau;x,t)  
&=&
- \,\partial_x j_x(\tau;x,t) + 
\frac{\kappa}{2\pi}\,
\partial_x \big[ \,\mu_+ \, V_+(u_+) - \mu_- \, V_-(u_-) \,\big]
\\
\rule{0pt}{.6cm}
&=&
- \,\partial_x \left(
j_x(\tau;x,t) -
\frac{\kappa}{2\pi}\,
 \big[ \,\mu_+ \, V_+(u_+) - \mu_- \, V_-(u_-) \,\big]
\right) .
\eea
This observation, or (\ref{cp4-mu}), leads us to introduce instead the operators
\be 
\label{eden-mu}
\hat{\varrho} (\tau;x,t) \equiv  
\left( j_{+} (\tau, u_+)  + \frac{\kappa \mu_{+}}{2\pi} \right) + \left( j_{-} (\tau, u_{-})  + \frac{\kappa \mu_{-}}{2\pi} \right) 
%\;\;\;\qquad\;\;\;
%u_\pm \equiv x \pm t
\ee
and
\be 
\label{ecurr-mu}
\hat{j} _x(\tau;x,t) 
\equiv 
-\,V_+(u_+)\left( j_{+} (\tau, u_+)  + \frac{\kappa \mu_{+}}{2\pi} \right) 
+ V_-(u_-) \left( j_{-} (\tau, u_{-})  + \frac{\kappa \mu_{-}}{2\pi} \right) 
\ee
which satisfy the following continuity equation also when $\mu_\pm \neq 0$
\be 
\label{ecr0-mu}
\partial_\tau \hat{\varrho}(\tau;x,t) = - \,\partial_x \hat{j}_x(\tau;x,t) \,.
\ee
In Sec.\,\ref{sec-rep-finite-density} we will see that the operators (\ref{eden-mu}) and (\ref{ecurr-mu}) have
vanishing mean values in the finite density representation;
hence they do not provide the operators employed in \cite{Liguori:1999tw, Hollands:2016svy, Bernard:2013aru, Bernard:2016nci} 
when $\tau=0$ and $V_+(u_+) = V_{-}(u_{-})=1$ identically.
Thus, in our analysis we mainly adopt the definitions (\ref{eden}) and (\ref{ecurr}).

In the following, 
also for the helicity, the energy and the momentum
we can introduce operators 
having vanishing mean values in the finite density representation
and playing the same role of (\ref{eden-mu}) and (\ref{ecurr-mu}).
The above considerations are straightforwardly adapted to these operators.

By employing (\ref{j-t-def}), 
the continuity relation (\ref{ecr-time}) takes the form 
\be
\label{ecr-cov-form1}
\partial_\tau \varrho(\tau;x,t) 
=
 -\, \partial_x j_x(\tau;x,t)  + \partial_t j_t(\tau;x,t)   
\ee
which can be rewritten also in following covariant form 
\be
\label{ecr-cov-form2}
\partial_\tau \varrho(\tau;x,t) = \partial_\mu j^\mu(\tau;x,t) 
\ee
via the Minkowski metric $\eta_{\mu\nu} \equiv  \textrm{diag}(1,-1)$ in the spacetime parameterised by $x^\mu = (t,x)$,
that allows to raise the index of the vector $j_\mu \equiv (j_t,j_x)$, finding  $j^\mu \equiv (j_t, \, -j_x)$.
%\\
%\textcolor{red}{\bf [IMPORTANT: is $(j_t, \, -j_x)$ a vector or one-form, or a mixed thing? $j_t$ and $j_x$ seem to have different nature, which becomes the same
%after taking the mean values]}

%
We remark that the 
arbitrary additive constant shifts in (\ref{ecurr}) and (\ref{j-t-def})
can be fixed also by requiring the vanishing of these quantities 
at the vertices $P_a$, $P_b$, $P_{+\infty}$ and $P_{-\infty} $ of the spacetime diamond $\mathcal{D}_A$ in (\ref{diamond}),
i.e. for $u_\pm \in \{a,b\}$.

From (\ref{eden}), we can define the total electric charge in $\mathcal{D}_A$ as
\be
\label{echarge}
Q_{A} 
\equiv 
\int_{\mathcal{D}_A} \!\! \varrho (\tau; x,t )\, \rd x\, \rd t
\,=\,
\frac{b-a}{2}
\left(\,
\int_a^b j_+(\tau, u_+)\, \rd u_+ 
+ 
\int_a^b j_-(\tau, u_-)\, \rd u_-
\right)
\ee
which is  independent of $x$ and $t$ by construction. 
The independence of $\tau$ in (\ref{echarge}), i.e. the condition 
$\partial_\tau Q_{A} = 0$ corresponding to the conservation of $Q_A$ in $\mathcal{D}_A$,
follows from the equations of motion for the currents in (\ref{mem4j}) 
combined with (\ref{cond}).

From (\ref{ecurr}) and (\ref{j-t-def}), notice that $\partial_\tau j _x$ is non vanishing, 
while $\partial_\tau j _t = 0$, being $j_t$ independent of $\tau$.
In particular, by employing (\ref{mem4j-bis}) we obtain
\bea
\label{ode-dtau-jx}
\partial_\tau j_x(\tau;x,t)  
&=&
-\, \partial_{u_+} \! \big[ \,V_+(u_+)^2 \, j_+(\tau,u_+)  \, \big]
- \partial_{u_-} \! \big[ \,V_-(u_-)^2 \, j_+(\tau,u_-)  \, \big]
\nn
\\
\rule{0pt}{.8cm}
& &
+\, \frac{ \partial_{u_+}\! V_+(u_+)^2 }{2}   \; j_+(\tau,u_+)
+ \frac{ \partial_{u_-}\! V_-(u_-)^2 }{2}   \; j_-(\tau,u_-)
\\
\rule{0pt}{.6cm}
& &
-\; \frac{\kappa}{4\pi} \Big[\, \mu_+  \, \partial_{u_+} \! V_+(u_+)^2 + \mu_-  \, \partial_{u_-} \! V_-(u_-)^2  \,\Big]
\nn
\eea
which implies that $j_x(\tau; x,t )$ does not correspond to the density of a conserved quantity in the diamond $\mathcal{D}_A$.
This happens because of the terms in the second line of (\ref{ode-dtau-jx});
indeed the first line and the third line can be written as total derivatives e.g. in $x$ and $t$ respectively.

The helicity can be investigated in a similar way.
Since $\phi_+$ and $\phi_-$ have opposite helicities, 
the commutators (\ref{cft6ab}) lead us to introduce the following density
\be 
\chi (\tau;x,t) \equiv   j_+(\tau,u_+) - j_-(\tau,u_-)  \,.
\label{hden}
\ee
From  (\ref{mem4j}), we find the continuity relation 
\be
\label{hel1}
\partial_\tau \chi(\tau;x,t) 
=
\partial_t k_t(\tau;x,t)  -\, \partial_x k_x(\tau;x,t)  = \partial_\mu k^\mu (\tau;x,t)
\ee
where the components of the helicity current are defined as 
\bea
\label{hcurr}
k_x(\tau;x,t) 
& \equiv & 
- \, V_+(u_+) \,j_+(\tau,u_+) - V_-(u_-) \, j_-(\tau,u_-) 
\\
\rule{0pt}{.6cm}
\label{k-xt-def}
k_t(\tau;x,t)   
& \equiv & 
\frac{\kappa}{2\pi}\,\big[ \,\mu_+ \, V_+(u_+) - \mu_- \, V_-(u_-) \,\big]
\eea
which vanish for $u_\pm \in \{a,b\}$.
The helicity charge in $\mathcal{D}_A$ following from (\ref{hden}) is
\be
\label{h-charge-def}
\widetilde{Q}_{A} 
\equiv 
\int_{\mathcal{D}_A} \!\! \chi (\tau; x,t )\, \rd x\, \rd t
\,=\,
\frac{b-a}{2}
\left(\,
\int_a^b j_+(\tau, u_+)\, \rd u_+ 
-
\int_a^b j_-(\tau, u_-)\, \rd u_-
\right)  .
\ee
This quantity is independent of $x$, $t$ and also conserved, i.e. independent of $\tau$, 
as we can find from (\ref{mem4j}) and (\ref{cond}).

\subsection{Energy and momentum}
\label{subsec-energy-continuity}

The energy density can be introduced from (\ref{me3b}) as follows
\be
\label{eden-E}
\CE(\tau ;x,t) \equiv  V_+(u_+) \, \T_+(\tau,u_+) + V_-(u_-) \, \T_-(\tau,u_-) 
+ f_+(u_{+}) + f_-(u_{-})
\ee
and $f_\pm(u_{\pm})$ are real functions that can be exploited to fix the mean value of this operator
but they do not influence the following analysis because they are independent of $\tau$.

The equations of motion (\ref{met3}) lead to 
\bea
\label{en-cons-0}
\partial_\tau \CE(\tau;x,t) 
&=&
\partial_{u_+}  \big[ \,V_+(u_+)^2\, \T_+(\tau,u_+) \,\big] 
- \frac{\kappa\, \mu_+^2}{2\pi}\, \partial_{u_+} \big[V_+(u_+)^2\big]
-\frac{c}{24 \pi}\,  V_+(u_+)\,\partial_{u_+}^3 V_+(u_+)
\nonumber
\\
& &-\,
\partial_{u_-}  \big[ \,V_-(u_-)^2\, \T_-(\tau,u_-) \,\big] 
+ \frac{\kappa \,\mu_-^2}{2\pi}\, \partial_{u_-} \big[V_-(u_-)^2\big]
+ \frac{c}{24 \pi}\,  V_-(u_-)\,\partial_{u_-}^3 V_-(u_-)  \,.
\nonumber
\\
& &
\eea
By using the following identity 
\be
\label{id-V2}
\partial_u \big( V(u)^2\,\CV[V](u) \big) = \, V(u)\, \partial_u^3 V(u)
\ee
where
\be 
\CV[V](u) \equiv
\frac{V''(u)}{V(u)} 
- \frac{1}{2} \left [ \frac{V' (u)}{V (u)} \right ]^2 
\label{CV}
\ee
and the fact that $\partial_x = \partial_{u_+} +\partial_{u_-}$ and $\partial_t = \partial_{u_+} -\partial_{u_-}$,
the differential equation (\ref{en-cons-0}) can be written also in the form of the continuity relation  
\be
\label{en-cons-2}
\partial_\tau \CE(\tau; x,t) = \partial_t \CJ_t(\tau;x,t)  - \partial_x \CJ_x(\tau;x,t)
= \partial_\mu \CJ^\mu(\tau; x,t)
\ee
where
\bea
\label{curl-J-x-def}
\CJ_x(\tau; x,t) &\equiv & - \,V_+(u_+)^2 \, \T_+(\tau,u_+) + V_-(u_-)^2 \, \T_-(\tau,u_-)
\\
\rule{0pt}{.8cm}
\label{curl-J-t-def}
\rule{0pt}{.7cm}
\CJ_t(\tau; x,t) 
& \equiv &
- \,\frac{\kappa}{4\pi}\, \Big\{ \mu^2_+ \,V_+(u_+)^2 + \mu^2_- \, V_-(u_-)^2 \Big\}
\nonumber
\\
\rule{0pt}{.55cm}
& &
- \, \frac{c}{24\pi}\, \Big\{ V_+(u_+)^2 \,\CV[V_+](u_+) + V_-(u_-)^2\,\CV[V_-](u_-) \Big\} 
+ C_{\CJ}  \,.
\eea
We remark that (\ref{curl-J-x-def}) satisfies
\be
\CJ_x(\tau=0; x,t) \big|_{V_{+}=V_{-}=1} = 
- \, \T_+(u_+) + \T_-(u_-)
\ee
whose expectation value in the finite density representation (see (\ref{fd1})) 
agrees with the corresponding result found in \cite{Hollands:2016svy} at zero temperature. 
In (\ref{curl-J-t-def}) we have introduced the constant $C_{\CJ}$ 
because, while $\CJ_x$ vanishes for $u_\pm \in \{a,b\}$,
this condition is not guaranteed for $\CJ_t$.

The energy density (\ref{eden-E}) leads us to define the total energy in $\mathcal{D}_A$ as follows
\bea
\label{toten}
E_{A} 
& \equiv &
\int_{\mathcal{D}_A} \!\! \mathcal{E}(\tau; x,t )\, \rd x\, \rd t
\\
\rule{0pt}{.7cm}
&=&
\frac{b-a}{2}
\left(\,
\int_a^b \! \Big[ V_+(u_+) \, \T_+(\tau,u_+) + f_+(u_+) \Big] \rd u_+ 
+ 
\int_a^b \! \Big[ V_-(u_-) \, \T_-(\tau,u_-) + f_-(u_-) \Big] \rd u_-
\right)
\nn
\eea
which is constant, i.e. independent of $x$, $t$ and $\tau$.
Its conservation along the $\tau$-evolution, i.e. the fact that $\partial_\tau E_{A}  = 0$,
is obtained from the continuity equation (\ref{en-cons-0}), the identity (\ref{id-V2}) and the condition (\ref{cond}).

By employing (\ref{met3-bis}), 
from (\ref{curl-J-t-def})  and (\ref{curl-J-x-def}) we obtain respectively 
$\partial_\tau \mathcal{J} _t = 0$ and 
\bea
\label{ode-dtau-cal-jx}
\partial_\tau \mathcal{J}_x(\tau;x,t)  
&=&
-\, \partial_{u_+} \! \big[ \,V_+(u_+)^3 \, \mathcal{T}_+(\tau,u_+)  \, \big]
- \partial_{u_-} \! \big[ \,V_-(u_-)^3 \, \mathcal{T}_+(\tau,u_-)  \, \big]
\\
\rule{0pt}{.8cm}
& &
+ \,\frac{ \partial_{u_+}\! V_+(u_+)^3 }{3}   \; \mathcal{T}_+(\tau,u_+)
+ \frac{ \partial_{u_-}\! V_-(u_-)^3 }{3}   \; \mathcal{T}_-(\tau,u_-)
\nn
\\
\rule{0pt}{.6cm}
& &
\hspace{-2.6cm}
+\, 
\frac{\kappa}{6\pi} \Big[\, \mu_+^2  \, \partial_{u_+} \! V_+(u_+)^3 + \mu_-^2  \, \partial_{u_-} \! V_-(u_-)^3  \,\Big]
+ 
\frac{c}{24\pi} \Big[\,V_+(u_+)^2 \,\partial_{u_+}^3 \! V_+(u_+) +  V_-(u_-)^2 \,\partial_{u_-}^3 \! V_-(u_-)  \,\Big]
\nn
\eea
which tells us that $\mathcal{J}_x(\tau;x,t)  $
is not the density of a conserved quantity in the diamond. 
%\\
%\textcolor{red}{\bf $\bullet$ [can we find $F$ such that $\partial_u F = V^2 \partial_u^3 V$, somehow extending (\ref{id-V2})?]}
%\\
%\textcolor{red}{\bf [it is not straightforward (Mathematica fails)]}

We find it worth introducing also the following generalized momentum density
\be
\widetilde{\CE}(\tau ;x,t) \equiv  V_+(u_+) \, \T_+(\tau,u_+) - V_-(u_-) \, \T_-(\tau,u_-)
+ f_+(u_{+}) - f_-(u_{-})  
\label{eden-bis}
\ee
where for the real functions $f_\pm(u_{\pm})$ 
we can repeat the considerations made below (\ref{eden-E}).
The equations of motion (\ref{met3}) give
\bea
\label{en-cons-0-bis}
\partial_\tau \widetilde{\CE}(\tau;x,t) 
&=&
\partial_{u_+}  \big[ \,V_+(u_+)^2\, \T_+(\tau,u_+) \,\big] 
- \frac{\kappa \mu_+^2}{2\pi}\, \partial_{u_+} \big[V_+(u_+)^2\big]
-\frac{c}{24 \pi}\,  V_+(u_+)\,\partial_{u_+}^3 V_+(u_+)
\nonumber
\\
& &+\,
\partial_{u_-}  \big[ \,V_-(u_-)^2\, \T_-(\tau,u_-) \,\big] 
- \frac{\kappa \mu_-^2}{2\pi}\, \partial_{u_-} \big[V_-(u_-)^2\big]
- \frac{c}{24 \pi}\,  V_-(u_-)\,\partial_{u_-}^3 V_-(u_-)
\nonumber
\\
& &
\eea
or, equivalently,
\be
\label{tilde-en-cons-2}
\partial_\tau \widetilde{\CE}(\tau; x,t) 
= \partial_t \widetilde{\CJ}_t(\tau;x,t)  - \partial_x \widetilde{\CJ}_x(\tau;x,t)  = 
\partial_\mu \widetilde{\CJ}^\mu(\tau; x,t)
\ee
where 
\bea
\label{tilde-curl-J-x-def}
\widetilde{\CJ}_x(\tau; x,t) &\equiv & -\, V_+(u_+)^2 \, \T_+(\tau,u_+) - V_-(u_-)^2 \, \T_-(\tau,u_-)
\\
\rule{0pt}{.8cm}
\label{tilde-curl-J-t-def}
\rule{0pt}{.7cm}
\widetilde{\CJ}_t(\tau; x,t) 
& \equiv &
- \, \frac{\kappa}{4\pi}\, \Big\{ \mu^2_+ \,V_+(u_+)^2 - \mu^2_- \, V_-(u_-)^2 \Big\}
\nonumber
\\
\rule{0pt}{.55cm}
& &
- \, \frac{c}{24\pi}\, \Big\{ V_+(u_+)^2 \,\CV[V_+](u_+) - V_-(u_-)^2\,\CV[V_-](u_-) \Big\} 
+ C_{\widetilde{\CJ}}  \,.
\eea

In analogy with (\ref{toten}),  
the momentum density (\ref{eden-bis}) leads us to introduce the total momentum in $\mathcal{D}_A$ as follows
\bea
\label{totmom}
\widetilde{E}_{A} 
& \equiv &
\int_{\mathcal{D}_A} \!\! \widetilde{\mathcal{E}}(\tau; x,t )\, \rd x\, \rd t
\\
\rule{0pt}{.7cm}
&=&
\frac{b-a}{2}
\left(\,
\int_a^b \! \Big[ V_+(u_+) \, \T_+(\tau,u_+) + f_+(u_+) \Big] \rd u_+ 
-
\int_a^b \! \Big[ V_-(u_-) \, \T_-(\tau,u_-) + f_-(u_-) \Big] \rd u_-
\right)
\nn
\eea
which is independent of $x$, $t$ and also of $\tau$,
as it can be found from the corresponding continuity equation (\ref{en-cons-0-bis}), 
the identity (\ref{id-V2}) and the condition (\ref{cond}).

The application of the above analysis to the chiral primaries is discussed in the Appendix\;\ref{app-current-phi}.

%%%%%%%%%%%%%%%%%%%%%%%%%%%%%%%%%%%%%%%%%%%%%%
\subsection{Transformation generated by the generalized momentum} 
\label{sec-momentum-evo}
%%%%%%%%%%%%%%%%%%%%%%%%%%%%%%%%%%%%%%%%%%%%%%

The operator (\ref{eden-bis}) provides the following evolution operator
\be 
P  \equiv 
\int_{-\infty}^\infty \!  \Big[ V_+(u_+) \, \T_+(\tau,u_+) + f_+(u_+) \Big]  \rd u_+
-
\int_{-\infty}^\infty \!  \Big[ V_-(u_-) \, \T_-(\tau,u_-) + f_-(u_-) \Big]  \rd u_-  \,.
\label{me3b-bis}
\ee
The evolution of a primary $\phi_\pm$ generated by (\ref{me3b-bis}) reads
\be 
\widetilde{\phi}_\pm (\lambda ,u) \equiv \e^{\ri \lambda P }\, \phi_\pm(u)\, \e^{-\ri \lambda P }
\label{nme1-bis}
\ee
and it can be studied by slightly modifying the analysis described for (\ref{nme1}). 
This leads to the following equation for (\ref{nme1-bis})
\be
\partial_\lambda \widetilde{\phi}_\pm(\lambda ,u) 
= 
V_\pm (u) \big(\partial_u \pm \ri \mu_\pm \big) \widetilde{\phi}_\pm(\lambda, u) 
+ h_\pm \big[\partial_u V_\pm(u)\big] \widetilde{\phi}_\pm (\lambda, u) 
\ee  
with initial condition $\widetilde{\phi}_\pm(0 ,u) = \phi_\pm(u) $,
which can be solved through $\zeta_\pm (\lambda,x)$ satisfying 
\be 
\partial_\lambda \zeta_\pm (\lambda,u) =  V_\pm (u)\, \partial_u \zeta_\pm (\lambda,u) 
\label{me2a-zeta}
\ee 
with the initial condition 
\be 
\zeta_\pm (0,u) = u  \,.
\label{me2b-zeta}
\ee 
which can be compared with the differential equation (\ref{me2a}) and its initial condition (\ref{me2b}) respectively. 
The solution of (\ref{me2a-zeta}) and (\ref{me2b-zeta}) can expressed 
for  $u \in [a,b]$ between two consecutive zeros of $V_\pm$ 
in terms of (\ref{s1}) as follows
\be 
\label{zeta-def-lambda}
\zeta_\pm (\lambda,u) \equiv w_\pm^{-1}\big(w_\pm(u) + \lambda  \big)  
\ee 
which is slightly different from (\ref{s2}).
The expressions (\ref{zeta-def-lambda}) satisfy
\be 
\partial_\lambda \zeta_\pm (\lambda,u) = 
V_\pm(\zeta_\pm (\lambda,u)) 
\;\;\; \qquad\;\;\; 
\partial_u \zeta_\pm (\lambda,u) = 
\frac{V_\pm (\zeta_\pm (\lambda,u))}{V_\pm(u)}  \;.
\ee
Comparing the functions in (\ref{zeta-def-lambda}) with the ones in (\ref{s2}), we observe that 
$\zeta_{+} (\lambda,u) = \xi_{+}(\lambda,u) $, while $\zeta_{-} (\lambda,u) = \xi_{-}(-\lambda,u) $.
The infinitesimal transformation is obtained by expanding (\ref{zeta-def-lambda}) for small $\lambda$
and this gives (see also the remark 4.2 of \cite{Moosavi:2019fas})
\be
\zeta_\pm (\lambda,u)  = u + V_\pm(u)\, \lambda + O(\lambda^2)  \,.
\ee 
For constant velocities $V_\pm$, we have   $\zeta_\pm (\lambda,u)  = u + V_\pm \lambda$ for (\ref{zeta-def-lambda}), 
i.e. the spatial translations.

By using (\ref{zeta-def-lambda}), for the evolution (\ref{nme1-bis}) we find 
\be 
\widetilde{\phi}_\pm (\lambda,u) = \e^{\pm \ri \mu_\pm [\zeta_\pm (\lambda,u) - u]}\, \big[\partial_u \zeta_\pm (\lambda,u)\big]^{h_\pm} \, \phi _\pm\big(\zeta_\pm (\lambda,u)\big) 
\label{me9a-zeta}
\ee
which satisfies the initial condition $\widetilde{\phi}_\pm(0,u) = \phi_\pm (u) $, as expected. 
Notice that the r.h.s.'s of (\ref{me9a}) and (\ref{me9a-zeta}) are formally identical. 
Thus, as for the evolutions in $\lambda$ of $j_\pm$ and of $\T_\pm$, i.e.
\be
\widetilde{j}_\pm (\lambda ,u) \equiv \e^{\ri \lambda \widetilde{H}  }\, j_\pm(u)\, \e^{-\ri \lambda \widetilde{H} }
\;\;\qquad\;\;
\widetilde{\T}_\pm (\lambda ,u) \equiv \e^{\ri \lambda \widetilde{H}  }\, \T_\pm(u)\, \e^{-\ri \lambda \widetilde{H} }
\ee
we find that they are given by the r.h.s.'s of (\ref{cp4}) and (\ref{met4}) respectively, 
with $\xi_\pm (\tau,u)$ replaced by $\zeta_\pm (\lambda,u)$.

Finally,  from (\ref{zeta-def-lambda}) one introduces the trajectories in $\mathcal{D}_A$ generated through the evolution 
governed by (\ref{me3b-bis}), as done in (\ref{mod-traj-tau}) for (\ref{s2}),
whose spacetime coordinates read
\be
\label{mod-traj-lambda}
x(\lambda) = \frac{\zeta_+(\lambda, u_+) + \zeta_-(\lambda, u_-) }{2}
\;\;\;\qquad\;\;\;
t(\lambda) = \frac{\zeta_+(\lambda, u_+) - \zeta_-(\lambda, u_-) }{2}
\ee
where $\lambda \in \RR$ and $(u_+, u_-)$ are the light-cone coordinates of the point corresponding to $\lambda =0$.

\subsection{Heat} 
\label{sec-heat-density}

Any linear combinations of the above conserved currents defines a conserved current as well. 
From non-equilibrium thermodynamics \cite{callen-book},
it is known that  the heat density in the system is described by 
\be 
\Q (\tau; x, t) = \E (\tau; x, t)- \mu_c \, \varrho (\tau; x, t) - \mu_h \, \chi (\tau; x, t)
\label{heat1}
\ee
defined through (\ref{eden}), (\ref{hden}) and (\ref{eden-E}),
where the electric charge and helicity chemical potentials are the combinations  
\be 
\mu_e \equiv \mu_+ + \mu_- 
\;\;\; \qquad \;\;\;
\mu_h \equiv \mu_+ - \mu_-    \,.
\label{heat2}
\ee
 Accordingly, the heat flow in the system is generated by the currents
\bea
\label{heat3-1}
q_x (\tau; x, t)  &=&  \CJ_x (\tau; x, t) - \mu_e\, j_x (\tau; x, t) - \mu_h\, k_x (\tau; x, t) 
\\
\label{heat3-2}
q_t (\tau; x, t) &=& \,\CJ_t (\tau; x, t) - \mu_e\, j_t (\tau; x, t) - \mu_h\, k_t (\tau; x, t)   \,.
\eea
Combining (\ref{heat1}) with (\ref{echarge}), (\ref{h-charge-def}) and  (\ref{toten}),
for the total heat in the diamond $\mathcal{D}_A$ we have
\be 
\Q_A = E_A - \mu_e\, Q_A - \mu_h\, {\widetilde Q}_A 
\label{heat4}
\ee
which is independent of $x$, $t$ and $\tau$.

%%%%%%%%%%%%%%%%%%%%%%%%%%%%%%%%%%%%%%%%%%%%%%
%\newpage
\section{Infinite volume}
\label{sec-line}
%%%%%%%%%%%%%%%%%%%%%%%%%%%%%%%%%%%%%%%%%%%%%%

In this section we consider the finite density representation of the CFT on the line,
which is characterised by the mean values reported in Sec.\,\ref{sec-rep-finite-density},
and the modular Hamiltonian corresponding to the bipartition provided by an interval.
The results of Sec.\,\ref{sec-mod-evolution} are employed
to discuss the modular evolution of the fields (Sec.\,\ref{subsec-mod-H-vacuum}), the modular conjugation and its
geometric action (Sec.\,\ref{subsec-mod-J-vacuum}). The modular correlators are considered in Sec.\,\ref{sec-mod-corr-line}.

\subsection{Finite density representation on the line}
\label{sec-rep-finite-density}

In the finite density representation of the CFT on the line, 
the mean values of the basic chiral fields described in Sec.\,\ref{subsec-commutators} are  
\be 
\langle \phi_\pm(u)\rangle_{\mu_\pm} = 0 
\;\;\qquad \;\;
\langle j_\pm(u)\rangle_{\mu_\pm}  =  - \frac{\kappa \mu_\pm}{2\pi} 
\;\;\qquad \;\;
\langle \T_\pm(u)\rangle_{\mu_\pm} =  \frac{\kappa \mu^2_\pm}{4\pi}  \,.
\label{fd1}
\ee
These expressions are obtained from the corresponding ones 
in the fundamental (ground state) representation $\pi_0$, where all these one-point function vanish
(see the Appendix\;\ref{app-comm-checks})
through an automorphism $\gamma_\mu =  \gamma_{\mu_{+}} \otimes \gamma_{\mu_{-}}$, 
as discussed in the Appendix\;\ref{app-reps}.
This automorphism provides the state $\Omega_{\mu_\pm}$
from the ground state $\Omega_0$, which corresponds to $\mu_\pm = 0$.
The finite density representation is given by $\pi_0 \circ \gamma_\mu$.
In the following we denote $\pi_0 \circ \gamma_\mu[\mathcal{O}_\pm]$ just by $\mathcal{O}_\pm$,
with a slight abuse of notation. 

This procedure can be employed to find also the two-point functions of these fields
in the finite density representation listed below, 
that are obtained from the ones in the fundamental representation, 
as explained in the Appendix\;\ref{app-reps}).
The two-point functions of $\phi_\pm$ are
(see e.g. \cite{Rehren:1987iy} for $\mu_\pm =0$)
\be 
\langle \phi_\pm^*(u) \, \phi_\pm (v)\rangle_{\mu_\pm}^{\textrm{\tiny con}} 
= 
\frac{ \e^{\mp \ri \pi h_\pm}\,\e^{\pm \ri \mu_\pm (u-v)}}{2\pi\, (u-v \mp \ri \varepsilon )^{2h_\pm}} 
\;\;\;\qquad\;\;\;
\langle \phi_\pm(u) \, \phi^*_\pm (v)\rangle_{\mu_\pm}^{\textrm{\tiny con}} 
= 
\frac{ \e^{\mp \ri \pi h_\pm}\,\e^{\mp \ri \mu_\pm (u-v)}}{2\pi\, (u-v \mp \ri \varepsilon )^{2h_\pm}} \,.
\label{fd5}
\ee 
%where the normalisation constants have been fixed like in \cite{Hollands:2019hje}.
The two-point functions of $j_\pm$ contain the constant $\kappa$ and read
\be 
\langle j_\pm(u) \, j_\pm (v)\rangle_{\mu_\pm}^{\textrm{\tiny con}} 
= 
\frac{\kappa}{4\pi^2} \, \frac{1}{(u-v \mp \ri \varepsilon )^2} 
\label{fd4}
\ee
while the two-point functions of $\T_\pm$ depend on the central charge $c$ as follows
\be 
\langle \T_\pm(u) \, \T_\pm (v)\rangle_{\mu_\pm}^{\textrm{\tiny con}} 
= 
\frac{c}{8\pi^2} \, \frac{1}{(u-v \mp \ri \varepsilon )^4}  \,.
%+ 
%\frac{\kappa\mu_\pm^2}{4\pi^2} \, \frac{1}{(x-y \mp \ri \varepsilon )^2} 
\label{fd3}
\ee
The positivity of (\ref{fd5})-(\ref{fd3}) imply $h_\pm \geqslant 0$ (see the Appendix\;\ref{app-michele-modular}), $\kappa \geqslant 0$ and $c \geqslant 0$
respectively.
Furthermore, we have that
\be
\langle \T_\pm(u) \, j_\pm (v)\rangle_{\mu_\pm}^{\textrm{\tiny con}} = 
-\,\mu_\pm\, \langle j_\pm(u) \, j_\pm (v)\rangle_{\mu_\pm}^{\textrm{\tiny con}}  \,.
\ee
The connected mixed correlators involving fields having different chiralities vanish identically. 

When $t \pm x$ are chosen as light-cone coordinates, 
the $\mp \ri \varepsilon$ must be replaced with $- \ri \varepsilon$
in all the chiral two-point functions occurring in this manuscript.  
%\\
%\textcolor{red}{\bf $\bullet$ 
%Do we need also the three-point function 
%\be 
%\label{3-point-j-line}
%\langle j_\pm(u_1) \, j_\pm (u_2)\, j_\pm (u_3)\rangle_{\mu_\pm}^{\textrm{\tiny con}} 
%= 
%\dots
%\ee
%which contains the structure constants, to define the representation? 
%\\
%And what about the higher-point functions?}

\subsection{Modular Hamiltonian}
\label{subsec-mod-H-vacuum}

The results discussed in Sec.\,\ref{sec-mod-evolution} 
can be employed to investigate
some modular Hamiltonians in two-dimensional CFT
corresponding to the spatial bipartition provided by an interval
\cite{Hislop:1981uh, Casini:2011kv, Wong:2013gua, Cardy:2016fqc}.

In the following we consider a CFT on a line and in the finite density state 
characterised by the correlators reported in Sec.\,\ref{sec-rep-finite-density}.
The spatial bipartition of the line is given by 
the interval $A=[a,b]$ and its complement $B=(-\infty,a)\cup(b,+\infty)$.
In this setup, the modular Hamiltonian of the interval $A$ reads
\cite{Wong:2013gua, Mintchev:2022xqh}
\be 
\label{KA-def-u-pm}
K_A  
= 
\int_A V(u_+) \left[\, {\cal T}_+(u_+) -  \frac{\kappa \mu^2_{+}}{4\pi}  \,\right] \rd u_+
+ 
\int_A V(u_-) \left[\, {\cal T}_-(u_-) -  \frac{\kappa \mu^2_{-}}{4\pi}  \,\right]  \rd u_-
\ee
in terms of the chiral operators (\ref{cal-T-pm-def}), with 
\be
\label{velocity_fund}
V(u) =2\pi \, \frac{(b-u)(u-a)}{b-a}=\frac{1}{w'(u)} 
\;\;\;\qquad\;\;\;
u \in A
\ee
where 
\be
\label{w-function-def}
w(u) \equiv \frac{1}{2\pi}\,\log \!\left(\! - \frac{u-a}{u-b} \right)  .
\ee
%\be
%\label{KA_fund_def}
%K_A = \int_{A}  v(x) \, \big[  \mathcal{T}_+(x) + \mathcal{T}_-(x)  \big] \rd x
%\ee 
%(in this case $v_+(x) = v_-(x) = v(x)$)
%where
%\be
%\label{velocity_fund}
%v(x) =2\pi \, \frac{(b-x)(x-a)}{b-a}=\frac{1}{w'(x)} 
%\ee
The modular Hamiltonian (\ref{KA-def-u-pm}) can be obtained by applying 
the automorphism $\gamma_{\mu}^{-1}$ described in the Appendix\;\ref{app-reps}
to the corresponding modular Hamiltonian in the fundamental representation (see (\ref{gamma-minus1-on-T})).
Notice that the integrand in (\ref{KA-def-u-pm}) corresponds to the special case of (\ref{eden-E})
where $V_\pm(u_\pm) = V(u_\pm)$ and $f_{\pm}(u_\pm) = -  \tfrac{\kappa \mu^2_{\pm}}{4\pi}  V(u_\pm) $.

The modular Hamiltonian of the complement $B$ can be found by exchanging $a$ and $b$ in (\ref{velocity_fund}) and (\ref{w-function-def})
(in the latter equation also the global sign in the argument of the logarithm must be inverted in order to have a well defined function); 
hence $V$ changes its global sign. Notice that $-V(u) \geqslant 0$ when $u\in B$.
This leads to 
\be 
\label{KB-def-u-pm}
K_B
= 
\int_B \big(\! -V(u_+)\big)  \left[\, {\cal T}_+(u_+) -  \frac{\kappa \mu^2_{+}}{4\pi}  \,\right] \rd u_+
+ 
\int_B \big(\! -V(u_-)\big)  \left[\, {\cal T}_-(u_-) -  \frac{\kappa \mu^2_{-}}{4\pi}  \,\right]  \rd u_-  \,.
\ee
We remark that the weight functions in (\ref{KA-def-u-pm}) and (\ref{KB-def-u-pm}), 
which are given by $V(u)$ and $-V(u)$ respectively, 
are positive in the corresponding domains. 
The modular Hamiltonians (\ref{KA-def-u-pm}) and (\ref{KB-def-u-pm}) 
provide the full modular Hamiltonian
for the bipartition and the finite density state $\Omega_{\mu_\pm}$ that we are considering, which reads
\bea
\label{fmh}
K
& \equiv &
K_{A} \otimes \boldsymbol{1}_B - \boldsymbol{1}_A \otimes K_{B} 
\\
\rule{0pt}{.7cm}
& = & 
\int_{-\infty}^{\infty} V(u_+)  \left[\, {\cal T}_+(u_+) -  \frac{\kappa \mu^2_{+}}{4\pi}  \,\right]  \rd u_+
+ 
\int_{-\infty}^{\infty} V(u_-)  \left[\, {\cal T}_-(u_-) -  \frac{\kappa \mu^2_{-}}{4\pi}  \,\right]  \rd u_- \,.
\nn
\eea
We remark that, because of the choice $f_{\pm}(u_\pm) = -  \tfrac{\kappa \mu^2_{\pm}}{4\pi}  V(u_\pm) $, 
this full modular Hamiltonian satisfies $\langle K \rangle_{\mu} =0$,
which is a straightforward consequence of the constraint 
$K \,\Omega_{\mu_\pm} = 0$ (see Eq.\,(V.2.7) of \cite{Haag:1992hx}).
The full modular Hamiltonian (\ref{fmh}) corresponds to the special case of (\ref{me3b}) 
where $V_+(u) = V_-(u) = V(u)$ is  the velocity (\ref{velocity_fund})
(which vanishes only at the endpoints of $A$)
and $\CT_\pm(u_\pm)$ are the operators in the finite density representation 
 introduced in Sec.\,\ref{sec-rep-finite-density}.

The modular evolution generated by the full modular Hamiltonian (\ref{fmh})
can be investigated by specialising the results discussed in Sec.\,\ref{sec-mod-evolution} 
to $V_+(u) = V_-(u) = V(u)$ given by (\ref{velocity_fund}).
In this case (\ref{s2}) becomes 
\be
\label{xi-map-fund}
\xi_\pm(\tau,u) = \xi(\pm\tau,u) 
\;\;\;\qquad\;\;\;
\xi(\tau,u) 
%= w^{-1}\left(2\pi\tau+w(x;a,b)\right) 
\equiv \frac{(b-u)\,a+(u-a)\,b\,\mathrm{e}^{2\pi \tau}}{(b-u)+(u-a)\,\mathrm{e}^{2\pi \tau}}  \,.
%\;\;\;\qquad\;\;\;
%u \in A
\ee

Although this expression has been obtained for $u\in A$, it can extended to $u\in B$;
indeed, it is invariant under the transformation that exchanges $a$ and $b$ and replaces $\tau$ with $-\,\tau$.
Another way to find this result is to consider $\tilde{w}(u) =  \tfrac{1}{2\pi} \log \! \big( \tfrac{u-b}{u-a} \big)$ for $u\in B$ introduced above
and observe that the expression $\tilde{w}^{-1}\big(\tilde{w}(u)-\tau\big)$ obtained from such $\tilde{w}(u)$ 
(see (\ref{s2}) combined with (\ref{fmh})) coincides with the one given in (\ref{xi-map-fund}).
Notice that $\xi(\tau,u) $ in (\ref{xi-map-fund}) can be written in the form $\xi(\tau,u) =\tfrac{\alpha u+ \beta}{ \gamma u + \delta}$,
with 
$\alpha = \tfrac{b \,\e^{2\pi \tau} -a}{b-a}$,
$\beta = \tfrac{a \, b \,(1-\e^{2\pi \tau}) }{b-a}$,
$\gamma = \tfrac{\e^{2\pi \tau} -1 }{b-a}$ and
$\delta = \tfrac{b - a \,\e^{2\pi \tau}}{b-a}$;
hence it is not a transformation of $SL(2,\mathbb{R})$ for any value of $\tau$
because $\alpha \delta - \beta \gamma = \e^{2\pi \tau}$,
which is different from one for $\tau \neq 0$.

For every $\tau \in \mathbb{R}$, we have that $ \xi(\tau,u) \in A$ when $u \in A$ and $\xi(\tau,u)  \in B$ when $u \in B$.
Moreover, given $u \in \mathbb{R}$, one observes that $\xi(\tau,u)  \rightarrow b$ as ${\tau\rightarrow+\infty}$ and $\xi(\tau,u)  \rightarrow a$ as ${\tau\rightarrow-\infty}$.
When $u \in B$, a finite value $\tau_{\infty}(u)$ for $\tau$ occurs such that $ \xi(\tau,u)  \rightarrow \pm \infty$ when $\tau \rightarrow \tau_{\infty}(u)^{\pm}$.
It corresponds to the zero of the denominator of (\ref{xi-map-fund}) and its explicit expression reads  \cite{Mintchev:2022fcp}
\be
\label{tau_infty_def}
\tau_{\infty}(u)
\equiv 
\frac{1}{2\pi} \,
\log \! \left(\frac{u-b}{u-a}\right)  .
%\hspace{2cm}
%x\in B
\ee
which is $\tau_{\infty}(u)<0$ when $u>b$ and $\tau_{\infty}(u)>0$ when $u<a$.

The modular evolutions of $\phi_\pm$, $j_\pm$ and $\T_\pm$
are obtained by plugging (\ref{xi-map-fund}) into (\ref{me9a}), (\ref{cp4}) and (\ref{met4}) respectively.
%Notice that, since (\ref{xi-map-fund}) is a transformation of $SL(2,\mathbb{R})$ in terms of $u$ with parameters depending on $\tau$,
We remark that the Schwarzian derivative (\ref{cft3}) for (\ref{xi-map-fund}) vanishes identically,
i.e. $\mathcal{S}_u [\xi ] (\tau,u)  = 0$.

From  (\ref{mod-traj-tau}) and (\ref{xi-map-fund}), we get
the modular trajectories in the spacetime as follows
\be
\label{mod-traj-tau-line}
x(\tau) = \frac{\xi(\tau, u_+) + \xi(-\tau, u_-) }{2}
\;\;\;\qquad\;\;\;
t(\tau) = \frac{\xi(\tau, u_+) - \xi(-\tau, u_-) }{2}
\ee
where $(u_+, u_-)$ are the light-cone coordinates of the spacetime point corresponding to $\tau=0$.

\begin{figure}[t!]
\vspace{-0.cm}
\hspace{.8cm}
\includegraphics[width=.9\textwidth]{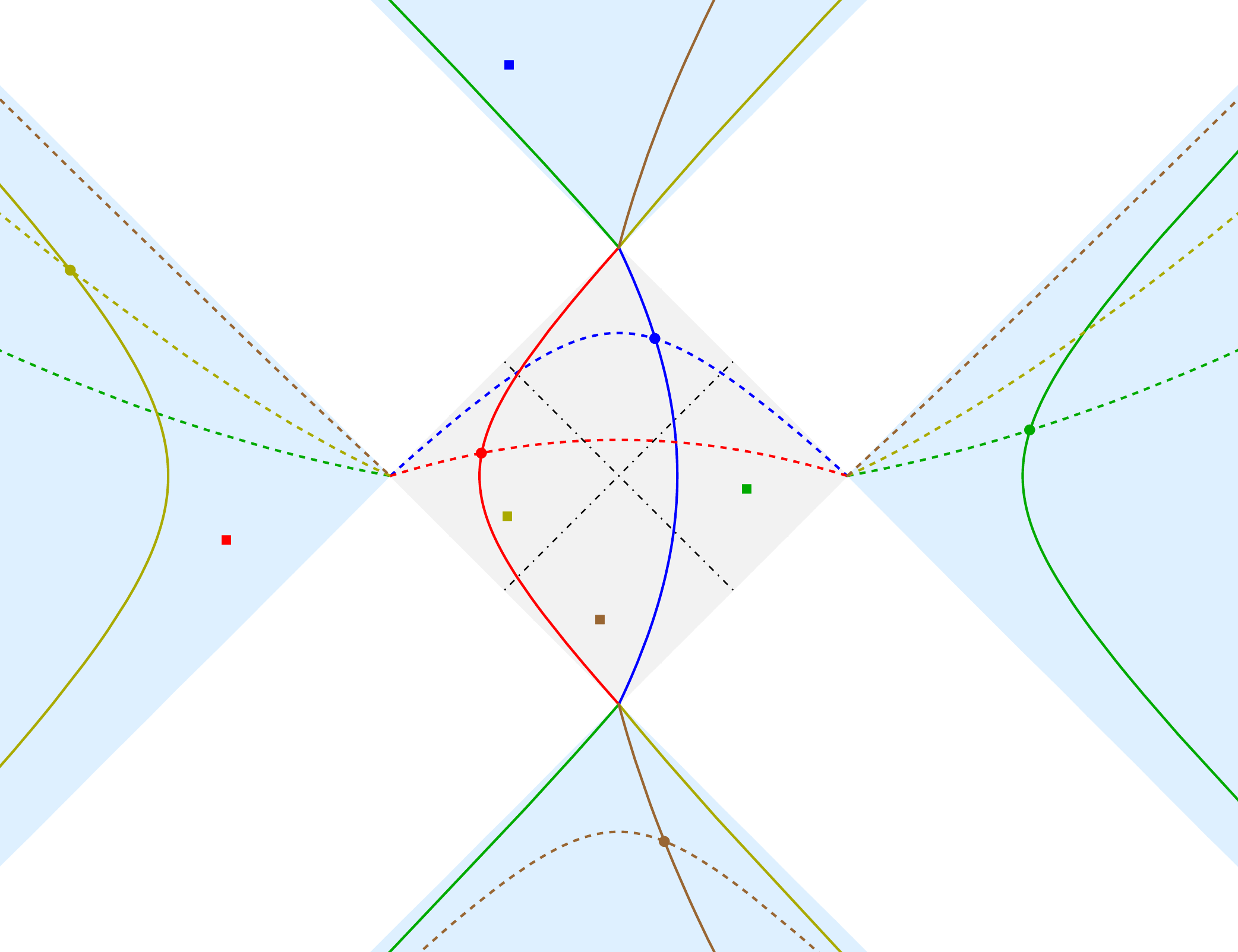}
\vspace{.3cm}
\caption{
Modular trajectories generated by 
either the modular Hamiltonian (\ref{fmh}) (solid lines)
or the modular momentum (\ref{full-mod-momentum}) (dashed lines),
obtained from (\ref{mod-traj-tau-line}) and (\ref{mod-traj-mom-gs}) respectively. 
The coloured squares denote the images 
through the modular conjugation (\ref{inversion-xt})
of the spacetime points corresponding to the dots having the same colour.
The dot dashed segments identify the partition 
$\mathcal{D}_A = \mathcal{D}_{\textrm{\tiny R}} \cup \mathcal{D}_{\textrm{\tiny L}} \cup \mathcal{D}_{\textrm{\tiny F}} \cup \mathcal{D}_{\textrm{\tiny P}} $
of the diamond $\mathcal{D}_A$.
}
\label{fig:diamond-AB}
\end{figure}

The domain of dependence $\mathcal{D}_A$ of the interval $A$ is the diamond
made by the points with light-cone coordinates $u_\pm \in A$
and it corresponds to the grey rhombus in Fig.\,\ref{fig:diamond-AB}.
Its vertices $P_a$, $P_b$, $P_{+\infty}$ and $P_{-\infty}$
have light-cone coordinates $(u_{+}, u_{-}) \in \big\{ (a,a)\,, (b,b)\,,  (b,a)\,, (a,b)  \big\}$
respectively.
Let us introduce the partition of $\mathcal{D}_A$ provided by the light rays from its center $P_c$, 
whose light-cone coordinates are $u_+ = u_- = \tfrac{a+b}{2}$
(see the dot-dashed segments in $\mathcal{D}_A$ in Fig.\,\ref{fig:diamond-AB}).
This gives $\mathcal{D}_A = \mathcal{D}_{\textrm{\tiny R}} \cup \mathcal{D}_{\textrm{\tiny L}} \cup \mathcal{D}_{\textrm{\tiny F}} \cup \mathcal{D}_{\textrm{\tiny P}} $,
where $\mathcal{D}_i$ belongs to the right wedge, the left wedge, the future cone and the past cone of $P_c$
for $i \in \big\{ \textrm{R} , \textrm{L} , \textrm{F} , \textrm{P} \big\}$ respectively.
A modular trajectory (\ref{mod-traj-tau-line}) whose initial point corresponding to $\tau=0$ belongs to $\mathcal{D}_A$
entirely stays within $\mathcal{D}_A$, for any real value of $\tau$.
In particular, its endpoints are the vertices $P_{+\infty}$ and $P_{-\infty}$, 
which are reached as $\tau\to +\infty$ and $\tau\to -\infty$ respectively. 
In Fig.\,\ref{fig:diamond-AB} 
the red and blue solid lines are the modular trajectories
whose initial points are the red and blue dots respectively.
The modular trajectories whose initial point has light-cone coordinates $(u_{+}, u_{-}) $ such that $u_\pm \in A$
span the diamond $\mathcal{D}_A$.

As for region $\mathcal{R}_A$ made by the points with light-cone coordinates $u_\pm \in B$
(light blue region in Fig.\,\ref{fig:diamond-AB}),
it can be naturally partitioned into four regions, i.e.
$\mathcal{R}_A = \mathcal{R}_{\textrm{\tiny R}} \cup \mathcal{R}_{\textrm{\tiny L}} \cup \mathcal{R}_{\textrm{\tiny F}} \cup \mathcal{R}_{\textrm{\tiny P}} $,
where $\mathcal{R}_i$ is the infinite region that shares a vertex of $\mathcal{D}_A$ with $\mathcal{D}_i$
and belongs to the right wedge, the left wedge, the future cone and the past cone of $P_c$
for $i \in \big\{ \textrm{R} , \textrm{L} , \textrm{F} , \textrm{P} \big\}$ respectively.
The modular trajectories (\ref{mod-traj-tau-line})
with initial points at $\tau=0$ in $\mathcal{R}_A$ 
span the entire domain $\mathcal{R}_A$
(in Fig.\,\ref{fig:diamond-AB}, see e.g. the green, yellow and brown solid lines, whose initial points 
are the dots having the corresponding color).
A modular trajectories
which does not belong to the vertical line passing through $P_{+\infty}$ and $P_{-\infty}$
has a non trivial intersection with $ \mathcal{R}_{\textrm{\tiny P}} $, $ \mathcal{R}_{\textrm{\tiny F}} $ 
and either $ \mathcal{R}_{\textrm{\tiny R}} $ or $ \mathcal{R}_{\textrm{\tiny L}} $,
depending on whether the $x$-coordinate of its initial point at $\tau=0$ 
(whose light-cone coordinate are $u_\pm$)
is either $x> (a+b)/2$ or $x < (a+b)/2$ respectively
(instead, the single modular trajectory belonging to the vertical line passing through $P_{+\infty}$ and $P_{-\infty}$
intersects $ \mathcal{R}_{\textrm{\tiny P}} $ and $ \mathcal{R}_{\textrm{\tiny F}} $ only).
The two transitions from $ \mathcal{R}_{\textrm{\tiny P}} $ to $ \mathcal{R}_i$ 
and from $ \mathcal{R}_i$ to $ \mathcal{R}_{\textrm{\tiny F}} $, where $i \in \big\{ \textrm{L} , \textrm{R}  \big\}$,
occur at two finite values of $\tau$ given by 
$\tau_{\infty}(u_+)$ and $\tau_{\infty}(u_-)$,
in terms of $\tau_{\infty}(u)$ introduced in (\ref{tau_infty_def})  \cite{Mintchev:2022fcp}.

We remark that
all the modular trajectories arrive to $P_{+\infty}$ and $P_{-\infty}$ 
as $\tau \to + \infty$ and $\tau \to -\infty$ respectively,
independently of whether the initial point is  either in  $\mathcal{D}_A$  or in $\mathcal{R}_A$.

The modular momentum operator is obtained by specialising (\ref{me3b-bis}) to the case that we are investigating, finding 
\be 
\label{full-mod-momentum}
P  
\equiv 
\int_{-\infty}^{\infty} V(u_+)  \left[\, {\cal T}_+(u_+) -  \frac{\kappa \mu^2_{+}}{4\pi}  \,\right]  \rd u_+
-
\int_{-\infty}^{\infty} V(u_-)  \left[\, {\cal T}_-(u_-) -  \frac{\kappa \mu^2_{-}}{4\pi}  \,\right]  \rd u_- 
\ee
where $V(u)$ has been introduced in (\ref{velocity_fund}).
This operator provides a transformation of the fields that can be  written
by specialising the results of Sec.\,\ref{sec-momentum-evo}
to $V_{+}(u)=V_{-}(u)=V(u)$.
The corresponding modular trajectories in the spacetime
are obtained from (\ref{mod-traj-lambda}) in the same way
and the result reads
\be
\label{mod-traj-mom-gs}
x(\lambda) = \frac{\zeta(\lambda, u_+) + \zeta(\lambda, u_-) }{2}
\;\;\;\qquad\;\;\;
t(\lambda) = \frac{\zeta(\lambda, u_+) - \zeta(\lambda, u_-) }{2}
\ee
where $\lambda \in \RR$ and $(u_+, u_-)$ are the light-cone coordinates of the initial point at $\lambda =0$.
The initial point can be either in $\mathcal{D}_A$ or in $\mathcal{R}_A$  
and the entire modular trajectory (\ref{mod-traj-mom-gs}) belongs to the same region
for all finite real values of $\lambda$,
reaching  $P_a$ and $P_b$ as $\lambda \to - \infty$ and $\lambda \to + \infty$
respectively. 
In Fig.\,\ref{fig:diamond-AB} the dashed curves we show
some modular trajectories generated by the momentum operator,
whose initial points are the dots with the same colour.

The above discussion is based on the fact that $V_+(u) = V_{-}(u)$, given by (\ref{velocity_fund}).
Since the assumption $V_+ = V_{-}$ has not been employed throughout Sec.\,\ref{sec-algebraic-1},
it is straightforward to extend our analysis to the boosted interval,
which is characterised by two different bipartitions along the chiral directions,
determined by the interval $(a_+, b_+)$ and $(a_-, b_-)$ for $u_+$ and $u_-$ directions respectively. 
When the CFT$_2$ is at finite densities, the modular Hamiltonian to consider is (\ref{me3b})
with the following weight functions
\be
\label{Vpm-boosted-interval}
V_\pm(u) \equiv 2\pi \, \frac{(b_\pm - u )( u - a_\pm)}{b_\pm - a_\pm} =\frac{1}{w_\pm'(u)} 
\;\;\;\qquad\;\;\;
u \in (a_\pm , b_\pm)
\ee
where
\be
\label{w-function-def-boosted}
w_\pm(u) \equiv \frac{1}{2\pi} \log \!\left(\! - \frac{u-a_\pm}{u-b_\pm} \right)  .
\ee

From Sec.\,\ref{sec-mod-evolution}, we have that the modular evolution of a primary chiral field is given by (\ref{me9a}),
where $\xi_\pm(\tau,u) $ is obtained by specifying (\ref{s2}) to this case. This leads to
\be
\label{xi-map-fund-boosted}
%\xi_\pm(\tau,u) = \xi(\pm\tau,u) 
%\;\;\;\qquad\;\;\;
\xi_\pm(\tau,u) 
%= w^{-1}\left(2\pi\tau+w(x;a,b)\right) 
=\frac{(b_\pm-u)\,a_\pm+(u-a_\pm)\,b_\pm\,\mathrm{e}^{\pm 2\pi \tau}}{(b_\pm-u)+(u-a_\pm)\,\mathrm{e}^{\pm2\pi \tau}}  
%\;\;\;\qquad\;\;\;
%u \in A
\ee
which reduce to (\ref{xi-map-fund}) when $a_+ = a_- $ and $b_+ = b_-$, as expected.
As for the modular evolution generated by the modular generalised momentum (\ref{me3b-bis}) 
defined through the weight functions (\ref{Vpm-boosted-interval}) characterising the boosted interval,
for a primary chiral field we find (\ref{me9a-zeta}) with
\be
\label{zeta-map-fund-boosted}
\zeta_\pm (\lambda,u) 
=
\frac{(b_\pm-u)\,a_\pm+(u-a_\pm)\,b_\pm\,\mathrm{e}^{2\pi \lambda}}{(b_\pm-u)+(u-a_\pm)\,\mathrm{e}^{2\pi \lambda}}  \,.
\ee
From (\ref{xi-map-fund-boosted}) and (\ref{zeta-map-fund-boosted}) it is straightforward to 
obtain the generalisation of Fig.\,\ref{fig:diamond-AB} corresponding to a region of spacetime determined 
by two different intervals $(a_+, b_+)$ and $(a_-, b_-)$ 
along the chiral directions parameterised by $u_+$ and $u_-$ respectively.

Let us conclude this discussion with a brief comment on the relation 
between the operator (\ref{fmh}) and the Tomita-Takesaki modular theory. Referring for details to \cite{Jovanovic:2025mwe}, 
here we observe  that the $K$-flow generated by the operator (\ref{fmh}) is well defined for conformal fields with any dimension $h_\pm\geqslant 0$. 
With some abuse of terminology, such flow is usually called modular flow, although strictly speaking it can be associated 
with the Tomita-Takesaki theory only for quantum fields which are local, i.e. that either commute or anticommute 
at spacelike distances. 
In CFT, this is certainly the case for fields with dimensions  
$h_\pm \in \mathbb Z$ and $h_\pm \in {\mathbb Z} +\frac{1}{2}$. 
Hence, according to Sec.\,\ref{sec-algebraic-1}, the modular theory 
applies for the currents describing the transport of electric charge and helicity ($h_\pm=1$) 
and of energy ($h_\pm=2$). With the exception of Appendix\;\ref{app-current-phi}, 
in the following we focus on the modular properties of these currents.

\subsection{Modular conjugation}
\label{subsec-mod-J-vacuum}

The modular theory of Tomita and Takesaki 
 \cite{takesaki-book-70, Haag:1992hx, Brattelli2, Borchers:2000pv, takesaki-book-03}
is constructed through the modular operator $e^{-K}$,
written in terms of the full modular Hamiltonian, and the modular conjugation $J$,
an antiunitary operator which leaves the state invariant 
and satisfies $J = J^\ast = J^{-1}$.
For the bipartition and the state of the CFT we are considering, 
the modular conjugation has a geometric action implemented by the real function 
$\mathsf{j} :\mathbb{R}\rightarrow\mathbb{R}$ 
which can be obtained by setting $\tau=\pm\, \ri/2$ in (\ref{xi-map-fund})
and reads \cite{Haag:1992hx}
\be
\label{j0-map-def}
\mathsf{j}(u)\equiv\frac{a+b}{2}+\frac{\left(\frac{b-a}{2}\right)^{2}}{u-\frac{a+b}{2}}
\ee
which is a bijective and idempotent function sending $A$ onto $B$.
The map (\ref{j0-map-def}) is invariant under $a \leftrightarrow b$ and satisfies $\mathsf{j}'(u) < 0$.
We remark  that (\ref{xi-map-fund}) and (\ref{j0-map-def}) commute, namely
\be
\label{id-xi-j-line}
\mathsf{j}\big( \xi(\tau, u) \big)
=
\xi\big(\tau,\mathsf{j}(u)\big)   \,.
\ee
Notice that $\mathsf{j}(u)$ in (\ref{j0-map-def}) becomes $\mathsf{j}(u)=2a - u$ as $b \to +\infty$
and $\mathsf{j}(u)=2b - u$ as $a \to -\infty$.

In Minkowski spacetime, the geometric action of the modular conjugation $J$
associated to the state and the bipartition we are investigating 
can be written through (\ref{j0-map-def}) as follows
\be
\label{inversion-xt}
\tilde{x}(x,t) \equiv \frac{\mathsf{j}(u_+) + \mathsf{j}(u_-)}{2}
\;\;\;\qquad\;\;\;
\tilde{t}(x,t) \equiv \frac{\mathsf{j}(u_+) - \mathsf{j}(u_-)}{2}  \;.
\ee

In Fig.\,\ref{fig:diamond-AB} the points labelled by coloured squares 
are the images through (\ref{inversion-xt}) of the points labelled by dots having the same colour.
Considering the partitions of $\mathcal{D}_A$ and $\mathcal{R}_A$ introduced in Sec.\,\ref{subsec-mod-H-vacuum},
we have that,  for any assigned $i \in \big\{ \textrm{R} , \textrm{L} , \textrm{F} , \textrm{P} \big\}$,
the idempotent map (\ref{inversion-xt}) sends $\mathcal{D}_i$ onto $\mathcal{R}_i$ in a bijective way.

The image of the modular trajectories (\ref{mod-traj-tau-line}) through (\ref{inversion-xt}) 
has been studied e.g. in \cite{Mintchev:2022fcp}.
Within the context of the gauge/gravity correspondence, 
the holographic dual of (\ref{inversion-xt}) tor $t=0$ has been discussed \cite{Mintchev:2022fcp, Caggioli:2024uza}
by employing the geodesic bit threads \cite{Freedman:2016zud, Agon:2018lwq}.

Considering a modular trajectory (\ref{mod-traj-tau-line})  in  $\mathcal{D}_A$
with initial point $P \in \mathcal{D}_A$ having light-cone coordinates $(u_+, u_-)$,
its image under (\ref{inversion-xt}) belongs to $\mathcal{R}_A$ and 
the spacetime coordinates of its generic point read
\cite{Mintchev:2022fcp}
\be
\label{mod-traj-conj-lambda-line}
\tilde{x}(\tau) = \frac{ \mathsf{j}\big(\xi(\tau, u_+)\big) + \,\mathsf{j}\big(\xi(-\tau, u_-)\big) }{2}
\;\;\;\qquad\;\;\;
\tilde{t}(\tau) = \frac{ \mathsf{j}\big(\xi(\tau, u_+)\big) -  \,\mathsf{j}\big(\xi(-\tau, u_-) \big)}{2}
\ee
which can be written in an equivalent form by employing (\ref{id-xi-j-line}).

\begin{figure}[t!]
\vspace{-0.cm}
\hspace{.8cm}
\includegraphics[width=.9\textwidth]{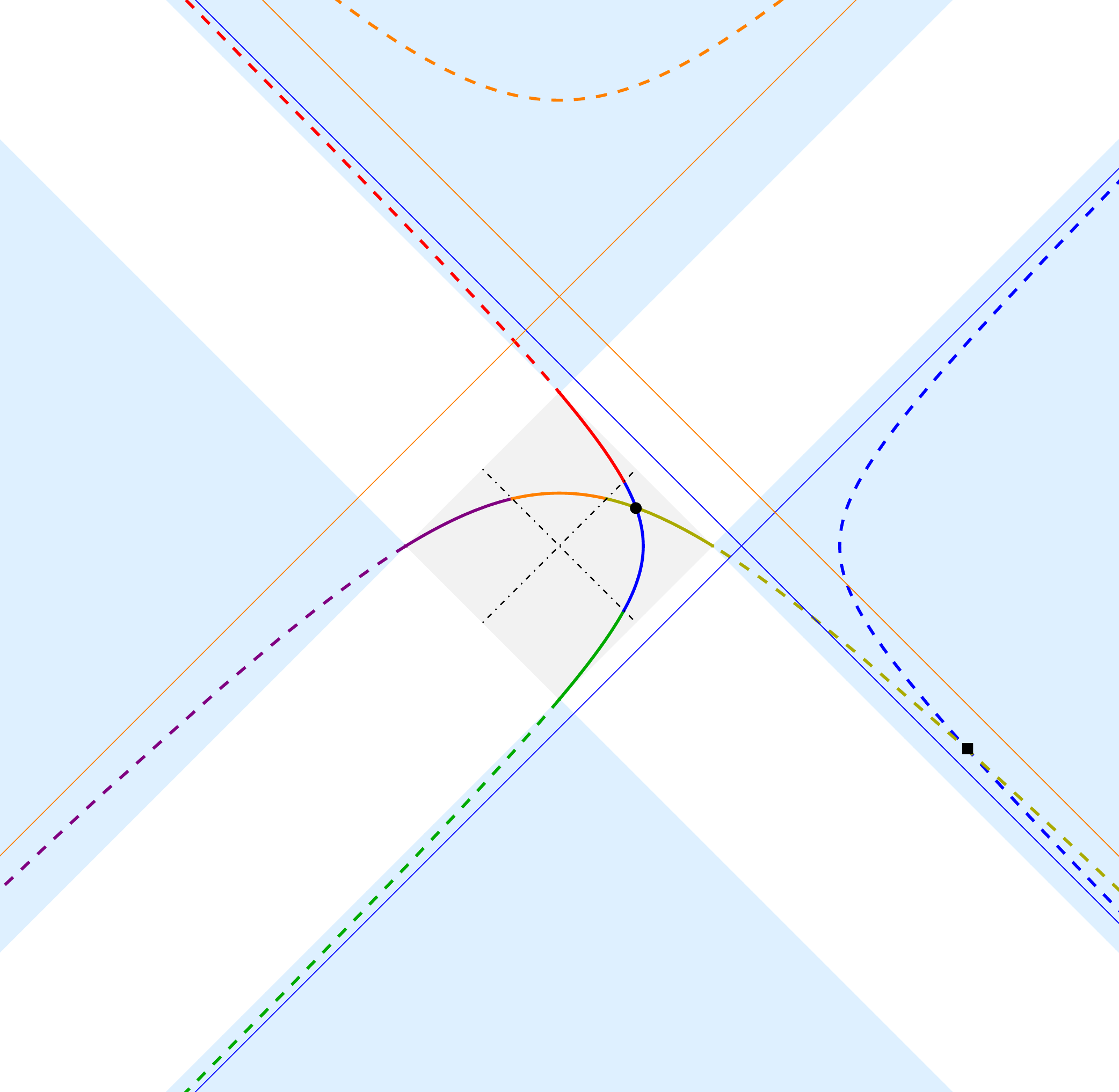}
\vspace{.3cm}
\caption{
The modular hyperbolae $\mathcal{I}_{_P}$ in (\ref{mod-hyper})
correspond to the red, blue and green arcs,
while 
the modular hyperbolae $\widetilde{ \mathcal{I} }_{_P}$ in(\ref{mod-hyper-P-evo})
is made by the purple, orange and dark yellow arcs.
The asymptotes of $\mathcal{I}_{_P}$ and $\widetilde{ \mathcal{I} }_{_P}$
are the orange and blue thin straight lines respectively. 
}
\label{fig:diamond-mod-hyperbolae}
\end{figure}

%\noindent 
%$\bullet$ 
%\textcolor{blue}{\bf [Modular hyperbolae ($K$ evolution)]}

In \cite{Mintchev:2022fcp, Jovanovic:2025mwe} it has been observed that 
the union of  the modular trajectory (\ref{mod-traj-tau-line}) in $\mathcal{D}_A$
and of the corresponding curve in $\mathcal{R}_A$ obtained through (\ref{mod-traj-conj-lambda-line})
provides the  hyperbola $\mathcal{I}_{_P}$ defined as follows
\be
\label{mod-hyper}
\big[ x(\tau) - x_0 \big]^2 - t(\tau)^2 = \kappa^2
\;\;\;\qquad\;\;\;
\big[ \tilde{x}(\tau) - x_0 \big]^2 -\tilde{t}(\tau)^2 = \kappa_0^2
\ee
whose parameters are 
\be
\label{mod-hyper-parameters}
x_0 \equiv \frac{u_+ \, u_- - a\, b}{ u_+ + u_- - (a + b) }
\;\;\;\qquad\;\;\;
\kappa_0 \equiv
\frac{\sqrt{ (b-u_+)(u_+ - a) \, (b-u_-)(u_- - a)  } }{ u_+ + u_- - (a + b)}
\ee
in terms of the light-cone coordinates of the initial point $P$.

In Fig.\,\ref{fig:diamond-mod-hyperbolae}, the point $P$ corresponds to the black dot
and its modular trajectory has been partitioned into the 
the green, blue and red solid arcs,
corresponding to the intersection of the modular trajectory with different $\mathcal{D}_i$,
with  $i \in \big\{ \textrm{R} , \textrm{L} , \textrm{F} , \textrm{P} \big\}$.
The black square is the image of $P$ under (\ref{inversion-xt}) 
and the image of each arc through (\ref{inversion-xt}) 
is indicated by the dashed curve with the same colour. 
The asymptotes of the hyperbolae (\ref{mod-hyper}) are $t= x - x_0$ and $t=  -x + x_0$
(see  the blue solid thin straight lines in Fig.\,\ref{fig:diamond-mod-hyperbolae}).
%
%As $b \to +\infty$, 
%from (\ref{mod-hyper-parameters}) 
%we have that $x_0 \to a$ and $\kappa \to 1$;
%hence the asymptotes of the hyperbolae become the 
%boundaries of $\mathcal{D}_A$ and $\mathcal{R}_A$.
%%
%These observations extend the result of \cite{Mintchev:2022fcp}, 
%where $\mathcal{I}_{_P}$ has been considered only for initial points $P$ at $t=0$,
%which correspond to the special case given by $u_{+} =u_{-}$
%in (\ref{mod-hyper-parameters}).

%\noindent 
%$\bullet$ 
%\textcolor{blue}{\bf [Modular hyperbolae ($P$ evolution)]}

It is worth considering also the modular trajectory (\ref{mod-traj-mom-gs})
provided by the evolution generated by the modular momentum operator (\ref{full-mod-momentum})
having the same point $P \in \mathcal{D}_A$ employed above
as initial point at $\lambda =0 $, where  $\lambda \in \RR$ is the modular parameter.
In Fig.\,\ref{fig:diamond-mod-hyperbolae}, 
this modular trajectory 
has been partitioned into the purple, orange and dark yellow solid arcs,
which correspond to the intersection of the modular trajectory with different $\mathcal{D}_i$,
with  $i \in \big\{ \textrm{R} , \textrm{L} , \textrm{F} , \textrm{P} \big\}$.
The image of the modular trajectory (\ref{mod-traj-mom-gs}) through (\ref{inversion-xt}) 
belongs to $\mathcal{R}_A$ and reads
\be
\label{mod-traj-mom-gs-conj}
\tilde{x}(\lambda) = \frac{ \mathsf{j}\big(\zeta(\lambda, u_+)\big) +\, \mathsf{j}\big( \zeta(\lambda, u_-)\big) }{2}
\;\;\;\qquad\;\;\;
\tilde{t}(\lambda) = \frac{ \mathsf{j}\big(\zeta(\lambda, u_+) \big) -\, \mathsf{j}\big( \zeta(\lambda, u_-)\big) }{2}
\ee
In Fig.\,\ref{fig:diamond-mod-hyperbolae}, 
the images under the inversion (\ref{inversion-xt}) 
of the solid arcs in purple, orange and dark yellow partitioning the modular trajectory obtained from (\ref{mod-traj-mom-gs})
are the dashed arcs having the corresponding colour. 

By adapting the construction of the hyperbola $\mathcal{I}_{_P}$ to the
evolution generated by the modular momentum operator (\ref{full-mod-momentum}),
here we observe that  the union of the modular trajectory (\ref{mod-traj-mom-gs})
and of its image under the map (\ref{inversion-xt}) described by (\ref{mod-traj-mom-gs-conj})
provide the hyperbola $\widetilde{ \mathcal{I} }_{_P}$ given by
\be
\label{mod-hyper-P-evo}
\big[ \,t(\lambda) - t_0 \,\big]^2 - \left[ \, x(\lambda) - \frac{a+b}{2} \,\right]^2  \! =  \lambda^2
\;\;\qquad\;\;
\big[ \,\tilde{t}(\lambda) - t_0 \,\big]^2 - \left[ \, \tilde{x}(\lambda) - \frac{a+b}{2} \,\right]^2  \! = \tilde{\kappa}_0^2
\phantom{xx}
\ee
where the parameters are
\be
\label{mod-hyper-parameters-P-evo}
t_0 \equiv 
\frac{ \big( u_+ - \tfrac{a+b}{2}\big) \big( u_- - \tfrac{a+b}{2}\big) -\big( \tfrac{b-a}{2} \big)^2}{  u_- - u_+}
\;\;\qquad\;\;
\tilde{\kappa}_0
\equiv
\frac{\sqrt{ (b-u_+)(u_+ - a) \, (b-u_-)(u_- - a)  } }{ u_+ - u_- }
\ee
in terms of  the light-cone coordinates of the initial point $P$.
These expressions are well defined whenever $u_+ \neq u_-$;
indeed, for $u_+ = u_-$ the hyperbola  $\widetilde{ \mathcal{I} }_{_P}$ becomes the horizontal line $t=0$.
The two hyperbolas $\mathcal{I}_{_P}$ and $\widetilde{ \mathcal{I} }_{_P}$ 
intersect at $P \in \mathcal{D}_A$ and at its image under (\ref{inversion-xt}) in $\mathcal{R}_A$.

It is worth describing the action of the modular conjugation on the fields of the CFT considered in Sec.\,\ref{sec-algebraic-1}.
In the literature, this transformation has been considered e.g. in \cite{Casini:2010bf, Papadodimas:2013jku}.
Since the geometric action of the modular conjugation is obtained from 
(\ref{xi-map-fund}) at $\tau=\pm\, \ri/2$, 
the action of $J$ on the basic fields of the CFT  can be found 
by combining the fact that $J$ is antiunitary
with (\ref{me9a}), (\ref{me9b}), (\ref{cp4}) and (\ref{met4}).
Thus, for the primaries we have 
\bea
\label{J-on-phi-1}
J\,\phi_{\pm}(u)\,J 
\,=\, 
\mathrm{e}^{\mp\mathrm{i}\mu_{\pm} (\,\mathsf{j}(u)-u )}
\, \mathsf{j}'(u)^{h_{\pm}} 
\, \phi^\ast_{\pm} (\,\mathsf{j}(u))
%\left|\mathsf{j}'(u)\right|^{h}\phi_{\pm}\left(\mathsf{j}(u)\right)
\\
\rule{0pt}{.5cm}
\label{J-on-phi-2}
J\,\phi^\ast_{\pm}(u)\,J 
\,=\, 
\mathrm{e}^{\pm \mathrm{i}\mu_{\pm} (\,\mathsf{j}(u)-u )}
\, \mathsf{j}'(u)^{h_{\pm}} 
\, \phi_{\pm} (\,\mathsf{j}(u))
\eea
where, from (\ref{j0-map-def}), we have that $\mathsf{j}'(u) <0$.
We remark that $\mathsf{j}'(u) \to -1$ as either $b \to +\infty$ or $a \to -\infty$.
As for the currents and the operators (\ref{cal-T-pm-def}) containing the energy-momentum tensor, 
which are hermitean operators, we find respectively
\be
\label{J-on-j}
J \, j_{\pm}(u)\, J 
\,=\, 
\mathsf{j}'(u)
\, j_{\pm} (\,\mathsf{j}(u))
-\frac{\kappa\mu_{\pm}}{2\pi}
\left[ \, 1-  \mathsf{j}'(u) \,\right]
\ee
and 
\be
\label{J-on-T}
J\,\mathcal{T}_{\pm}(u) \, J 
\,=\, 
\mathsf{j}'(u)^2 \, \mathcal{T}_{\pm} (\,\mathsf{j}(u))
+\frac{\kappa\mu_{\pm}^{2}}{4\pi}
\left[ \,1-\mathsf{j}'(u)^{2} \,\right]
\ee
where in the last expression
we used that the Schwarzian derivative (see (\ref{cft3})) of (\ref{j0-map-def}) vanishes identically,
i.e. $\mathcal{S}_u[\,\mathsf{j}\,](u) = 0$.

The transformation rule (\ref{J-on-T}) can be employed to observe that
\bea
\label{JKJ-chiral}
J \left(\,\int_a^b V(u)\, \mathcal{T}_\pm(u) \, \rd u \right) J
& = &
\int_a^b \mathsf{j}'(u) \, V(u)\, \mathcal{T}_\pm(\,\mathsf{j}(u)) \, \mathsf{j}'(u) \, \rd u + \textrm{const}
\\
& = &
 - \left[ \,
\int_{-\infty}^a \!\! V(u)\, \mathcal{T}_\pm(u) \, \rd u
+
\int_b^{+\infty} \!\! V(u)\, \mathcal{T}_\pm(u) \, \rd u
\,\right]
+ \textrm{const}
\hspace{1cm}
\nn
\eea
where we used that (\ref{velocity_fund}) and (\ref{j0-map-def}) satisfy the following identity 
\be
\mathsf{j}'(u) \, V(u) = V(\, \mathsf{j}(u)) \,.
\ee
By applying (\ref{JKJ-chiral}) to (\ref{KA-def-u-pm}), it is straightforward to find that (\ref{KB-def-u-pm}) can be written as $K_B = J\, K_{A} \, J $;
hence the full modular Hamiltonian (\ref{fmh}) can be equivalently expressed in the following suggestive form
\be
\label{fmh-J}
K
\equiv
K_{A} \otimes \boldsymbol{1}_B - \boldsymbol{1}_A \otimes \big( J\, K_{A} \, J \big)
%= 
%\int_{-\infty}^{\infty} V(u_+)\, {\cal T}_+(u_+)\, \rd u_+
%+ 
%\int_{-\infty}^{\infty} V(u_-)\, {\cal T}_-(u_-)\, \rd u_-
%+
%C_K
%=
%\int_{\mathbb{R}}dx\,\mathsf{V}(x;a,b)\,T(x)\,. 
\ee

Since $J\, \Omega_{\mu_\pm} = \Omega_{\mu_\pm}$,
taking the mean values of 
(\ref{J-on-phi-1}), (\ref{J-on-phi-2}), (\ref{J-on-j}) and (\ref{J-on-T})
in the l.h.s.'s one finds $\langle \phi_\pm(u)\rangle_{\mu_\pm} $, $\langle \phi^\ast_\pm(u)\rangle_{\mu_\pm} $, 
$\langle j_\pm(u)\rangle_{\mu_\pm} $ and $\langle \T_\pm(u)\rangle_{\mu_\pm} $ respectively.
By using (\ref{fd1}) in the corresponding r.h.s.'s, consistency is observed;
indeed, the r.h.s.'s of the expressions in (\ref{fd1}) are obtained.

The modular conjugation $J$ allows us to investigate the modular evolution in $\mathcal{R}_A$.
Indeed, considering a chiral operator $\mathcal{O}_{\pm}(u)$ placed at $u \in B$,
i.e. in the domain complementary to $A$ on the line, 
its modular evolution $\mathcal{O}_{\pm}(\tau,u)$ can be written as 
\be
\label{mod-evo-O-in-B}
\mathcal{O}_{\pm}(\tau,u)
\,=\,
\e^{\mathrm{i} K \tau} \, \mathcal{O}_{\pm}(u) \, \e^{-\mathrm{i} K \tau}
\,=\,
J J \, \e^{\mathrm{i} K \tau} \, \mathcal{O}_{\pm}(u) \, \e^{-\mathrm{i} K \tau} J J 
\,=\,
J  \left[ \, \e^{\mathrm{i} K \tau} \big( J \,\mathcal{O}_{\pm}(u) \,J \,\big) \, \e^{-\mathrm{i} K \tau} \, \right] J 
%=
%J\e^{\mathrm{i} K \tau}J\mathcal{O}_{\pm}(u)J\e^{-\mathrm{i} K \tau}J
\ee
where we used that $J$ is idempotent and that $J \, \e^{\mathrm{i} K \tau} J = \e^{\mathrm{i} K \tau}$
(see e.g. Eq.\,(V.2.9) of \cite{Haag:1992hx}).
In the last expression of (\ref{mod-evo-O-in-B}), the operator $J \,\mathcal{O}_{\pm}(u) \,J $ is located in $A$;
hence the same holds for its modular evolution, which corresponds to the operator within the 
square brackets in the last expression of (\ref{mod-evo-O-in-B}).
In the Appendix\;\ref{app-evolution-B-region} we have obtained explicit expressions for \eqref{mod-evo-O-in-B}
when $\mathcal{O}_{\pm}$ is either $\phi_\pm$ or $j_\pm$ or $\CT_\pm$, 
finding that the expressions for the modular evolutions given by
\eqref{me9a}, \eqref{me9b}, \eqref{cp4} and \eqref{met4} with $\xi_\pm(\tau,u)$ reported in  \eqref{xi-map-fund} 
hold also when $u \in B$.

\subsection{Modular correlators}
\label{sec-mod-corr-line}

The two-point functions of the primaries, of the currents and of the energy-momentum tensor
along the modular evolution generated by the full modular Hamiltonian (\ref{fmh})
can be written by combining the results discussed in Sec.\,\ref{sec-mod-evolution}
with the expressions reported in Sec.\,\ref{sec-rep-finite-density}.

As for the one-point functions, 
from (\ref{fd1}) we can take the mean values of 
\eqref{me9a}, \eqref{me9b}, \eqref{cp4} and \eqref{met4} with $\xi_\pm(\tau,u)$ given by \eqref{xi-map-fund}, 
finding that they are independent of $\tau$,
i.e. 
$\langle \phi_\pm(u)\rangle_{\mu_\pm} = \langle \phi_\pm (\tau,u)  \rangle_{\mu_\pm} $ for the primaries,
$\langle j_\pm(u)\rangle_{\mu_\pm} = \langle j_\pm (\tau,u)  \rangle_{\mu_\pm} $ for the currents
and $\langle \T_\pm(u)\rangle_{\mu_\pm} = \langle \T_\pm (\tau,u)  \rangle_{\mu_\pm} $ for the operators (\ref{cal-T-pm-def}).

In order to investigate the modular two-point functions, 
we first observe that, 
when $u \neq v$ and $\tau_1 \neq \tau_2$ are real, for (\ref{xi-map-fund}) 
we have 
\be
\label{identity-mod-corr-zero-temp}
\frac{\partial_{u} \xi(\tau_1, u) \, \partial_{v} \xi(\tau_2, v) }{\big[ \xi(\tau_1, u) - \xi(\tau_2, v)\big]^2}
=
\bigg(
\frac{R(\tau_{12}; u, v) }{ u-v }
\bigg)^2 
\ee
where we have introduced
\be
\label{R-factor-def}
R(\tau; u, v) 
\equiv 
\frac{ \e^{2\pi w(u)} - \e^{2\pi w(v)}  }{  \e^{2\pi w(u) +\pi \tau} - \e^{2\pi w(v) - \pi \tau}}
\,=\,
\frac{  (u-a)(v-b) - (u-b) (v-a)}{   (u-a)(v-b)\,\e^{\pi \tau } - (u-b)(v-a)\,\e^{- \pi \tau }  } 
\ee
which satisfies
\be
R(\tau = 0; u, v) = 1
\;\qquad\;
R(-\tau; v, u) = R(\tau; u, v) 
\;\qquad\;
R(\tau + \ri\,; u, v) = - R(-\tau; v, u) 
\ee
In the derivation of (\ref{identity-mod-corr-zero-temp}) we used that
\be
\label{xi12-tau-chiral-tau12}
\xi(\tau_1,u)  - \xi(\tau_2,v)
%= 
%\frac{(b-a)^2 \, \e^{\pi (\tau_1 +\tau_2)} }{ q(\tau_1,u)\, q(\tau_2,v)\; R(\tau_{12} ; u,v)     }
%\; (u - v)
= 
\frac{ p(\tau_1 ,u) \, p(\tau_2 ,v) }{ R(\tau_{12} ; u,v)     }
\; (u - v)
\ee
in terms of (\ref{R-factor-def}) and of
\be
\label{pA-qA-def}
p(\tau,u) \equiv \frac{(b-a)\, \e^{\pi \tau} }{ b-u +(u-a)\, \e^{2\pi \tau}   } 
\;\;\;\qquad\;\;\;
q(\tau,u)  \equiv b-u +(u-a)\, \e^{2\pi \tau} \,.
%\;\;\;\qquad\;\;\;
%\partial_u \xi(\tau,u) = p(\tau,u)^2
\ee
%which simplifies to (\ref{xi12-tau-chiral}) when $\tau_1= \tau_2$, as expected. 
The dependence on  $\tau_{12}$ in (\ref{identity-mod-corr-zero-temp}),
which is not evident in the l.h.s.,
occurs because the product $q(\tau_1,u)\, q(\tau_2,v)$ simplifies in the ratio
(see (\ref{xi12-tau-chiral-tau12})). 
Notice that $\e^{2\pi w(u)}$ in (\ref{identity-mod-corr-zero-temp}) is well defined for $u\in \RR$,
although $w(u)$ in (\ref{w-function-def}) holds for $u\in A$.
Moreover, we remark that (\ref{identity-mod-corr-zero-temp}) is invariant under the simultaneous
exchange $a \leftrightarrow b$ and $\tau_1  \leftrightarrow  \tau_2$.

%\textcolor{red}{[The signs in the $\varepsilon$ prescription in the following must be understood]}
For the primaries $\phi_\pm$,  
from (\ref{me9a}), (\ref{me9b}), (\ref{fd1}), (\ref{fd5}) and (\ref{identity-mod-corr-zero-temp}), 
one finds the following modular correlators
\bea
\label{mod-corr-phi-mu}
\langle \phi_\pm^*(\tau_1, u) \,\phi_\pm (\tau_2, v)\rangle_{\mu_\pm}^{\textrm{\tiny con}} 
& =&
 \nonumber
\\
\rule{0pt}{.5cm}
& & \hspace{-3.5cm}
=\,
\e^{\mp \ri \mu_\pm [\xi_\pm (\tau_1,u) - \xi_\pm (\tau_2,v)  - u+ v]}
\big[ 
\partial_{u} \xi_{\pm}(\tau_1, u) \, \partial_{v} \xi_{\pm}(\tau_2, v) 
\big]^{h_\pm}
 \langle 
 \phi_\pm^*\big( \xi_{\pm}(\tau_1, u)  \big) \,\phi_\pm \big( \xi_{\pm}(\tau_2, v)  \big)
 \rangle_{\mu_\pm}
 \nonumber
 \\
\rule{0pt}{.5cm}
& & \hspace{-3.5cm}
=\,
\frac{ \e^{\pm \ri \mu_\pm (u-v)} }{2\pi \, \e^{\pm \ri \pi h_\pm}}\;
W_\pm(\pm \tau_{12}; u, v)^{2h_{\pm}}
\eea
and
\be
\label{mod-corr-phi-mu-2}
\langle \phi_\pm(\tau_1, u) \,\phi^*_\pm (\tau_2, v)\rangle_{\mu_\pm}^{\textrm{\tiny con}} 
=\,
\frac{ \e^{\mp \ri \mu_\pm (u-v)} }{2\pi \, \e^{\pm \ri \pi h_\pm}}\;
W_\pm(\pm \tau_{12}; u, v)^{2h_{\pm}}
\ee
where we have introduced 
\be
\label{cap-W-def}
W_\pm(\tau; u, v)
\,\equiv \,
\frac{ \e^{2\pi w(u)} - \e^{2\pi w(v)}  }{u-v  }\;
\frac{1}{  \e^{2\pi w(u) +\pi \tau} - \e^{2\pi w(v) - \pi \tau} \mp \ri \varepsilon}  \;.
\ee
When either $u \neq v$ or $\tau \neq 0$, we can set $\varepsilon = 0$ and $W_\pm(\tau; u, v)$ becomes 
\be
\label{cap-W-eps0-def}
W(\tau; u, v)
\equiv 
\frac{ R(\tau; u, v) }{u-v  }
\ee
in terms of (\ref{R-factor-def}).
Notice that the Hilbert space structure implies that (\ref{mod-corr-phi-mu}) satisfies
\be
\label{id-mod-corr-J}
\langle \phi_\pm^*(\tau_1, u) \,\phi_\pm (\tau_2, v)\rangle_{\mu_\pm}^{\textrm{\tiny con}} 
=
\overline{
\langle \phi^\ast_\pm(\tau_2, v) \,\phi_\pm (\tau_1, u)\rangle_{\mu_\pm}^{\textrm{\tiny con}} 
}
\ee
where the overline denotes the complex conjugation.

As for the currents $j_\pm$, 
from (\ref{cp4}),  (\ref{fd1}), (\ref{fd4}) and (\ref{identity-mod-corr-zero-temp}), 
their modular correlators read
\bea
\label{mod-corr-j-mu}
\langle j_\pm(\tau_1, u) \,j_\pm (\tau_2,v)\rangle_{\mu_\pm}^{\textrm{\tiny con}}  
&=&
\big[ 
\partial_{u} \xi_{\pm}(\tau_1, u) \, \partial_{v} \xi_{\pm}(\tau_2, v) 
\big] \,
 \langle 
 j_\pm\big( \xi_{\pm}(\tau_1, u)  \big) \,j_\pm \big( \xi_{\pm}(\tau_2, v)  \big)
 \rangle_{\mu_\pm}
 \nonumber
\\
\rule{0pt}{.5cm}
&=&
\frac{\kappa}{4\pi^2} \,W_\pm(\pm \tau_{12}; u, v)^{2}   \,.
\eea
Similarly, for the modular correlators of the operators (\ref{cal-T-pm-def}), 
which contains the energy-momentum tensor, 
from (\ref{met4}),  (\ref{fd1}),  (\ref{fd3}) and (\ref{identity-mod-corr-zero-temp}), we find
\bea
\label{mod-corr-emtensor-mu}
\langle \T_\pm(\tau_1, u) \,\T_\pm (\tau_2,v)\rangle_{\mu_\pm}^{\textrm{\tiny con}}  
&=&
\big[ 
\partial_{u} \xi_{\pm}(\tau_1, u) \, \partial_{v} \xi_{\pm}(\tau_2, v) 
\big]^2\,
 \langle 
 \T_\pm\big( \xi_{\pm}(\tau_1, u)  \big) \,\T_\pm \big( \xi_{\pm}(\tau_2, v)  \big)
 \rangle_{\mu_\pm}
 \nonumber
\\
\rule{0pt}{.5cm}
&=&
\frac{c}{8\pi^2} \, W_\pm(\pm \tau_{12}; u, v)^{4}  \,.
\eea
We remark that these modular correlators are functions of $\tau_{12}$
which depend on $t_1$ and $t_2$ separately;
hence the modular energy is conserved along the modular evolution
(see Sec.\,\ref{subsec-energy-continuity}), 
while the conventional energy is not. 
Furthermore, according to the discussion reported at the end of Sec.\,\ref{subsec-mod-J-vacuum}
and in the Appendix\;\ref{app-evolution-B-region}, 
we have that the expressions 
(\ref{mod-corr-phi-mu}), (\ref{mod-corr-phi-mu-2}), (\ref{mod-corr-j-mu}) and (\ref{mod-corr-emtensor-mu}) 
for the modular correlators hold for any $u,v \in \RR$.

Since for (\ref{cap-W-def}) the following property holds
\be
\label{KMS}
W_\pm(\tau + \ri ; u, v) = W_\pm(\tau  - \ri ; u, v) = W_\pm( - \tau  ; v, u)
\ee
the modular correlators (\ref{mod-corr-phi-mu}), (\ref{mod-corr-phi-mu-2}), (\ref{mod-corr-j-mu}) and (\ref{mod-corr-emtensor-mu}) 
satisfy the KMS condition with modular inverse temperature $\tilde{\beta} = 1$.
This is a characterising feature of the modular correlators
that exposes the thermal nature of the modular evolution. 
Furthermore, the validity of this condition confirms the expression (\ref{KA-def-u-pm}) for the modular Hamiltonian
(this criterion has been adopted e.g. in \cite{Longo:2009mn} to confirm the expression of the 
modular Hamiltonian of disjoint intervals on the line for the massless Dirac field in the ground state, 
found in \cite{Casini:2009vk}).

In the Appendix\;\ref{app-michele-modular} we also provide 
a consistency check for (\ref{mod-corr-phi-mu})
based on the modular reflection positivity property (see e.g. \cite{Casini:2010bf}).

Let us highlight that, when $\tau \neq 0$, 
the limit $v \to u$ of (\ref{cap-W-def}) is well defined and reads
\be
\label{cap-W-equal-point}
\lim_{v \to u} 
W_\pm(\tau; u, v)
=
\frac{\pi }{ V (u) \, \sinh(\pi \tau\mp \ri \varepsilon) }   \;.
\ee
This observation will be employed in Sec.\,\ref{subsec-noise} in a crucial way.

From the properties of the modular conjugation, for a chiral operator $\mathcal{O}_\pm$ we have 
\be
\label{id-mod-corr-J}
\langle \mathcal{O}_\pm^*(\tau_1, u) \,\mathcal{O}_\pm (\tau_2, v)\rangle_{\mu_\pm}^{\textrm{\tiny con}} 
=
\langle \big[ J \, \mathcal{O}^\ast_\pm(\tau_2, v) \, J \big] \big[ J \, \mathcal{O}_\pm (\tau_1, u)\, J \big] \rangle_{\mu_\pm}^{\textrm{\tiny con}} 
%=
%\overline{
%\langle \phi^\ast_\pm(\tau_2, v) \,\phi_\pm (\tau_1, u)\rangle_{\mu_\pm}^{\textrm{\tiny con}} 
%}
\ee
By employing the following identity 
\be 
W_\pm (\tau, u, v) ^2
= 
\mathsf{j}^\prime(u)\, \mathsf{j}^\prime(v)\, W_\pm (-\tau, \mathsf{j}(v), \mathsf{j}(u))^2
\ee
we checked that the r.h.s.'s of  
\eqref{J-on-phi-1}, \eqref{J-on-phi-2}
\eqref{J-on-j} and \eqref{J-on-T} 
are consistent with  (\ref{id-mod-corr-J}).

The analyses discussed above can be extended to investigate 
the correlators of the fields whose evolution is determined by the momentum operator
(see (\ref{full-mod-momentum})-(\ref{mod-traj-mom-gs}) and Sec.\,\ref{sec-momentum-evo}).
It is straightforward to adapt the expressions (\ref{mod-corr-phi-mu}), (\ref{mod-corr-phi-mu-2}),
(\ref{mod-corr-j-mu}) and (\ref{mod-corr-emtensor-mu}) to these flows,
finding correlators that satisfy the corresponding KMS condition, 
whose validity is based on the property (\ref{KMS}) for (\ref{cap-W-def}).

\section{Modular transport and fluctuations}
\label{sec-transport-line}
%%%%%%%%%%%%%%%%%%%%%%%%%%%%%%%%%%%%%%%%%%%%%%%%%%

%
%In this section we consider the modular evolution 
%$\O (\tau; x,t) \equiv \e^{\ri \tau K} \O(x,t) \, \e^{-\ri \tau K}$ of some observables $\O$ 
%of the CFT in the finite density representation and for the bipartition given by an interval and its complement on the line. 
%The non-trivial transport properties are described by the following expectation values 
%\be 
%\langle \O (\tau; x,t) \rangle_{\mu} 
%\;\;\;\qquad \;\;\;
%\langle \O (\tau_1 ; x,t) \,\O (\tau_2 ; x,t) \rangle_{\mu}^{\textrm{\tiny con}}   
%\label{a2}
%\ee 
%which provide the mean values and the quadratic quantum fluctuations respectively.  
%%
%%In (\ref{a2}) 
%The notation $\langle \, \dots \rangle_{\mu} $ denotes that both 
%$\langle \, \dots \rangle_{\mu_+} $ and $\langle \, \dots \rangle_{\mu_-} $ are employed.
%%
%For the operators that will be considered in the following, 
%$\langle \O (\tau; x,t) \rangle_{\mu} $ is independent of $\tau$,
%as discussed in Sec.\,\ref{sec-mod-corr-line}.
%\\
%\textcolor{red}{[{\bf (MICHELE)} comment about the transport as probability distribution from all the correlators (add a ref)]}
%\\

In this section we consider the modular evolution 
$\O (\tau; x,t) \equiv \e^{\ri \tau K} \O(x,t) \, \e^{-\ri \tau K}$ of some observables $\O$ (essentially currents and densities) 
of the CFT in the finite density representation and for the bipartition given by an interval on the line. 
In unitary quantum field theory, the sequence of connected correlation functions 
$\big\{ \langle \O (\tau_1 ; x_1,t_1) \, ...\,  \O (\tau_n ; x_n,t_n) \rangle_{\mu}^{\textrm{\tiny con}}  \big\}  $
provides the cumulants of a probability distribution (see e.g.\;\cite{Haag:1992hx}).  
Hereafter the notation $\langle \, \dots \rangle_{\mu} $ denotes that both 
$\langle \, \dots \rangle_{\mu_+} $ and $\langle \, \dots \rangle_{\mu_-} $ 
are employed.
When $\O$ is a current, this distribution fully describes the microscopic transport properties  
of the associated charge. For $n=1$ and $n=2$ one gets respectively the mean value and the quadratic 
fluctuations (quantum noise) of the current. 
In the following we consider operators whose one-point function 
$\langle \O (\tau; x,t) \rangle_{\mu} $ is independent of $\tau$,
as discussed in Sec.\,\ref{sec-mod-corr-line}.

\subsection{Charge and helicity transport} 
\label{sec-charge-helicity-trans}

%\noindent
%\textcolor{blue}{\bf $\bullet$ Charge}
%\\
The mean values of the charge currents in (\ref{ecurr}) and (\ref{j-t-def})
in the finite density representation of the CFT on the line
can be written by using (\ref{fd1})
and specialising the velocities to $V_{+}(u)=V_{-}(u)$ given by (\ref{velocity_fund}).
The result is
\bea
\label{mc1x}
\langle j_x (\tau;x,t) \rangle_{\mu}  
&=&
\frac{\kappa}{2\pi} \,\big[ \mu_+ V(u_+) - \mu_- V(u_-)  \big] 
\\
\rule{0pt}{.6cm}
\label{mc1t}
\langle j_t (\tau;x,t) \rangle_{\mu}  
&=&
\frac{\kappa}{2\pi} \,\big[ \mu_+ V(u_+)  + \mu_- V(u_-) \big]   \,.
\eea
We remark that, instead, the mean values of the operators (\ref{eden-mu}) and (\ref{ecurr-mu}) vanish identically. 

%\noindent \textcolor{blue}{\bf $\bullet$ Helicity} \\
The mean values of the helicity currents (\ref{k-xt-def})
are obtained in the same way and read
\bea
\label{mc10x}
\langle k_x (\tau;x,t) \rangle_{\mu}  
&=&
 \frac{\kappa}{2\pi} \,\big[\mu_+ V(u_+) + \mu_- V(u_-)  \big] 
 \\
 \label{mc10t}
 \rule{0pt}{.6cm}
 \langle k_t (\tau;x,t) \rangle_{\mu}  
&=&
 \frac{\kappa}{2\pi} \,\big[\mu_+ V(u_+) - \mu_- V(u_-)  \big]   \,.
\eea

%\noindent
%\textcolor{red}{\bf $\bullet$ Conductivity [to be understood]}
%\\
%\noindent
%\textcolor{blue}{\bf $\longrightarrow$ Can we define a conductivity? It should be a matrix in our problem.}
%\\
%\textcolor{blue}{\bf $\longrightarrow$ Since we are considering a two-dimensional vector field, the transport is non trivial also when $\mu-=\mu_+$.}
%\\
%\textcolor{red}{
%{\bf [the following part must be revised]}
%This tells us that, when $\mu_+ \neq \mu_-$, 
%there is a non-trivial charge transport along the modular evolution of the system
%and the mean 
%value of the current is controlled by the modular velocity and vanishes at the endpoints of $A$. 
%By assuming without loss of generality that $\Delta \mu \equiv \mu_--\mu_+ >0$,
%we have that  the current is positive in $A$ and negative in the spatial complement $B$ 
%with respect to the standard orientation of the real line $\RR$. 
%This leads us to interpret the endpoint $u=a$ as a source and the endpoint $u=b$ as a sink. 
%The mean current (\ref{mc1}) satisfies Ohm's law with conductivity  
%\be 
%\sigma \equiv \frac{\rd}{\rd \Delta \mu}\, \langle j (\tau,x) \rangle_{\mu_\pm} = 
%\dots
%%\frac{\kappa}{2\pi} v(x) 
%\label{mc2}
%\ee 
%which is proportional to the modular velocity.} 
%\\

%\noindent
%\textcolor{blue}{\bf $\bullet$ Vector fields:}

\begin{figure}[t!]
\vspace{-.6cm}
\hspace{-1.3cm}
%\rule{0pt}{8cm}
\includegraphics[width=1.2\textwidth]{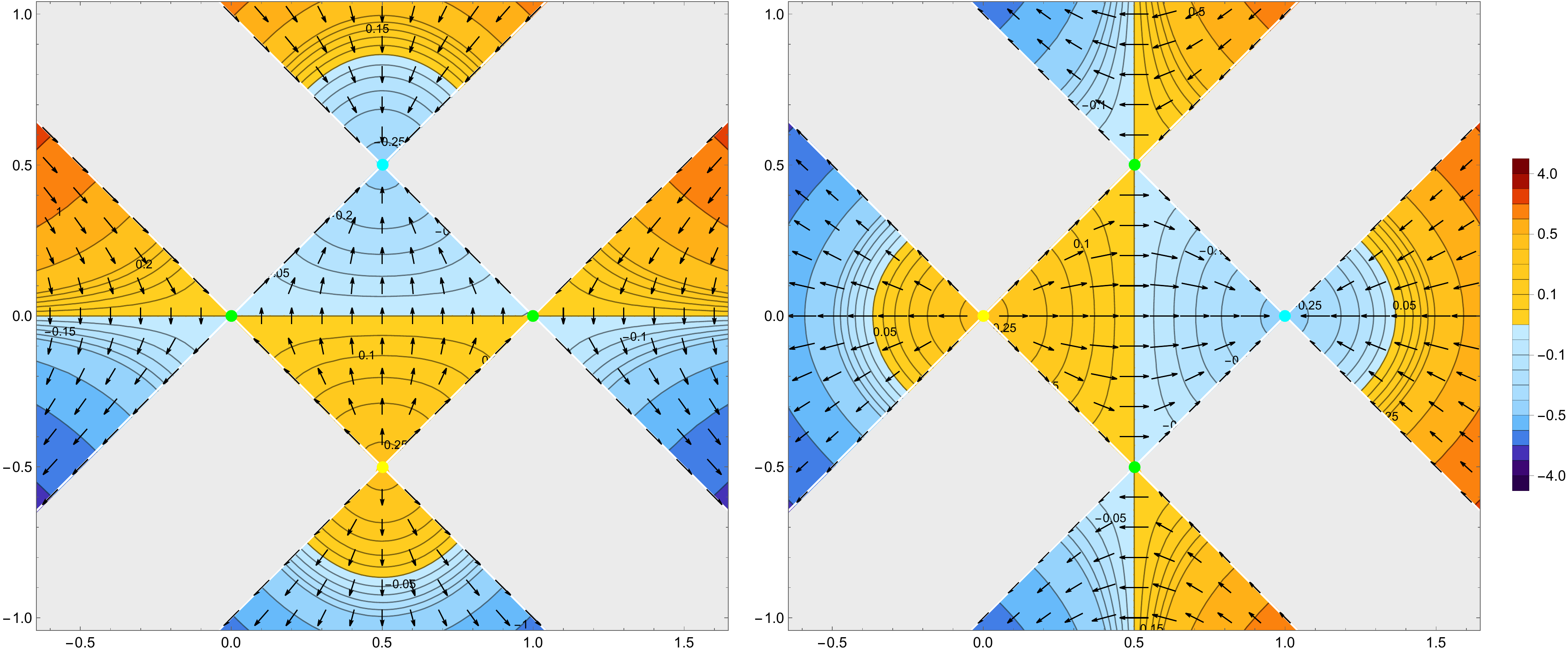}
\\
\rule{0pt}{8.3cm}
\vspace{-.3cm}
\hspace{-1.35cm}
\includegraphics[width=1.2\textwidth]{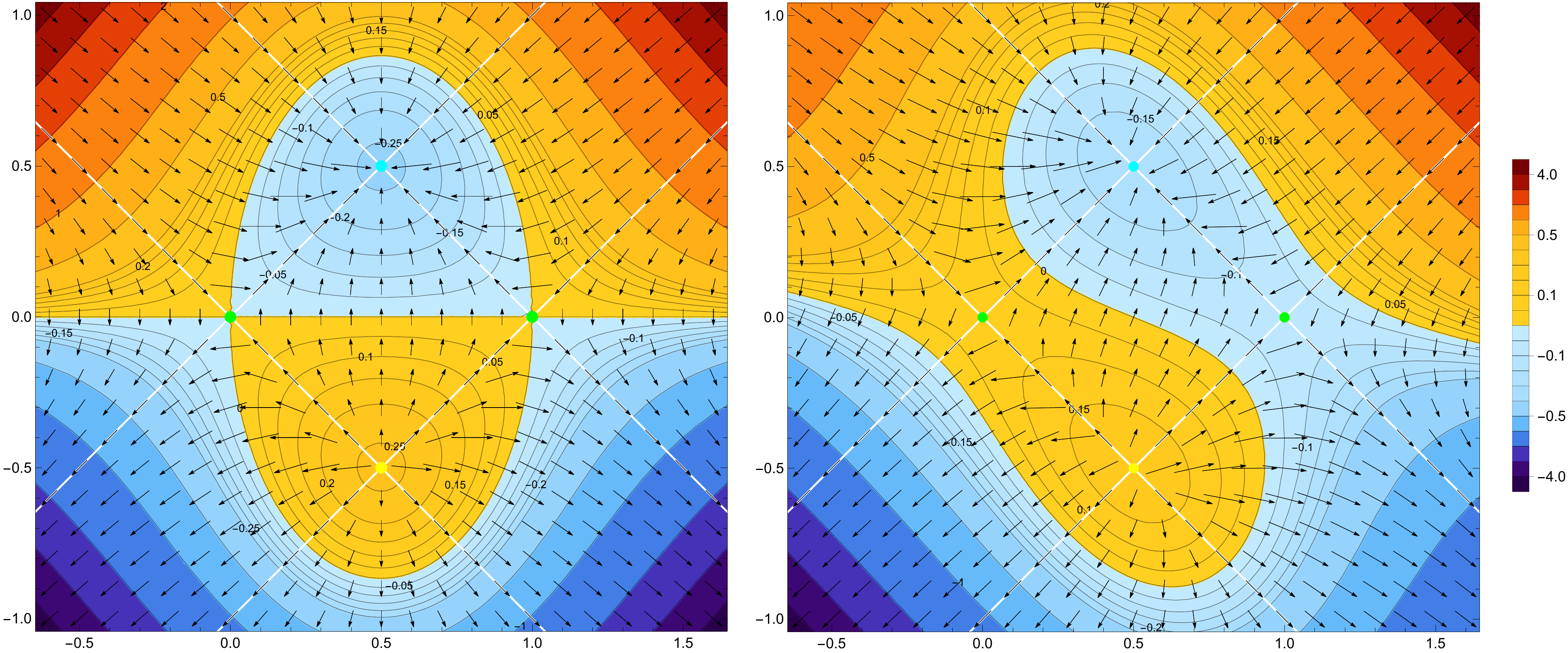}
\vspace{.0cm}
\caption{Vector fields for the mean values of the charge currents  (\ref{mc1x}) and (\ref{mc1t}),
whose potential is the first expressions in (\ref{potentials-W-line}).
The interval is $A=[0,1]$ on the line and the CFT has $c=1$, $\kappa = 3$
and either equal chemical potentials $\mu_{+} = \mu_{-} = 0.52$ (left panels)
or different chemical potentials  $\mu_{+} =  0.52$ and $\mu_{-} =  0.22$ (right panels).
In the top panels the initial point $(x,t) \in \mathcal{D}_A \cup \mathcal{R}_A$, 
while in the bottom panels the vector field $\boldsymbol{j} (x,t)$ 
is extended to the entire Minkowski spacetime. 
}
\label{fig:j-curr}
\end{figure}

\begin{figure}[t!]
\vspace{-.6cm}
\hspace{-1.3cm}
\includegraphics[width=1.2\textwidth]{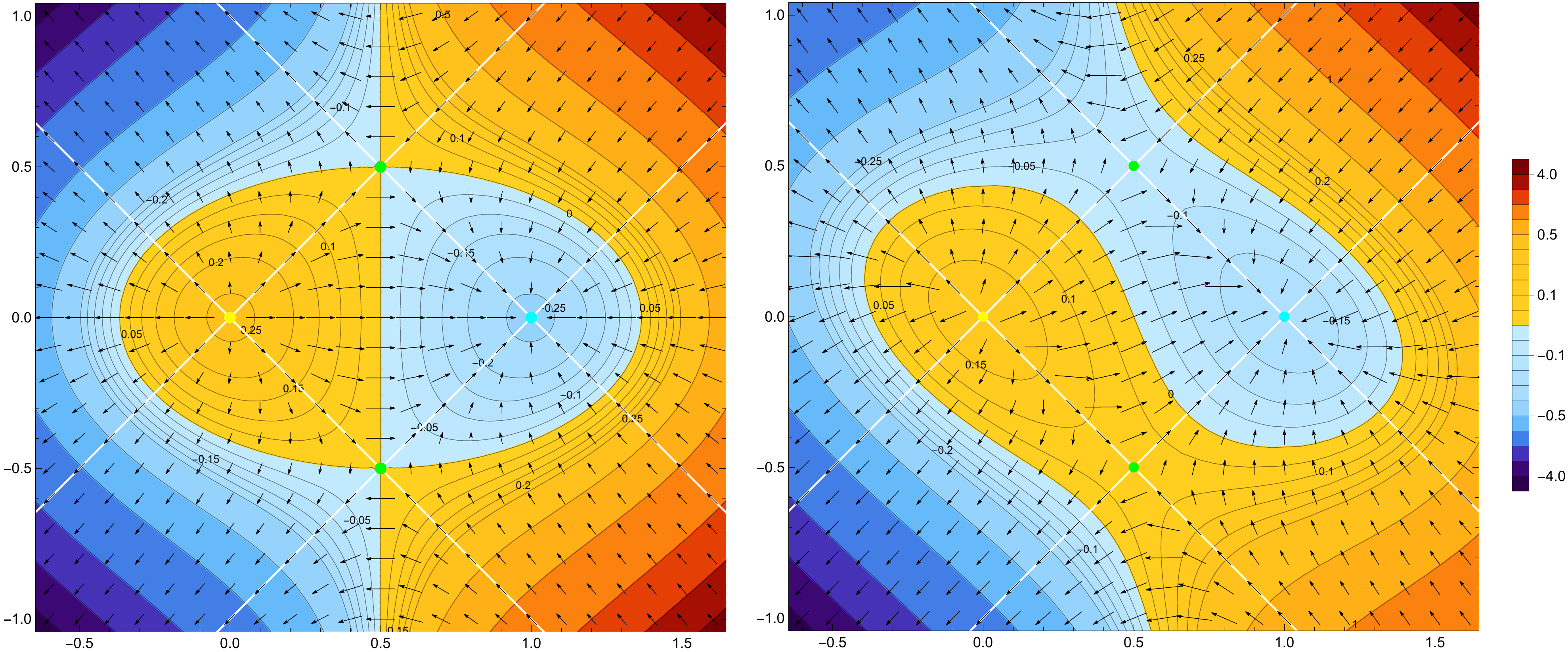}
\vspace{-.4cm}
\caption{Vector fields for the mean values of the helicity currents  (\ref{mc10x}) and (\ref{mc10t}),
whose potential is the second expressions in (\ref{potentials-W-line}),
for either equal (left panel) or different (right panel) chemical potentials,
in the same setup of Fig.\,\ref{fig:j-curr}.
}
\label{fig:k-curr}
\end{figure}

These mean values for the charge currents and for the helicity currents  
%with the corresponding potentials (\ref{potentials-W-line}),
are independent of $\tau$ and the event with spacetime coordinates $(x,t)$
corresponds to the initial point of the modular evolution; hence $(x,t) \in \mathcal{D}_A \cup \mathcal{R}_A$.
We find it worth introducing  the smooth planar vector fields 
$\boldsymbol{j} (x,t) \equiv \big( \langle j_x (\tau;x,t) \rangle_{\mu} \, , \langle j_t (\tau;x,t) \rangle_{\mu}   \big) $
and
$\boldsymbol{k} (x,t)\equiv \big( \langle k_x (\tau;x,t) \rangle_{\mu}  \, , \langle k_t (\tau;x,t) \rangle_{\mu}   \big) $
for $(x,t) \in \mathcal{D}_A \cup \mathcal{R}_A$.
Despite the fact that these vector fields are defined in $\mathcal{D}_A \cup \mathcal{R}_A$,
it is natural to extend them to the entire Minkowski spacetime and in the following 
such extension will be mainly explored. 
In Fig.\,\ref{fig:j-curr} and Fig.\,\ref{fig:k-curr} the vector fields $\boldsymbol{j} (x,t)$ and $\boldsymbol{k} (x,t)$
are displayed for a specific choice of the parameters (see the caption of Fig.\,\ref{fig:j-curr}).
In particular, in the left panels $\mu_{+} = \mu_{-}$, while $\mu_{+} \neq \mu_{-}$ in the right panels.
Moreover, in Fig.\,\ref{fig:j-curr} the top panels show $\boldsymbol{j} (x,t)$ for $(x,t) \in \mathcal{D}_A \cup \mathcal{R}_A$,
while in the bottom panels the corresponding extensions to the entire Minkowski spacetime are displayed.

%
%\noindent
%\textcolor{red}{
%$\bullet$ 
%[we do not know whether is more physical the vector or the one-form.
%Here we show the one-form because: (a) it provides the sink/source interpretation; (b) there is no flux through the diamond;
%(c) inside the diamond, it follows qualitatively the modular trajectories]
%}

Two-dimensional vector fields have been largely explored \cite{arnold-book, milnor-book}.
For instance, it is insightful to consider 
the critical points (also called singular points in the mathematical literature) of the vector field, 
i.e. the points where it vanishes. 
By construction, the critical points of 
both the vector fields $\boldsymbol{j} (x,t)$ and $\boldsymbol{k} (x,t)$
extended to the entire Minkowski spacetime are isolated points located
in the vertices of the diamond $\mathcal{D}_A$, 
i.e. $P_a$, $P_b$, $P_{+\infty}$ and $P_{-\infty}$,
whose light-cone coordinates are 
$(u_{+}, u_{-}) \in \big\{ (a,a)\,, (b,b)\,,  (b,a)\,, (a,b)  \big\}$ respectively 
(see the coloured dots in Fig.\,\ref{fig:j-curr} and Fig.\,\ref{fig:k-curr}),
which are obtained by combining the zeros of the function $V(u)$ in (\ref{velocity_fund}).
Since $a$ and $b$ are zeros of  (\ref{velocity_fund}) of order $1$,
all these critical points have multiplicity $1$.

%\textcolor{blue}{
%We find it worth highlighting that the qualitative behaviour \dots resembles the modular flow\dots
%}
%\textcolor{red}{[{\bf (MICHELE)}
%comment on the suggestive qualitative behaviour of the vector field
%in Fig.\,\ref{fig:j-curr},
%from the south pole to the north pole of the charges, like the modular flow in $\tau$]}

The Poincar\'e index is a useful tool to study smooth vector fields $\boldsymbol{v}\equiv (v_x, v_t)$ with isolated zeros. 
Given a smooth closed curve $\gamma$ where the vector field is not vanishing,
the Poincar\'e index of $\gamma$ relative to the vector field $\boldsymbol{v}$ can be computed as follows
\be
\label{poincare-index-gamma}
\mathcal{I}[\boldsymbol{v}](\gamma) \equiv \frac{1}{2\pi} \oint_\gamma 
\frac{v_t \, \rd v_x - v_x \, \rd v_t}{v_x^2 + v_t^2}
%\rd \varphi
\ee
%where
%\be
%\varphi = \arctan(v_t/v_x)
%\;\;\;\qquad\;\;\;
%\rd \varphi = \frac{v_t \, \rd v_x - v_x \, \rd v_t}{v_x^2 + v_t^2}
%\ee
and it corresponds to the number of rotations in the positive (counterclockwise)
direction that the vector field performs when we go around $\gamma$ once.
The index $\mathcal{I}_P$ of a critical point $P$ is the Poincar\'e index of a closed smooth curve that encloses only $P$;
hence it is determined by the behaviour of the vector field nearby $P$.
Depending on such behaviour, $P$ is a critical point of certain type
(e.g. either a node or a center or a focus or a saddle or something else).  
A node contains all the nearby trajectories
and it is either stable or unstable, depending on whether
all the trajectories move away from the point.
A saddle has two transversal trajectories called separatices,
one of which is ingoing and the other outgoing, 
while the other trajectories behave like a family
of hyperbolas whose asymptotes are the separatrices.
A node has Poincar\'e index  $+1$ (like a focus and a center),
while a saddle has Poincar\'e index $-1$.
The vector fields $\boldsymbol{j} (x,t)$ and $\boldsymbol{k} (x,t)$
display two nodes, one stable and one unstable, and two saddles
(in the bottom panels of Fig.\,\ref{fig:j-curr} and Fig.\,\ref{fig:k-curr}, 
see the cyan dot, the yellow dot and the green dots respectively).

Various theorems about the sum of the indices for smooth vector fields with isolated zeros
have been established \cite{arnold-book}.
For instance, the index of a closed curve is equal to 
the sum of the indices of the critical points enclosed by the curve. 
A fundamental result in this context is the Poincar\'e-Hopf theorem, 
which claims that the sum of the indices of all the isolated critical points of a vector field
on a two-dimensional compact manifold is independent of the vector field 
and equal to the Euler characteristic of the manifold.
For the vector fields $\boldsymbol{j} (x,t)$ and $\boldsymbol{k} (x,t)$,
which are defined on the plane, the sum of the indices of all the critical points is zero.
These vector fields can be mapped to the Riemann sphere 
through the stereographic projection 
and the resulting vector fields on this compact manifold
must have a critical point with Poincar\'e index $+2$ at the north pole,
which corresponds to infinity on the plane.

%\noindent
%\textcolor{red}{\bf $\bullet$ [CONDUCTIVITIES \& BALLISTIC TRANSPORT for $\boldsymbol{j}$ and $\boldsymbol{k}$]}

By employing the definitions in (\ref{heat2}),
from (\ref{mc1x})-(\ref{mc1t}) and (\ref{mc10x})-(\ref{mc10t}) 
we introduce the local conductivities respectively as 
\bea
\boldsymbol{\sigma}_{j, e} 
&\equiv &
 \partial_{\mu_e} \boldsymbol{j} (x,t)
=
\frac{\kappa}{4\pi}\, 
\Big( V(u_+) - V(u_-) \, , \, V(u_+) + V(u_-) \Big)
\\
\rule{0pt}{.7cm}
\boldsymbol{\sigma}_{j, h} 
&\equiv &
 \partial_{\mu_h} \boldsymbol{j} (x,t)
=
\frac{\kappa}{4\pi}\, 
\Big( V(u_+) + V(u_-) \, , \, V(u_+) - V(u_-) \Big)
\eea
and
\be
\boldsymbol{\sigma}_{k, e} 
\equiv
 \partial_{\mu_e} \boldsymbol{k} (x,t)
=
\boldsymbol{\sigma}_{j, h} 
\;\;\;\qquad\;\;\;
\boldsymbol{\sigma}_{k, h} 
\equiv 
 \partial_{\mu_h} \boldsymbol{k} (x,t)
=
\boldsymbol{\sigma}_{j, e} \,.
\ee
%where we have restored the role of the charge $e$, that has been fixed to one in Sec.\,\ref{subsec-commutators}.
%\textcolor{red}{\bf [mettere carica uguale a 1]}
%
%The conductivities from (\ref{mc1t}) and (\ref{mc10t}) are defined in a similar way, 
%finding that 
%$\sigma_{j_t, e}  =  \sigma_{k_x, e} $, $\sigma_{j_t, h} = \sigma_{k_x, h}$,
%$\sigma_{k_t, e}  = \sigma_{j_x, e}$ and $\sigma_{k_t, h} =  \sigma_{j_x, h}$.
%
These four local conductivities are independent of $\tau$;
hence their Fourier transform gives only a Dirac delta term.

In order to understand the nature of the modular transport, let us consider
\be
\label{ballistic-j-k}
\int_0^\tau \boldsymbol{j} (x,t)\, \rd \tilde{\tau} = \boldsymbol{j} (x,t)\, \tau
\;\;\;\qquad\;\;\;
\int_0^\tau \boldsymbol{k} (x,t)\, \rd \tilde{\tau} = \boldsymbol{k} (x,t)\, \tau \,.
\ee
These linear growths tell us that the modular transport is ballistic (see e.g. \cite{Ilievski:2018esb}).
We remark that the linear response approximation is not employed here.

A vector field $\boldsymbol{v}$ has vanishing curl when its components satisfy $\partial_t v_x - \partial_x v_t = 0$ 
and this feature implies that it can be written 
as the gradient of the potential $\mathsf{W}$, namely $v_\mu = - \,\partial_\mu \mathsf{W}$.

Both the vector fields $\boldsymbol{j} (x,t)$ and $\boldsymbol{k} (x,t)$ 
have vanishing curl and therefore the corresponding potentials 
$\mathsf{W}_j(x,t) $ and $\mathsf{W}_k(x,t) $ respectively can be obtained. 
Indeed, we find that
\be
\label{ham-vec-j}
\langle j_x (\tau;x,t) \rangle_{\mu}  = -\,\partial_x \mathsf{W}_j(x,t) 
\;\;\;\qquad \;\;\;
\langle j_t (\tau;x,t) \rangle_{\mu}   = -\,\partial_t \mathsf{W}_j(x,t) 
\ee
and 
\be
\label{ham-vec-k}
\langle k_x (\tau;x,t) \rangle_{\mu}  = -\,\partial_x \mathsf{W}_k(x,t) 
\;\;\;\qquad \;\;\;
\langle k_t (\tau;x,t) \rangle_{\mu}   = -\,\partial_t \mathsf{W}_k(x,t) 
\ee
where the potentials $\mathsf{W}_j$ and $\mathsf{W}_k$ are defined respectively as
\be
\label{potentials-W-line}
\mathsf{W}_j(x,t) \equiv \frac{\kappa}{2\pi} \big[\mu_+ g(u_+) - \mu_- g(u_-) \big]
\;\;\;\qquad\;\;\;
\mathsf{W}_k(x,t) \equiv \frac{\kappa}{2\pi} \big[\mu_+ g(u_+) + \mu_- g(u_-) \big]
\ee
in terms of 
\be 
\label{potential}
g(u)\equiv
\frac{2\pi}{3(b-a)}
 \left( u - \frac{a+b}{2} \right) 
\left( u^2 - (a+b)\, u - \frac{a^2 -4a b + b^2}{2} \, \right)  .
\ee 
The arbitrary additive constants in (\ref{potentials-W-line})
have been fixed by imposing the vanishing condition at the center of $\mathcal{D}_A$
for both these potentials; indeed (\ref{potential}) has a zero at $u=(a+b)/2$.
We remark that (\ref{mc1x})-(\ref{mc1t}), (\ref{mc10x})-(\ref{mc10t})
are consistent with (\ref{ham-vec-j}) and (\ref{ham-vec-k})  because 
\be
\label{id-der-g-V}
- \,\partial_u g(u) = V(u)  \,.
\ee

Given a curve $\gamma$ (not necessarily closed) parameterised by $s$,
let us denote the line integral of the vector field $ \boldsymbol{v} $ along $\gamma$ 
and the flux of $ \boldsymbol{v} $ through $\gamma$ respectively as
\be
\mathcal{L}[\boldsymbol{v}](\gamma) \equiv \int_\gamma \boldsymbol{v} \cdot \hat{\boldsymbol{\tau}}\, \rd s
\;\;\;\qquad\;\;\;
\mathcal{F}[\boldsymbol{v}](\gamma) \equiv \int_\gamma \boldsymbol{v} \cdot \hat{\boldsymbol{n}}\, \rd s
\ee
where $\hat{\boldsymbol{\tau}}$ and $\hat{\boldsymbol{n}}$ are the unit vectors which are respectively tangent and normal to $\gamma$.

We highlight the vanishing of the fluxes of the vector fields $\boldsymbol{j} (x,t)$ and $\boldsymbol{k} (x,t)$
through the straight white lines in Fig.\,\ref{fig:j-curr} and Fig.\,\ref{fig:k-curr},
which identify the diamond $\mathcal{D}_A$ and the region $\mathcal{R}_A$
(i.e. the grey and light blue regions in Fig.\,\ref{fig:diamond-AB}).
This can be shown by observing that 
the absolute value of the ratio of theirs components 
is equal to one along these lines. 
This absence of flux naturally suggests to consider the total charges in the diamond $\mathcal{D}_A$.
In the finite density representation, by using (\ref{fd1}) and (\ref{heat2}),
for the mean values of (\ref{echarge}) and (\ref{h-charge-def}) we find respectively
\be
\label{QA-tQA-mean}
\langle Q_A \rangle_{\mu} = \,-\frac{\kappa}{4\pi}\,\mu_e \,\ell^2
\;\;\;\qquad\;\;\;
\langle \widetilde{Q}_A \rangle_{\mu} = \,-\frac{\kappa}{4\pi}\,\mu_h \,\ell^2  \,.
\ee

A crucial feature of $\boldsymbol{j} (x,t)$ and $\boldsymbol{k} (x,t)$ is that they are curl free vector fields.
The Green theorem implies that any line integral of a curl free vector field along a closed curve vanishes;
hence, all the line integrals along curves anchored to the same endpoints provide the same result
given by the difference between the values of the potential at these endpoints. 
In our case, we find it worth considering the lines anchored to the opposite vertices of $\mathcal{D}_A$
and, among them, the convenient representatives to choose are
the horizontal segment $A=\{ (x,t=0)\,;\, a\leqslant x \leqslant b \}$ along the real axis,
whose endpoints are $P_a$ and $P_b$,
and the vertical segment  $\tilde{A}= \{ (x=\tfrac{a+b}{2},t)\,;\,  - \tfrac{b-a}{2} \leqslant t \leqslant \tfrac{b-a}{2} \}$ on the imaginary axis,
whose endpoints are $P_{-\infty}$ and $P_{+\infty}$.
Given a smooth curves $\gamma(P_1 \to P_2)$ starting in $P_1$ and ending in $P_2$
and a vector field $\boldsymbol{v} (x,t)$,
let us denote by $\mathcal{L}[\boldsymbol{v}]\big(\gamma(P_1 \to P_2)\big) $
the line integral of $\boldsymbol{v} (x,t)$ along $\gamma(P_1 \to P_2)$
and by $\mathcal{F}[\boldsymbol{v}]\big(\gamma(P_1 \to P_2)\big) $
its flux through $\gamma(P_1 \to P_2)$.
When $\boldsymbol{v} (x,t)$ is curl free, 
$\mathcal{L}[\boldsymbol{v}]\big(\gamma(P_1 \to P_2)\big) $
depends only on the endpoints $P_1$ and $P_2$.
For the vector field $\boldsymbol{j} (x,t)$ in (\ref{mc1x})-(\ref{mc1t}),
these line integrals can be written in terms of the mean values of total charges (\ref{QA-tQA-mean}) as follows
\bea
\label{flow-j-t-A}
\mathcal{L}[\boldsymbol{j}]\big(\gamma(P_a \to P_b)\big) 
%\int_A \langle j_x (\tau;x,0) \rangle_{\mu_\pm}  \rd x 
&=& \mathsf{W}_j\big|_{P_a} - \mathsf{W}_j\big|_{P_b} 
\,=\,
-\frac{2\pi}{3}\, \langle \widetilde{Q}_A \rangle_{\mu}
\\
\label{flow-j-x-At}
\rule{0pt}{.7cm}
\mathcal{L}[\boldsymbol{j}]\big(\gamma(P_{-\infty} \to P_{+\infty})\big) 
% \int_{\tilde{A}} \langle j_t (\tau; \tfrac{a+b}{2} ,t) \rangle_{\mu_\pm}  \rd t
&=& \mathsf{W}_j\big|_{P_{-\infty}}  - \mathsf{W}_j \big|_{P_{+\infty}} 
=\,
-\frac{2\pi}{3}\, \langle Q_A \rangle_{\mu}
\eea
and, similarly, for the vector field $\boldsymbol{k} (x,t)$ in (\ref{mc10x})-(\ref{mc10t})  we have that
\bea
\label{flow-k-t-A}
\mathcal{L}[\boldsymbol{k}]\big(\gamma(P_a \to P_b)\big) 
%\int_A \langle k_x (\tau;x,0) \rangle_{\mu_\pm}  \rd x 
&=& \mathsf{W}_k\big|_{P_a} - \mathsf{W}_k\big|_{P_b} 
\,=\,
-\frac{2\pi}{3}\, \langle Q_A \rangle_{\mu}
\\
\label{flow-k-x-At}
\rule{0pt}{.7cm}
\mathcal{L}[\boldsymbol{k}]\big(\gamma(P_{-\infty} \to P_{+\infty})\big) 
%\int_{\tilde{A}} \langle k_t (\tau; \tfrac{a+b}{2} ,t) \rangle_{\mu_\pm}  \rd t
&=& \mathsf{W}_k\big|_{P_{-\infty}}  - \mathsf{W}_k \big|_{P_{+\infty}} 
=\,
-\frac{2\pi}{3}\, \langle \widetilde{Q}_A \rangle_{\mu}  \,.
\eea
When $\mu_+ = \mu_-$, the line integrals in (\ref{flow-j-t-A}) and (\ref{flow-k-x-At}) vanish,
as one can observe also from the left panel of Fig.\,\ref{fig:j-curr} and Fig.\,\ref{fig:k-curr} respectively. 

We find it worth studying also the fluxes of the vector fields
$\boldsymbol{j} (x,t)$ and $\boldsymbol{k} (x,t)$ through the above mentioned curves.
However, since the divergence of these vector fields does not vanish,
these fluxes depend on the curve.
Among the curves starting in $P_a$ and ending in $P_b$,
let us consider the spacelike curve $\gamma_{t_0}$
given by \eqref{mod-traj-mom-gs} with $u_{\pm}=\frac{a+b}{2}\pm t_0$ and $-\ell /2 \leqslant t_0 \leqslant \ell /2$.
%(see e.g. the dashed lines in the grey diamond in Fig.\,\ref{fig:diamond-AB}).
Analogously, in the class of curves starting in $P_{-\infty}$ and ending in $P_{+\infty}$,
we choose the modular trajectories 
$\gamma_{x_0}$ given by (\ref{mod-traj-tau-line})
with $u_{\pm}=  x_0$ and $a\leqslant x_0 \leqslant b$.
The fluxes of $\boldsymbol{j} (x,t)$ through these curves
can be written in terms of the mean values of total charges (\ref{QA-tQA-mean}) as follows
\bea
\label{F-j-gamma-t0}
\mathcal{F}[\boldsymbol{j}] \big(\gamma_{t_0}(P_a \to P_b)\big) 
&=&
\frac{2\pi}{3}\, \langle Q_A \rangle_{\mu} \, f(2t_0/\ell)
\\
\label{F-j-gamma-x0}
\rule{0pt}{.7cm}
\mathcal{F}[\boldsymbol{j}] \big(\gamma_{x_0}(P_{-\infty} \to P_{+\infty})\big) 
&=&
\frac{2\pi}{3}\, \langle \widetilde{Q}_A \rangle_{\mu} \, f \big(2(x_0 -\tfrac{a+b}{2})/\ell\big)
\eea
and, similarly, for the fluxes of $\boldsymbol{k} (x,t)$ we get
\bea
\label{F-k-gamma-t0}
\mathcal{F}[\boldsymbol{k}] \big(\gamma_{t_0}(P_a \to P_b)\big) 
&=&
\frac{2\pi}{3}\, \langle \widetilde{Q}_A \rangle_{\mu} \,  f(2t_0/\ell)
\\
\label{F-k-gamma-x0}
\rule{0pt}{.7cm}
\mathcal{F}[\boldsymbol{k}] \big(\gamma_{x_0}(P_{-\infty} \to P_{+\infty})\big) 
&=&
\frac{2\pi}{3}\, \langle Q_A \rangle_{\mu} \, f \big(2(x_0 -\tfrac{a+b}{2})/\ell\big)
\eea
where the function $f(y)$ is defined for $|y| < 1$ as
\be
f(y) \equiv  \frac{3 \big(1-y^{2}\big)^{2}}{8\,y^{3}} \left[\left(1+y^{2} \right) \log\!\left(\frac{1+y}{1-y}\right)-2y\,\right]  .
\ee
Notice that this function is even, vanishes as $|y| \to 1$ and takes its maximum value equal to $1$ at $y=0$.
%\\
%{\color{red}
%$\Longrightarrow$ [Is the non trivial modular transport even when $\mu_+ = \mu_{-}$
%(in contrast with the standard case of \cite{Liguori:1999tw, Bernard:2013aru, Hollands:2016svy, Bernard:2016nci, Cappelli:2001mp})
%related to the fact that we have two times?]
%}

Focussing on the case $\mu_+ = \mu_{-}$ considered in the left panel Fig.\,\ref{fig:j-curr} for simplicity,
which has already non trivial modular transport properties,
a heuristic physical picture for the charge transport in the diamond $\mathcal{D}_A$ is the following. 
The critical points $P_{-\infty}$ and $P_{+\infty}$ play the role of a source and a sink respectively:
the charged excitations are emitted by $P_{-\infty}$ and absorbed by $P_{+\infty}$ with vanishing velocity. 
Since the total charge in $\mathcal{D}_A$ is conserved (see (\ref{QA-tQA-mean})),
the amount of charge emitted in $P_{-\infty}$ is equal to the one absorbed in $P_{+\infty}$. 
After the emission, in the lower part of $\mathcal{D}_A$, where $t<0$, 
the charges are accelerated, arriving at the segment $A$ on the $x$-axes with maximal velocity.
The flux through $A$, which is given by (\ref{F-j-gamma-t0}) for $t_0 = 0$, is maximal.  
In the upper part of $\mathcal{D}_A$, where $t>0$, the charges decelerate until they reach $P_{+\infty}$, where they are absorbed. 
When $\mu_+ \neq \mu_-$ the situation is similar, but deformed as shown 
in the right panel of Fig.\,\ref{fig:j-curr} for a specific setup.
In this case the maximum flux should be reached along the curve separating the 
light blue and the orange regions, where the potential vanishes. 

A similar heuristic picture for the helicity transport (see Fig.\,\ref{fig:k-curr})
is obtained  by adapting the above observations properly.
For instance, in this case the vertices $P_a$ and $P_b$ of the diamond $\mathcal{D}_A$
play the role of  the source and the sink respectively
and, when $\mu_{+} = \mu_{-}$, the maximal flux corresponds to $\tilde{A}$.

We find it worth mentioning that 
the charge transport displayed in the left panel of Fig.\,\ref{fig:j-curr} 
resembles the transport of electrons in the vacuum tube called triode in the context of electromagnetism.
In this analogy, $P_{-\infty}$ and $P_{+\infty}$ play the role of the cathode and anode respectively. 
Indeed, the cathode emits electrons, which are accelerated by the electric field in the vacuum. 
The control grid of the triode is represented by the segment $A$ at $t=0$
and its potential is tuned in such a way that the electrons produced by the cathode in $P_{-\infty}$ 
reach the anode in $P_{+\infty}$ with vanishing velocity.

\subsection{Energy and momentum transport} 
\label{sec-energy-momentum-trans}

The analysis discussed in Sec.\,\ref{sec-charge-helicity-trans} can be adapted
to the energy and momentum currents. 

%\noindent
%\textcolor{blue}{\bf $\bullet$ Energy \& Momentum}
%\\
%From (\ref{fd1}) one finds the mean value of the energy density (\ref{eden-E}) 
%\\
%\textcolor{red}{\bf [Should we report the following? NO. It is influenced by $f_\pm$]}
%\be
%\label{mc11a}
%\langle \CE (\tau;x,t) \rangle_{\mu_\pm}  
%\,=\, 
%\frac{\kappa}{4\pi} \,\big[\,\mu^2_+ V(u_+) + \mu^2_- V(u_-) \, \big] 
%\ee

The mean values of the energy currents (\ref{curl-J-x-def})-(\ref{curl-J-t-def}) 
specialised to $V_{+}(u)=V_{-}(u)$ given by (\ref{velocity_fund})
are obtained by employing (\ref{fd1}) and the results read respectively
\bea
\label{mc11x}
\langle \CJ_x (\tau;x,t) \rangle_{\mu}  
&=&
 - \frac{\kappa}{4\pi} \,\big[ \, \mu^2_+ V(u_+)^2 - \mu^2_- V(u_-)^2 \,\big] 
 \\
 \label{mc11t}
 \rule{0pt}{.6cm}
 \langle \CJ_t (\tau;x,t) \rangle_{\mu}  
&=&
 -\frac{\kappa}{4\pi} \,\big[ \, \mu^2_+ V(u_+)^2 + \mu^2_- V(u_-)^2 \,\big]
\eea
where we fixed the constant $C_{\mathcal{J}} = -\,\pi c/6$ to get the last expression.
This constant has been determined 
by observing that for (\ref{velocity_fund}) we have
\be
\label{id-Nu-on-V}
V(u)^2 \,\CV[V](u) \,=\, -\, 2\pi^2
\ee
which implies that the term in the second line of (\ref{curl-J-t-def}) 
drastically simplifies to
\be
\label{identity-V-zero-temp}
- \frac{c}{24\pi}\, \Big\{ V(u_+)^2 \,\CV[V](u_+) + V(u_-)^2\,\CV[V](u_-) \Big\}
=  \frac{\pi\,c}{6}  
\ee
and imposing the vanishing of the final expression at all the vertices of $\mathcal{D}_A$.
The mean values (\ref{mc11x}) and (\ref{mc11t}) provide the components of the planar vector field $\boldsymbol{\CJ} (x,t) $.
\\
Similarly, we introduce the planar vector field $\boldsymbol{\widetilde{\CJ}} (x,t)$ whose components are 
the mean values of the momentum currents  (\ref{tilde-curl-J-x-def}) and (\ref{tilde-curl-J-t-def}),
which are obtained through the same steps\footnote{In this case $C_{\widetilde{\mathcal{J}}} = 0$ since, 
from (\ref{id-Nu-on-V}) we have that $V(u_+)^2 \,\CV[V](u_+) - V(u_-)^2\,\CV[V](u_-)  =  0 $,
which occurs because the central charge is the same for the two chiralities (see (\ref{cft1})). 
} 
and read respectively
\bea
\label{mc11tilde-x}
\langle \widetilde{\CJ}_x (\tau;x,t) \rangle_{\mu}  
&=&
 - \frac{\kappa}{4\pi} \,\big[ \, \mu^2_+ V(u_+)^2 + \mu^2_- V(u_-)^2 \,\big] 
 \\
 \label{mc11tilde-t}
 \rule{0pt}{.6cm}
 \langle \widetilde{\CJ}_t (\tau;x,t) \rangle_{\mu}  
&=&
 -\frac{\kappa}{4\pi} \,\big[ \, \mu^2_+ V(u_+)^2 - \mu^2_- V(u_-)^2 \,\big]  \,.
\eea

\begin{figure}[t!]
\vspace{-.6cm}
\hspace{-1.3cm}
\includegraphics[width=1.2\textwidth]{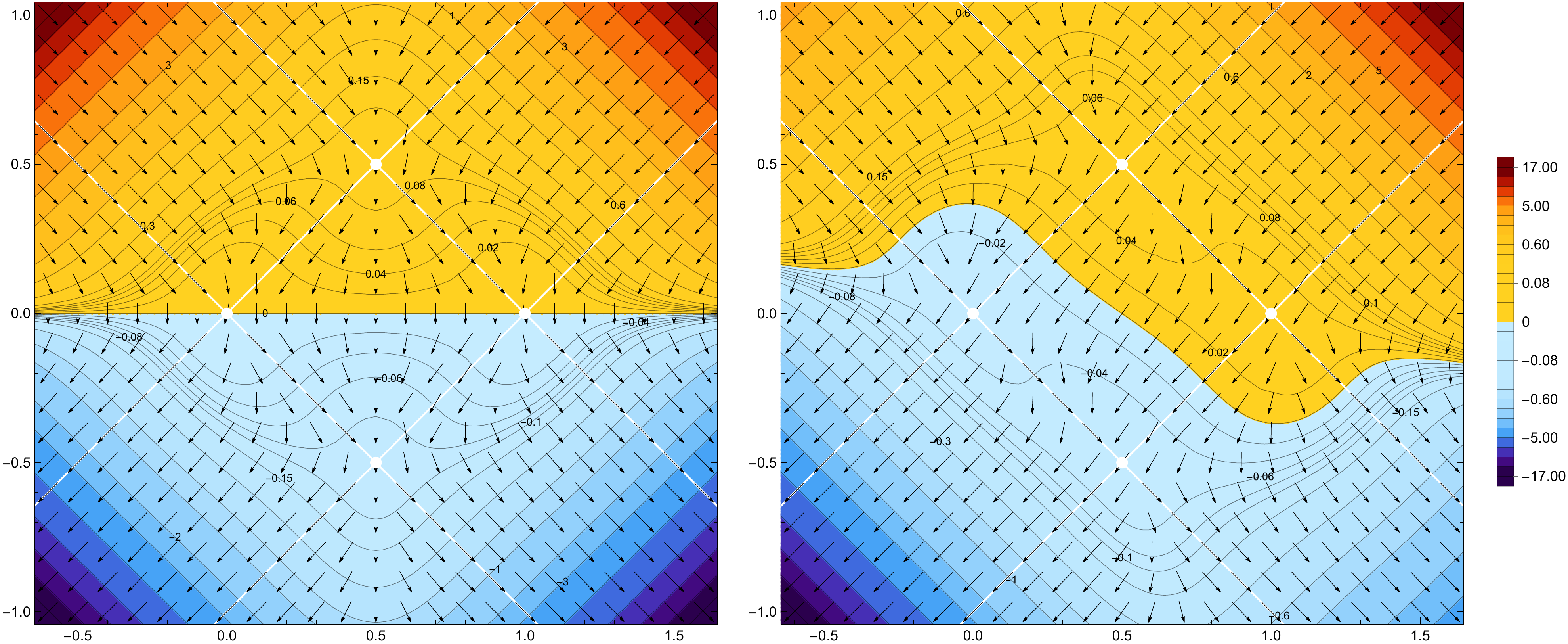}
\vspace{-.4cm}
\caption{Vector fields for the mean values of the energy density currents  (\ref{mc11x}) and (\ref{mc11t}),
whose potential is the first expressions in (\ref{potential-W-energy}),
for either equal (left panel) or different (right panel) chemical potentials,
in the same setup of Fig.\,\ref{fig:j-curr}.
}
\label{fig:E-curr}
\end{figure}

\begin{figure}[t!]
\vspace{-.6cm}
\hspace{-1.3cm}
\includegraphics[width=1.2\textwidth]{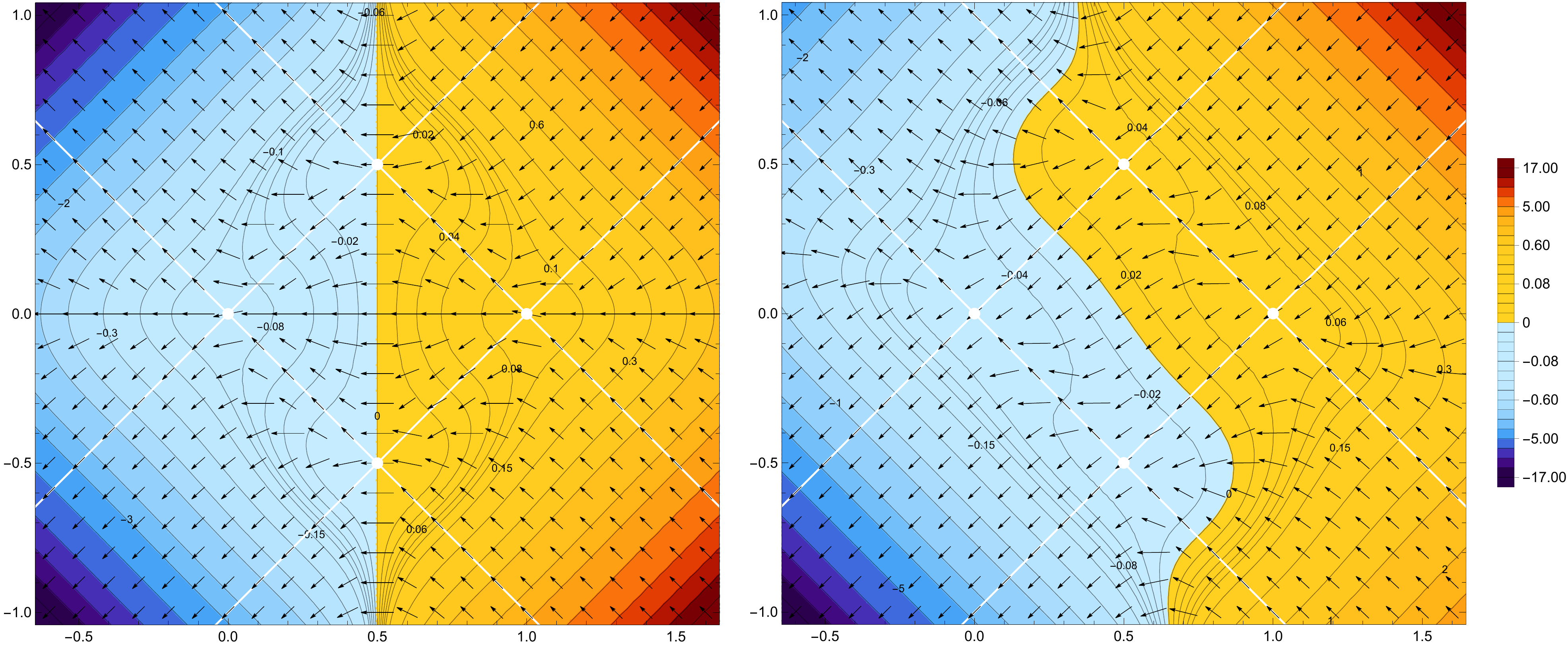}
\vspace{-.4cm}
\caption{Vector fields for the mean values of the momentum density currents  (\ref{mc11tilde-x}) and (\ref{mc11tilde-t}),
whose potential is the second expressions in (\ref{potential-W-energy}),
for either equal (left panel) or different (right panel) chemical potentials,
in the same setup of Fig.\,\ref{fig:j-curr}.
}
\label{fig:tE-curr}
\end{figure}

The planar vector fields $\boldsymbol{\CJ} (x,t) $ and $\boldsymbol{\widetilde{\CJ}} (x,t)$
%whose components are defined by (\ref{mc11x})-(\ref{mc11t}) and (\ref{mc11tilde-x})-(\ref{mc11tilde-t}) respectively,
%have the same critical points corresponding to the vertices of the diamond $\mathcal{D}_A$.
%The planar curl free vector fields defined by the mean values 
%of the energy currents  (\ref{mc11x})-(\ref{mc11t})
%and of the momentum currents  (\ref{mc11tilde-x})-(\ref{mc11tilde-t})
%with the corresponding potentials (\ref{potential-W-energy}),
are displayed in Fig.\,\ref{fig:E-curr} and Fig.\,\ref{fig:tE-curr}
for the choice parameters reported in the caption of Fig.\,\ref{fig:j-curr}.
In particular, $\mu_{+} = \mu_{-}$ in the left panels, 
while $\mu_{+} \neq \mu_{-}$ in the right panels.
These vector fields have the same isolated critical points 
which are given by the vertices of the diamond $\mathcal{D}_A$.
All of them have multiplicity $2$ and Poincar\'e index $0$.
By applying the Poincar\'e-Hopf theorem (see Sec.\,\ref{sec-charge-helicity-trans}),
we find consistency with the fact that 
these vector fields can me mapped (through the stereographic projection)
to vector fields on the Riemann sphere
which have a critical point of index $2$ at the north pole.

The fluxes of the vector fields $\boldsymbol{\CJ} (x,t) $ and $\boldsymbol{\widetilde{\CJ}} (x,t)$
through the straight white lines in Fig.\,\ref{fig:E-curr} and Fig.\,\ref{fig:tE-curr} vanish. 
Indeed, the absolute value of the ratios of their components is equal to 1 along these lines. 
Unfortunately, this analytic result is not properly shown in Fig.\,\ref{fig:E-curr} and Fig.\,\ref{fig:tE-curr}
because of a graphical failure in the displaying of these vector fields around their critical points. 
Another similar failure occurs at the vertices of $\mathcal{D}_A$, where arrows are displayed, 
while they should not because they are critical points of the vector fields. 
These failures could be related to the multiplicity of the critical points;
indeed they do not occur for the vector fields shown in Fig.\,\ref{fig:j-curr} and Fig.\,\ref{fig:k-curr}, 
whose critical points have multiplicity 1.

The local conductivities corresponding to 
(\ref{mc11x})-(\ref{mc11t}) and (\ref{mc11tilde-x})-(\ref{mc11tilde-t}) 
are respectively 
\bea
\boldsymbol{\sigma}_{\CJ \! ,e} 
&\equiv &
\partial_{\mu_e} \boldsymbol{\CJ} (x,t)
=
-\frac{\kappa}{4\pi}\, 
\Big( \mu_+ V(u_+)^2 - \mu_- V(u_-)^2  \, , \, \mu_+ V(u_+)^2 + \mu_- V(u_-)^2 \Big)
\hspace{.8cm}
\\
\rule{0pt}{.7cm}
\boldsymbol{\sigma}_{\CJ\! , h} 
&\equiv &
\partial_{\mu_h}\boldsymbol{\CJ} (x,t)
=
-\frac{\kappa}{4\pi}\, 
\Big( \mu_+ V(u_+)^2 + \mu_- V(u_-)^2  \, , \, \mu_+ V(u_+)^2 - \mu_- V(u_-)^2 \Big)
\eea
and
\be
\boldsymbol{\sigma}_{\widetilde{\CJ} \! ,e} 
\equiv
\partial_{\mu_e} \boldsymbol{\widetilde{\CJ}} (x,t)
=
\boldsymbol{\sigma}_{\CJ \! ,h} 
\;\;\;\qquad\;\;\;
\boldsymbol{\sigma}_{\widetilde{\CJ} \! ,h} 
\equiv 
 \partial_{\mu_h} \boldsymbol{\widetilde{\CJ}} (x,t)
=
\boldsymbol{\sigma}_{\CJ \! ,e} 
\ee
which are independent of $\tau$;
hence their Fourier transform gives only a Dirac delta term. 

The ballistic nature of the modular transport observed through (\ref{ballistic-j-k})
is confirmed by the linear growth in $\tau$ of the following quantities
\be
\int_0^\tau \boldsymbol{\CJ} (x,t)\, \rd \tilde{\tau} = \boldsymbol{\CJ}  (x,t)\, \tau
\;\;\;\qquad\;\;\;
\int_0^\tau \boldsymbol{\widetilde{\CJ}} (x,t)\, \rd \tilde{\tau} = \boldsymbol{\widetilde{\CJ}} (x,t)\, \tau \,.
\ee

Both the vector fields defined by the energy and momentum currents are curl free
and can be written as the gradient some potentials as follows
\be
\label{ham-vec-E}
\langle \mathcal{J}_x (\tau;x,t) \rangle_{\mu}  = -\,\partial_x \mathsf{W}_{\mathcal{J}}(x,t) 
\;\;\;\qquad \;\;\;
\langle \mathcal{J}_t (\tau;x,t) \rangle_{\mu}   = -\,\partial_t \mathsf{W}_{\mathcal{J}}(x,t) 
\ee
and 
\be
\label{ham-vec-E-tilde}
\langle \widetilde{\mathcal{J}}_x (\tau;x,t) \rangle_{\mu}  = -\,\partial_x \mathsf{W}_{\widetilde{\mathcal{J}}}(x,t) 
\;\;\;\qquad \;\;\;
\langle \widetilde{\mathcal{J}}_t (\tau;x,t) \rangle_{\mu}   = -\,\partial_t \mathsf{W}_{\widetilde{\mathcal{J}}}(x,t)   
\ee
For the potentials $\mathsf{W}_{\mathcal{J}}$ and $\mathsf{W}_{\widetilde{\CJ}}$, we find   respectively
\be
\label{potential-W-energy}
\mathsf{W}_{\mathcal{J}}(x,t) \equiv \frac{\kappa}{4\pi} \big[\mu_+^2 G(u_+) - \mu_-^2 G(u_-) \big] 
\;\qquad\;
\mathsf{W}_{\widetilde{\CJ}}(x,t) \equiv \frac{\kappa}{4\pi} \big[\mu_+^2 G(u_+) + \mu_-^2 G(u_-) \big] 
\ee
being $G(u)$ defined as 
\bea
\label{potential-E}
G(u) 
&\equiv &
\frac{2 \pi^2}{15\,(b-a)^2}
\left( u -\frac{a+b}{2} \right)
\Big[\,
6\, u^4
- 12(a+b)\, u^3
+ 4\big(a^2 + 7 a b + b^2\big)\, u^2
\\
& &\hspace{3.5cm}
+\,2 \big(a^3 - 7a^2 b- 7 a b^2+b^3\big)\, u
+ a^4 -6 a^3 b+16 a^2 b^2 - 6 a b^3 + b^4
\,\Big]
\nonumber
\eea
whose additive constant has been fixed by imposing the condition $G(\tfrac{a+b}{2}) =0$.
The expressions  (\ref{mc11x})-(\ref{mc11t}), (\ref{mc11tilde-x})-(\ref{mc11tilde-t}), 
(\ref{ham-vec-E}) and (\ref{ham-vec-E-tilde}) 
are consistent because (\ref{potential-E}) and (\ref{velocity_fund}) are related as follows
\be
\label{id-GE-from-Vsq-line}
\partial_u G(u) = V(u)^2 \,.
\ee

%\noindent
%\textcolor{blue}{\bf $\bullet$ Vector fields and their singular  (critical)  points}
%\textcolor{red}{\bf [TO DO]}
%\\
%{\color{blue}
%$\Longrightarrow$ {\bf Poincar\'e index}: 
%Each vector field has four critical points with multiplicity $2$ with Poincar\'e index $0$.
%\\
%$\Longrightarrow$ \dots
%}
%\\
% The  light-cone coordinates of the singular points $(u_{0,+}, u_{0,-})$ of the vector field
%given by (\ref{mc11a})-(\ref{mc11x}) are 
%\be
%\label{sing-points-E}
%u_{0,\pm } = \frac{a+b}{2} + \eta_1 \,\frac{b-a}{2} \;
%\sqrt{1+ \eta_2\, \frac{2 \,\sqrt{c/\kappa}}{\sqrt{3}\, | \mu_\pm | (b-a)}}
%\;\;\;\qquad\;\;
%\eta_1, \eta_2 \in \{-1\,, +1 \}
%\ee
%hence we can have at most $16$ singular points. 
%\\
%$\longrightarrow$ The four points corresponding to $\eta_2 =+1$ are always real, while whenever $\eta_2 =-1$ we have to impose the reality condition.
%This is obtained by studying the sign of the expression under the square root in (\ref{sing-points-E}).
%The critical case corresponds to the following condition
%\be
%\label{critical-ratio}
%\mu_\pm^2 (b-a)^2 = \frac{4\, c}{3\, \kappa}
%\ee
%\\
%$\longrightarrow$ In the limit $c/\kappa \to 0$ we recover the singular points (\ref{sing-points-jk}), as expected. 
%\\

Finally, since the vector fields $\boldsymbol{\CJ} (x,t) $ and $\boldsymbol{\widetilde{\CJ}} (x,t)$
%defined by the mean values 
%of the energy currents  (\ref{mc11x})-(\ref{mc11t})
%and of the momentum currents  (\ref{mc11tilde-x})-(\ref{mc11tilde-t}) 
are curl free, 
their line integrals along a curve 
depend only on the endpoints of the curve.
Let us consider the curves anchored to the opposite vertices of $\mathcal{D}_A$, 
as done in Sec.\,\ref{sec-charge-helicity-trans}.
By employing (\ref{ham-vec-E}) and (\ref{ham-vec-E-tilde}), we find
\bea
\label{flow-Jx-A}
\mathcal{L}[\boldsymbol{\CJ}]\big(\gamma(P_a \to P_b)\big) 
%\int_A \langle \mathcal{J}_x (\tau;x,0) \rangle_{\mu_\pm}  \rd x
&=&
\mathsf{W}_{\mathcal{J}}\big|_{P_a} - \mathsf{W}_{\mathcal{J}}\big|_{P_b} 
\,=\,
-\frac{4\pi}{5}\, \widetilde{ \mathsf{E}}_A
%- \frac{\pi \kappa \big(\mu_{+}^2 - \mu_{-}^2\big)}{30}\; \ell^3
\hspace{1.4cm}
\\
\label{flow-Jt-At}
\rule{0pt}{.6cm}
\mathcal{L}[\boldsymbol{\CJ}]\big(\gamma(P_{-\infty} \to P_{+\infty})\big) 
%\int_{\tilde{A}} \langle \mathcal{J}_t (\tau; \tfrac{a+b}{2} ,t) \rangle_{\mu_\pm}  \rd t
&=&
\mathsf{W}_{\mathcal{J}}\big|_{P_{-\infty}} - \mathsf{W}_{\mathcal{J}}\big|_{P_{+\infty}} 
\,=\,
-\frac{4\pi}{5}\,  \mathsf{E}_A
%-\frac{\pi \kappa \big(\mu_{+}^2 + \mu_{-}^2\big)}{30}\; \ell^3
\hspace{1cm}
\eea
and
\bea
\label{flow-Jx-A-tilde}
\mathcal{L}[\boldsymbol{\widetilde{\CJ}} ]\big(\gamma(P_a \to P_b)\big) 
%\int_A \langle \widetilde{\CJ}_x (\tau;x,0) \rangle_{\mu_\pm}  \rd x
&=&
\mathsf{W}_{\widetilde{\CJ}}\big|_{P_a} - \mathsf{W}_{\widetilde{\CJ}}\big|_{P_b} 
\,=\,
-\frac{4\pi}{5}\, \mathsf{E}_A
%- \frac{\pi \kappa \big(\mu_{+}^2 + \mu_{-}^2\big)}{30}\; \ell^3
\hspace{1.4cm}
\\
\label{flow-Jt-At-tilde}
\rule{0pt}{.6cm}
\mathcal{L}[\boldsymbol{\widetilde{\CJ}}]\big(\gamma(P_{-\infty} \to P_{+\infty})\big) 
%\int_{\tilde{A}} \langle \widetilde{\CJ}_t (\tau; \tfrac{a+b}{2} ,t) \rangle_{\mu_\pm}  \rd t
&=&
\mathsf{W}_{\widetilde{\CJ}}\big|_{P_{-\infty}} - \mathsf{W}_{\widetilde{\CJ}}\big|_{P_{+\infty}} 
\,=\,
-\frac{4\pi}{5}\, \widetilde{ \mathsf{E}}_A
%-\frac{\pi \kappa \big(\mu_{+}^2 - \mu_{-}^2\big)}{30}\; \ell^3
\hspace{1cm}
\eea
where we have introduced 
\be
\label{EA-tEA-mean}
 \mathsf{E}_A  \equiv  \frac{\kappa}{24} \big( \mu_{+}^2 + \mu_{-}^2\big) \,\ell^3
\;\;\;\qquad\;\;\;
\widetilde{ \mathsf{E}}_A  \equiv \frac{\kappa}{24} \big( \mu_{+}^2 - \mu_{-}^2\big) \,\ell^3  
\ee
which correspond respectively to 
the mean value of the total energy (\ref{toten}) and of the total momentum (\ref{totmom}) 
in the diamond $\mathcal{D}_A$ 
in the case where $f_{+}(u_+) = f_{-}(u_-) = 0$ are imposed.
Instead, from (\ref{KA-def-u-pm}) we have that 
the mean values of (\ref{toten}) and (\ref{totmom})  are $\langle E_A \rangle_{\mu} = \langle \widetilde{E}_A \rangle_{\mu} = 0$.

The vanishing of the line integrals in  (\ref{flow-Jt-At}) and (\ref{flow-Jt-At-tilde}) for $\mu_+ = \mu_-$
can be observed also from the left panel of Fig.\,\ref{fig:E-curr} and Fig.\,\ref{fig:tE-curr} respectively.

We find it worth considering also the fluxes of $\boldsymbol{\CJ} (x,t)$ and $\boldsymbol{\widetilde{\CJ}} (x,t)$
through  the curves $\gamma_{t_0}$ and $\gamma_{x_0}$ introduced in Sec.\,\ref{sec-charge-helicity-trans}.
They can be written as
\bea
\label{F-E-gamma-t0}
\mathcal{F}[\boldsymbol{\CJ}] \big(\gamma_{t_0}(P_a \to P_b)\big) 
&=&
-\frac{4\pi}{5} \, \mathsf{E}_A   \, F(2t_0/\ell)
\\
\label{F-E-gamma-x0}
\rule{0pt}{.6cm}
\mathcal{F}[\boldsymbol{\CJ}] \big(\gamma_{x_0}(P_{-\infty} \to P_{+\infty})\big) 
&=&
-\frac{4\pi}{5} \, \widetilde{ \mathsf{E}}_A\, F\big(2(x_0 -\tfrac{a+b}{2})/\ell\big)
\eea
and 
\bea
\label{F-tE-gamma-t0}
\mathcal{F}[\boldsymbol{\widetilde{\CJ}}] \big(\gamma_{t_0}(P_a \to P_b)\big) 
&=&
-\frac{4\pi}{5} \, \widetilde{ \mathsf{E}}_A\, F(2t_0/\ell)
\\
\label{F-tE-gamma-x0}
\rule{0pt}{.6cm}
\mathcal{F}[\boldsymbol{\widetilde{\CJ}}] \big(\gamma_{x_0}(P_{-\infty} \to P_{+\infty})\big) 
&=&
-\frac{4\pi}{5} \, \mathsf{E}_A  \, F\big(2(x_0 -\tfrac{a+b}{2})/\ell\big)
\eea
in terms of the mean values of total charges (\ref{EA-tEA-mean}),
where $F(y) $ is defined  for $|y| < 1$ as follows
\be
\label{cal-F-def}
F(y) \equiv  \frac{5\left(1-y^{2}\right)^{2}}{64\,y^{5}}
\left[\,
2y\big(3y^{4}-2y^{2}+3\big)
-3\left(1-y^{2}\right)^{2}\left(1+y^{2}\right)\log\!\left(\frac{1+y}{1-y}\right)
\right] .
\ee
This function is even, vanishes as $|y| \to 1$ and takes its maximum value equal to $1$ at $y=0$.

%\DP{We have
%\be
%\int_{\gamma_{t}}\bar{\mathcal{J}}\cdot\hat{n}\,d\gamma_{t} 
%=  
%-\frac{4\pi}{5}\, \langle E_A \rangle_{\mu_\pm}
%%-\frac{\pi\ell^{3}}{30}\kappa(\text{\ensuremath{\mu}}_{+}^{2}+\text{\ensuremath{\mu}}_{-}^{2}) 
%\,\mathcal{F}(2t/\ell)
%\ee
%where the function $\mathcal{F}(y) $ is defined  for $|y| < 1$ as
%\be
%\mathcal{F}(y) \equiv  \frac{5\left(1-y^{2}\right)^{2}}{16y^{5}}
%\left[\,
%2y\big(3y^{4}-2y^{2}+3\big)
%-3\left(1-y^{2}\right)^{2}\left(1+y^{2}\right)\log\!\left(\frac{1+y}{1-y}\right)
%\right]
%\ee
%where $\gamma_{t}$ is the spacelike curve attached to $a$ and $b$
%given by equation \eqref{mod-traj-lambda} with $u_{\pm}=\frac{b+a}{2}\pm t$, and $-\ell/2\leqslant t\leqslant\ell/2$.
%}

\subsection{Quantum noise} 
\label{subsec-noise}

Quantum effects along the conventional temporal evolution of any observable $\O$ 
generate non-trivial fluctuations around its mean value $\langle \O \rangle$. 
This is the case for the modular evolution as well. 
For instance, the quadratic fluctuations of the current $j_x(\tau;x,t)$ 
describe the quantum noise produced by the transport of charged particles. 
It is known \cite{Landauer1998, Kane94} that, 
besides spoiling the charge propagation and detection, 
the noise carries also useful information because 
it provides the experimental basis of noise spectroscopy. 

The basic quantity of interest is the (modular) noise power 
generated by the charge current (\ref{ecurr})
at frequency $\omega$ in the point $(x,t)$ of the spacetime,
defined as follows
\be 
P_j(\omega; x,t) 
\,\equiv\, 
- \int_{-\infty}^\infty  
\langle j_x(\tau_1;x,t) \,j_x(\tau_2;x,t)\rangle_{\mu}^{\textrm{\tiny con}}   
\; \e^{\ri \omega \tau_{12} }\,\rd \tau_{12}  \,.
%\;\;\qquad \;\;
%\tau_{12} \equiv \tau_1-\tau_2 
\label{no1}
\ee
Since $\tau_{12}$ is dimensionless, the corresponding frequency $\omega$ is dimensionless as well. 
Notice that the noise power is generated only by $j_x$ 
because $\langle j_t(\tau_1;x,t) \,j_t(\tau_2;x,t)\rangle_{\mu}^{\textrm{\tiny con}} $ vanishes identically.
Indeed, $j_t$ introduced in (\ref{j-t-def})
is proportional to the identity operator, being generated only by the 
central extension $\kappa$ in the r.h.s. of (\ref{cft7}). 

Considering (\ref{ecurr-mu}), 
%which has vanishing mean value, 
we remark that $\langle j_x(\tau_1;x,t) \,j_x(\tau_2;x,t)\rangle_{\mu}^{\textrm{\tiny con}}   =
\langle \hat{j}_x(\tau_1;x,t) \, \hat{j}_x(\tau_2;x,t)\rangle_{\mu}^{\textrm{\tiny con}}   $;
indeed $j_x(\tau;x,t)$ and $\hat{j}_x(\tau;x,t)$ differ by a real function.

By using (\ref{ecurr}), the modular noise power can be  expressed in terms of 
the two-point functions of the chiral currents $j_\pm(\tau, u)$
%\\
%\textcolor{red}{
%{\bf [missing term $V_{+} V_{-}$ below? Same issue in (\ref{no8}) and (\ref{no11}).}
%No because of (\ref{j-2point-mix-app})]}
%\\
%\textcolor{red}{ [ {\bf do we have a Drude weight term?} No because of (\ref{j-2point-mix-app})]}
%\\
%\textcolor{red}{ \bf [why Moosavi finds a non trivial Drude weight in (4.11a) of \cite{Moosavi:2019fas}?)]}
\bea
\label{no3}
P_j(\omega;x,t) 
&=&
-\int_{-\infty}^\infty  
\Big[\,
V(u_+)^2 \,\langle j_+(\tau_1,u_+) \, j_+(\tau_2,u_+)\rangle_{\mu_+}^{\textrm{\tiny con}} 
\\
& & \hspace{1.5cm}
 + \, V(u_-)^2 \,\langle j_-(\tau_1,u_-)\, j_-(\tau_2,u_-)\rangle_{\mu_-}^{\textrm{\tiny con}}  
\,\Big]
\e^{\ri \omega \tau_{12} }\,\rd \tau_{12} \,.
\nonumber
\eea
By employing (\ref{mod-corr-j-mu}) and the limit given in (\ref{cap-W-equal-point}), 
we have that the dependence on $V(u_+)$ and $V(u_-)$ drops out and 
therefore (\ref{no3}) becomes
(see (\ref{app-noise-int-current}) for the evaluation of the integral)
\bea
P_j(\omega;x,t) 
&=&
 -\frac{\kappa}{4} \,\lim_{\varepsilon \to 0}\,
  \int_{-\infty}^\infty  
\left [ \,
\frac{1}{\sinh^2 (\pi \tau  - \ri \varepsilon )} 
+ 
\frac{1}{\sinh^2 (- \pi \tau + \ri \varepsilon )}  
\,\right ] 
\e^{\ri \omega \tau } \,  \rd \tau 
\nn
\\
\label{no4}
  \rule{0pt}{.7cm}
&=&
 -\frac{\kappa}{2} \,\lim_{\varepsilon \to 0}\,
  \int_{-\infty}^\infty  
\frac{\e^{\ri \omega \tau }}{\sinh^2 (\pi \tau  - \ri \varepsilon )} 
 \,  \rd \tau 
\,=\,
\frac{\kappa}{2\pi}
 \; \omega \coth (\omega / 2) 
 + \frac{\kappa}{2\pi}\, \omega
\eea
which is indepedent of $x$ and $t$.
The zero frequency limit of (\ref{no4}) reads
\be 
P_j(0;x,t) = \lim_{\omega \to 0} P_j(\omega; x,t ) = \frac{\kappa}{\pi}   \;.
\label{no5}
\ee

It is worth comparing this result with the Johnson-Nyquist law \cite{Johnson1928, Nyquist1928}.
In a CFT on the line at finite inverse temperature $\beta$ and vanishing chemical potential,
the two-point functions of the chiral currents read
\be 
\langle j_\pm(u) \,j_\pm (v)\rangle_{_{\textrm{\tiny $\beta$}}}^{\textrm{\tiny con}}   
=
\frac{\kappa}{4\pi^2} 
\bigg(
\frac{\pi }{\beta\,\sinh \! \big[ \pi (u-v \mp \ri \varepsilon)/\beta\big] }
\bigg)^{2} .
\label{ftd4}
\ee
Since $\varrho(x,t) = j_{+}(u_+) + j_{-}(u_{-})$, $j_{x}(x,t) = j_{+}(u_+)- j_{-}(u_{-})$ 
and $\langle j_{\pm}(u) \,j_{\mp} (v)\rangle_{_{\textrm{\tiny $\beta$}}}^{\textrm{\tiny con}}   =0$,
by using (\ref{app-noise-int-current}) 
 the noise power $P_{\textrm{\tiny JN}}(\omega )  $
at frequency $\omega$ in the point $x$ of the space 
for $\varrho$ and $j_{x}$ coincide and read
\bea
\label{P-JN-current}
P_{\textrm{\tiny JN}}(\omega ) 
&\equiv &
- \int_{-\infty}^\infty  \!
\langle \varrho(x,t_1) \, \varrho(x,t_2)\rangle_{\beta}^{\textrm{\tiny con}}   
\; \e^{\ri \omega t_{12} } \, \rd t_{12}
\,=\,
- \int_{-\infty}^\infty  \!
\langle j_{x}(x,t_1) \, j_{x}(x,t_2)\rangle_{\beta}^{\textrm{\tiny con}}   
\; \e^{\ri \omega t_{12} } \, \rd t_{12}
\nn
\\
\label{P-JN-current-exp}
  \rule{0pt}{.5cm}
&=&
- \frac{\kappa}{4 \, \beta^2} \, 
\lim_{\varepsilon \to 0} \,
\int_{-\infty}^\infty 
\bigg( \,
\frac{1}{\sinh^2 \! \big[ \pi ( t - \ri \varepsilon)/\beta\big] } 
+ 
\frac{1}{\sinh^2 \! \big[ \pi ( -t + \ri \varepsilon)/\beta\big] } 
\,\bigg)\,
  \e^{\ri \omega t } \, \rd t
  \nn
  \hspace{1cm}
  \\
  \rule{0pt}{.8cm}
\label{P-JN-current-ris}
&=&
- \frac{\kappa}{2 \, \beta^2} \, 
\lim_{\varepsilon \to 0} \,
\int_{-\infty}^\infty 
\frac{\e^{\ri \omega t }}{\sinh^2 \! \big[ \pi ( t - \ri \varepsilon)/\beta\big] } 
   \, \rd t
\,=\,
\frac{\kappa}{2\pi}
 \; \omega \coth (\beta\omega / 2) 
 + \frac{\kappa}{2\pi}\, \omega
\eea
which is independent of $x$ because of the translation invariance of the CFT.
The zero frequency limit of (\ref{P-JN-current-ris}) gives the Johnson-Nyquist law
\be 
P_{\textrm{\tiny JN}}(0) = \frac{\kappa}{\pi \beta}   
\label{no6}
\ee
Comparing (\ref{no4}) with (\ref{P-JN-current-ris})
and also the corresponding zero frequency limits given by (\ref{no5}) with (\ref{no6})
confirms that the modular evolution has a thermal 
character with inverse temperature $\tilde \beta =1$, which appears in the KMS condition (\ref{KMS}) 
satisfied by the modular correlation functions (\ref{mod-corr-phi-mu}), (\ref{mod-corr-phi-mu-2}),
(\ref{mod-corr-j-mu}) and (\ref{mod-corr-emtensor-mu}). 
We remark that the noise power (\ref{no5}) provides a physical observable \cite{Landauer1998, Kane94}
where in principle the modular temperature can be measured.

We can also introduce the modular noise power $P_k(\omega; x,t) $
at frequency $\omega$ in the point $(x,t)$ of the spacetime
generated by the helicity current (\ref{k-xt-def}),
defined by the r.h.s. (\ref{no1}) with $j_x$ replaced by $k_x$.
From the explicit expression of 
$\langle k_x(\tau_1;x,t) \,k_x(\tau_2;x,t)\rangle_{\mu}^{\textrm{\tiny con}} $
and (\ref{j-2point-app})-(\ref{j-2point-mix-app}),
it is straightforward to obtain the r.h.s. of (\ref{no3}) for $P_k(\omega; x,t) $;
hence $P_j(\omega;x,t) = P_k(\omega;x,t) $.
Indeed, the differences due to the diverse relative signs in (\ref{ecurr}) and (\ref{hcurr}) 
do not lead to a relevant result in this computation 
because the mixed connected correlators vanish.

Similarly, it is worth investigating 
the noise generated by the energy current (\ref{curl-J-x-def}), namely
\be 
{\cal P_J}(\omega; x,t ) \equiv 
\int_{-\infty}^\infty  \!
\langle \CJ_x(\tau_1; x,t) \, \CJ_x(\tau_2; x,t)\rangle_{\mu}^{\textrm{\tiny con}}   
\; \e^{\ri \omega \tau_{12} } \, \rd \tau_{12} \,.
\label{no7}
\ee
From (\ref{mod-corr-emtensor-mu}) and the identity (\ref{cap-W-equal-point}),
one finds (see (\ref{app-noise-int-energy}) for the evaluation of the integral)
\bea
\label{no8}
{\cal P_J}(\omega; x,t ) 
&=&
\frac{\pi^2 c}{8}\, \lim_{\varepsilon \to 0} \,
\int_{-\infty}^\infty 
\left [ \,\frac{1}{\sinh^4 (\pi \tau  - \ri \varepsilon )} 
+ 
\frac{1}{\sinh^4 (-\pi \tau + \ri \varepsilon )}  
\,\right ] 
  \e^{\ri \omega \tau } \, \rd \tau
 \\
 \label{no8-bis}
 \rule{0pt}{.7cm}
&=&
\frac{\pi^2 c}{4}\, \lim_{\varepsilon \to 0} \,
\int_{-\infty}^\infty 
\frac{ \e^{\ri \omega \tau }}{\sinh^4 (\pi \tau  - \ri \varepsilon )} 
  \, \rd \tau
\,=
 \frac{c}{24\pi } \, \big( 4 \pi^2 + \omega^2\big) \,\big[ \, \omega \coth (\omega / 2) +\omega\, \big]
 \hspace{1cm}
\eea
which is indepedent of $x$ and $t$,
and whose zero frequency limit gives
\be 
{\cal P_J}(0 ; x,t ) = \lim_{\omega \to 0} {\cal P_J}(\omega; x,t  ) = \frac{ \pi  \,c}{3}   \;.
\label{no9}
\ee

It is worth comparing these results based on the modular evolution with the 
corresponding ones based on the standard temporal evolution. 
In a CFT on the line at finite temperature and vanishing chemical potential, we have 
\be 
\langle T_\pm(u) \,T_\pm (v)\rangle_{_{\textrm{\tiny $\beta$}}}^{\textrm{\tiny con}}   
= 
\frac{c}{8\pi^2} 
\bigg(\frac{\pi }{\beta \,\sinh \! \big[ \pi (u-v \mp \ri \varepsilon)/\beta\big] }\bigg)^{4}  .
%+ \frac{\kappa\mu_\pm^2}{4\pi^2} \left [\frac{\pi }{\beta_\pm \sinh \left (\frac{\pi}{\beta_\pm }(x-y \mp \ri \varepsilon)\right ) }\right ]^2 
\label{ftd3}
\ee
Since $T_{tt}(x,t) = T_{+}(u_+) + T_{-}(u_{-})$, $T_{xt}(x,t) = T_{+}(u_+)- T_{-}(u_{-})$ 
and $\langle T_\pm(u) \,T_\mp (v)\rangle_{_{\textrm{\tiny $\beta$}}}^{\textrm{\tiny con}}   =0$,
the noise power $\mathcal{P}_{\textrm{\tiny JN}}(\omega )  $
at frequency $\omega$ in the point $x$ of the space
of $T_{tt}$ and $T_{xt}$ coincide. 
It reads
\bea
\label{P-JN-energy}
\mathcal{P}_{\textrm{\tiny JN}}(\omega ) 
&\equiv &
 \int_{-\infty}^\infty  \!
\langle T_{xt}(x,t_1) \, T_{xt}(x,t_2)\rangle_{\beta}^{\textrm{\tiny con}}   
\; \e^{\ri \omega t_{12} } \, \rd t_{12}
\,=\,
 \int_{-\infty}^\infty  \!
\langle T_{tt}(x,t_1) \, T_{tt}(x,t_2)\rangle_{\beta}^{\textrm{\tiny con}}   
\; \e^{\ri \omega t_{12} } \, \rd t_{12}
\nn
\\
\label{P-JN-energy-exp}
  \rule{0pt}{.6cm}
&=&
 \frac{\pi^2 c}{8 \, \beta^4} \, 
\lim_{\varepsilon \to 0} \,
\int_{-\infty}^\infty 
\bigg( \,
\frac{1}{\sinh^4 \! \big[ \pi ( t - \ri \varepsilon)/\beta\big] } 
+ 
\frac{1}{\sinh^4 \! \big[ \pi ( -t + \ri \varepsilon)/\beta\big] } 
\,\bigg)
\,  \e^{\ri \omega t } \, \rd t
  \hspace{1cm}
  \\
  \label{P-JN-energy-ris0}
  \rule{0pt}{.8cm}
&=&
 \frac{\pi^2 c}{4 \, \beta^4} \, 
\lim_{\varepsilon \to 0} \,
\int_{-\infty}^\infty 
\frac{\e^{\ri \omega t } }{\sinh^4 \! \big[ \pi ( t - \ri \varepsilon)/\beta\big] } 
\, \rd t
  \\
  \rule{0pt}{.6cm}
\label{P-JN-energy-ris}
&=&
 \frac{c}{24\pi \, \beta^2} \, \big[\,4 \pi^2 + (\beta \omega)^2\,\big] \,\big[ \, \omega \coth (\beta\omega / 2) +\omega\, \big]
\eea
where (\ref{app-noise-int-energy}) has been employed and whose zero frequency limit is
\be 
\mathcal{P}_{\textrm{\tiny JN}}(0 ) = \frac{\pi \,c}{3 \,\beta^3}   \,.
\ee
Comparing (\ref{P-JN-energy-ris}) with the modular noise (\ref{no8-bis})
provides another consistency check 
for the fact that the modular evolution has a thermal 
character with inverse temperature $\tilde \beta =1$.

Thus, while $P_j(0; x,t)$ provides the coefficient $\kappa$ of the central term in (\ref{cft7}) through (\ref{no5}),
${\cal P_J}(0; x,t) $ gives the central charge $c$  through (\ref{no9}). 
Notice that in the derivation of (\ref{no8-bis}), the expression (\ref{identity-mod-corr-zero-temp}) has been employed, 
which holds for the specific velocity given in (\ref{velocity_fund}).
%

%\noindent
%\textcolor{red}{\bf [A $C$-theorem from (\ref{no8-bis}) or (\ref{no9})?]}
%\\
%\textcolor{red}{ [$\Longrightarrow$ what is $\mathcal{J}$ outside the critical point? ]}
%\\

The noise ${\cal P}_{\widetilde{\CJ}}(\omega; x,t )$ generated by the momentum current (\ref{tilde-curl-J-x-def})
is defined by the r.h.s. of (\ref{no7}) with $\CJ_x$ replaced by $\widetilde{\CJ}_x$.
By adapting the observations made above to get $P_j(\omega;x,t) = P_k(\omega;x,t) $
and using (\ref{T-2point-app})-(\ref{T-2point-mix-app}),
one finds that ${\cal P}_{\widetilde{\CJ}}(\omega; x,t )={\cal P}_{\CJ}(\omega; x,t )$.

The previous analysis shows that 
the modular noise power generated by the charge and energy currents 
is uniform in space and time, 
despite the fact that the translation invariance is broken by the bipartition of the system. 
This peculiar feature does not hold for the quadratic fluctuations of a generic observable. 
Indeed, consider for instance the modular noise power relative to the charge density (\ref{eden}), 
namely
\be
\label{no10}
P_\varrho(\omega;x,t) 
\,\equiv\, 
- \int_{-\infty}^\infty  \!
\langle \varrho (\tau_1;x,t) \, \varrho(\tau_2;x,t)\rangle_{\mu}^{\textrm{\tiny con}}   
\; \e^{\ri \omega \tau_{12} }\,\rd \tau_{12} \,.
\ee
By using (\ref{mod-corr-j-mu}), (\ref{cap-W-equal-point}) and (\ref{app-noise-int-current}), we find
\bea 
\label{no11}
P_\varrho (\omega;x,t) 
&=&
 -\frac{\kappa}{4} \,\lim_{\varepsilon \to 0}\,
  \int_{-\infty}^\infty  
\left [ \,
\frac{1}{V(u_+)^2 \,\sinh^2 (\pi \tau  - \ri \varepsilon )} 
+ \frac{1}{V(u_-)^2 \,\sinh^2 (-\pi \tau + \ri \varepsilon )}  
\,\right ] 
\e^{\ri \omega \tau } \,  \rd \tau  
\nn
\\
\rule{0pt}{.7cm}
&=&
 -\frac{\kappa}{4} 
 \left [ \frac{1}{V^2(u_+)} + \frac{1}{V^2(u_-)} \right ] 
 \lim_{\varepsilon \to 0}\,
  \int_{-\infty}^\infty  
\frac{\e^{\ri \omega \tau }}{\sinh^2 (\pi \tau  - \ri \varepsilon )} 
 \,  \rd \tau  
 \nn
 \\
 \rule{0pt}{.7cm}
\label{no12}
&=&
 \frac{\kappa}{4 \pi } 
 \left [ \,\frac{1}{V^2(u_+)} + \frac{1}{V^2(u_-)} \,\right ] 
 \big[ \, \omega \coth (\omega / 2)  + \omega \, \big]
\eea
that  depends on the frequency and on the position in spacetime. 
The zero frequency limit of (\ref{no12}) gives
\be
P_\varrho (0;x,t) = 
\frac{\kappa}{2\pi}  \left [ \frac{1}{V(u_+)^2} + \frac{1}{V(u_-)^2} \right ]
\label{no13}
\ee
which is qualitatively different from (\ref{no5}) because of 
the occurrence of a non trivial dependence on the spacetime position.
Notice that, setting $V(u) =1$ identically in (\ref{no13})
one recovers (\ref{no6}) in the special case of $\beta =1$.

We can introduce also the noise $P_\chi (\omega;x,t) $ generated by $\chi$ in (\ref{hden})
as the Fourier transform in $\tau_{12}$ of the connected modular two-point function of $\chi$ at coincident points,
as done in (\ref{no10}) for $P_\varrho (\omega;x,t)$.
Comparing this computation with the one reported above for $P_\varrho (\omega;x,t)$,
we observe again that the differences due to the different relative sign in (\ref{eden}) and (\ref{hden})
do not play any role because the mixed connected correlators vanish;
hence $P_\chi (\omega;x,t) = P_\varrho (\omega;x,t)$.

%\textcolor{red}{\bf [adapt the above computations to the PRIMARIES (new appendix?)]}

We can perform a non trivial consistency check of the above results by considering 
the Fourier transform of the anticommutator
\be
\label{anti-comm-modular-def}
\mathcal{A}[\mathcal{O}](\omega; x,t)
\,\equiv\,
\frac{1}{2}
\int_{-\infty}^\infty  
 \langle \,\big\{ \, \mathcal{O}(\tau_1;x,t) \,,  \mathcal{O}(\tau_2;x,t)\, \big\} \,\rangle_{\mu}^{\textrm{\tiny con}}   
\, \e^{\ri \omega \tau_{12} }\,\rd \tau_{12}  
\ee
and the Fourier transform of the commutator
\be
\label{comm-modular-def}
\rule{0pt}{.7cm}
\mathcal{C}[\mathcal{O}](\omega; x,t)
\,\equiv\,
\frac{1}{2}
\int_{-\infty}^\infty  
 \langle \,\big[ \, \mathcal{O}(\tau_1;x,t) \,, \mathcal{O}(\tau_2;x,t)\, \big] \,\rangle_{\mu}^{\textrm{\tiny con}}   
\, \e^{\ri \omega \tau_{12} }\,\rd \tau_{12}  
\ee
for the modular correlators of the operators $\mathcal{O} \in \big\{ j_x, k_x, \CJ_x, \widetilde{\CJ}_x, \rho, \chi \big\}$
considered above. 
Since the Fourier transform $F(\omega) \equiv \int_{-\infty}^{+\infty} f(t) \, \e^{\ri \omega t} \rd t$
of a generic function $f(t)$
satisfies the property
 $\tfrac{1}{2}\int_{-\infty}^{+\infty} \! \big[ f(t) \pm f(-t)\big]  \e^{\ri \omega t} \rd t = \tfrac{1}{2}\big[ F(\omega) \pm F(-\omega)\big] $,
by employing (\ref{no4}), (\ref{no8-bis}) and (\ref{no12}),
for (\ref{anti-comm-modular-def}) and (\ref{comm-modular-def})
we find that the following modular fluctuation-dissipation relation
\be
\mathcal{A}[\mathcal{O}](\omega; x,t)
\,=\,
\coth \! \left( \frac{\omega}{2}\right)\,
\mathcal{C}[\mathcal{O}](\omega; x,t)
\ee
which corresponds to the fluctuation-dissipation relation 
\cite{Kubo_1966, Kubo-book, MARCONI2008}
with inverse temperature given by $\tilde \beta =1$.
In the case of two-dimensional CFT, 
this result further confirms the thermal nature of the modular evolution with inverse temperature $\tilde \beta =1$,
in agreement with the KMS condition discussed in Sec.\,\ref{sec-mod-corr-line}.

%%%%%%%%%%%%%%%%%%%%%%%%%%%%%%%%%%%%%%%%%%
%\newpage 
\section {Finite volume} 
\label{sec-finite-volume}
%%%%%%%%%%%%%%%%%%%%%%%%%%%%%%%%%%%%%%%%%%

In this section the analyses of  Sec.\,\ref{sec-line} are extended
to a two-dimensional CFT at finite density and finite volume
by compactifying each  chiral direction on the circle of length $L$,
hence the resulting spacetime $\mathbb{M}_{\diamond}$ has the topology of the torus. 
%
%\textcolor{red}{[{\bf is $\mathbb{M}_{\diamond}=\overline {\mathbb M}$?} No, because $\overline {\mathbb M}$ comes from the unit circle, hence it does not contain $L$]}
In Sec.\,\ref{sec-rep-finite-density-circle} the relevant chiral correlators are discussed.
The modular Hamiltonian associated 
to the bipartition of each chiral direction provided by the interval $A=[a,b]$
and the corresponding modular correlators
are explored in Sec.\,\ref{sec-mod-ham-circle} and Sec.\,\ref{sec-mod-corr-circle} respectively.

%%%%%%%%%%%%%%%%%%%%%%%%%%%%%%%%%%%%%%%%%%
\subsection {Finite density representation on the circle} 
\label{sec-rep-finite-density-circle}

The finite density and finite volume representation of a CFT on a circle of length $L$
can be obtained from its ground state representation in the infinite line,
by employing first the conformal transformation $u \mapsto \e^{2\pi \ri u/L}$ 
and then the automorphism described in the Appendix\;\ref{app-reps}.

In this representation, the one-point functions are obtained
by first applying the conformal transformation $u \mapsto \e^{2\pi \ri u/L}$
to the one-point functions in the ground state and on the line,
which are given by (\ref{fd1}) with $\mu_+=\mu_-=0$, 
and then employing the automorphism discussed in the Appendix\;\ref{app-reps}.
The result is 
\be 
\langle \phi_\pm(u)\rangle_{_{\textrm{\tiny $L,\mu_\pm$}}} \! =\,  0 
\;\;\qquad \;\;
\langle j_\pm(u)\rangle_{_{\textrm{\tiny $L,\mu_\pm$}}} \! =\,  -\frac{\kappa \mu_\pm}{2\pi} 
\;\;\qquad \;\;
 \langle \T_\pm(u)\rangle_{_{\textrm{\tiny $L,\mu_\pm$}}} \! = \, \frac{\kappa \mu^2_\pm}{4\pi} - \frac{\pi c}{12 L^2} \;\;
\label{ftd1-L}
\ee
where we used that the Schwarzian derivative (\ref{cft3}) of the conformal map $u \mapsto \e^{2\pi \ri u/L}$ is equal to $2\pi^2/L^2$.
In the infinite volume limit $L \to + \infty$, the one-point functions (\ref{ftd1-L}) become the ones in (\ref{fd1}), as expected. 

The connected two-point functions in the finite density and finite volume representation can be written through the same procedure, 
starting from the connected two-point functions in the ground state representation.
From (\ref{fd5}), for the connected two-point expectation values of $\phi_\pm$ we obtain
%\textcolor{red}{\bf [REFORMULATE: $\tilde{\mu}_\pm$ function of $\mu_\pm$ and $h_\pm$]}
%\textcolor{red}{[is it expected that $f(u,v)$ depends also on $u+v$? See also the discussion about the Cayley transform in \cite{Jovanovic:2025mwe}]}
\be 
\langle \phi_\pm^*(u)\, \phi_\pm (v)\rangle_{_{\textrm{\tiny $L,\mu_\pm$}}}^{\textrm{\tiny con}}  
\! =
\frac{ \e^{\pm \ri \mu_\pm (u-v)} }{2\pi\, \e^{\pm \ri \pi h_\pm} }
\bigg(\frac{\pi }{L\,\sin \! \big[ \pi (u-v \mp \ri \varepsilon)/L\big] }\bigg)^{2h_\pm}  
\;\;\qquad\;\;
\frac{L \mu_\pm}{2\pi} \in \mathbb{Z}
\label{ftd5}
\ee 
%with the same multiplicative constants introduced in (\ref{fd5}) and the quantisation condition on $\mu_\pm$ guarantees that the 
whose r.h.s. is periodic, as expected. 
In the Appendix\;\ref{subsec-app-mu}, a consistency check of (\ref{ftd5}) is discussed
by considering the special case of a free fermion with anti-periodic boundary conditions 
and obtaining the corresponding two-point function through the Fermi-Dirac distribution. 
\\
%where the periodic function  $f(u,v) \equiv \e^{2\pi \ri u/L} - \e^{2\pi \ri v/L} = 2\ri \,\e^{\pi \ri (u+v)/L} \sin[\pi (u-v)/L] $
%occurs in the exponent. 
%
As for the two-point functions of $j_\pm$ and $\T_\pm$, 
from (\ref{fd4}) and (\ref{fd3}) we find  respectively
\be 
\langle j_\pm(u) \,j_\pm (v)\rangle_{_{\textrm{\tiny $L,\mu_\pm$}}}^{\textrm{\tiny con}}   
=
\frac{\kappa}{4\pi^2} 
\bigg(\frac{\pi }{L\,\sin \! \big[ \pi (u-v \mp \ri \varepsilon)/L\big] }\bigg)^{2} 
\label{ftd4}
\ee
and 
\be 
\langle \T_\pm(u) \,\T_\pm (v)\rangle_{_{\textrm{\tiny $L,\mu_\pm$}}}^{\textrm{\tiny con}}   
= 
\frac{c}{8\pi^2} 
\bigg(\frac{\pi }{L \,\sin \! \big[ \pi (u-v \mp \ri \varepsilon)/L\big] }\bigg)^{4}  \,.
%+ \frac{\kappa\mu_\pm^2}{4\pi^2} \left [\frac{\pi }{\beta_\pm \sinh \left (\frac{\pi}{\beta_\pm }(x-y \mp \ri \varepsilon)\right ) }\right ]^2 
\label{ftd3}
\ee
We also have that
\be
\langle \T_\pm(u) \, j_\pm (v)\rangle_{_{\textrm{\tiny $L,\mu_\pm$}}}^{\textrm{\tiny con}}    
= 
-\,\mu_\pm\, \langle j_\pm(u) \, j_\pm (v)\rangle_{_{\textrm{\tiny $L,\mu_\pm$}}}^{\textrm{\tiny con}}    \,.
\ee
The connected mixed correlators involving fields having different chiralities vanish identically. 
Notice that
the infinite volume limit $L \to + \infty$ of the two-point correlators 
(\ref{ftd5}), (\ref{ftd4}) and (\ref{ftd3}) 
gives the two-point correlators on line (\ref{fd5}),  (\ref{fd4}) and (\ref{fd3}) respectively, 
as expected.

%%%%%%%%%%%%%%%%%%%%%%%%%%%%%%%%%%%%%%%%%%
%\newpage
\subsection {Modular Hamiltonian and modular conjugation} 
\label{sec-mod-ham-circle}
%%%%%%%%%%%%%%%%%%%%%%%%%%%%%%%%%%%%%%%%%%

\begin{figure}[t!]
\vspace{-.8cm}
\hspace{-1.1cm}
\includegraphics[width=1.14\textwidth]{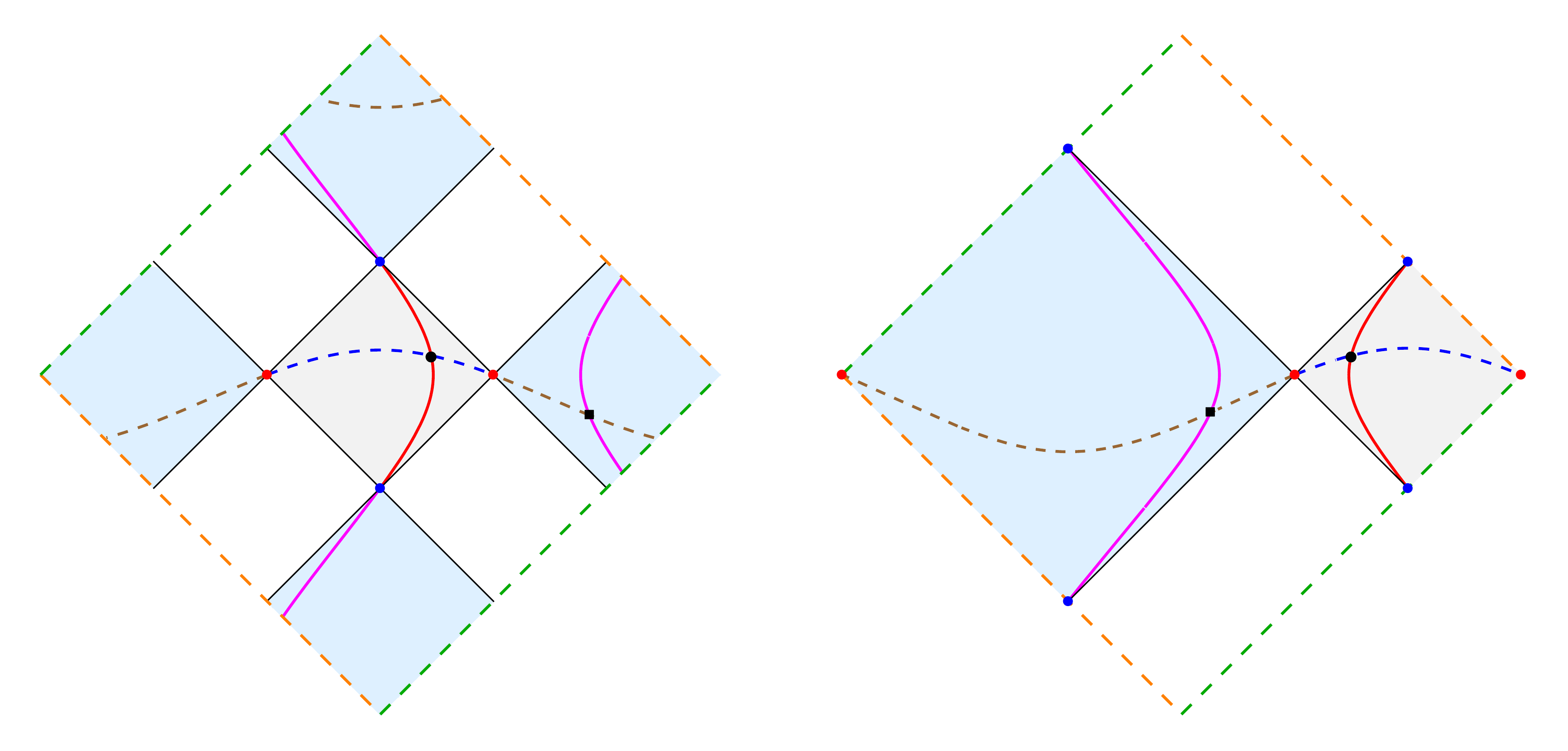}
\vspace{-.5cm}
\caption{
Modular trajectories generated 
either by the modular Hamiltonian (\ref{fmh-circle}) (solid lines)
or by the modular momentum (dashed lines)
for the CFT on the circle. 
Since periodic boundary conditions are imposed, 
the green (orange) dashed straight lines must be identified.
The solid (dashed) curves having different colours are related through 
the geometric action of the modular conjugation,
constructed from (\ref{j0-map-circle-def}).
}
\label{fig:diamond-circle}
\end{figure}

In the following we consider the portion of Minkowski spacetime described by the light-cone coordinates $(u_+, u_-)$
when periodic boundary conditions are imposed along both the chiral directions with the same period equal to $L$.
The resulting spacetime $\mathbb{M}_{\diamond}$ has the topology of a torus
and it is shown is Fig.\,\ref{fig:diamond-circle}, 
where  $u_\pm \in (-L/2\,, L/2)$ and the dashed segments having the same colour are identified.
We consider a two-dimensional CFT in the finite density state on this spacetime.
Moreover, each chiral direction is  partitioned through the interval $A =[a,b]$ with length $\ell \equiv b-a$
and its complement; hence the diamond $\mathcal{D}_A$ can be introduced.
The two panels of Fig.\,\ref{fig:diamond-circle} describe the same setup in two equivalent ways
and $\mathcal{D}_A$ corresponds to the grey region in each panel.

A standard way to compactify a chiral direction exploits the Cayley map,
which relates the real line to the unit circle $ \mathbb{S}$ with one point removed (see e.g. \cite{Hollands:2019hje})
and reads $u \mapsto z=\tfrac{1+ \ri \, v}{1- \ri \, v}$ where $v \in \RR$ and $z\in \mathbb{S}\setminus \{P_0\}$,
being $P_0$ the point at $\theta=\pi$ on $ \mathbb{S}$. 
By employing the complex number $z=\e^{2\pi \ri u/L} $ with  $u \sim u+L$ to parameterise $ \mathbb{S}$, 
where $L$ corresponds to the compactification parameter, 
the Cayley map and its inverse read respectively
\be
\label{cayley-one-dim}
\e^{2\pi \ri u/L} = \frac{1+ \ri \, v}{1- \ri \, v}
\;\;\;\; \qquad\;\;\;\;
v \,=\, \ri\; \frac{1 - \e^{2\pi \ri u/L} }{ 1 + \e^{2\pi \ri u/L} } \,=\, \tan(\pi u/L)\,\equiv \, \mathcal{C}(u) \,.
\ee
Alternatively \cite{Ginsparg:1988ui},
one first introduces the periodic identification $u \sim u+L$ on the real line 
and then uses the exponential map $u \mapsto \e^{2\pi \ri u/L}$.
%as done in the standard approach a two-dimensional CFT in Euclidean signature 

%

%\noindent 
%$\bullet$ 
%\textcolor{blue}{\bf [(Full) Modular Hamiltonian]} 

%The full modular Hamiltonian corresponding to a CFT on the circle in the finite density state
%and to the spatial bipartition determined by an interval $A =[a,b]$
%can be studied through the general results described in Sec.\,\ref{sec-mod-evolution}.
%%
%The spacetime is the surface of an infinite vertical cylinder
%which can be parameterised by either $(x,t)$ or $(u_+, u_{-})$.
%%
%It is shown in Fig.\,\ref{fig:diamond-circle}, 
%for two different choices of the generatrix of the cylinder along which the cylinder has been opened
%(see the vertical dashed thick lines, that must be identified),
%whose $x$-coordinate is the middle point of $B$ in the left panel and $b$ in the right panel. 
%The grey region is the diamond $\mathcal{D}_A$, like in Fig.\,\ref{fig:diamond-AB}.
%The periodicity in $u_\pm$ provides the partition the spacetime 
%into infinitely many fundamental regions 
%identified by the dashed green segments in Fig.\,\ref{fig:diamond-circle}.

By adapting the general results described in Sec.\,\ref{sec-mod-evolution},
in this CFT setup the modular Hamiltonian of $A$  
and the corresponding full modular Hamiltonian read respectively  
\cite{Wong:2013gua, Cardy:2016fqc}
\be 
\label{KA-def-u-pm-circle}
K_A  
= 
\int_A  V_L(u_+) \left[\, {\cal T}_+(u_+) - \frac{\kappa \mu^2_{+} }{4\pi} + \frac{\pi c}{12 L^2}\,  \right]  \!\rd u_+
+ 
\int_A  V_L(u_-) \left[\, {\cal T}_-(u_-) - \frac{\kappa \mu^2_{-} }{4\pi} + \frac{\pi c}{12 L^2}\,  \right]  \! \rd u_-
\ee
and 
\bea
\label{fmh-circle}
K
\equiv
K_{A} \otimes \boldsymbol{1}_B - \boldsymbol{1}_A \otimes K_{B} 
&=&
\int_{-L/2}^{L/2} V_L(u_+) \left[\, {\cal T}_+(u_+) - \frac{\kappa \mu^2_{+} }{4\pi} + \frac{\pi c}{12 L^2}\,  \right]   \rd u_+
\nn
\\
\rule{0pt}{.7cm}
& &
+ 
\int_{-L/2}^{L/2}  V_L(u_-) \left[\, {\cal T}_-(u_-) - \frac{\kappa \mu^2_{-} }{4\pi} + \frac{\pi c}{12 L^2}\,  \right]   \rd u_-
%+
%C_{K,L} 
%=
%\int_{\mathbb{R}}dx\,\mathsf{V}(x;a,b)\,T(x)\,. 
\eea
where 
%$C_{K,L} $ is a constant and  
the velocity $V_L(u)$ is
\be
\label{velocity_fund-circle}
V_L(u) =2L\, \frac{\sin[\pi(b-u)/L]\, \sin[\pi(u-a)/L]}{\sin[\pi(b-a)/L]}=\frac{1}{w_L'(u)} 
\;\;\;\qquad\;\;\;
u \in A
\ee
being $w_L(u)$ defined as follows
\be
\label{w-function-def-circle}
w_L(u) \equiv \frac{1}{2\pi}\,\log \!\left(\! - \frac{\sin[\pi(u-a)/L]}{\sin[\pi(u-b)/L]} \right)  .
\ee
The weight function (\ref{velocity_fund-circle}) can be obtained from (\ref{velocity_fund}) 
as follows
\be
\label{VL-from-V}
V_L(v) 
= \frac{\widetilde{V}\big( \e^{2\pi \ri v/L}\big)}{\partial_v\big( \e^{2\pi \ri v/L}\big)}
= \frac{\widehat{V}  \big( \mathcal{C}(v)\big)}{ \mathcal{C}'(v) }
\ee
where $\widetilde{V}(v)$ is defined as (\ref{velocity_fund}) with $a$ and $b$ replaced by $\e^{2\pi \ri a/L}$ and $\e^{2\pi \ri b/L}$ respectively,
while in the last expression the Cayley map (\ref{cayley-one-dim}) is employed
and $\widehat{V}(v)$ is given by  (\ref{velocity_fund}) with  $a$ and $b$ replaced by $\mathcal{C}(a)$ and $\mathcal{C}(b)$ respectively. 
%The velocity (\ref{velocity_fund-circle}) is related to the corresponding velocity on the line given in (\ref{velocity_fund}) as follows
%\be
%\label{VL-from-V}
%V_L(u) = \frac{\widetilde{V}\big( \e^{2\pi \ri u/L}\big)}{\partial_u\big( \e^{2\pi \ri u/L}\big)}
%\ee
%where $\widetilde{V}(u)$ is defined as (\ref{velocity_fund}) where $a$ and $b$ 
%are replaced by $\e^{2\pi \ri a/L}$ and $\e^{2\pi \ri b/L}$ respectively. 
%
The full modular Hamiltonian (\ref{fmh-circle}) 
corresponds (\ref{me3b}) specialised to $V_+(u) = V_-(u) = V_L(u)$ given by  (\ref{velocity_fund-circle}),
which vanishes only at the endpoints of $A$.
Notice that a vanishing additive constant has been chosen in (\ref{KA-def-u-pm-circle}).

%\begin{figure}[t!]
%\vspace{-.3cm}
%\hspace{-.9cm}
%\includegraphics[width=1.1\textwidth]{fig_DAB_mod_circle}
%\vspace{-.2cm}
%\caption{
%\textcolor{red}{\bf [Is this the right figure? See paper with Dobrica (appendix)]}
%\\
%Modular trajectories generated 
%either by the modular Hamiltonian (\ref{fmh-circle}) (solid lines)
%or by the modular momentum (dashed lines)
%for the CFT on the circle. 
%%
%The solid (dashed) lines having different colours are related through 
%the geometric action of the modular conjugation,
%constructed from (\ref{j0-map-circle-def}).
%}
%\label{fig:diamond-circle}
%\end{figure}

The modular evolution generated by (\ref{fmh-circle}) can be studied
by applying the results discussed in Sec.\,\ref{sec-mod-evolution} to $V_+(u) = V_-(u) = V_L(u)$
and $w_+(u) = w_{-}(u) =w_L(u)$, 
introduced in (\ref{velocity_fund-circle}) and (\ref{w-function-def-circle}) respectively. 
In this case (\ref{s2}) becomes \cite{Mintchev:2020uom}
\be
\label{xi-map-fund-circle}
\xi_\pm(\tau,u) = \xi_L(\pm\tau,u) 
\;\;\;\qquad\;\;\;
\xi_L(\tau,u) 
%= w^{-1}\left(2\pi\tau+w(x;a,b)\right) 
\equiv \frac{L}{2\pi \, \ri} \, \log\! \bigg( \frac{ \e^{\pi \ri  (b+a)/L} + \e^{ 2\pi \ri  b/L}\, \e^{2\pi w_L(u)+2\pi \tau} }{ \e^{\pi \ri (b-a)/L} +  \e^{2\pi w_L(u)+2\pi \tau}}   \bigg)
%\;\;\;\qquad\;\;\;
%u \in A
\ee
whose infinite volume limit $L \to +\infty$ gives (\ref{xi-map-fund}), as expected. 
From (\ref{xi-map-fund-circle}) we have
\bea
\label{exp-xi-L-1}
\e^{2\pi \ri  \xi_L(\tau,u)  /L} 
&=&
\frac{
\big( \e^{ 2\pi \ri  b/L} - \e^{ 2\pi \ri  u/L}  \big)\,\e^{ 2\pi \ri  a/L} + \big(\e^{ 2\pi \ri  u/L} - \e^{ 2\pi \ri  a/L} \big)\,\e^{ 2\pi \ri  b/L} \,\mathrm{e}^{2\pi \tau}
}{
\big( \e^{ 2\pi \ri  b/L}  - \e^{ 2\pi \ri  u/L} \big)+ \big( \e^{ 2\pi \ri  u/L} - \e^{ 2\pi \ri  a/L} \big)\, \mathrm{e}^{2\pi \tau}} 
\\
\label{exp-xi-L-2}
\rule{0pt}{.75cm}
&=&
\e^{\pi \ri (a+b)/L}\;
\frac{ \e^{\pi \ri a/L} \sin\!\big[\pi(b-u)/L\big] + \e^{\pi \ri b/L} \sin\!\big[\pi(u-a)/L\big]\, \e^{2\pi \tau} 
}{ 
\e^{\pi \ri b/L} \sin\!\big[\pi(b-u)/L\big] + \e^{\pi \ri a/L} \sin\!\big[\pi(u-a)/L\big]\, \e^{2\pi \tau} 
 }
 \hspace{1cm}
\eea
where the r.h.s. of (\ref{exp-xi-L-1}) 
corresponds to the r.h.s. of $\xi(\tau,u)$ in  (\ref{xi-map-fund})
with $a$, $b$ and $u$ replaced by $\e^{ 2\pi \ri  a/L} $, $\e^{ 2\pi \ri  b/L} $ and $\e^{ 2\pi \ri  u/L} $ respectively. 

From $\xi_L(\tau,u) $ in  (\ref{xi-map-fund-circle}) we obtain
the spacetime coordinates of
the modular trajectory in the diamond $\mathcal{D}_A$
whose initial point at $\tau=0$ has light-cone coordinates $(u_+, u_-)$,
namely
\be
\label{mod-traj-tau-circle}
x(\tau) = \frac{\xi_L(\tau, u_+) + \xi_L(-\tau, u_-) }{2}
\;\;\;\qquad\;\;\;
t(\tau) = \frac{\xi_L(\tau, u_+) - \xi_L(-\tau, u_-) }{2}
\ee
which correspond to (\ref{mod-traj-tau-line}) with $\xi(\tau,u) $ replaced by $\xi_L(\tau,u) $.
In Fig.\,\ref{fig:diamond-circle}, the solid red curve is a modular trajectory 
whose  initial point is the black dot.

The modular evolutions of the operators $\phi_\pm$, $j_\pm$ and $\CT_\pm$
are obtained by specialising (\ref{me9a}), (\ref{cp4}) and (\ref{met4}) respectively 
to (\ref{xi-map-fund-circle}).
It is worth writing explicitly the result for $\CT_\pm$
\be 
\label{T-evolved-circle}
\T_\pm (\tau,u) 
= 
\big[\partial_u \xi_L (\pm \tau,u)\big]^2 \, \T _\pm\big(\xi_L (\pm \tau,u)\big) 
+ \left( \frac{\kappa \mu^2_\pm}{4\pi} - \frac{\pi c}{12 L^2} \right)
\Big\{1- \big[\partial_u \xi_L (\pm \tau,u)\big]^2\Big\} 
\ee
where we used that\footnote{Another interesting result about (\ref{xi-map-fund-circle}) is
\be
\mathcal{S}_u\big[\e^{2\pi \ri \,\xi_L(\tau, u)/ L}\big](\tau,u) = \mathcal{S}_u\big[\e^{2\pi \ri u/ L}\big](u) = \frac{2\pi^2}{L^2}  \,.
\nn
\ee}
\be
\label{id-sch-L}
\mathcal{S}_u[\xi_L](\tau, u)  
=\,
\frac{2\pi^2}{L^2}\,
\Big\{
1-\big[ \partial_u  \xi_L (\tau, u) \big]^2 
\Big\}  \;.
\ee

The modular momentum operator can be introduced as (\ref{me3b-bis}) 
specialised to $V_+(u) = V_-(u) = V_L(u)$ given by (\ref{velocity_fund-circle}),
finding (\ref{full-mod-momentum}) with $V(u)$ replaced by $V_L(u)$.
The coordinates of the corresponding modular trajectories are (\ref{mod-traj-mom-gs})
with $\zeta(\lambda, u)$ replaced by $\zeta_L(\lambda, u)$,
which is defined by specialising (\ref{zeta-def-lambda}) 
to $w_+(u) = w_{-}(u) =w_L(u)$ in (\ref{w-function-def-circle}).
In Fig.\,\ref{fig:diamond-circle}, the dashed blue curve in $\mathcal{D}_A$ 
is a modular trajectory generated by the momentum operator 
whose initial point is the black dot.

The modular conjugation $J$ for the state and the bipartition of the circle that we are considering 
displays a geometric action in the spacetime
characterised by the map $( x,t) \to \big(\tilde{x}(x,t), \tilde{t}(x,t) \big)$ given by (\ref{inversion-xt})
where $\mathsf{j}(u)$ is replaced by the function $\mathsf{j}_L :\mathbb{R}\rightarrow\mathbb{R}$ 
defined as
\be
\label{j0-map-circle-def}
\mathsf{j}_L(u)\equiv 
\frac{L}{2\pi \ri} \, \log
\!\left(
\frac{ \e^{2\pi \ri b/L} + \e^{2\pi \ri a/L} }{2} 
+ \frac{  \big[ \big( \e^{2\pi \ri b/L} - \e^{2\pi \ri a/L} \big)/2 \big]^2 }{ \e^{2\pi \ri u/L} - \big( \e^{2\pi \ri b/L} + \e^{2\pi \ri a/L} \big)/2 }
\right)
%\frac{a+b}{2}+\frac{\left(\frac{b-a}{2}\right)^{2}}{u-\frac{a+b}{2}}
\ee
which is a bijective and idempotent function sending $A$ onto $B$
with negative derivative
\be
\label{der-j0-map-circle}
\mathsf{j}'_L(u)
=\,
-\, \frac{\sin^2[\pi (b-a)/L] }{ \big| \sin[\pi (b-u)/L] - \e^{\pi \ri (b-a)/L}  \sin[\pi (u-a)/L] \, \big|^2 }  \;.
\ee
Notice that the maps (\ref{xi-map-fund-circle}) and (\ref{j0-map-circle-def}) commute, namely they satify
\be
\label{id-xi-j-circle}
\mathsf{j}_L \big( \xi_L(\tau, u) \big)
=
\xi_L\big(\tau,\mathsf{j}_L (u)\big)
\ee
whose infinite volume limit gives (\ref{id-xi-j-line}).
In Fig.\,\ref{fig:diamond-circle} the solid and dashed curves in the light blue region
are obtained from the corresponding ones in $\mathcal{D}_A$
through the above mentioned map 
providing the geometric action of the modular conjugation.
These curves  also the modular trajectories generated by 
either the modular Hamiltonian or the modular momentum 
whose initial point is labelled by the black square,
which is the image of the black dot in $\mathcal{D}_A$
through the geometric action of modular conjugation.

The field transformations of the basic CFT fields 
are obtained by adapting the observations 
made in Sec.\,\ref{subsec-mod-J-vacuum} to the finite volume case 
we are considering. 
This leads us to conclude that the action of $J$ on 
$\phi_\pm$, $\phi^\ast_\pm$ and $j_\pm$ is given by 
(\ref{J-on-phi-1}), (\ref{J-on-phi-2}) are (\ref{J-on-j}) respectively, 
with $\mathsf{j}(u)$ replaced by $\mathsf{j}_L(u)$ defined in (\ref{j0-map-circle-def}).
As for $\CT_\pm$, the non trivial term due to the Schwarzian derivative must be taken into account. 
The result is obtained by setting $\tau=\pm \ri/2$ in (\ref{T-evolved-circle}) and reads
\be
\label{J-on-T-circle}
J\,\mathcal{T}_{\pm}(u) \, J 
\,=\, 
\mathsf{j}'_L(u)^2 \, \mathcal{T}_{\pm} (\,\mathsf{j}_L(u))
+\left( \frac{\kappa \mu^2_\pm}{4\pi} - \frac{\pi c}{12 L^2} \right)
\left[ \,1-\mathsf{j}'_L(u)^{2} \,\right]  .
\ee

By adapting (\ref{JKJ-chiral}) to the finite volume case, 
this transformation rule combined with the fact that (\ref{velocity_fund-circle}) and (\ref{j0-map-circle-def}) satisfy
\be
\mathsf{j}_L'(u) \, V_L(u) = V_L(\, \mathsf{j}_L(u)) 
\ee
leads to write the full modular Hamiltonian in the form given in (\ref{fmh-J}).

At finite volume, we performed a consistency check of these field transformations rules
by taking their mean values, employing the fact that $J$ leaves the state invariant and using (\ref{ftd1-L}),
as done in the end of Sec.\,\ref{subsec-mod-J-vacuum} for the interval in the infinite line.
%From this analysis consistency has been found. 
For instance, in the case of $\CT_\pm$ we have that
$ \langle \T_\pm(u)\rangle_{_{\textrm{\tiny $L,\mu_\pm$}}} = \langle J\,\T_\pm(u)\,J\rangle_{_{\textrm{\tiny $L,\mu_\pm$}}} $
and we found that the r.h.s. of (\ref{J-on-T-circle}) is consistent with the last expression in (\ref{ftd1-L}).

Following the analysis reported in the final part of Sec.\,\ref{subsec-mod-J-vacuum},
we computed the modular evolution of an operator belonging to the complementary region $B$, 
from \eqref{mod-evo-O-in-B} specialised to either $\phi_\pm$ or $j_\pm$ or $\CT_\pm$.
In the Appendix\;\ref{app-evolution-B-region},
by employing also (\ref{id-xi-j-circle}) and (\ref{J-on-T-circle}), 
we have found that the expressions \eqref{me9a}, \eqref{me9b}, \eqref{cp4} and \eqref{met4} 
with $\xi_\pm(\tau,u)$ given by \eqref{xi-map-fund-circle}  for the modular evolution 
hold also for $u \in B$.

We remark that the compact manifold $\mathbb{M}_{\diamond}$ considered above does not coincide with the 
compactification $\overline {\mathbb M} = ({\mathbb S} \times {\mathbb S} ) /{\mathbb Z}_2$ of the 
two-dimensional Minkowski spacetime $\mathbb M$
(often called Dirac-Weyl compactification)
discussed in \cite{Segal:1971aa, Segalbook, Luscher:1974ez, Todorov:1978rf, Brunetti:1992zf},
where ${\mathbb S}$ is the unit circle. 
Since $\overline {\mathbb M}$ is not causally orientable, its universal covering 
${\widetilde {\mathbb M}} = {\mathbb S}  \times  {\mathbb R}$ is employed
to define a consistent CFT on the cylinder \cite{Segal:1971aa, Luscher:1974ez, Brunetti:1992zf}.
However, from the group theoretical point of view, 
the time $t_c \in \RR$ on ${\widetilde {\mathbb M}}$ is associated to the conformal Hamiltonian $\frac{1}{2}(P_0 + K_0)$ 
rather than to the Hamiltonian $P_0$ in ${\mathbb M}$, 
where $K_0$ is  the generator of the special conformal transformations \cite{Segalbook, Todorov:1978rf}. 

\subsection {Modular correlators} 
\label{sec-mod-corr-circle}
%%%%%%%%%%%%%%%%%%%%%%%%%%%%%%%%%%%%%%%%%%

The modular evolutions of $\phi_\pm$, $j_\pm$ and $\CT_\pm$
can be written by specialising (\ref{me9a}), (\ref{cp4}) and (\ref{met4}) to (\ref{xi-map-fund-circle}),
as already mentioned above in Sec.\,\ref{sec-mod-ham-circle}.
The result for $\CT_\pm$ has been reported explicitly in (\ref{T-evolved-circle}).
Taking the mean values of the resulting expressions and using (\ref{ftd1-L}),
we find one-point functions that are independent of $\tau$,
i.e. $\langle \phi_\pm(u)\rangle_{_{\textrm{\tiny $L,\mu_\pm$}}} = \langle \phi_\pm (\tau,u)  \rangle_{_{\textrm{\tiny $L,\mu_\pm$}}} $ for the primaries,
$\langle j_\pm(u)\rangle_{_{\textrm{\tiny $L,\mu_\pm$}}} = \langle j_\pm (\tau,u)  \rangle_{_{\textrm{\tiny $L,\mu_\pm$}}} $ for the currents
and $\langle \T_\pm(u)\rangle_{_{\textrm{\tiny $L,\mu_\pm$}}} = \langle \T_\pm (\tau,u)  \rangle_{_{\textrm{\tiny $L,\mu_\pm$}}} $ 
for the operators (\ref{cal-T-pm-def})
(in the latter case also (\ref{id-sch-L}) has been used).

As for the modular two-point correlators at finite volume, 
when $u \neq v$, $\tau_1 \neq \tau_2$ and
for the velocity (\ref{velocity_fund-circle}) providing (\ref{xi-map-fund-circle}),
one finds the following identity
\be
\label{identity-mod-corr-circle}
\frac{ \partial_{u} \xi_L(\tau_1, u) \, \partial_{v} \xi_L(\tau_2, v) }{ \Big( \frac{L}{\pi} \sin \! \big[ \frac{\pi}{L} \big( \xi_L(\tau_1, u) - \xi_L(\tau_2, v)\big) \big]\Big)^2}
\,=
\left(
\frac{R_L(\tau_{12}; u,v)}{ \frac{L}{\pi} \sin \!\big[ \frac{\pi}{L}\,(u-v) \big]}
\right)^2
\ee
where
\bea
\label{R-L-factor-def}
R_L(\tau; u,v)
&\equiv &
\frac{ \e^{2\pi w_L(u)} - \e^{2\pi w_L(v)} }{  \e^{2\pi w_L(u) +\pi \tau } - \e^{2\pi w_L(v) - \pi \tau } } 
\\
\rule{0pt}{.7cm}
& = &
\frac{  
\sin\!\big[ \tfrac{\pi}{L}(u-a) \big] \, \sin\!\big[ \tfrac{\pi}{L}(v-b) \big] - \sin\!\big[ \tfrac{\pi}{L}(u-b) \big] \, \sin\!\big[ \tfrac{\pi}{L}(v-a) \big]
}{   
\sin\!\big[ \tfrac{\pi}{L}(u-a) \big] \, \sin\!\big[ \tfrac{\pi}{L}(v-b) \big] \, \e^{\pi \tau } - \sin\!\big[ \tfrac{\pi}{L}(u-b) \big] \, \sin\!\big[ \tfrac{\pi}{L}(v-a) \big]\, \e^{- \pi \tau }  
} 
\hspace{1cm}
\nn
\eea
which satisfies
\be
\label{RL-ids}
R_L(\tau = 0; u, v) = 1
\;\qquad\;
R_L(-\tau; v, u) = R_L(\tau; u, v) 
\;\qquad\;
R_L(\tau + \ri\,; u, v) = - R_L(-\tau; v, u) \,.
\ee
The infinite volume limit of (\ref{identity-mod-corr-circle}) and (\ref{R-L-factor-def})
gives (\ref{identity-mod-corr-zero-temp}) and (\ref{R-factor-def}) respectively, as expected. 
The r.h.s. of (\ref{identity-mod-corr-circle}) has been obtained by observing that
\be
\label{sin-xi12-circle-tau12}
\sin\!\big[ \tfrac{\pi}{L} \big( \xi_L(\tau_1,u)  - \xi_L(\tau_2,v) \big)\big]
= 
\frac{ p_L (\tau_1 ,u) \, p_L (\tau_2 , v) }{ R_L(\tau_{12} ; u , v)  }
\, \sin\!\big[ \tfrac{\pi}{L}(u - v)\big]
\ee
in terms of (\ref{R-L-factor-def}) and of 
\be
p_L (\tau , u) \equiv \frac{
2\ri\, \e^{\pi \ri (a+b)/L} \sin[\pi(b-a)/L] \, \e^{\pi \tau} \, \e^{\pi \ri u/L}
}{ 
\big[ \, \e^{ 2\pi \ri  b/L} - \e^{ 2\pi \ri  u/L}  + \big(\e^{ 2\pi \ri  u/L} - \e^{ 2\pi \ri  a/L} \big)  \,\mathrm{e}^{2\pi \tau} \,\big] \, \e^{\pi \ri \xi_L(\tau,u)  /L}  } 
\ee
which satisfies $p_L (\tau=0 , u)=1$ and  is a real function when $u \in A$, indeed we notice that its square can be written as
\bea
\label{pL-squared}
p_L (\tau , u)^2 
&=& 
\partial_u \xi_L(\tau,u) 
\\
&=& 
\frac{\sin^2[\tfrac{\pi}{L}(b-a)]
}{
\sin^2[\tfrac{\pi}{L}(b-a)]
+
\big(\e^{2\pi \tau} -1 \big) \sin^2[\tfrac{\pi}{L}(u-a)]
+
\big(\e^{-2\pi \tau} -1 \big)  \sin^2[\tfrac{\pi}{L}(b-u)]
}
\hspace{1cm}
\eea
which is positive for $u\in A$ and any $\tau \in \RR$.
From (\ref{pL-squared}), we have that $p_L (\tau , u)$ does not vanish for any finite value of $\tau$;
hence, since $p_L (\tau=0 , u)=1$, we conclude that $p_L (\tau , u) >0$.
%\textcolor{red}{\bf [checked numerically]}
%\be
%\label{sin-xi12-circle-tau12}
%\sin\!\big[ \tfrac{\pi}{L} \big( \xi_L(\tau_1,u)  - \xi_L(\tau_2,v) \big)\big]
%= 
%\frac{ \big(\sin\!\big[\tfrac{\pi}{L} (b - a)\big] \big)^2 \, \e^{\pi (\tau_1 +\tau_2)} }{ \sqrt{\tilde{q}_L(\tau_1,u)\, \tilde{q}_L(\tau_2,v) } \; R_L(\tau_{12} ; u,v)     }
%\; \sin\!\big[ \tfrac{\pi}{L}(u - v)\big]
%\ee
%\textcolor{red}{[{\bf some issues here:} numerically, I find that if $b-a$ is too large the sign changes, for given $u_1$ and $u_2$. It also depends on the signs of $\tau_1$ and $\tau_2$]}
%\\
%in terms of (\ref{R-L-factor-def}) and of the positive quantity $\tilde{q}_L(\tau,u) $ defined as follows
%\bea
%\label{q-L-def-A}
%\tilde{q}_L(\tau,u) 
%& \equiv &
%\frac{1}{4} \,\Big| \,\e^{ 2\pi \ri  b/L} - \e^{ 2\pi \ri  u/L}  + \big(\e^{ 2\pi \ri  u/L} - \e^{ 2\pi \ri  a/L} \big)  \,\mathrm{e}^{2\pi \tau} \Big|^2
%\\
%\rule{0pt}{.6cm}
%&=&
%\big( \sin\!\big[\tfrac{\pi}{L} (u - a)\big] \big)^2 \, \e^{4\pi \tau} 
%+ \big( \sin\!\big[\tfrac{\pi}{L} (b - u)\big] \big)^2 
%\nn
%\\
%& & +\, 2 \cos\!\big[\tfrac{\pi}{L} (b - a)\big]  \sin\!\big[\tfrac{\pi}{L} (b - u)\big]  \sin\!\big[\tfrac{\pi}{L} (u-a)\big] \, \e^{2\pi \tau}
%\eea
The infinite volume limit of (\ref{sin-xi12-circle-tau12}) gives (\ref{xi12-tau-chiral-tau12}), as expected. 
%In this limit we also have that $L^2\, \tilde{q}_L(\tau,u) \to \big[ \pi \,q(\tau, u)\big]^2$
%(see (\ref{pA-qA-def})).
%

%We remark that, 
%since $s_{12}(\tau) \equiv - \big[ \xi_L(\tau, u_{1,+}) - \xi_L(\tau, u_{2,+}) \big] \big[ \xi_L(-\tau, u_{1,-}) - \xi_L(-\tau, u_{2,-}) \big]$ 
%and  $\sin\!\big[ \tfrac{\pi}{L} \big( \xi_L(\tau,u_{1,+})  - \xi_L(\tau,u_{2,+}) \big)\big]\, \sin\!\big[ \tfrac{\pi}{L} \big( \xi_L(-\tau,u_{1,-})  - \xi_L(-\tau,u_{2,-}) \big)\big]$
%have the same sign, 
%by setting $\tau_1 = \tau_2$ in (\ref{sin-xi12-circle-tau12}) 
%and employing (\ref{RL-ids}) combined with the fact that $p_L (\tau , u) >0$,
%one observes that the modular evolution in $\mathcal{D}_A$ preserves the relativistic causality
%also in this finite volume setup.
%%
%From (\ref{pL-squared}), we have that $p_L (\tau , u)  \to 0^+$ as $\tau \to \pm \infty$,
%and this implies that $s_{12}(\tau) $ vanishes as $\tau \to \pm \infty$, 
%as also suggested from the modular trajectories. 

The modular correlators can be written by adapting the procedure described in Sec.\,\ref{sec-mod-corr-line}.
Thus, from the expressions in (\ref{me9a}), (\ref{me9b}), (\ref{cp4}) and (\ref{met4}),
the correlators (\ref{ftd5}), (\ref{ftd4}) and  (\ref{ftd3}) 
and the identity (\ref{identity-mod-corr-circle}),
for the connected modular correlators of the primary $\phi_\pm$ we get 
%\textcolor{red}{\bf [(\ref{mod-cor-phi-circ})-(\ref{mod-cor-phi-circ-2}) seem wrong if we employ the new version of (\ref{ftd5}).
%Do (\ref{me9a}) and (\ref{me9b}) hold also at finite volume?]}
 \bea
 \label{mod-cor-phi-circ}
\langle \phi_\pm^*(\tau_1, u) \,\phi_\pm (\tau_2, v)\rangle_{_{\textrm{\tiny $L,\mu_\pm$}}}^{\textrm{\tiny con}} 
&=&
\frac{ \e^{\pm \ri \mu_\pm (u-v)} }{2\pi \, \e^{\pm \ri \pi h_\pm}}\; W_{L,\pm}(\pm \tau_{12}; u, v)^{2h_{\pm}}
 \\
  \label{mod-cor-phi-circ-2}
  \rule{0pt}{.78cm}
\langle \phi_\pm(\tau_1, u) \,\phi^*_\pm (\tau_2, v)\rangle_{_{\textrm{\tiny $L,\mu_\pm$}}}^{\textrm{\tiny con}} 
&=&
\frac{ \e^{\mp \ri \mu_\pm (u-v)} }{2\pi \, \e^{\pm \ri \pi h_\pm}}\; W_{L,\pm}(\pm  \tau_{12}; u, v)^{2h_{\pm}}
\eea
and for the connected modular correlators of 
the current $j_\pm$ and the energy-momentum tensor
one obtains respectively
\bea
  \label{mod-cor-j-circ}
\rule{0pt}{.6cm}
\langle j_\pm(\tau_1, u) \,j_\pm (\tau_2,v)\rangle_{_{\textrm{\tiny $L,\mu_\pm$}}}^{\textrm{\tiny con}}  
&=&
\frac{\kappa}{4\pi^2} \,W_{L,\pm}(\pm  \tau_{12}; u, v)^{2}
 \\
  \label{mod-cor-T-circ}
\rule{0pt}{.6cm}
\langle \T_\pm(\tau_1, u) \,\T_\pm (\tau_2,v)\rangle_{_{\textrm{\tiny $L,\mu_\pm$}}}^{\textrm{\tiny con}}  
&=&
\frac{c}{8\pi^2} \, W_{L,\pm}(\pm  \tau_{12}; u, v)^{4}
\eea
where $W_{L,\pm}$ is defined in terms of (\ref{w-function-def-circle}) as follows \cite{Mintchev:2020uom}
\be
\label{cap-W-def-circle}
W_{L,\pm}(\tau ; u, v)
\,\equiv \,
\frac{ \e^{2\pi w(u)} - \e^{2\pi w(v)}  }{\tfrac{L}{\pi} \,\sin \!\big[ \tfrac{\pi}{L} (u-v) \big] }\;
\frac{1}{  \e^{2\pi w(u) +\pi \tau} - \e^{2\pi w(v) - \pi \tau } \mp \ri \varepsilon} 
\ee
which becomes (\ref{cap-W-def}) in the infinite volume limit $L \to + \infty$.

For (\ref{cap-W-def-circle}) one finds the following property
\be
\label{KMS-0}
W_{L,\pm}(\tau + \ri ; u, v) = W_{L,\pm}(\tau - \ri ; u, v) = W_{L,\pm}( - \tau  ; v, u)
\ee
which implies that the modular correlators (\ref{mod-cor-phi-circ}), (\ref{mod-cor-phi-circ-2}), (\ref{mod-cor-j-circ}) and (\ref{mod-cor-T-circ}) 
satisfy the KMS condition with modular inverse temperature $\tilde{\beta} = 1$.

Furthermore, when $\tau \neq 0$, 
the limit $v \to u$ of (\ref{cap-W-def}) is well defined and given by 
\be
\label{cap-W-equal-point-circle}
\lim_{v \to u} 
W_{L,\pm}(\tau; u, v)
=
\frac{\pi }{ V_L (u) \, \sinh(\pi \tau \mp \ri \varepsilon) } 
\ee
in terms of (\ref{velocity_fund-circle}),
which becomes (\ref{cap-W-equal-point}) as $L\to+\infty$
and will be employed in Sec.\,\ref{sec-noise-circle}.

Finally, by employing the following identity for (\ref{cap-W-def-circle}) and (\ref{j0-map-circle-def})
\be 
W_{L,\pm} (\tau, u, v) ^2
= 
\mathsf{j}_L^\prime(u)\; \mathsf{j}_L^\prime(v)\, W_{L,\pm} (-\tau, \mathsf{j}_L(v), \mathsf{j}_L(u))^2
\ee
we checked that the expressions reported in the r.h.s.'s of  \eqref{J-on-phi-1}, \eqref{J-on-phi-2}, \eqref{J-on-j} 
and \eqref{J-on-T-circle},
with $\mathsf{j}$ replaced by $\mathsf{j}_L$,
are consistent with (\ref{id-mod-corr-J}) written for finite volume and finite density representation.

%%%%%%%%%%%%%%%%%%%%%%%%%%%%%%%%%%%%%%%%%%%%%%%%%%%%%%
%\newpage
\section {Modular transport and fluctuations at finite volume} 
\label{sec-noise-circle}
%%%%%%%%%%%%%%%%%%%%%%%%%%%%%%%%%%%%%%%%%%%%%%%%%%%%%%

In this section, 
the analyses discussed in Sec.\,\ref{sec-transport-line} are extended 
to the CFT at finite density and finite volume described in Sec.\,\ref{sec-finite-volume}.

The mean values of the charge currents $\langle j_x (\tau;x,t) \rangle_{_{\textrm{\tiny $L,\mu$}}} $ 
and $\langle j_t (\tau;x,t) \rangle_{_{\textrm{\tiny $L,\mu$}}} $
are obtained by employing (\ref{ecurr}), (\ref{j-t-def}) and (\ref{ftd1-L}).
This gives the r.h.s.'s of (\ref{mc1x})-(\ref{mc1t})
with the velocity $V(u)$ replaced by $V_L(u)$ introduced in  (\ref{velocity_fund-circle}),
which provide the components of the vector field $\boldsymbol{j} (x,t)$.
Similarly, the mean values of the helicity currents
$\langle k_x (\tau;x,t) \rangle_{_{\textrm{\tiny $L,\mu$}}} $ 
and $\langle k_t (\tau;x,t) \rangle_{_{\textrm{\tiny $L,\mu$}}} $,
which are the components of the vector field $\boldsymbol{k} (x,t)$,
can be written from (\ref{k-xt-def}) and (\ref{ftd1-L}),
finding the r.h.s.'s of (\ref{mc10x})-(\ref{mc10t}) with $V(u)$ replaced by $V_L(u)$.

\begin{figure}[t!]
\vspace{-.6cm}
\hspace{-1.7cm}
\includegraphics[width=1.23\textwidth]{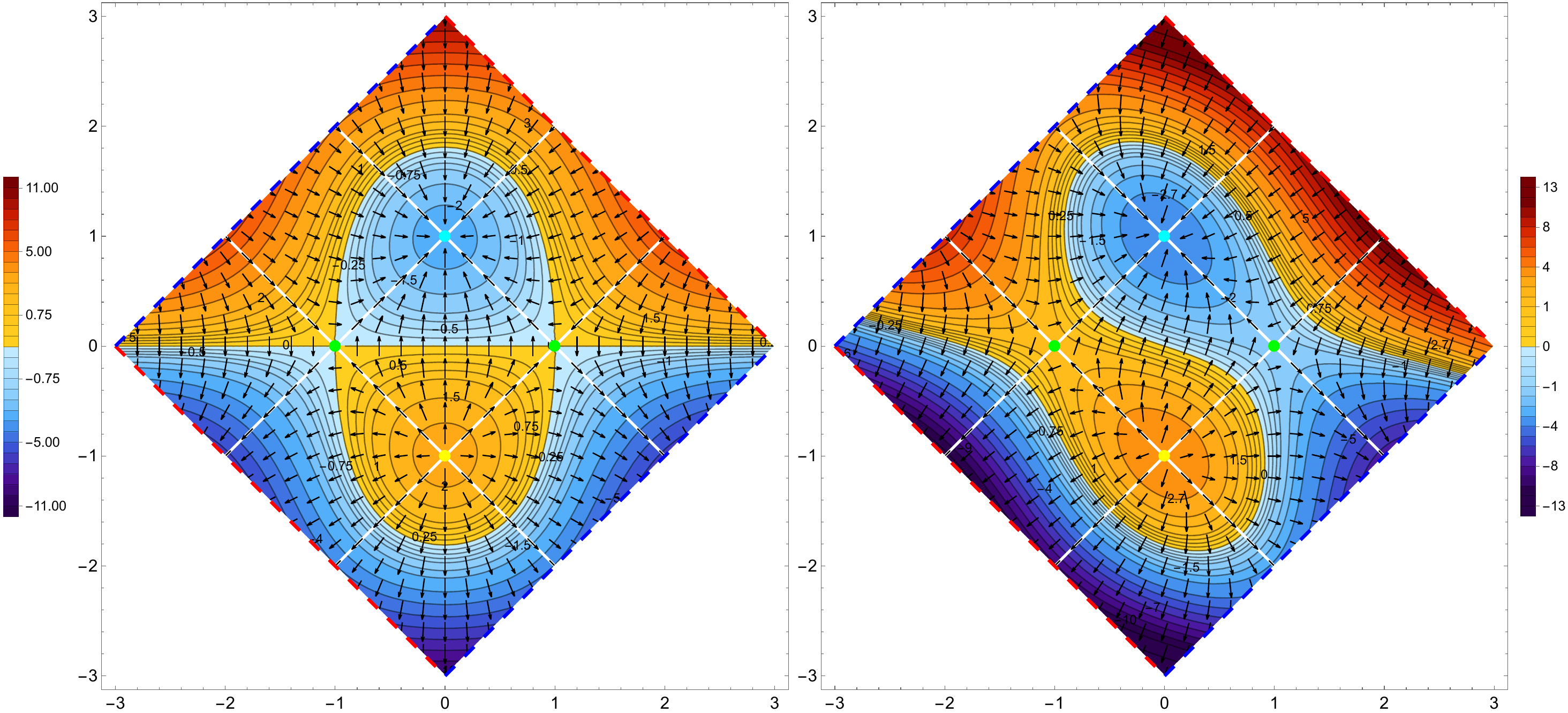}
\vspace{-.4cm}
\caption{Vector fields for the mean values of the charge currents (\ref{ham-vec-j-0-circle}) in $\mathbb{M}_{\diamond}$,
whose potential is the first expressions in (\ref{potentials-circle-W}).
The CFT has $c=1$, $\kappa = 3$
and either equal chemical potentials $\mu_{+} = \mu_{-} = 2\pi/L$ (left panel)
or different chemical potentials  $\mu_{+} =  4\pi/L$ and $\mu_{-} =  2\pi/L$ (right panel).
Here $L=6$ and  $\ell = 2$.
}
\label{fig:j-L-curr}
\end{figure}

\begin{figure}[t!]
\vspace{-.6cm}
\hspace{-1.7cm}
\includegraphics[width=1.23\textwidth]{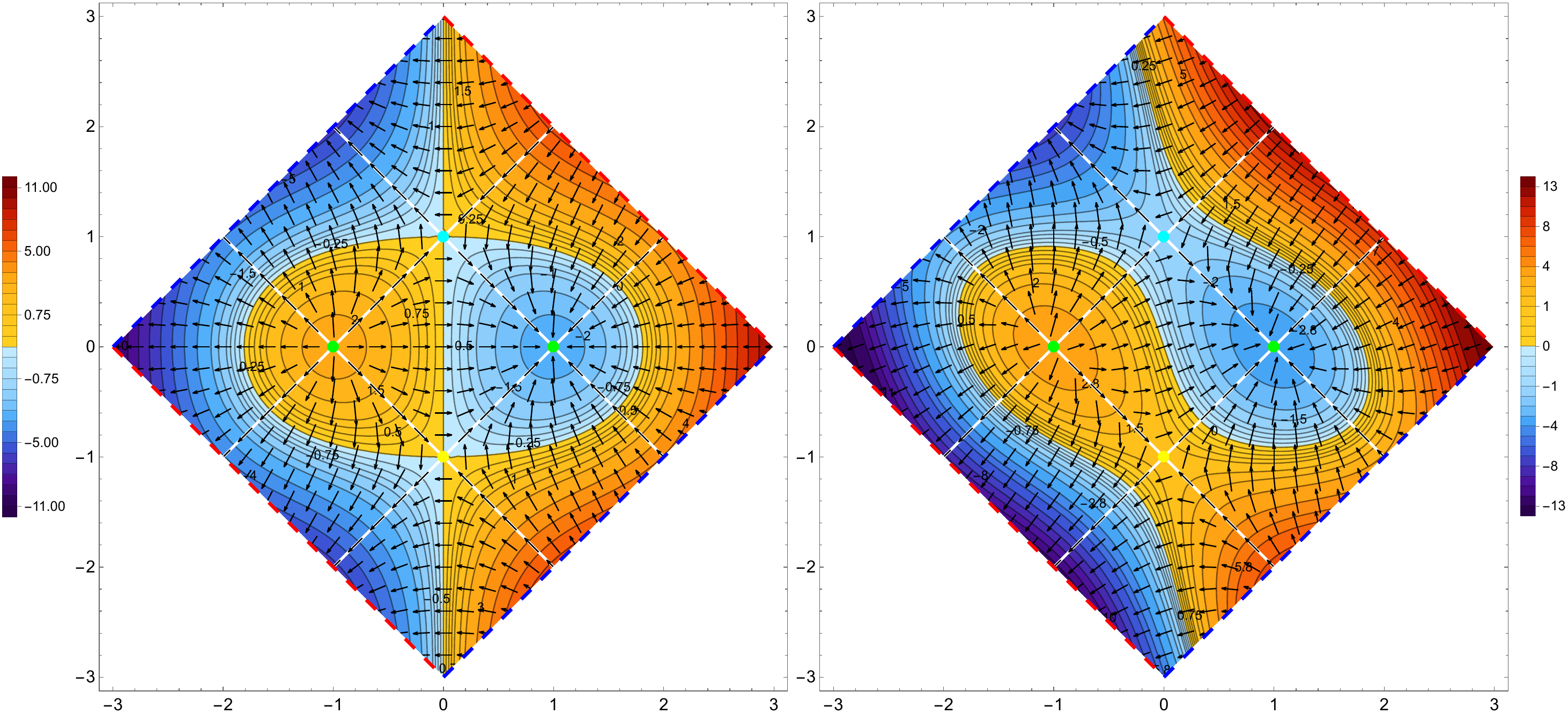}
\vspace{-.4cm}
\caption{Vector fields for the mean values of the helicity currents (\ref{ham-vec-k-0-circle}) in $\mathbb{M}_{\diamond}$,
whose potential is the second expressions in (\ref{potentials-circle-W}),
for either equal (left panel) or different (right panel) chemical potentials,
in the same setup of Fig.\,\ref{fig:j-L-curr}.
}
\label{fig:k-L-curr}
\end{figure}

The smooth planar vector fields $\boldsymbol{j} (x,t)$ and $\boldsymbol{k} (x,t)$ in $\mathbb{M}_{\diamond}$
are shown in Fig.\,\ref{fig:j-L-curr} and Fig.\,\ref{fig:k-L-curr} 
%we shown these currents and their potentials (\ref{potentials-circle-W})
for the choice of the parameters described in the caption of Fig.\,\ref{fig:j-L-curr}.
In all the figures of this section $\mathcal{D}_A$ and $\mathbb{M}_{\diamond}$
have been represented like in the left panel of Fig.\,\ref{fig:diamond-circle}.
Moreover, 
the extension of the vector fields to the entire spacetime $\mathbb{M}_{\diamond}$
have been displayed, 
as discussed in Sec.\,\ref{sec-charge-helicity-trans} for the case of the Minkowski spacetime
(see also Fig.\,\ref{fig:j-curr}).

Both the vector fields $\boldsymbol{j} (x,t)$ and $\boldsymbol{k} (x,t)$ 
have the same critical points (by construction)
and all of them have multiplicity $1$.
In particular, four critical points occur in $\mathbb{M}_{\diamond}$:
two nodes (one stable and one unstable) and two saddles,
which are denoted through the same notation adopted in the figures of Sec.\,\ref{sec-transport-line}.
We recall that the nodes have Poincar\'e index $+1$, while the saddles have Poincar\'e index $-1$.
Thus, the sum of the Poincar\'e indices of all the isolated critical points in a fundamental region vanishes.
This is consistent with the Poincar\'e-Hopf theorem mentioned in Sec.\,\ref{sec-charge-helicity-trans};
indeed $\mathbb{M}_{\diamond}$ has the topology of the torus, whose Euler characteristic is equal to zero.

The vector fields $\boldsymbol{j} (x,t)$ and $\boldsymbol{k} (x,t)$  are curl free and satisfy
\bea
\label{ham-vec-j-0-circle}
& &
\langle j_x (\tau;x,t) \rangle_{_{\textrm{\tiny $L,\mu$}}}  = -\,\partial_x \mathsf{W}_{L,j}(x,t) 
\;\;\;\qquad \;\;\;
\langle j_t (\tau;x,t) \rangle_{_{\textrm{\tiny $L,\mu$}}}   = -\,\partial_t \mathsf{W}_{L,j}(x,t) 
\hspace{1cm}
\\
\rule{0pt}{.5cm}
\label{ham-vec-k-0-circle}
& &
\langle k_x (\tau;x,t) \rangle_{_{\textrm{\tiny $L,\mu$}}}  = -\,\partial_x \mathsf{W}_{L,k}(x,t) 
\;\;\;\qquad \;\;\,
\langle k_t (\tau;x,t) \rangle_{_{\textrm{\tiny $L,\mu$}}}   = -\,\partial_t \mathsf{W}_{L,k}(x,t) 
\eea
where the potentials read respectively
\be
\label{potentials-circle-W}
\mathsf{W}_{L,j}(x,t) \equiv \frac{\kappa}{2\pi} \big[\mu_+ \,g_{_{\textrm{\tiny $L$}}}\! (u_+) - \mu_- \,g_{_{\textrm{\tiny $L$}}}\! (u_-) \big]
\;\qquad\;
\mathsf{W}_{L,k}(x,t) \equiv \frac{\kappa}{2\pi} \big[\mu_+ \,g_{_{\textrm{\tiny $L$}}}\! (u_+) + \mu_- \,g_{_{\textrm{\tiny $L$}}}\! (u_-) \big]
\hspace{.2cm}
\ee
being the function $g_{_{\textrm{\tiny $L$}}}\! (u)$ defined as follows
\be
\label{potential-circle}
g_{_{\textrm{\tiny $L$}}}\!(u)
\equiv
\frac{L}{\tan\!\big[\tfrac{\pi}{L}(b-a)\big]}  \bigg(u-\frac{a+b}{2} \bigg)
-
\frac{L^2}{2\pi \sin\!\big[\tfrac{\pi}{L}(b-a)\big]} \,\sin\!\bigg[\frac{2\pi}{L} \bigg(u-\frac{a+b}{2} \bigg) \bigg]     \,.
\ee
Notice that, 
although (\ref{potential-circle}) is not a periodic function of $u$
and therefore the potentials in (\ref{potentials-circle-W}) are not periodic $\mathbb{M}_{\diamond}$ as well,
the corresponding vector fields (\ref{ham-vec-j-0-circle}) and (\ref{ham-vec-k-0-circle}) are periodic $\mathbb{M}_{\diamond}$.
Thus, the potentials in (\ref{potentials-circle-W}) are well defined on an open subset of $\mathbb{M}_{\diamond}$;
hence the potentials displayed in Fig.\,\ref{fig:j-L-curr} and Fig.\,\ref{fig:k-L-curr} 
are not defined in a neighbourhood of boundary made by the union of the dashed straight segments.

Consistency between (\ref{ham-vec-j-0-circle})-(\ref{potential-circle}) and 
the mean values of the currents occurs because 
(\ref{potential-circle}) and  (\ref{velocity_fund-circle}) are related as follows
\be
\label{der-gL-VL}
- \,\partial_u \,g_{_{\textrm{\tiny $L$}}}\!(u) = V_L(u)  \,.
\ee

Combining (\ref{id-der-g-V}), (\ref{VL-from-V}) and (\ref{der-gL-VL}), 
it is straightforward to find that the functions (\ref{potential}) and (\ref{potential-circle}) are related as follows
\be
\partial_u g_L(u) = \frac{ \partial_u \tilde{g}\big( \e^{2\pi \ri u/L}\big)}{\big[ \partial_u\big( \e^{2\pi \ri u/L}\big)\big]^2}
\ee
where $\tilde{g}(u)$ is defined as (\ref{potential}) where $a$ and $b$ 
are replaced by $\e^{2\pi \ri a/L}$ and $\e^{2\pi \ri b/L}$ respectively. 
Moreover, 
the infinite volume limit $L \to +\infty$ of (\ref{potential-circle}) gives (\ref{potential});
hence the potentials (\ref{potentials-W-line}) obtained for the CFT on the line
are the infinite volume limit of (\ref{potentials-circle-W}).

The vector fields $\boldsymbol{j} (x,t)$ and $\boldsymbol{k} (x,t)$,
which are well defined in $\mathbb{M}_{\diamond}$,
have vanishing fluxes through the solid white lines in Fig.\,\ref{fig:j-L-curr} and Fig.\,\ref{fig:k-L-curr};
%(which corresponds to the solid black lines in the left panel of Fig.\,\ref{fig:diamond-circle}).
indeed,  the absolute value of the ratio of theirs components is equal to one along these lines. 
These vanishing fluxes lead us to consider the total charges in the diamond $\mathcal{D}_A$.
In the finite volume and finite density representation, 
from (\ref{ftd1-L}) and (\ref{QA-tQA-mean}),
for the mean values of (\ref{echarge}) and (\ref{h-charge-def}) we find respectively
\be
\label{QA-tQA-mean-L}
\langle Q_A \rangle_{_{\textrm{\tiny $L,\mu$}}} \! = \langle Q_A \rangle_{\mu}
\;\;\;\qquad\;\;\;
\langle \widetilde{Q}_A \rangle_{_{\textrm{\tiny $L,\mu$}}} \! = \langle \widetilde{Q}_A \rangle_{\mu}   \,.
\ee

It is worth considering the line integrals of the curl free vector fields $\boldsymbol{j} (x,t)$ and $\boldsymbol{k} (x,t)$ 
along curves anchored to the opposite vertices of $\mathcal{D}_A$,
as done in Sec.\,\ref{sec-charge-helicity-trans}.
The results read
\bea
\label{flow-j-t-A-circle}
\mathcal{L}[\boldsymbol{j}]\big(\gamma(P_a \to P_b)\big) 
%\int_A \langle j_x (\tau;x,0) \rangle_{_{\textrm{\tiny $L,\mu_\pm$}}}  \rd x \,
&=& 
\, \mathsf{W}_{L,j}\big|_{P_a} - \mathsf{W}_{L,j}\big|_{P_b} 
=
%\frac{\kappa (\mu_{+} - \mu_{-})}{6} 
-\frac{2\pi}{3}\, \langle \widetilde{Q}_A \rangle_{\mu} \, M(\pi \ell/L) 
%\left( 1 -\frac{\pi \ell}{L} \, \cot(\pi \ell/L) \right) \hspace{.8cm}
\\
\label{flow-j-x-At-circle}
\rule{0pt}{.8cm}
\mathcal{L}[\boldsymbol{j}]\big(\gamma(P_{-\infty} \to P_{+\infty})\big) 
%\int_{\tilde{A}} \langle j_t (\tau; \tfrac{a+b}{2} ,t) \rangle_{_{\textrm{\tiny $L,\mu_\pm$}}}  \rd t \,
&=&
\, \mathsf{W}_{L,j}\big|_{P_{-\infty}}  - \mathsf{W}_{L,j} \big|_{P_{+\infty}} 
=
-\frac{2\pi}{3}\, \langle Q_A \rangle_{\mu} \, M(\pi \ell/L) 
%\frac{\kappa (\mu_{+} + \mu_{-})}{6}\; L^2 M(\ell/L)
\eea
and
\bea
\label{flow-k-t-A-circle}
\mathcal{L}[\boldsymbol{k}]\big(\gamma(P_a \to P_b)\big) 
%\int_A \langle k_x (\tau;x,0) \rangle_{_{\textrm{\tiny $L,\mu_\pm$}}}  \rd x \,
&=&
\, \mathsf{W}_{L,k}\big|_{P_a} - \mathsf{W}_{L,k}\big|_{P_b} 
=
-\frac{2\pi}{3}\, \langle Q_A \rangle_{\mu} \, M(\pi \ell/L) 
\\
\label{flow-k-x-At-circle}
\rule{0pt}{.8cm}
\mathcal{L}[\boldsymbol{k}]\big(\gamma(P_{-\infty} \to P_{+\infty})\big) 
%\int_{\tilde{A}} \langle k_t (\tau; \tfrac{a+b}{2} ,t) \rangle_{_{\textrm{\tiny $L,\mu_\pm$}}}  \rd t \,
&=& 
\, \mathsf{W}_{L,k}\big|_{P_{-\infty}}  - \mathsf{W}_{L,k} \big|_{P_{+\infty}} 
=
-\frac{2\pi}{3}\, \langle \widetilde{Q}_A \rangle_{\mu} \, M(\pi \ell/L) 
%\frac{\kappa (\mu_{+} - \mu_{-})}{6} \; L^2 M(\ell/L)
\eea
where
\be
\label{M-func-def}
M(y) \equiv \frac{3}{y^2}\Big( 1 - y \cot(y) \Big)   \,.
\ee
Since $M(\pi \ell/L) \to 1$ as $L/\ell \to +\infty$,
the line integrals (\ref{flow-j-t-A-circle})-(\ref{flow-j-x-At-circle}) and (\ref{flow-k-t-A-circle})-(\ref{flow-k-x-At-circle}) 
become respectively  (\ref{flow-j-t-A})-(\ref{flow-j-x-At}) and (\ref{flow-k-t-A})-(\ref{flow-k-x-At}) in the infinite volume limit.
When $\mu_+ = \mu_-$, the line integrals in (\ref{flow-j-t-A-circle}) and (\ref{flow-k-x-At-circle}) vanish
(see also the left panel of Fig.\,\ref{fig:j-L-curr} and Fig.\,\ref{fig:k-L-curr} respectively).

The mean values of the energy currents 
$\langle \mathcal{J}_x (\tau;x,t) \rangle_{_{\textrm{\tiny $L,\mu$}}} $ 
and $\langle  \mathcal{J}_t (\tau;x,t) \rangle_{_{\textrm{\tiny $L,\mu$}}} $
for the two-dimensional CFT in $\mathbb{M}_{\diamond}$
provide the components of the vector field $\boldsymbol{\CJ} (x,t)$.
From the expressions of the operators in (\ref{curl-J-x-def})-(\ref{curl-J-t-def})
with $C_{\mathcal{J}} = -\,\pi c/6$ (see also in the text below (\ref{mc11t})),
the mean values (\ref{ftd1-L}) 
and the velocity $V_L(u)$ in (\ref{velocity_fund-circle})
characterising the representation and the bipartition we are considering,
for the mean values of these energy currents we find 
\bea
\label{mc11x-circle}
\langle \CJ_x (\tau;x,t) \rangle_{_{\textrm{\tiny $L,\mu$}}} 
&=&
 - \left[ \left(  \frac{\kappa \mu^2_+ }{4\pi} -\frac{\pi c}{12 L^2} \right)  V_L(u_+)^2 - \left(  \frac{\kappa \mu^2_- }{4\pi} -\frac{\pi c}{12 L^2} \right) V_L(u_-)^2 \,\right] 
 \eea
 and 
\bea
 \label{mc11t-circle-0} 
 \rule{0pt}{.7cm}
\langle \CJ_t (\tau;x,t) \rangle_{_{\textrm{\tiny $L,\mu$}}} 
& \equiv &
- \frac{\kappa}{4\pi}\, \Big\{ \mu^2_+ \,V_L(u_+)^2 + \mu^2_- \, V_L(u_-)^2 \Big\}
\nonumber
\\
\rule{0pt}{.55cm}
& &
- \, \frac{c}{24\pi}\, \Big\{ V_L(u_+)^2 \,\CV[V_L](u_+) + V_L(u_-)^2\,\CV[V_L](u_-) \Big\} 
-\frac{\pi \,c}{6}
\\
 \label{mc11t-circle} 
 \rule{0pt}{.8cm}
&=&
 - \left[ \left(  \frac{\kappa \mu^2_+ }{4\pi} -\frac{\pi c}{12 L^2} \right)  V_L(u_+)^2 + \left(  \frac{\kappa \mu^2_- }{4\pi} -\frac{\pi c}{12 L^2} \right) V_L(u_-)^2 \,\right] 
\eea
where the non trivial function of the spacetime position 
occurring in the second line of (\ref{mc11t-circle-0}) 
can be simplified by observing that
\be
\label{id-Nu-on-V-circle}
V_L(u)^2 \,\CV[V_L](u) \,=\,  -\, 2\pi^2 - \frac{2\pi^2}{L^2}\,  V_L(u)^2
\ee
which becomes constant in the infinite volume limit (see (\ref{id-Nu-on-V})).
%

%\begin{figure}[t!]
%\vspace{-.6cm}
%\hspace{-1.3cm}
%\includegraphics[width=1.2\textwidth]{fig-E-L-vec-field}
%\vspace{-.3cm}
%\caption{Vector fields for the mean values of the energy density currents  (\ref{ham-vec-E-0})  on the vertical cylinder,
%whose potential is (\ref{potentials-circle-W-energy}),
%for either equal (left panel) or different (right panel) chemical potentials,
%in the same setup of Fig.\,\ref{fig:j-L-curr}.
%}
%\label{fig:E-L-curr}
%\end{figure}

In a similar way, the mean values 
$\langle \widetilde{\mathcal{J}}_x (\tau;x,t) \rangle_{_{\textrm{\tiny $L,\mu$}}} $ 
and $\langle  \widetilde{\mathcal{J}}_t (\tau;x,t) \rangle_{_{\textrm{\tiny $L,\mu$}}} $
of the momentum currents define 
the components of the vector field $\boldsymbol{\widetilde{\CJ}} (x,t)$ in $\mathbb{M}_{\diamond}$.
From the expressions of the operators in (\ref{tilde-curl-J-x-def})-(\ref{tilde-curl-J-t-def}) with $\widetilde{C}_{\mathcal{J}} = 0$,
the mean values (\ref{ftd1-L}) and the velocity $V_L(u)$ in (\ref{velocity_fund-circle}),
for the mean values of the operators (\ref{tilde-curl-J-x-def}) and (\ref{tilde-curl-J-t-def}) with $\widetilde{C}_{\mathcal{J}} = 0$
in the finite density representation in $\mathbb{M}_{\diamond}$
we get respectively
\be
\label{mc11x-tilde-circle}
\langle \widetilde{\CJ}_x (\tau;x,t) \rangle_{_{\textrm{\tiny $L,\mu$}}}
\,=\,
 - \left[ \left(  \frac{\kappa \mu^2_+ }{4\pi} -\frac{\pi c}{12 L^2} \right)  V_L(u_+)^2 + \left(  \frac{\kappa \mu^2_- }{4\pi} -\frac{\pi c}{12 L^2} \right) V_L(u_-)^2 \,\right] 
\ee
and
\bea
 \label{mc11t-tilde-circle-0} 
 \rule{0pt}{.7cm}
\langle \widetilde{\CJ}_t (\tau;x,t) \rangle_{_{\textrm{\tiny $L,\mu$}}}  
& \equiv &
- \frac{\kappa}{4\pi}\, \Big\{ \mu^2_+ \,V_L(u_+)^2 - \mu^2_- \, V_L(u_-)^2 \Big\}
\nonumber
\\
\rule{0pt}{.55cm}
& &
- \, \frac{c}{24\pi}\, \Big\{ V_L(u_+)^2 \,\CV[V_L](u_+) - V_L(u_-)^2\,\CV[V_L](u_-) \Big\}  
\\
 \label{mc11t-tilde-circle} 
 \rule{0pt}{.8cm}
&=&
 - \left[ \left(  \frac{\kappa \mu^2_+ }{4\pi} -\frac{\pi c}{12 L^2} \right)  V_L(u_+)^2 - \left(  \frac{\kappa \mu^2_- }{4\pi} -\frac{\pi c}{12 L^2} \right) V_L(u_-)^2 \,\right] .
\eea
The additive constants in 
(\ref{mc11x-circle})-(\ref{mc11t-circle-0}) and (\ref{mc11x-tilde-circle})-(\ref{mc11t-tilde-circle-0})
have been fixed by imposing that the resulting expressions 
vanish at the vertices of the diamond $\mathcal{D}_A$.

\begin{figure}[t!]
\vspace{-.6cm}
\hspace{-1.7cm}
\includegraphics[width=1.23\textwidth]{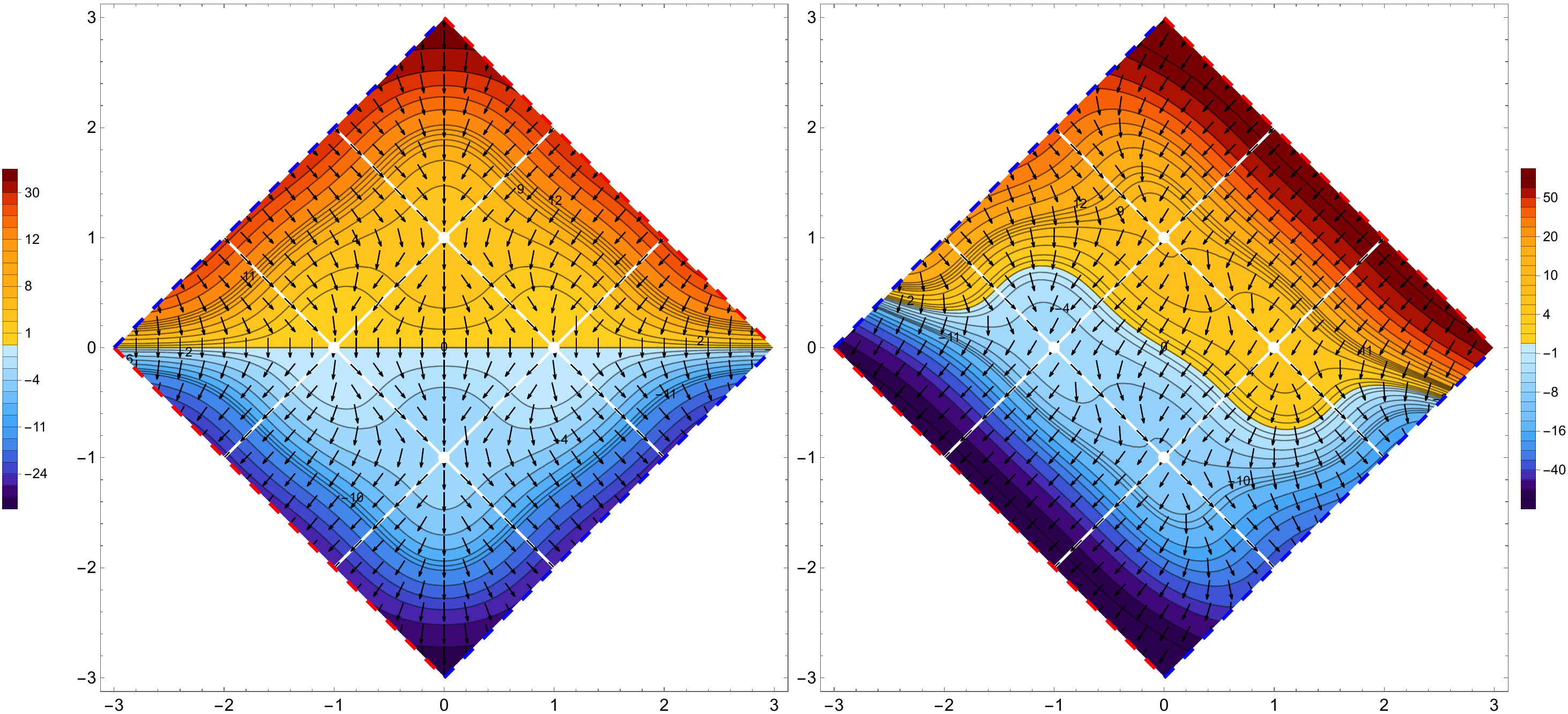}
\vspace{-.4cm}
\caption{Vector fields for the mean values of the energy density currents  (\ref{ham-vec-E-0}) in $\mathbb{M}_{\diamond}$,
whose potential is (\ref{potentials-circle-W-energy}),
for either equal (left panel) or different (right panel) chemical potentials,
in the same setup of Fig.\,\ref{fig:j-L-curr}.
}
\label{fig:E-L-curr}
\end{figure}

\begin{figure}[t!]
\vspace{-.6cm}
\hspace{-1.7cm}
\includegraphics[width=1.23\textwidth]{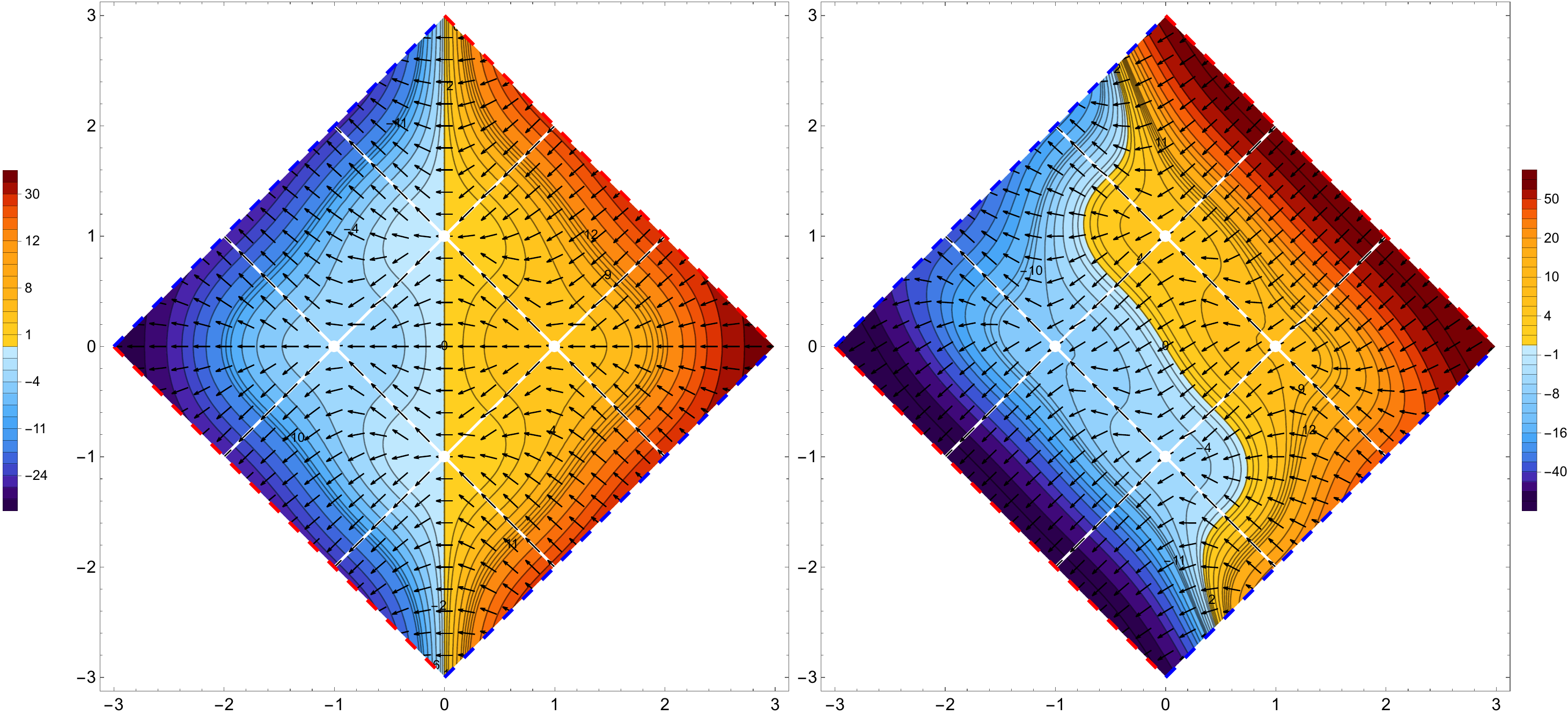}
\vspace{-.4cm}
\caption{Vector fields for the mean values of the momentum density currents  (\ref{ham-vec-tildeE-0})  in $\mathbb{M}_{\diamond}$,
whose potential is (\ref{potentials-circle-W-momentum}),
for either equal (left panel) or different (right panel) chemical potentials,
in the same setup of Fig.\,\ref{fig:j-L-curr}.
}
\label{fig:tE-L-curr}
\end{figure}

In Fig.\,\ref{fig:E-L-curr} and Fig.\,\ref{fig:tE-L-curr} we show
the vector fields $\boldsymbol{\CJ} (x,t)$ and $\boldsymbol{\widetilde{\CJ}} (x,t)$  in $\mathbb{M}_{\diamond}$
for the choice of the parameters described in the caption of Fig.\,\ref{fig:j-L-curr}.
These vector fields vanish at the same critical point,
which correspond to the vertices of $\mathcal{D}_A$.
All these isolated critical points have multiplicity $2$ and Poincar\'e index $0$.
Hence, the Poincar\'e-Hopf theorem can be checked also in these cases;
indeed, the sum of the Poinca\'re indices in $\mathbb{M}_{\diamond}$ is zero,
consistently with the fact that $\mathbb{M}_{\diamond}$ has the topology of the torus.

The vector fields $\boldsymbol{\CJ} (x,t)$ and $\boldsymbol{\widetilde{\CJ}} (x,t)$
are curl free and can be written respectively as 
\bea
\label{ham-vec-E-0}
& &
\langle \CJ_x (\tau;x,t) \rangle_{_{\textrm{\tiny $L,\mu$}}}  = -\,\partial_x \mathsf{W}_{L,\CJ}(x,t) 
\;\;\;\qquad \;\;\;
\langle \CJ_t (\tau;x,t) \rangle_{_{\textrm{\tiny $L,\mu$}}}   = -\,\partial_t \mathsf{W}_{L,\CJ}(x,t) 
\\
\rule{0pt}{.5cm}
\label{ham-vec-tildeE-0}
& &
\langle \widetilde{\CJ}_x (\tau;x,t) \rangle_{_{\textrm{\tiny $L,\mu$}}}  = -\,\partial_x \mathsf{W}_{L,\widetilde{\CJ}}(x,t) 
\;\;\;\qquad \;\;\;
\langle \widetilde{\CJ}_t (\tau;x,t) \rangle_{_{\textrm{\tiny $L,\mu$}}}   = -\,\partial_t \mathsf{W}_{L,\widetilde{\CJ}}(x,t) 
\eea
where the potentials $\mathsf{W}_{L,\CJ} $ and $\mathsf{W}_{L,\widetilde{\CJ}}$  read respectively
\bea
\label{potentials-circle-W-energy}
\mathsf{W}_{L,\CJ}(x,t) 
& \equiv & 
 \left(  \frac{\kappa \mu^2_+ }{4\pi} -\frac{\pi c}{12 L^2} \right)  G_L(u_+) - \left(  \frac{\kappa \mu^2_- }{4\pi} -\frac{\pi c}{12 L^2} \right) G_L(u_-) 
\\
\rule{0pt}{.7cm}
\label{potentials-circle-W-momentum}
\mathsf{W}_{L,\widetilde{\CJ}}(x,t) 
& \equiv & 
 \left(  \frac{\kappa \mu^2_+ }{4\pi} -\frac{\pi c}{12 L^2} \right)  G_L(u_+) + \left(  \frac{\kappa \mu^2_- }{4\pi} -\frac{\pi c}{12 L^2} \right) G_L(u_-) 
\eea
in terms of the function $G_L(u)$ defined as follows
\bea
\label{GL-def}
G_L(u)
& \equiv &
\frac{L^3}{2\pi \sin^2\!\big[\tfrac{\pi}{L}(b-a)\big]} \; 
\Bigg\{
\sin\!\left( \frac{2\pi (b-u)}{L} \right) 
-
\sin\!\left( \frac{2\pi (u-a)}{L} \right) 
\\
\rule{0pt}{.8cm}
& & \hspace{1.5cm}
+\,
 \frac{1}{4}  \sin\!\left( \frac{4\pi }{L} \!\left( u -\frac{a+b}{2}\right) \! \right)  
+
\frac{2\pi}{L}
\left[ 1+ \frac{1}{2} \, \cos\!\left( \frac{2\pi (b-a)}{L} \right) \right] \left( u -  \frac{a+b}{2} \right)
\Bigg\}
\nn
\eea
%\be
%\label{GL-def}
%G_L(u)
% \equiv 
%\frac{L^2}{\sin^2\!\big[\tfrac{\pi}{L}(b-a)\big]} \; 
%\left\{ 
%\,R_L(u)  
%+
%\left[ 1+ \frac{1}{2} \, \cos\!\left( \frac{2\pi (b-a)}{L} \right) \right] \left( u -  \frac{a+b}{2} \right)
%\right\}
%%\\
%%\label{GL-tilde-def}
%%\rule{0pt}{.7cm}
%%\widehat{G}_L(u)
%%& \equiv &
%%-\,\frac{2\pi^2}{\sin^2\!\big[\tfrac{\pi}{L}(b-a)\big]} \; 
%%\left\{ 
%%\,R_L(u)  
%%+
%%\frac{3}{2} \left( u -  \frac{a+b}{2} \right)
%%\right\}
%\ee
%where
%\be
%\label{RL-def}
%R_L(u)
%\equiv 
%\frac{L}{2\pi} \left[ \,
%\sin\!\left( \frac{2\pi (b-u)}{L} \right) 
%-
%\sin\!\left( \frac{2\pi (u-a)}{L} \right) 
% + 
% \frac{1}{4}  \sin\!\left( \frac{4\pi }{L} \!\left( u -\frac{a+b}{2}\right) \! \right)  
% \right]  .
%\ee
which satisfies $G_L\big(\frac{a+b}{2}\big)=0$; 
hence (\ref{potentials-circle-W-energy}) and (\ref{potentials-circle-W-momentum}) vanish in the center of the diamond $\mathcal{D}_A$.

Similarly to (\ref{potential-circle}) and (\ref{ham-vec-j-0-circle})-(\ref{ham-vec-k-0-circle}),
the function (\ref{GL-def}) is not periodic in $u$
and therefore the potentials (\ref{potentials-circle-W-energy}) and (\ref{potentials-circle-W-momentum})
are not well defined in the whole spacetime $\mathbb{M}_{\diamond}$
but only in a subset of $\mathbb{M}_{\diamond}$ where 
a neighbourhood of boundary made by the union of the dashed straight segments
in Fig.\,\ref{fig:E-L-curr} and Fig.\,\ref{fig:tE-L-curr} 
has been subtracted.

Consistency among the expressions reported  between (\ref{mc11x-circle}) and (\ref{GL-def}) 
occurs because (\ref{GL-def}) and $V_L(u)$ in (\ref{velocity_fund-circle}) are related as follows
\be
\label{der-GL-from-VLsq}
\partial_u G_L(u) = V_L(u)^2\,.
%\;\;\;\qquad\;\;\;
%\partial_u \widehat{G}_L(u) = V_L(u)^2 \, \CV[V_L](u)
\ee

From (\ref{id-GE-from-Vsq-line}), (\ref{VL-from-V}) and (\ref{der-GL-from-VLsq}), 
we observe that the functions in (\ref{potential-E}) and (\ref{GL-def}) are related as follows
\be
\partial_u G_L(u) = \frac{ \partial_u \widetilde{G}\big( \e^{2\pi \ri u/L}\big)}{\big[ \partial_u\big( \e^{2\pi \ri u/L}\big)\big]^3}
\ee
where $\widetilde{G}(u)$ is defined as (\ref{potential-E}) where $a$ and $b$ 
are replaced by $\e^{2\pi \ri a/L}$ and $\e^{2\pi \ri b/L}$ respectively. 
In the infinite volume limit $L \to +\infty$ of (\ref{GL-def}) gives (\ref{potential-E});
hence the potentials in
(\ref{potentials-circle-W-energy}) and (\ref{potentials-circle-W-momentum})
become respectively the first and the second potential
 in (\ref{potential-W-energy}) in this regime.

The fluxes of the vector fields $\boldsymbol{\CJ} (x,t)$ and $\boldsymbol{\widetilde{\CJ}} (x,t)$
through the straight white lines in Fig.\,\ref{fig:E-L-curr} and Fig.\,\ref{fig:tE-L-curr} vanish,
as we can show by observing that
the absolute value of the ratios of their components is equal to 1 along these lines. 
As already highlighted in Sec.\,\ref{sec-energy-momentum-trans} for the corresponding vector fields on the plane, 
also in this case this analytic result is not properly displayed  in the left panels
of Fig.\,\ref{fig:E-L-curr} and Fig.\,\ref{fig:tE-L-curr}
(see around the vertices of $\mathcal{D}_A$)
because of a failure in the graphical representation of the vector fields.
A similar issue occurs in both the panels of these figures at the vertices of $\mathcal{D}_A$, 
where arrows are displayed, despite the fact that these are critical points of the vector fields. 
Such failures do not occur for the vector fields $\boldsymbol{j} (x,t)$ and $\boldsymbol{k} (x,t)$  
in Fig.\,\ref{fig:j-L-curr} and Fig.\,\ref{fig:k-L-curr}, whose critical points have multiplicity 1.

%\textcolor{red}{\bf [In order to finalise the following part we need to clarify the issue of the mean values for the modular energy.
%Insights by Longo?]}

%The vanishing of these fluxes naturally leads us to consider
%the mean values of the total energy and of the total momentum in $\mathcal{D}_A$.
%In the finite density and finite volume representation, where (\ref{ftd1-L}) hold, 
%for the mean values of (\ref{toten}) and (\ref{totmom}) we find respectively
%\bea
%\label{EA-mean-circle}
% \mathsf{E}_{A,L} 
%&=&
%\left(  \frac{\kappa (\mu^2_{+} + \mu^2_{-})}{8\pi} - \frac{\pi c}{12 L^2} \right) \frac{\pi \ell^3}{3}\, M(\pi \ell/L) 
%\\
%\label{tEA-mean-circle}
%\rule{0pt}{.7cm}
% \widetilde{\mathsf{E}}_{A,L} 
%&=&
%\frac{\kappa (\mu^2_{+} - \mu^2_{-})}{8\pi}\;  \frac{\pi \ell^3}{3}\, M(\pi \ell/L) 
%\eea
%being $M(y)$ the function introduced in (\ref{M-func-def});
%hence these expressions become the ones introduced in (\ref{EA-tEA-mean}) in the infinite volume limit, as expected. 
%
%The observations reported in the text below (\ref{EA-tEA-mean}) about the additive constants in 
%the operators $K$ and $P$ can be adapted to (\ref{EA-mean-circle}) and (\ref{tEA-mean-circle}) 
%straightforwardly.

The line integrals of the curl free vector fields 
$\boldsymbol{\CJ} (x,t)$ and $\boldsymbol{\widetilde{\CJ}} (x,t)$
along curves anchored to the opposite vertices of $\mathcal{D}_A$
read
\bea
\label{flow-Jx-A-circle}
\mathcal{L}[\boldsymbol{\CJ}]\big(\gamma(P_a \to P_b)\big) 
%\int_A \langle \mathcal{J}_x (\tau;x,0) \rangle_{_{\textrm{\tiny $L,\mu_\pm$}}}   \rd x \,
&=&
\, \mathsf{W}_{L,\mathcal{J}}\big|_{P_a} - \mathsf{W}_{L,\mathcal{J}}\big|_{P_b} 
\,=\,
-\frac{4\pi}{5}\, \widetilde{\mathsf{E}}_{A,L} \;
% - \frac{\kappa (\mu^2_{+} - \mu^2_{-})}{8\pi}\; \ell^3 
\mathcal{M}(\pi\ell/L)
\\
\label{flow-Jt-At-circle}
\rule{0pt}{.6cm}
\mathcal{L}[\boldsymbol{\CJ}]\big(\gamma(P_{-\infty} \to P_{+\infty})\big) 
%\int_{\tilde{A}} \langle \mathcal{J}_t (\tau; \tfrac{a+b}{2} ,t) \rangle_{_{\textrm{\tiny $L,\mu_\pm$}}}  \rd t  \,
&=&
\, \mathsf{W}_{L,\mathcal{J}}\big|_{P_{-\infty}} - \mathsf{W}_{L,\mathcal{J}}\big|_{P_{+\infty}} 
=\,
-\frac{4\pi}{5}\,  \mathsf{E}_{A,L} \; \mathcal{M}(\pi\ell/L)
\eea
and
\bea
\label{flow-Jx-A-tilde-circle}
\mathcal{L}[\boldsymbol{\widetilde{\CJ}}]\big(\gamma(P_a \to P_b)\big) 
%\int_A \langle \widetilde{\CJ}_x (\tau;x,0) \rangle_{_{\textrm{\tiny $L,\mu_\pm$}}}   \rd x \,
&=&
\, \mathsf{W}_{L,\widetilde{\CJ}}\big|_{P_a} - \mathsf{W}_{L,\widetilde{\CJ}}\big|_{P_b} 
\,=\,
-\frac{4\pi}{5}\,  \mathsf{E}_{A,L} \;
%- \left(  \frac{\kappa (\mu^2_{+} + \mu^2_{-})}{8\pi} - \frac{\pi c}{12 L^2} \right) \ell^3 
\mathcal{M}(\pi\ell/L)
\\
\label{flow-Jt-At-tilde-circle}
\rule{0pt}{.6cm}
\mathcal{L}[\boldsymbol{\widetilde{\CJ}}]\big(\gamma(P_{-\infty} \to P_{+\infty})\big) 
%\int_{\tilde{A}} \langle \widetilde{\CJ}_t (\tau; \tfrac{a+b}{2} ,t) \rangle_{_{\textrm{\tiny $L,\mu_\pm$}}}   \rd t \,
&=&
\, \mathsf{W}_{L,\widetilde{\CJ}}\big|_{P_{-\infty}} - \mathsf{W}_{L,\widetilde{\CJ}}\big|_{P_{+\infty}} 
=\,
-\frac{4\pi}{5}\,  \widetilde{\mathsf{E}}_{A,L} \;
%- \frac{\kappa (\mu^2_{+} - \mu^2_{-})}{8\pi}\;\ell^3 
\mathcal{M}(\pi\ell/L)
%\hspace{1cm}
\eea
where we have introduced
\bea
\label{EA-mean-circle}
 \mathsf{E}_{A,L} 
&\equiv &
\left(  \frac{\kappa (\mu^2_{+} + \mu^2_{-})}{8\pi} - \frac{\pi c}{12 L^2} \right) \frac{\pi \ell^3}{3}\, M(\pi \ell/L) 
\\
\label{tEA-mean-circle}
\rule{0pt}{.7cm}
 \widetilde{\mathsf{E}}_{A,L} 
&\equiv &
\frac{\kappa (\mu^2_{+} - \mu^2_{-})}{8\pi}\;  \frac{\pi \ell^3}{3}\, M(\pi \ell/L) 
\eea
in terms of the function (\ref{M-func-def})
(hence these expressions become the ones defined in (\ref{EA-tEA-mean}) in the infinite volume limit, as expected)
and 
\be
\mathcal{M}(y) \equiv 
\frac{5\big[ 3 \cot(y) +y\big( 2-3 \csc(y)^2 \big) \big] }{ 4\, y\big[ y\cot(y) -1 \big] }   \;.
\ee
From  (\ref{ftd1-L}), 
we notice that (\ref{EA-mean-circle}) and (\ref{tEA-mean-circle}) provide
the mean values of the total energy (\ref{toten}) and of the total momentum (\ref{totmom}) in $\mathcal{D}_A$
in the finite density and finite volume representation
when  $f_{+}(u_+) = f_{-}(u_-) = 0$.
Instead, from (\ref{KA-def-u-pm-circle}),
for the mean values of (\ref{toten}) and (\ref{totmom})  
we have  $\langle E_A \rangle_{_{\textrm{\tiny $L,\mu$}}} = \langle \widetilde{E}_A \rangle_{_{\textrm{\tiny $L,\mu$}}} = 0$.

Since $ \mathcal{M}(\pi\ell/L) \to 1 $ in the infinite volume limit, 
 the line integrals (\ref{flow-Jx-A-circle})-(\ref{flow-Jt-At-circle}) and (\ref{flow-Jx-A-tilde-circle})-(\ref{flow-Jt-At-tilde-circle})
become respectively (\ref{flow-Jx-A})-(\ref{flow-Jt-At}) and (\ref{flow-Jx-A-tilde})-(\ref{flow-Jt-At-tilde}) in this regime. 
When $\mu_+ = \mu_-$, the integrals in  (\ref{flow-Jt-At-circle}) and (\ref{flow-Jt-At-tilde-circle}) vanish,
as one can realise also from the left panel of Fig.\,\ref{fig:E-L-curr} and Fig.\,\ref{fig:tE-L-curr} respectively.

By using the correlators in Sec.\,\ref{sec-rep-finite-density-circle},
also at finite volume we can introduce the modular noise power 
at frequency $\omega$ and in the spacetime point $(x,t) \in \mathbb{M}_{\diamond}$
generated by the various quantities as done in Sec.\,\ref{subsec-noise}
for the Minkowski spacetime. 
In $\mathbb{M}_{\diamond}$ one introduces
$P_{L,j}(\omega; x,t)$ for the charge current, 
$P_{L,k}(\omega; x,t)$ for the helicity current, 
$P_{L,\CJ}(\omega; x,t)$ for the energy current, 
$P_{L,\widetilde{\CJ}}(\omega; x,t)$ for the momentum current,
$P_{L,\varrho}(\omega; x,t)$ for the charge density 
and $P_{L,\chi}(\omega; x,t)$ for the density (\ref{hden}).
Since in the computation of these quantities the coincident points limit must be considered, 
the only difference with respect to the expressions reported in Sec.\,\ref{subsec-noise}
is due to the fact that $V(u)$ must be replaced by $V_L(u)$ in (\ref{velocity_fund-circle}).
Thus, $P_{L,j}(\omega; x,t) = P_{j}(\omega; x,t)$,  $P_{L,k}(\omega; x,t) = P_{k}(\omega; x,t)$, 
$P_{L,\CJ}(\omega; x,t) = P_{\CJ}(\omega; x,t)$ and $P_{L,\widetilde{\CJ}}(\omega; x,t) = P_{\widetilde{\CJ}}(\omega; x,t)$,
while in $P_{L,\varrho}(\omega; x,t)$ and $P_{L,\chi}(\omega; x,t)$
do not coincide with $P_{\varrho}(\omega; x,t)$ and $P_{\chi}(\omega; x,t)$ respectively
because the velocity explicitly occurs in their expressions (see (\ref{no11})).

These observations lead to introduce 
$\mathcal{A}_L[\mathcal{O}](\omega; x,t)$ and $\mathcal{C}_L[\mathcal{O}](\omega; x,t)$
as the r.h.s. of (\ref{anti-comm-modular-def}) and (\ref{comm-modular-def}) respectively
with $ \langle \dots \rangle_{\mu}^{\textrm{\tiny con}}$ replaced by 
$ \langle \dots \rangle^{\textrm{\tiny con}}_{_{\textrm{\tiny $L,\mu$}}}$,
finding that they satisfy the following modular fluctuation-dissipation relation
\be
\mathcal{A}_L[\mathcal{O}](\omega; x,t)
\,=\,
\coth \! \left( \frac{\omega}{2}\right)\,
\mathcal{C}_L[\mathcal{O}](\omega; x,t)
\ee
which encodes the fact that the modular evolution has a thermal nature 
with inverse temperature $\tilde \beta =1$,
in agreement with the KMS condition for the modular correlators
(see Sec.\,\ref{sec-mod-corr-circle}).

Finally, we emphasise that the heuristic picture for the transport described in the final part of Sec.\,\ref{sec-charge-helicity-trans}
can be adapted to the finite volume case in a straightforward way.

%%%%%%%%%%%%%%%%%%%%%%%%%%%%%%%%%%%%%%%%%%%
%\newpage
\section{Conclusions}
\label{sec_conclusions}
%%%%%%%%%%%%%%%%%%%%%%%%%%%%%%%%%%%%%%%%%%%

%\textcolor{red}{\bf [citare \cite{Fraenkel:2023abm}]}

We investigated the modular quantum transport 
in a two-dimensional CFT at finite density and zero temperature
for the bipartition given by an interval
either on the line or on the circle.

The modular flows of the operators that we have considered
(primaries, currents and energy-momentum tensor) 
are generated by modular Hamiltonians 
which depend also on the chemical potentials
(see (\ref{me3b}), (\ref{fmh}) and (\ref{fmh-circle})) 
\cite{Wong:2013gua, Mintchev:2022xqh}.
Their explicit expressions can be written by specialising
(\ref{me9a}), (\ref{me9b}), (\ref{cp4}) and (\ref{met4})
to the modular evolutions corresponding to 
(\ref{xi-map-fund}) for the interval on the infinite line 
and to (\ref{xi-map-fund-circle}) for the interval on the circle,
as discussed in Sec.\,\ref{subsec-mod-H-vacuum} 
and Sec.\,\ref{sec-mod-ham-circle} respectively. 
From these modular flows, we have found modular continuity equations 
(see (\ref{ecr-cov-form1}), (\ref{hel1}), (\ref{en-cons-2}) and (\ref{tilde-en-cons-2})
specialised to $V_{+}(u) = V_{-}(u)$, which is equal to 
(\ref{velocity_fund}) for the interval on the line
and to (\ref{velocity_fund-circle}) for the interval on the circle)
and the corresponding conserved quantities along the modular evolution
(see (\ref{echarge}), (\ref{h-charge-def}), (\ref{toten}), (\ref{totmom}) 
and also (\ref{heat4}) for the modular heat, in the special cases just mentioned),
where the dependence on the representation occurs
through the expression of the velocities. 

In the finite density representations, either on the line or on the circle, 
the mean values of the modular currents that we have introduced 
naturally provide two-dimensional curl free vector fields
that describe the modular quantum transport in the spacetime
(see Sec.\,\ref{sec-charge-helicity-trans} and Sec.\,\ref{sec-energy-momentum-trans} 
for the CFT on the line and in Sec.\,\ref{sec-noise-circle} for the CFT on the circle).
This modular quantum transport  
is different from the one discussed in  \cite{Czech:2017zfq, Czech:2019vih, Banerjee:2022jnv},
based on the Berry phase.

Finally, we have investigated the modular quantum noise power generated by various currents
for a CFT either on the line or on the circle 
(see Sec.\,\ref{subsec-noise} and Sec.\,\ref{sec-noise-circle} respectively).
A modular Johnson-Nyquist law (see (\ref{no5}), which holds also at finite volume)
for the modular noise power generated by the charge current
and the modular analogue of the fluctuation-dissipation relation are obtained. 
These results confirm the thermal nature of the modular evolution with inverse temperature $\tilde{\beta} = 1$,
in agreement with the KMS condition 
for the modular correlators (see Sec.\,\ref{sec-mod-corr-line} and Sec.\,\ref{sec-mod-corr-circle}).
While the modular noise power generated by the charge or by the helicity current
contains the coefficient occurring in the central term in (\ref{cft7}),
the modular noise power generated by the energy or by the momentum current 
contains the central charge of the CFT. 
Furthermore, while the modular noise power generated by these currents 
is independent of the spacetime position (see (\ref{no5}) and (\ref{no9})), 
the modular noise power generated by the charge density 
depends on the position in the spacetime (see (\ref{no13})).

The results discussed in this manuscript can be extended in various directions. 
In two spacetime dimensions, it is worth studying the modular transport properties 
for CFT at finite temperature,
for either bi-local or non-local modular Hamiltonians 
\cite{Casini:2009vk, Longo:2009mn, Arias:2018tmw, Blanco:2019xwi, Fries:2019ozf, Bostelmann:2022yvj, Abate:2022dyw,Abate:2023ldj},
including the ones corresponding to systems with boundaries or defects \cite{Mintchev:2020uom, Mintchev:2020jhc, Rottoli:2022plr},
for modular Hamiltonians in inhomogeneous systems \cite{Tonni:2017jom}
and in non-relativistic field theories \cite{Spitzer-14, Mintchev:2022xqh, Mintchev:2022yuo}.
It is relevant to explore the modular quantum transport also in higher dimensional quantum field theories
 \cite{Hislop:1981uh, Casini:2011kv}
and to investigate possible gravitational duals of these results, maybe by using \cite{Ryu:2006bv, Ryu:2006ef, Hubeny:2007xt}.
Another direction concerns the connections between our results and the previous analyses in inhomogeneous CFT
\cite{Ramirez:2015yfa, Allegra:2015hxc, Dubail:2016tsc, Gawedzki:2017woc, Langmann:2017dyy, Langmann:2018skr, Moosavi:2019fas, Fan:2019upv, Fan:2020orx, deBoer:2023lrd}.
Finally, it is worth exploring the modular evolution and the corresponding transport properties
also in lattice models \cite{peschel-03,Casini:2009sr, EislerPeschel:2009review, Arias:2016nip,Eisler:2017cqi, Eisler:2018aaa,
Eisler:2019rnr, DiGiulio:2019cxv, Eisler:2020lyn, Eisler:2022rnp, Javerzat:2021hxt}.

%\noindent
%$\bullet$ \textcolor{red}{\bf Further future directions?}
%\\
%{\color{blue}
%$\longrightarrow$  non-abelian current algebra in 2D CFT \cite{Knizhnik:1984nr}
%\\
%$\longrightarrow$ modular inequalities? E.g. \cite{Blanco:2017akw}, ANEC papers, etc.
%}

\vskip 20pt 
\centerline{\bf Acknowledgments} 
\vskip 5pt

We are grateful to Bruno Bertini, Andrea Gambassi, Stefano Lepri, Per Moosavi, Domenico Seminara
and especially Stefan Hollands, Antonio Lerario, Hong Liu and Roberto Longo
for useful discussions or correspondence. 
ET acknowledges
Center for Theoretical Physics at MIT (Boston), University of Florence and NORDITA (Stockholm) 
for warm hospitality and financial support during part of this work. 
DP has been supported by ERC under Consolidator grant number 771536 (NEMO)
during his last year in SISSA.

\vskip 20pt

%%%%%%%%%%%%%%%%%%%%%%%%%%%%%%%%%%%%%%%%%%%%%%%%%%%%%
%\newpage
\appendix

%%%%%%%%%%%%%%%%%%%%%%%%%%%%%%%%%%%%%%%%%%%%%%%%%%%%%
%\newpage
\section{Basic correlators in the fundamental representation}
\label{app-comm-checks}
%%%%%%%%%%%%%%%%%%%%%%%%%%%%%%%%%%%%%%%%%%%%%%%%%%%%%

In this appendix we summarise the one-point functions and the two-point functions of the chiral fields on the line
considered in Sec.\,\ref{subsec-commutators} in the fundamental representation. 

A characteristic feature of this representation is the vanishing of the one-point functions
\be 
\langle \phi_\pm(u)\rangle_{_{0}} = \langle j_\pm(u)\rangle_{_{0}}  = \langle T_\pm(u)\rangle_{_{0}} =  0
\label{App1}
\ee
hence the corresponding two-point functions coincide with their connected parts. 
By using the normalisation adopted in \cite{Rehren:1987iy, Hollands:2019hje}, by conformal invariance, for the primaries we have
\be 
\langle \phi_\pm^*(u) \, \phi_\pm (v)\rangle_{_{0}}
=
\langle \phi_\pm(u) \, \phi_\pm^* (v)\rangle_{_{0}} 
= 
\frac{ \e^{\mp \ri \pi h_\pm}  }{2\pi\, (u-v \mp \ri \varepsilon )^{2h_\pm}}   
\,.
\label{App2}
\ee 

The normalisation of the two-point functions of $j_\pm(u)$ and $T_\pm(u)$ follow from (\ref{cft7}) and 
(\ref{cft1}) respectively. 
This can be shown by considering the following well known distribution 
\be
\label{distrib-0-def}
\frac{1}{u\pm \ri \varepsilon} = \frac{1}{u} \mp \ri \pi \delta (u) 
\ee
and the ones obtained by taking its first, second and third derivative, 
that read respectively
\be
\label{distrib-0-deriv}
\frac{1}{(u\pm \ri \varepsilon)^2} = \frac{1}{u^2} \pm \ri \pi \,\delta^\prime (u) 
\;\qquad\;
\frac{1}{(u\pm \ri \varepsilon)^3} = \frac{1}{u^3} \mp \frac{\ri \pi}{2} \,\delta^{\prime \prime} (u) 
\;\qquad\;
\frac{1}{(u\pm \ri \varepsilon)^4} = \frac{1}{u^4} \pm \frac{\ri \pi}{6} \,\delta^{\prime \prime \prime} (u)   \,.
\ee
By conformal invariance \cite{luschernotes}, we have that
\be 
\langle j_\pm(x_1 \pm t_1) \, j_\pm (x_2\pm t_2)\rangle_{_{0}}
\,=\, 
\frac{C_{j_\pm}}{(x_{12} \pm t_{12} \mp \ri \varepsilon )^2} 
\label{f2}
\ee
where we remind that $x_{12} \equiv x_1 - x_2$ and $t_{12} \equiv t_1 - t_2$. 
In order to fix $C_{j_\pm}$, 
from (\ref{f2}) we evaluate  the expectation value of the commutator 
\bea 
\label{com-j-app-test}
\langle \big[\,  j_\pm(x_1\pm t)\, , \,   j_\pm (x_2\pm t) \, \big]\rangle_{_{0}}
&=&
C_{j_\pm} \! \left [\frac{1}{(x_{12} \mp \ri \varepsilon )^2} - \frac{1}{(x_{21} \mp \ri \varepsilon )^2} \right ]
\\
\rule{0pt}{.6cm}
&=&
C_{j_\pm} \! \left [ \frac{1}{(x_{12} \mp \ri \varepsilon )^2} - 
\frac{1}{(x_{12} \pm \ri \varepsilon )^2} \right ]
\,=\, 
\mp \, 2 \pi \ri\, C_{j_\pm}\, \delta^{\prime} (x_{12})
\nonumber
\eea 
where the first distribution in (\ref{distrib-0-deriv}) has been employed. 
Comparing this result with the expectation value of (\ref{cft7}), one finds 
\be 
C_{j_\pm} = \frac{\kappa}{4\pi^2}  \;.
\label{App3}
\ee
Hence, the positivity of (\ref{f2}) implies $\kappa \geqslant 0$. 

A similar analysis can be performed for $T_\pm (u)$. In this case the two-point function is
\be 
\langle \,T_\pm(x_1 \pm t_1) \,T_\pm (x_2\pm t_2) \, \rangle_{_{0}}
= 
 \frac{C_{T_\pm}}{(x_{12} \pm t_{12} \mp \ri \varepsilon )^4} 
\label{f3}
\ee 
where the constant $C_{T_\pm}$ has to be fixed. 
From (\ref{f3}), one obtains 
\bea 
\label{com-T-app-test}
\langle \big[\, T_\pm(x_1\pm t)\, ,\,  T_\pm (x_2\pm t)\, \big]\rangle_{_{0}}
&=&
C_{T_\pm}\!\left [\frac{1}{(x_{12} \mp \ri \varepsilon )^4} - 
\frac{1}{(x_{21} \mp \ri \varepsilon )^4}\right ]
\\
\rule{0pt}{.6cm}
&=&
C_{T_\pm} \! \left [ \frac{1}{(x_{12} \mp \ri \varepsilon )^4} - 
\frac{1}{(x_{12} \pm \ri \varepsilon )^4} \right ]= 
\mp \, \frac{\ri \pi}{3} \, C_{T_\pm} \delta^{\prime \prime \prime} (x_{12})
\nonumber 
\eea
which can be  compared with the expectation value of (\ref{cft1}), finding that consistency leads to
\be 
C_{T_\pm} = \frac{c}{8\pi^2}  \;.
\label{App4}
\ee
In this case, positivity of (\ref{f3}) implies the well known constraint $c\geqslant 0$ for unitary CFT.

%%%%%%%%%%%%%%%%%%%%%%%%%%%%%%%%%%%%%%%%%%%%%%%%%%%%%
%\newpage
\section{Currents involving the chiral primaries}
\label{app-current-phi}
%%%%%%%%%%%%%%%%%%%%%%%%%%%%%%%%%%%%%%%%%%%%%%%%%%%%%

In this appendix we apply to the chiral primaries
the analyses discussed in Sec.\,\ref{subsec-charge-continuity} and Sec.\,\ref{subsec-energy-continuity}
about the continuity equations and the conservation laws involving
the electric charge, the helicity, the energy and the momentum.

Multiplying (\ref{mem4}) by $V_\pm(u_\pm)^{h_\pm -1}$ first 
and then taking either the sum or the difference of the resulting equations, 
we obtain
\bea
\label{ecr-0-ph-plus}
\partial_\tau \Phi(\tau;x,t) 
&=&
\big( \partial_{u_+}  + \ri\, \mu_+ \big)\big[ \,V_+(u_+)^{h_+}\, \phi_+(\tau,u_+) \,\big] 
%+ \ri\, \mu_+ \, V_+(u_+)^{h_+}\, \phi_+(\tau,u_+) 
-\big( \partial_{u_-}  - \ri\, \mu_- \big) \big[ \,V_-(u_-)^{h_-}\, \phi_-(\tau,u_-) \,\big] 
%+ \ri\, \mu_- \, V_-(u_-)^{h_-}\, \phi_-(\tau,u_-) 
\nonumber \\
& &
\\
\label{ecr-0-ph-minus}
\partial_\tau \widetilde{\Phi}(\tau;x,t) 
&=&
\big( \partial_{u_+}  + \ri\, \mu_+ \big)\big[ \,V_+(u_+)^{h_+}\, \phi_+(\tau,u_+) \,\big] 
%+ \ri\, \mu_+ \, V_+(u_+)^{h_+}\, \phi_+(\tau,u_+) 
 + \big( \partial_{u_-}  - \ri\, \mu_- \big) \big[ \,V_-(u_-)^{h_-}\, \phi_-(\tau,u_-) \,\big] 
%+ \ri\, \mu_- \, V_-(u_-)^{h_-}\, \phi_-(\tau,u_-) 
\nonumber \\
& &
\eea
where we have introduced
\bea
\label{Phi-def-plus}
\Phi(\tau;x,t)  & \equiv & V_+(u_+)^{h_+ -1} \,\phi_+(\tau, u_+) + V_-(u_-)^{h_- -1} \,\phi_-(\tau, u_-)
\\
\label{Phi-def-minus}
\widetilde{\Phi}(\tau;x,t)  & \equiv & V_+(u_+)^{h_+ -1} \,\phi_+(\tau, u_+) - V_-(u_-)^{h_- -1} \,\phi_-(\tau, u_-)  \,.
\eea

%\textcolor{red}{\bf [why do we need to impose $\mu_\pm =0$ to get a continuity equation?]}
%When $\mu_\pm =0$, 
%\be
%\partial_\tau \Phi(\tau;x,t) 
%\,=\,
%\partial_{x}  \big[ \,V_+(u_+)^{h_+}\, \phi_+(\tau,u_+) - V_-(u_-)^{h_-}\, \phi_-(\tau,u_-) \,\big] 
%\ee

By using $\partial_{u_\pm} = \tfrac{1}{2} \big( \partial_{x} \pm \partial_t\big)$ 
and renaming (\ref{heat2}) as  $(\mu_x, \mu_t) \equiv (\mu_e , \mu_h)$,
%as $\mu_\pm \equiv \tfrac{1}{2}\big(\mu_x \pm \mu_t \big)$, 
%\textcolor{red}{\bf [uniformize the notation with (\ref{heat2})]}
one finds that (\ref{ecr-0-ph-plus}) and (\ref{ecr-0-ph-minus})
can be written respectively as the following continuity equations
\be
\label{eom-Psi-Psi}
\partial_\tau \Phi(\tau;x,t)  = \big( \partial_\nu +\ri \, \mu_\nu \big) \Psi^\nu(\tau;x,t) 
\;\;\;\qquad\;\;\;
\partial_\tau \widetilde{\Phi}(\tau;x,t)  = \big( \partial_\nu +\ri \, \mu_\nu \big) \widetilde{\Psi}^\nu(\tau;x,t) 
\ee
where we have introduced
\bea
\Psi^x(\tau;x,t) &\equiv & \frac{1}{2} \big[\,V_+(u_+)^{h_+}\, \phi_+(\tau,u_+) - V_-(u_-)^{h_-}\, \phi_-(\tau,u_-) \,\big]
\\
\rule{0pt}{.65cm}
\Psi^t(\tau;x,t) &\equiv & \frac{1}{2} \big[\,V_+(u_+)^{h_+}\, \phi_+(\tau,u_+) + V_-(u_-)^{h_-}\, \phi_-(\tau,u_-) \,\big]
\eea
and 
\be
\widetilde{\Psi}^x(\tau;x,t)  \equiv  \Psi^t(\tau;x,t) 
\;\;\;\qquad\;\;\;
\widetilde{\Psi}^t(\tau;x,t)  \equiv  \Psi^x(\tau;x,t)  \,.
\ee
The differential equations in (\ref{eom-Psi-Psi}) can be expressed more conveniently in the following form
\bea
\label{eom-Psi-Psi-exp-0}
\partial_{\tau} \big[ \e^{\ri \mu_{\alpha}x^{\alpha}}\Phi(\tau;x,t) \big]
&=& 
\partial_{\nu}\big[ \e^{\ri \mu_{\alpha}x^{\alpha}}\Psi^{\nu}(\tau;x,t)\big]
\\
\label{eom-Psi-Psi-exp-1}
\partial_{\tau} \big[ \e^{\ri \mu_{\alpha}x^{\alpha}} \widetilde{\Phi}(\tau;x,t) \big]
&=& 
\partial_{\nu}\big[ \e^{\ri \mu_{\alpha}x^{\alpha}} \widetilde{\Psi}^{\nu}(\tau;x,t)\big]
\eea
where $\mu_{\alpha}x^{\alpha} = \mu_{x}x+\mu_{t}t = \mu_{+}u_{+}+\mu_{-}u_{-}\,$.
Considering (\ref{eom-Psi-Psi-exp-0}), its r.h.s. can be written as 
\bea
\partial_{\nu}\left[e^{i\mu_{\alpha}x^{\alpha}}\Psi^{\nu}(\tau;x,t)\right] 
= & &
\\
& & \hspace{-3cm}
 =\,
\e^{\ri \mu_{-}u_{-}} \, \partial_{u_{+}}
\big[ \e^{i\mu_{+}u_{+}} V_{+}(u_{+})^{h_{+}}\phi_{+}(\tau,u_{+})\big]
-
\e^{\ri \mu_{+}u_{+}} \, \partial_{u_{-}}
\big[ \e^{\ri \mu_{-}u_{-}}V_{-}(u_{-})^{h_{-}}\phi_{-}(\tau,u_{-})\big] 
\nn
\eea
hence 
\begin{eqnarray}
\int_{\mathcal{D}_{A}} \!\!
\partial_{\nu}\big[ \e^{\ri \mu_{\alpha}x^{\alpha}} \Psi^{\nu}(\tau;x,t)\big]\rd x \,\rd t 
& = & 
C_{a,b}(\mu_{-}) 
\left(\,\int_{a}^{b}\partial_{u_{+}} \big[ \e^{\ri \mu_{+}u_{+}}V_{+}(u_{+})^{h_{+}}\phi_{+}(\tau,u_{+})\big] du_{+}\right)
\hspace{1.4cm}
\\
 &   & 
+\; C_{a,b}(\mu_{+}) 
 \left(\,\int_{a}^{b}\partial_{u_{-}}\big[\e^{\ri \mu_{-}u_{-}}V_{-}(u_{-})^{h_{-}}\phi_{-}(\tau,u_{-})\big]
 du_{-}\right) 
 =0
 \nonumber
\end{eqnarray}
where we have introduced
\be
C_{a,b}(\mu) 
\equiv 
\frac{\e^{- \ri \mu a} - \e^{- \ri \mu b} }{2\mu}
% - \frac{\cos(\mu b) - \cos(\mu a)}{2\mu} + \ri \, \frac{\sin(\mu b) - \sin(\mu a)}{2\mu}
\ee
and (\ref{cond}) has been used. 
This result and the corresponding one obtained by performing a similar analysis for (\ref{eom-Psi-Psi-exp-1})
imply that 
\bea
\label{phi-charge}
F_{A} 
\equiv 
\int_{\mathcal{D}_{A}} \!\! \e^{\ri \mu_{\alpha}x^{\alpha}}\Phi(\tau;x,t) \, \rd x\, \rd t
&=&
C_{a,b}(\mu_{-}) 
\int_a^b \e^{\ri \mu_{+}u_{+}} V_+(u_+)^{h_+ -1 }\, \phi_+(\tau,u_+) \, \rd u_+ 
\\
& &
+ \, C_{a,b}(\mu_{+}) 
\int_a^b \e^{\ri \mu_{-}u_{-}} V_-(u_-)^{h_- - 1}\, \phi_-(\tau,u_-) \, \rd u_-
\hspace{1cm}
\nonumber 
\eea
and
\bea
\label{phi-charge-tilde}
\widetilde{F}_{A} 
\equiv 
\int_{\mathcal{D}_{A}} \!\! \e^{\ri \mu_{\alpha}x^{\alpha}} \,\widetilde{\Phi}(\tau;x,t) \, \rd x\, \rd t
&=&
C_{a,b}(\mu_{-}) 
\int_a^b \e^{\ri \mu_{+}u_{+}} V_+(u_+)^{h_+ -1 }\, \phi_+(\tau,u_+) \, \rd u_+ 
\\
& &
- \, C_{a,b}(\mu_{+}) 
\int_a^b \e^{\ri \mu_{-}u_{-}} V_-(u_-)^{h_- - 1}\, \phi_-(\tau,u_-) \, \rd u_-
\hspace{1cm}
\nonumber 
\eea
are conserved, i.e. independent of $\tau$.

%%%%%%%%%%%%%%%%%%%%%%%%%%%%%%%%%%%%%%%%%%%%%%%%%%%%%
%\newpage
\section{Representations and automorphisms}
\label{app-reps}
%%%%%%%%%%%%%%%%%%%%%%%%%%%%%%%%%%%%%%%%%%%%%%%%%%%%%

In this appendix we describe the construction of the finite density representation of a chiral 
CFT on either the line $\mathbb R$ or on the circle $\mathbb S$ 
(see Sec.\,\ref{sec-rep-finite-density} and Sec.\,\ref{sec-rep-finite-density-circle} respectively)
using a specific automorphism
$\gamma_\mu \equiv \gamma_{\mu_{+}} \otimes \gamma_{\mu_{-}}$ 
and the corresponding fundamental representation. 

We begin by considering the line, where 
the automorphism $\gamma_\mu \equiv \gamma_{\mu_{+}} \otimes \gamma_{\mu_{-}}$ is defined as follows
\cite{Araki:1977iz, Liguori:1999tw, Camassa:2011wk, Bernard:2013aru, Hollands:2016svy, Bernard:2016nci, Moosavi:2019fas}
\be
\label{auto-gamma-def}
\gamma_{\mu_{\pm}} \! ( \mathcal{O}_\pm )
\,\equiv\,
\mathrm{e}^{\mathrm{i}D_{\mu_{\pm}}}\,  \mathcal{O}_\pm\, \mathrm{e}^{-\mathrm{i}D_{\mu_{\pm}}}
\ee
with 
\be
\label{D-operator-shift}
D_{\mu_{\pm}}
\,\equiv\, 
\pm \,\mu_{\pm} \! \int_{-\infty}^\infty \! u\,j_{\pm}(u)\, \rd u  \,.
\ee 
By applying (\ref{auto-gamma-def}) to the fields $\phi_\pm$, $j_\pm$ and $T_\pm$, we find respectively
\begin{eqnarray}
\label{gamma-pm-phi}
\gamma_{\mu_{\pm}} \! \big(\phi_{\pm}(u)\big)
& = & 
\e^{\mp \ri\mu_{\pm}u} \, \phi_{\pm}(u)
\\
\label{gamma-pm-j}
\rule{0pt}{.7cm}
\gamma_{\mu_{\pm}}\! \big( j_{\pm}(u)\big)
& = & 
j_{\pm}(u)-\frac{\kappa\mu_{\pm}}{2\pi}
\\
\label{gamma-pm-T}
\rule{0pt}{.6cm}
\gamma_{\mu_{\pm}} \! \big( T_{\pm}(u)\big)
& = & 
T_{\pm}(u)+\mu_{\pm} \, j_{\pm}(u)-\frac{\kappa\mu_{\pm}^{2}}{4\pi}  \;.
\end{eqnarray}
Thus, for (\ref{cal-T-pm-def}) we obtain
\be
\label{gamma-pm-cal-T}
\gamma_{\mu_{\pm}} \! \big(\mathcal{T}_{\pm}(u)\big)
\,=\,
%\gamma_{\mu_{\pm}}\left[T_{\pm}(x)-\mu_{\pm}j_{\pm}(x)\right]
%\,=\, 
\gamma_{\mu_{\pm}}\! \big(T_{\pm}(u)\big) - \mu_{\pm} \, \gamma_{\mu_{\pm}}\! \big( j_{\pm}(u)\big)
%\nonumber \\
% & = & 
% T_{\pm}(x)+\mu_{\pm}j_{\pm}(x)-\frac{\kappa\mu_{\pm}^{2}}{4\pi}-\mu_{\pm}\left(j_{\pm}(x)-\frac{\kappa\mu_{\pm}}{2\pi}\right)
 \,=\, T_{\pm}(u)+\frac{\kappa\mu_{\pm}^{2}}{4\pi}
\ee
where in the r.h.s. $T_{\pm}(u)$ occurs. 
The inverse of (\ref{gamma-pm-phi}), (\ref{gamma-pm-j}), (\ref{gamma-pm-T}) and (\ref{gamma-pm-cal-T})
read respectively
\begin{eqnarray}
\gamma_{\mu_{\pm}}^{-1}  \big(\phi_{\pm}(u)\big)
& = & 
\e^{\pm \ri\mu_{\pm} u}\phi_{\pm}(u)
\\
\rule{0pt}{.7cm}
\gamma_{\mu_{\pm}}^{-1} \big( j_{\pm}(u)\big)
& = & 
j_{\pm}(u)+\frac{\kappa\mu_{\pm}}{2\pi}
\\
\label{gamma-minus1-on-T}
\rule{0pt}{.6cm}
\gamma_{\mu_{\pm}}^{-1} \big( T_{\pm}(u)\big)
& = & 
T_{\pm}(u)-\mu_{\pm}j_{\pm}(u)-\frac{\kappa\mu_{\pm}^{2}}{4\pi}  
\end{eqnarray}
and 
\be
\label{gamma-pm-cal-T-inv}
\gamma^{-1}_{\mu_{\pm}}  \big(\mathcal{T}_{\pm}(u)\big)
\,=\,
\CT_{\pm}(u)-\mu_{\pm}j_{\pm}(u) - \frac{3\kappa\mu_{\pm}^{2}}{4\pi}  \;.
%\gamma_{\mu_{\pm}}\left[T_{\pm}(x)-\mu_{\pm}j_{\pm}(x)\right]
%\,=\, 
\ee
We remark that (\ref{gamma-pm-phi})-(\ref{gamma-pm-T}) 
preserve the commutation relations (\ref{cft1})-(\ref{cft7}).
%Another consistency check for the expressions above is that e.g. 
%$\gamma_{\mu_{\pm}}^{-1}\big( \gamma_{\mu_{\pm}}\! \big(T_{\pm}(u)\big) \big)  = T_{\pm}(u)$.

 The $n$-point function of a generic operator $\mathcal{O}$ in the finite density representation 
 can be constructed through the automorphism (\ref{auto-gamma-def}) and 
 the corresponding $n$-point function in the fundamental representation as follows
\begin{equation}
\label{app-B-one-point}
\langle  \mathcal{O}(u_1) \dots \mathcal{O}(u_n)\rangle_{\textrm{\tiny $\mu_\pm$}}
\! =
\langle \gamma_{\mu_{\pm}} ( \mathcal{O}(u_1)) \dots \gamma_{\mu_{\pm}} ( \mathcal{O}(u_n) ) \rangle_{_{0}}
\end{equation}
The one-point functions at finite density in (\ref{fd1})  
are straightforwardly obtained by combining (\ref{app-B-one-point}) 
with the fact that  for a CFT in its ground state and on the line we have
\be
\langle \phi_\pm(u) \rangle_{_{0}}
=\,
\langle j_\pm(u) \rangle_{_{0}}
=\,
\langle \mathcal{T}_\pm(u) \rangle_{_{0}}
=\,
0
\ee

The prescription (\ref{app-B-one-point}) tell us that 
the automorphism $\gamma_{\mu_\pm}$ maps an operator in the finite density representation 
into the corresponding operator in the fundamental representation;
hence $\gamma_{\mu_\pm}^{-1}$ can be employed to construct operators in the finite density representation
from the corresponding ones in the fundamental representation. 
An important example is given by the modular Hamiltonians 
in (\ref{KA-def-u-pm}), (\ref{KB-def-u-pm}) and (\ref{fmh}),
which can be easily obtained by applying $\gamma_{\mu_\pm}^{-1}$ (see (\ref{gamma-minus1-on-T}))
to the corresponding modular Hamiltonians in the fundamental representation.

The action of the automorphism $\gamma_{\mu_\pm}$ described in (\ref{gamma-pm-phi})-(\ref{gamma-pm-cal-T})
provides also the two-point functions at finite density on the line reported in Sec.\,\ref{sec-rep-finite-density}.
The two-point function for the primaries in (\ref{fd5}) is obtained by employing (\ref{gamma-pm-phi})  and (\ref{fd1}) as follows
\be
\langle \phi_\pm^*(u) \, \phi_\pm (v)\rangle_{\mu_\pm}^{\textrm{\tiny con}} 
= 
\big\langle 
\gamma_{\mu_{\pm}}\!\big(\phi_\pm^*(u)\big) \, \gamma_{\mu_{\pm}}\!\big(\phi_\pm (v)\big)
\big\rangle_{_{0}}^{\textrm{\tiny con}} 
 =
\e^{\pm \ri\mu_{\pm}(u-v)}
\langle \phi_\pm^*(u) \, \phi_\pm (v)\rangle_{_{0}}^{\textrm{\tiny con}}   \,.
\ee
This procedure tells us that the correlators involving primaries with different chiralities vanish 
also at finite density. 
Similarly, from (\ref{gamma-pm-j})  and (\ref{fd1}) one finds 
the two-point function for the current given in (\ref{fd4}) as follows
\bea
\label{j-2point-app}
\langle j_\pm(u) \, j_\pm (v)\rangle_{\mu_\pm}^{\textrm{\tiny con}} 
&=&
\langle j_\pm(u) \, j_\pm (v)\rangle_{\mu_\pm}
-
\langle j_\pm(u) \rangle_{\mu_\pm}
\langle  j_\pm (v)\rangle_{\mu_\pm}
\\
 \rule{0pt}{.5cm}
&=&
\langle \gamma_{\mu_\pm} \! \big( j_\pm(u)\big) \,  \gamma_{\mu_\pm} \! \big( j_\pm (v) \big)\rangle_{_{0}}
-
\langle \gamma_{\mu_\pm} \! \big( j_\pm(u)\big) \rangle_{_{0}}\,
\langle  \gamma_{\mu_\pm} \! \big( j_\pm (v) \big)\rangle_{_{0}}
 \nn
\\
 \rule{0pt}{.6cm}
&=&
 \big\langle 
 \Big(j_{\pm}(u)-\frac{\kappa\mu_{\pm}}{2\pi}\Big)
 \Big(j_{\pm}(v)-\frac{\kappa\mu_{\pm}}{2\pi}\Big)
 \big\rangle_{_{0}}
 -
 \left(\frac{\kappa\mu_{\pm}}{2\pi}\right)^{2}
 =
 \langle j_\pm(u) \, j_\pm (v)\rangle_{_{0}}^{\textrm{\tiny con}}
 \nn
 \hspace{1cm}
\eea
and through similar steps also
\be
 \label{j-2point-mix-app}
 \langle j_\pm(u) \, j_\mp (v)\rangle_{\mu}^{\textrm{\tiny con}} 
\,=\,
 \big\langle 
 \Big(j_{\pm}(u)-\frac{\kappa\mu_{\pm}}{2\pi}\Big)
 \Big(j_{\mp}(v)-\frac{\kappa\mu_{\mp}}{2\pi}\Big)
 \big\rangle_{_{0}}
 -
\frac{\kappa^2 \mu_{+} \mu_{-}}{(2\pi)^2}
 =
 \langle j_\pm(u) \, j_\mp (v)\rangle_{_{0}}^{\textrm{\tiny con}}
 =\,0
\ee
which implies (see e.g. \cite{Moosavi:2019fas})
\be
 \langle j_\pm(u) \, j_\mp (v)\rangle_{\mu}
=
\kappa^2\,
\frac{ \mu_{+} \mu_{-}}{(2\pi)^2}\,.
\ee

From (\ref{gamma-pm-cal-T})  and (\ref{fd1}), the two-point function (\ref{fd3}) is obtained as follows
\bea
\label{T-2point-app}
\langle \mathcal{T}_\pm(u) \, \mathcal{T}_\pm (v)\rangle_{\mu_\pm}^{\textrm{\tiny con}} 
&=&
 \big\langle 
 \bigg(T_{\pm}(u)+\frac{\kappa\mu^2_{\pm}}{4\pi}\bigg)
 \bigg(T_{\pm}(u)+\frac{\kappa\mu^2_{\pm}}{4\pi}\bigg)
 \big\rangle_{_{0}}
 -
 \left(\frac{\kappa\mu^2_{\pm}}{4\pi}\right)^{2}
 =
 \langle T_\pm(u) \, T_\pm (v)\rangle_{_{0}}^{\textrm{\tiny con}} 
\nonumber
\\
& &
  \\
  \label{T-2point-mix-app}
 \langle \CT_\pm(u) \, \CT_\mp (v)\rangle_{\mu_\pm}^{\textrm{\tiny con}} 
&=&
 \big\langle 
 \bigg(T_{\pm}(u)+\frac{\kappa\mu^2_{\pm}}{4\pi}\bigg)
 \bigg(T_{\mp}(u)+\frac{\kappa\mu^2_{\mp}}{4\pi}\bigg)
 \big\rangle_{_{0}}
 -
\frac{\kappa^2 \mu^2_{+} \mu^2_{-}}{(4\pi)^2}
 =
 \langle T_\pm(u) \, T_\mp (v)\rangle_{_{0}}^{\textrm{\tiny con}} 
 =0  \,.
\nonumber
\\
& &
\eea

The above considerations can be extended to any chiral field theory on a circle of length $L$
by fixing the periodicity of the fields. 
This circle is obtained by imposing the periodicity condition on the line;
hence we can restrict to the the interval $[-L/2\, ,\, L/2]$,  imposing the following boundary conditions
\be 
\phi_\pm (L/2) = (-1)^{2h_\pm} \,\phi_\pm (-L/2) 
\qquad 
j_\pm (L/2) = j_\pm (-L/2)  
\qquad 
T_\pm (L/2) = T_\pm (-L/2) \,.
\label{nc1}
\ee  
In particular, $\phi_\pm$ are periodic for $h_\pm \in \mathbb Z$ and anti-periodic for $h_\pm \in \mathbb Z +\frac{1}{2}$. 
In this case, instead of the automorphism $D_{\mu_{\pm}}$ in (\ref{auto-gamma-def}),
one introduces 
\be
\label{D-operator-shift-circle}
D_{_{\textrm{\tiny $L, \mu_\pm$}}}
\,\equiv\, 
\pm \,\mu_{\pm} \! \int_{-L/2}^{L/2}\! u\,j_{\pm}(u)\, \rd u  
\ee 
where $j_\pm (u)$ satisfies (\ref{nc1}). 
Requiring that $\gamma_{\mu_{\pm}}$ preserves the periodicity condition (\ref{nc1}) 
leads to the following  constraint
\be 
\frac{L\,\mu_\pm}{2\pi}  = n_\pm 
\;\;\; \qquad \;\;\;
n_\pm \in \mathbb Z \,.
\label{nc2}
\ee 
Taking into account this condition, 
the finite density and the zero density correlators on the circle are related as follows
\begin{equation}
\label{app-B-n-point-circle}
\langle  \mathcal{O}(u_1) \dots \mathcal{O}(u_n)\rangle_{_{\textrm{\tiny $L, \mu_\pm$}}}
\! =\,
\langle \gamma_{\mu_{\pm}} ( \mathcal{O}(u_1))\dots \gamma_{\mu_{\pm}} ( \mathcal{O}(u_n) ) \rangle_{_{\textrm{\tiny $L,0$}}}\,.
\end{equation}
Thus, the one-point functions at finite density (\ref{ftd1-L}) 
can be written straightforwardly by combining (\ref{app-B-n-point-circle}) 
with the fact that  for a CFT in its ground state and on the circle  we have
%In this way we derive (\ref{KA-def-u-pm-circle})  and (\ref{fmh-circle}) from 
\be
\label{app-B-one-point-circle}
\langle \phi_\pm(u) \rangle_{_{L,0}}
=\,
\langle j_\pm(u) \rangle_{_{L,0}}
=\,
0
\;\;\;\qquad\;\;\;
\langle \mathcal{T}_\pm(u) \rangle_{_{L,0}} = \langle T_\pm(u)\rangle_{_{\textrm{\tiny $L,0$}}}
=\,
- \frac{\pi c}{12 L^2}  \;.
\ee
Analogously, the two-point function (\ref{ftd3}) is obtained 
from (\ref{ftd1-L}) and (\ref{gamma-pm-cal-T}) as follows
\bea
\langle \T_\pm(u) \,\T_\pm (v)\rangle_{_{\textrm{\tiny $L,\mu_\pm$}}}^{\textrm{\tiny con}}   
&=&
\langle \T_\pm(u) \,\T_\pm (v)\rangle_{_{\textrm{\tiny $L,\mu_\pm$}}}
-
\langle \T_\pm(u) \rangle_{_{\textrm{\tiny $L,\mu_\pm$}}}
\langle \T_\pm (v)\rangle_{_{\textrm{\tiny $L,\mu_\pm$}}}
 \nonumber
\\
&=&
 \big\langle 
 \bigg(T_{\pm}(u)+\frac{\kappa\mu^2_{\pm}}{4\pi}\bigg)
 \bigg(T_{\pm}(u)+\frac{\kappa\mu^2_{\pm}}{4\pi}\bigg)
 \big\rangle_{_{\textrm{\tiny $L,0$}}}
\! -
 \left( \frac{\kappa \mu^2_\pm}{4\pi} - \frac{\pi c}{12 L^2} \right)^2
 \nonumber
 \\
 \rule{0pt}{.6cm}
 &=&
 \langle T_\pm(u) \,T_\pm (v)\rangle_{_{\textrm{\tiny $L,0$}}}
-
 \left(  \frac{\pi c}{12 L^2} \right)^2
 =\,
  \langle T_\pm(u) \,T_\pm (v)\rangle_{_{\textrm{\tiny $L,0$}}}^{\textrm{\tiny con}}    \,.
\eea

Summarising, the fundamental input are the correlation functions 
$\langle \,\cdots  \rangle_{_{0}}$ in the ground state representation on the line. 
They generate, 
through the mapping $u \mapsto \e^{2\pi \ri u/L}$, the ground state correlation functions 
$\langle \,\cdots  \rangle_{_{\textrm{\tiny $L,0$}}}$ on the circle. 
The associated finite density correlators are obtained in turn through the 
automorphism $\gamma_{\mu_\pm}$ (see (\ref{app-B-one-point}) and (\ref{app-B-n-point-circle})). 
The allowed values for the chemical potential are  
$\mu_\pm \in \mathbb R$ on the line and $L\mu_\pm /(2\pi) \in \mathbb Z$ on the circle.

\section{Consistency checks for the correlators}
\label{app-michele-modular}
%%%%%%%%%%%%%%%%%%%%%%%%%%%%%%%%%%%%%%%%%%%%%%%%%%%%%

In this appendix we describe some consistency checks for both the correlation functions 
(see (\ref{fd5}) and (\ref{ftd5}))
and the modular correlators (\ref{mod-corr-phi-mu}).
These checks are based 
on the special case of free fermions (Sec.\,\ref{subsec-app-mu}),
the positivity (Sec.\,\ref{subsec-app-POS})
and the properties of the entanglement spectrum (Sec.\,\ref{subsec-app-ES}).
Since both chiralities can be analysed in the same way, 
we focus on the right movers $\phi_+(u)$, 
by setting $\phi_+(u) \equiv \phi(u)$, $h_+ \equiv h$ and $\mu_+ \equiv \mu$ 
throughout this appendix, to enlighten the notation.

\subsection{Fermionic correlators at finite density}
\label{subsec-app-mu}

In order to check the first expression in (\ref{fd5}) in a special case, 
let us consider the free chiral fermion on the line,
whose two-point function at finite density can be written through the 
Fermi-Dirac distribution as follows
\be
C(u,v) =
\int_\RR \frac{\rd p}{2\pi} \int_\RR \frac{\rd k}{2\pi} 
\; \e^{\ri (p u - k v)}  \; \frac{ 2\pi \, \delta(u-v) }{1+\e^{\beta(p -\mu)}} 
\,=\,
\int_\RR \frac{\rd p}{2\pi} \; \e^{\ri p (u - v)} \; \frac{1 }{1+\e^{\beta(p -\mu)}} 
\ee
which can be regularised at $p \to -\infty$ by introducing the infinitesimal $\varepsilon >0$ as follows
\be
C_{\varepsilon}(u,v) \equiv 
\int_\RR \frac{\rd p}{2\pi} \; \e^{\ri p (u - v- \ri \varepsilon)} \; \frac{1 }{1+\e^{\beta(p -\mu)}} 
\ee
the limit of this expression at zero temperature reads
\bea
\lim_{\beta \to +\infty} C_{\varepsilon}(u,v) 
&=&
\int_\RR \frac{\rd p}{2\pi} \; \e^{\ri p (u - v- \ri \varepsilon)} \; \theta(\mu - p)
\nn
\\
&=&
\e^{\ri \mu (u - v)}  \int_\RR \frac{\rd q}{2\pi} \; \e^{-\ri q (u - v- \ri \varepsilon)} \; \theta(q)
\,=\,
\frac{\e^{\ri \mu (u - v)} }{2\pi \ri \, (u-v -\ri \varepsilon)}
\eea
which corresponds to the first expression in (\ref{fd5}) for the right moving fields when $h_{+}=1/2$.

It is worth studying the analogue computation on the circle,  in order to check (\ref{ftd5}) in a special case. 
Considering the free fermion on a circle of length $L$ satisfying anti-periodic boundary conditions
we can employ the complete set of orthonormal functions $\big\{ \frac{1}{\sqrt{L}}\,\e^{\ri \,2\pi n \,u/L } \, ; n \in \widetilde{\mathbb{Z}} \,\big\}$,
where $\widetilde{\mathbb{Z}} $ denotes the set made by the half-integers.
The periodic boundary conditions are more subtle in the context of entanglement \cite{Klich:2015ina}.
The two-point function of this field can be expressed through the Fermi-Dirac distribution as follows
\be
C_L(u,v) 
=
\frac{1}{L} 
\sum_{m \,\in\, \widetilde{\mathbb{Z}}}  \, \sum_{n \,\in\, \widetilde{\mathbb{Z}} } \,
 \e^{2\pi \ri (m u - n v)/L}  \, \frac{\delta_{m,n} }{1+\e^{\beta(2\pi n/L  -\mu)}} 
\,=
\frac{1}{L}  \sum_{n \,\in\, \widetilde{\mathbb{Z}} } 
 \e^{2\pi \ri \,n(u - v)/L}  \, \frac{ 1 }{1+\e^{\beta(2\pi n/L  -\mu)}} 
\ee
which can be regularised as $n \to -\infty$ through the infinitesimal $\varepsilon >0$ as above, 
namely
\be
\label{CL-eps}
C_{L, \varepsilon}(u,v) 
\,=
\frac{1}{L}  \sum_{n \,\in\, \widetilde{\mathbb{Z}} } 
 \e^{2\pi \ri \,n(u - v- \ri \varepsilon)/L}  \, \frac{ 1 }{1+\e^{\beta(2\pi n/L  -\mu)}} 
\ee
which is convergent for both $n \to -\infty$ and $n \to +\infty$,
because of $\varepsilon >0 $ and $\beta >0$ respectively. 

The zero temperature limit of (\ref{CL-eps}) selects 
the values of $n \in \widetilde{\mathbb{Z}} $ in the series
such that $2\pi \, n/L - \mu \leqslant 0$, namely $n \leqslant L\mu /(2\pi)$.
Since  $L\mu /(2\pi) \in \mathbb{Z}$ (see Appendix\;\ref{app-reps}),
the values of $n$ providing a non-vanishing contribution to (\ref{CL-eps}) in the zero temperature limit 
are given by  $n = L\mu /(2\pi) + \tilde{n}$ with $\tilde{n} \in \widetilde{\mathbb{Z}}_{-} $,
being $\widetilde{\mathbb{Z}}_{-} $ made by the half-integer and negative numbers
(in particular, we have that  $n < L\mu /(2\pi) $).
Thus, the zero temperature limit of (\ref{CL-eps}) when $L\mu /(2\pi) \in \mathbb{Z}$ gives
\bea
\lim_{\beta \to +\infty} \! C_{L, \varepsilon}(u,v) 
&=&
\frac{1}{L} \!  \sum_{\tilde{n} \,\in\, \widetilde{\mathbb{Z}}_{-} } 
\! \! \e^{2\pi \ri (L\mu /(2\pi) + \tilde{n} ) (u - v- \ri \varepsilon)/L} 
\\
&=&
\frac{\e^{\ri \mu (u-v)}}{L} \!  \sum_{\tilde{n} \,\in\, \widetilde{\mathbb{Z}}_{-} } 
\! \! \e^{2\pi \ri \,\tilde{n}  (u - v- \ri \varepsilon)/L} 
\,=\,
\frac{\e^{\ri \mu (u-v)}}{L} \;
\frac{1}{2\ri\, \sin\!\big[ \pi (u - v- \ri \varepsilon)/L\big]}
\nn
\eea
which corresponds to (\ref{ftd5}) for the right moving fields 
and specialised to $h_{+}=1/2$.

%{\color{orange}
%Assuming that $L\mu /(2\pi) \notin \widetilde{\mathbb{Z}}$,
%consider the integer $m_\mu \in \mathbb{Z}$ such that $|  L\mu /(2\pi) - m_\mu |<1/2$.
%The values of $n$ providing a non-vanishing contribution to (\ref{CL-eps}) in the zero temperature limit 
%are given by  $n = m_\mu + \tilde{n}$ with $\tilde{n} \in \widetilde{\mathbb{Z}}_{-} $,
%being $\widetilde{\mathbb{Z}}_{-} $ made by the half-integer and negative numbers.
%It is worth introducing $\tilde{\mu}$ as $m_\mu = L \tilde{\mu} /(2\pi) $.
%Thus, the zero temperature limit of (\ref{CL-eps}) when $L\mu /(2\pi) \notin \widetilde{\mathbb{Z}}$ gives
%\bea
%\lim_{\beta \to +\infty} \! C_{L, \varepsilon}(u,v) 
%&=&
%\frac{1}{L} \!  \sum_{\tilde{n} \,\in\, \widetilde{\mathbb{Z}}_{-} } 
%\! \! \e^{2\pi \ri (L \tilde{\mu} /(2\pi) + \tilde{n} ) (u - v- \ri \varepsilon)/L} 
%\\
%&=&
%\frac{\e^{\ri \tilde{\mu} (u-v)}}{L} \!  \sum_{\tilde{n} \,\in\, \widetilde{\mathbb{Z}}_{-} } 
%\! \! \e^{2\pi \ri \,\tilde{n}  (u - v- \ri \varepsilon)/L} 
%\,=\,
%\frac{\e^{\ri \tilde{\mu} (u-v)}}{L} \;
%\frac{1}{2\ri\, \sin\!\big[ \pi (u - v- \ri \varepsilon)/L\big]}
%\nn
%\eea
%which corresponds to (\ref{ftd5}) for the right moving fields specialised to $h_{+}=1/2$.
%}
%\\
%\textcolor{red}{\bf [see Eq.\,(2.41) of \cite{TodorovFurlan-review} for the $\mu=0$ case]}
%\\
%\textcolor{red}{\bf [should we discuss also the cases where $L\mu /(2\pi) \in \widetilde{\mathbb{Z}}$?]}

\subsection{Positivity}
\label{subsec-app-POS}

Assuming that we are dealing with a unitary CFT, 
the positivity of the scalar product $(\cdot \, ,\cdot )$ in the state space 
implies that   
\be 
\big(\mathcal{O}(f)\,\Omega_{\mu}\, ,\, \mathcal{O}(f)\,\Omega_{\mu}\big) 
\geqslant 0 
\label{e1}
\ee 
where the state $\Omega_\mu$ characterises the finite density representation introduced  
in Sec.\,\ref{sec-rep-finite-density} and  
\be 
\mathcal{O}(f) \equiv \int_{-\infty}^\infty \! f(u) \,\mathcal{O}(u)\, \rd u
\label{m1}
\ee
is a chiral field smeared with a generic complex test function $f$.  
Choosing $\mathcal{O}(u) = \phi (u)$, the inequality (\ref{e1}) implies 
\be
\label{e2} 
\int_{-\infty}^\infty \! \rd u \int_{-\infty}^\infty \! \rd v \;
\overline{f(u)}\, \langle \phi^* (u) \phi (v) \rangle_{\mu}\, f(v) \geqslant 0
\ee
for any test function $f$. Plugging in (\ref{e2}) the explicit form (\ref{fd5}) of the correlation function one gets
\be
\label{e3} 
\frac{\e^{-\ri \pi h}}{2\pi}\int_{-\infty}^\infty \! \rd u \int_{-\infty}^\infty \! \rd v \;
\frac{\overline{f_{\mu}(u)} \, f_{\mu}(v)}{(u-v - \ri \varepsilon)^{2h}} \geqslant 0
\ee
where $f_{\mu}(u) \equiv \e^{-\ri \mu u} f(u)$ and the overline denotes complex conjugation. 

The inequality (\ref{e3}) implies a condition on 
the dimension $h$, which is easily obtained in momentum space. In fact, 
performing the Fourier transform (see for instance \cite{GelfandShilov}) (\ref{e3}) takes the form 
\be 
\label{e4}
\frac{1}{\Gamma (2h)}\int_{-\infty}^\infty \frac{\rd p}{2\pi} \,  \big| \widehat{f}_{\mu }(p) \big|^2 \, p_+^{2h -1} 
\,=\, 
\frac{1}{\Gamma (2h)}\int_{0}^\infty \frac{\rd p}{2\pi} \, \big| \widehat{f}_{\mu }(p) \big|^2 \, p^{2h -1} 
\, \geqslant  \, 0
\ee 
where the distribution $p_+^\sigma \equiv \theta (p) \, p^\sigma $ has been introduced. 
The bound (\ref{e4}) is satisfied provided that the integral converges. 
This is the case for large $p$, where the integrand is dominated by the 
exponential decay of $|\widehat{f}_{\mu }(p)|$. The convergence at $p=0$ implies that $h>0$. 
Summarising, for both chiralities the dimensions of the primary fields in unitary CFT satisfy 
\be
h_\pm >0 \,.
\label{e5}
\ee 

Let us discuss now the impact of positivity on the modular correlator (\ref{mod-corr-phi-mu}). 
Given the interval $A \equiv [a,b]$, we consider the chiral field (\ref{m1}) localised in $A$ assuming 
for the support of the test functions 
${\rm supp}(f) \subset A$. Let $S$ be the conjugate linear operator occurring in the Tomita-Takesaki theorem
(see Eq.\,(V.2.1) of \cite{Haag:1992hx}) which acts as follows
\be 
S \,\mathcal{O}(f) \,\Omega_{\mu} =  \, \mathcal{O}^*(f) \,\Omega_{\mu} \;\;\;\qquad \;\;\; {\rm supp}(f) \subset A\,.
\label{m2}
\ee 
The unique polar decomposition of $S$ reads
\be 
S= J \,\Delta^{1/2}
\label{m3} 
\ee 
defining the (antiunitary) modular conjugation $J$ studied in Sec.\,\ref{subsec-mod-J-vacuum} 
and the self-adjoint positive (in general unbounded) modular operator $\Delta$, which satisfy 
\be
J=J^* = J^{-1}
\;\;\qquad\;\;
\Delta\, \Omega_{\mu } = \Omega_{\mu}
\;\;\qquad\;\;
J\,\Omega_{\mu} = \Omega_{\mu}   \,.
\label{m3-bis} 
\ee
The modular operator $\Delta $ is expressed in terms of the full modular Hamiltonian $K$ in \eqref{fmh} 
by $\Delta \equiv  \e^{-K} $.  

Following \cite{Landsman:1986uw, Casini:2010bf, Papadodimas:2013jku} (see also (\ref{J-on-phi-1})-(\ref{J-on-T})),  
we introduce the reflected operator 
\be 
\mathcal{O}^{\textrm{\tiny ref}} (f) \equiv  J\,\mathcal{O} (f)\,J^* 
\label{m4}
\ee 
By employing \eqref{m3} and \eqref{m3-bis}, one finds 
\bea 
\label{m5}
\big(\Omega_{\mu}\, ,\, \mathcal{O}^{\textrm{\tiny ref}}(f) \,\mathcal{O}(f) \,\Omega_{\mu}\big) 
&=& 
\big( J^2\,\Omega_{\mu}\, ,\, J\mathcal{O}(f)J^* \,\mathcal{O}(f) \,\Omega_{\mu}\big) 
\nn
\\
& & \hspace{-1.5cm}
=\,
\big(\mathcal{O}(f)J^*  \mathcal{O}(f) \,\Omega_{\mu}\, ,\, J\,\Omega_{\mu}\big) 
= \big(\mathcal{O}(f)J^* \mathcal{O}(f) \,\Omega_{\mu}\, ,\, \Omega_{\mu}\big) \nn \\
& & \hspace{-1.5cm}
=\,
\big(\Omega_{\mu}\, ,\, \mathcal{O}^*(f)\,J \,\mathcal{O}^*(f)\,\Omega_{\mu}\big) 
=
\big(\Omega_{\mu}\, ,\, \mathcal{O}^*(f)\,J \,S \,\mathcal{O}(f)\,\Omega_{\mu}\big) 
\nonumber \\
& & \hspace{-1.5cm}
=\,
 \big(\Omega_{\mu}\, ,\, \mathcal{O}^*(f)\,J^2 \,\Delta^{1/2} \,\mathcal{O}(f)\,\Omega_{\mu}\big)  
=
\big(\Omega_{\mu}\, ,\, \mathcal{O}^*(f) \,\Delta^{1/2} \,\mathcal{O}(f)\,\Omega_{\mu}\big) 
\nonumber \\
& & \hspace{-1.5cm}
=\,
\big(\mathcal{O}(f)\,\Omega_{\mu}\, ,\, \Delta^{1/2} \,\mathcal{O}(f)\,\Omega_{\mu}\big) 
\geqslant 0 
\eea
which has been called modular reflection positivity \cite{Casini:2010bf}. 
Thus,
the modular reflection positivity can be derived from
the properties of $J$,
the positivity of the scalar product $(\cdot \, ,\cdot )$ 
and the positivity of the modular operator $\Delta$. 

For the field $\mathcal{O}(u)=\phi (u)$ in (\ref{m1}), 
the inequality (\ref{m5}) takes the form 
\be
\label{m52} 
\int_{-\infty}^\infty \! \rd u \int_{-\infty}^\infty \! \rd v \;
\overline{f(u)}\,  
\big(\phi (u) \,\Omega_\mu\, ,  \,\e^{-K/2} \,\phi (v) \,\Omega_{\mu } \big) \, f(v) \geqslant 0 \,.
\ee
On the other hand, 
by using $\e^{-\ri \tau K}\,\Omega_{\mu}=\Omega_{\mu}$, 
the modular evolution \eqref{nme1}
and the modular correlator \eqref{mod-corr-phi-mu} for the field $\phi (u)$, 
one gets 
\bea 
\label{app-fin-C-W}
\frac{\e^{\pm \ri \mu (u-v)}}{2 \pi \, \e^{i \pi h }} \,W_+(\tau_{12};u,v)^{2h}
&=&
\big(\phi (\tau_1,u) \,\Omega_{\mu}\, ,\, \phi (\tau_2, v) \,\Omega_{\mu} \big) 
\nonumber \\
&=&
\big(\e^{\ri \tau_1 K}\,\phi (u)\,\e^{-\ri \tau_1 K}\,\Omega_{\mu}\, , \,\e^{\ri \tau_2 K}\,\phi (v)\,\e^{-\ri \tau_2 K}\,\Omega_{\mu}\big) 
\nonumber \\
&=&
\big(\phi (u)\,\Omega_{\mu}\, ,\, \e^{-\ri \tau_{12} K}\, \phi (v)\,\Omega_{\mu}\big)  \,.
\label{m53}
\eea 
Combining (\ref{m53}) specialised to $\tau_{12} = -\ri/2$ with (\ref{m52}), 
one finds
\be 
 \int_{-\infty}^\infty \! \rd u \int_{-\infty}^\infty \!\rd v \;
 \overline{f_\mu(u) } \; \frac{W_+(-\ri/2;u,v)^{2h} }{2\pi\, \e^{i \pi h}} \; f_\mu (v)
\,\geqslant \,0 
\label{m6}
\ee
where $f_\mu$ has been defined in the text below (\ref{e3})
and we have that ${\rm supp} (f_\mu) = {\rm supp} (f) \subset A$.
The inequality (\ref{m6}) provides a non-trivial consistency condition 
for the explicit expression of $W_+(\tau;u,v)$ in (\ref{cap-W-def}).
Indeed, when $A$ has finite length
we have that (\ref{cap-W-def}) for $\varepsilon =0$ and $\tau_{12} = -\ri /2$ can be written as 
\be
\label{Wplus-with-s}
W_+(-\ri/2;u,v)^{2h} 
=
\Bigg( \frac{\ri}{\tfrac{b-a}{2}\, \big[ 1- s(u) \, s(v) \big]} \Bigg)^{2h }
\;\;\qquad\;\;
s(u) \equiv \frac{u - \tfrac{a+b}{2}}{\tfrac{b-a}{2}} \in (-1,1)   \,.
\ee 
Plugging this expression into (\ref{m6})
and employing the following Taylor series
\be
\frac{1}{\left(1-y\right)^{h}} 
=
\sum_{n=0}^{\infty}\alpha_{n}(h)\,y^{n}
\;\;\;\qquad\;\;\;
\alpha_{n}(h)  \equiv  \frac{1}{n!}\prod_{k=0}^{n-1}(h+k)>0
\ee
where $h>0$ and $|y |<1$, 
for an interval of finite length we find that 
the inequality (\ref{m6}) becomes equivalent to 
\be
 \int_{-\infty}^{\infty} \! \! \rd u
 \int_{-\infty}^{\infty} \! \! \rd v
 \,\overline{f_\mu (u)} 
 \left[ \, \sum_{n=0}^{\infty}\alpha_{n}(2h)\,s(u)^{n} s(v)^{n}\right]\!
 f_\mu(v) 
=
 \sum_{n=0}^{\infty}\alpha_{n}(2h)
 \left| \, \int_{-\infty}^{\infty}
 \! \! \rd u\,f_\mu(u)\,s(u)^{n}\right|^{2}\geqslant 0  
\ee
which is verified because $\alpha_{n}(2h) >0$.

In the special case of the Rindler wedge, i.e. in the limiting regime given by  $b \to +\infty$,
the inequality (\ref{m6}) becomes equivalent to 
\bea
& &
 \int_{-\infty}^{\infty} \! \! \rd u
 \int_{-\infty}^{\infty} \! \! \rd v
 \,\overline{f_\mu (u)} \,
 \left( \frac{1}{u-a + v-a}  \right)^h\!
 f_\mu(v) 
 \\
 \rule{0pt}{.7cm}
 & &
 =
\frac{1}{2}  \int_{-\infty}^{\infty} \! \! \rd u
 \int_{-\infty}^{\infty} \! \! \rd v
  \left( \frac{1}{u-a + v-a}  \right)^h
 \left[
 \,\overline{f_\mu (u)} \,
 f_\mu(v) 
 + \textrm{c.c.}
 \right]
\, \geqslant \,
   \frac{1}{M_f^h} 
  \left[ \, \int_{-\infty}^{\infty}
 \! \! \rd u\,f_\mu(u)\,\right]^{2}
 \geqslant \, 0
 \nn
% \\
%  \rule{0pt}{.7cm}
% & &
% \geqslant
% \frac{1}{2 M_f^h}  \int_{-\infty}^{\infty} \! \! \rd u
% \int_{-\infty}^{\infty} \! \! \rd v\;
% \left[
% \,\overline{f_\mu (u)} \,
% f_\mu(v) 
% + \textrm{c.c.}
% \right]
% =
%  \frac{1}{M_f^h} 
%  \left| \, \int_{-\infty}^{\infty}
% \! \! \rd u\,f_\mu(u)\,\right|^{2}
% \geqslant \, 0
\eea
where c.c. denotes the complex conjugate,
$M_f \equiv \textrm{max}\big\{ u-a + v-a> 0 \; ;\, u,v \in \textrm{supp}(f)\big\}$
and the crucial inequality has been obtained by assuming that
$f_\mu$ are real and positive functions with compact support properly included in $A = [a, +\infty)$.

The above analysis provides  a non-trivial consistency check between
the expression (\ref{cap-W-def}) of $W_+(\tau;u,v)$ 
and the modular reflection positivity condition (\ref{m5}).

\subsection{Entanglement spectrum}
\label{subsec-app-ES}

As further consistency check of the modular correlator in (\ref{mod-corr-phi-mu})-(\ref{cap-W-def}),
we show that
\be
\label{ineq-Delta-2pt}
\big(\mathcal{O}(f)\,\Omega_{\mu}\, , \,\mathcal{O}(f)\,\Omega_{\mu}\big) 
\geqslant 
\big(\mathcal{O}(f)\,\Omega_{\mu}\, ,\, \Delta^{1/2} \,\mathcal{O}(f)\,\Omega_{\mu}\big) 
\geqslant 
0
\ee
where the first inequality comes from the fact that the spectrum of $\Delta$
(i.e. the entanglement spectrum)  is a subset of $(0,1)$,
while the last one corresponds to the positivity of the modular operator $\Delta$ (see (\ref{m5})).
For non-coincident $u,v \in [a,b]$, from (\ref{R-factor-def}) we introduce
\be
\label{rA-def}
 r(u,v) 
 \equiv 
 -\,\ri\, R(-\ri/2;u,v)
 = \frac{(b-a) \, (u-v)}{2\big[ \tfrac{a+b}{2}(u+v) - (u\, v+ a\, b) \big]}
 \,\in\,[ -1, 1]
\ee
which satisfies (see also (\ref{cap-W-eps0-def}) and (\ref{Wplus-with-s}))
\be
\label{rA-from-W}
\frac{W_+(-\ri/2;u,v)^{2h} }{\e^{i \pi h}} = \frac{r(u,v)}{u-v} \,.
\ee
From (\ref{m53}) specialised to $\tau_{12} = -\ri/2$, (\ref{m6}) and (\ref{rA-from-W}), 
we obtain 
\bea
\label{ineq-app-spec}
\big(\phi (u)\,\Omega_{\mu}\, ,\, \Delta^{1/2}\, \phi (v)\,\Omega_{\mu}\big) 
&=&
 \int_{-\infty}^\infty \! \rd u \int_{-\infty}^\infty \!\rd v \;
 \overline{f_\mu(u) } \; \frac{W_+(-\ri/2;u,v)^{2h} }{2\pi\, \e^{i \pi h}} \; f_\mu (v)
 \nn
\\
\rule{0pt}{.75cm}
&=&
 \int_{-\infty}^\infty \! \rd u \int_{-\infty}^\infty \!\rd v \;
 \overline{f_\mu(u) } \; \frac{r(u,v)^{2h}}{2\pi\, (u-v)^{2h}} \; f_\mu (v)
 \nn
 \\
 \rule{0pt}{.75cm}
& \leqslant &
 \int_{-\infty}^\infty \! \rd u \int_{-\infty}^\infty \!\rd v \;
f_\mu(u)  \; \frac{1}{2\pi\, (u-v)^{2h}} \; f_\mu (v)
\eea
where the inequality originates from the fact that $r(u,v)^2 \in [0,1]$ for $u,v \in [a,b]$
and it has been obtained by assuming that $f_\mu$ are real and positive functions.
Finally, from (\ref{fd5}),
one realises that the inequality in (\ref{ineq-app-spec}) provides
the first inequality in (\ref{ineq-Delta-2pt}).

\section{Modular evolution in the complementary region}
\label{app-evolution-B-region}
%%%%%%%%%%%%%%%%%%%%%%%%%%%%%%%%%%%%%%%%%%%%%%%%%%%%%

The modular evolution of an operator localised in the region $B$ complementary to the interval $A$ 
can be studied by combining its modular evolution in $A$ and the modular conjugation,
as obtained in (\ref{mod-evo-O-in-B}).
In this appendix we derive the explicit expressions of these modular evolutions
for the fields $\phi_\pm$, $j_\pm$ and $\CT_\pm$.

In order to specify (\ref{mod-evo-O-in-B}) to the primary fields $\phi_\pm$,
by employing \eqref{me9a}, \eqref{me9b} and \eqref{J-on-phi-1}, let us consider 
%\be
%\label{J-on-phi-1-a}
%J \phi_{\pm}(u)J=\mathrm{e}^{\mp\mathrm{i}\mu_{\pm}(\mathsf{j}(u)-u)}(-\mathsf{j}'(u))^{h}\phi_{\pm}^{*}(\mathsf{j}(u)) 
%\ee
%which implies that
\bea
\label{exp-app-Bmod-phi}
& & \hspace{-.6cm}
\e^{\ri K \tau} \big( J\, \phi_{\pm}(u)\,J \,\big) \, \e^{-\ri K \tau} 
% \e^{\ri K \tau} \,\mathrm{e}^{\mp \ri\mu_{\pm}(\mathsf{j}(u)-u)}(-\mathsf{j}'(u))^{h}\phi_{\pm}^{*}(\mathsf{j}(u))\e^{-\ri K \tau}
%\\
% \rule{0pt}{.5cm}
% & &  
=\;
 \mathrm{e}^{\mp \ri\mu_{\pm}(\mathsf{j}(u)-u)} 
 \,\mathsf{j}'(u)^{h_\pm} 
\,\e^{\ri K \tau}\phi_{\pm}^{*}(\,\mathsf{j}(u))\, \e^{-\ri K \tau}
 \nn \\
  \rule{0pt}{.7cm}
 & &  =\,
 \mathrm{e}^{\mp \ri\mu_{\pm}(\mathsf{j}(u)-u)}
 \,\mathsf{j}'(u)^{h_\pm} 
 \Big[\,
 \mathrm{e}^{\mp\mathrm{i}\mu_{\pm}(\xi_{\pm}(\tau, \, \mathsf{j}(u))-\mathsf{j}(u))}
 \Big( \partial_{v}\xi_{\pm}(\tau,v)\big|_{v=\mathsf{j}(u)}\Big)^{h_\pm}
 \phi_{\pm}^{*}\big(\xi_{\pm}(\tau,\mathsf{j}(u))\big)
 \Big]
 \nonumber \\
   \rule{0pt}{.5cm}
 & &  =\,
 \mathrm{e}^{\mp\mathrm{i}\mu_{\pm} (\xi_{\pm}(\tau, \, \mathsf{j}(u))-u )}
 \big[ \partial_{u}\xi_{\pm}(\tau,\mathsf{j}(u)) \big]^{h_\pm} 
 \,\phi_{\pm}^{*}\big(\xi_{\pm}(\tau,\mathsf{j}(u))\big)  \,.
\eea
Then, 
the r.h.s. of (\ref{mod-evo-O-in-B}) is obtained by applying $J (\cdot) J$ to (\ref{exp-app-Bmod-phi});
hence, by using (\ref{J-on-phi-2}),  
for the primaries we find 
\be
\phi_{\pm}(\tau,u)
%=J\e^{\ri K \tau}J\phi_{\pm}(u)J\e^{-\ri K \tau}J
=\mathrm{e}^{\pm\mathrm{i}\mu_{\pm} (\tilde{\xi}_{\pm}(\tau,u)-u )}
\big(\partial_{u}\tilde{\xi}_{\pm}(\tau,u)\big)^{h_\pm}
\phi_{\pm}\big(\tilde{\xi}_{\pm}(\mathsf{\tau,}u))\big)
\ee
with $\tilde{\xi}_{\pm}(\tau,u)$ being defined as follows
\be
\label{app-xi-tilde-def}
\tilde{\xi}_{\pm}(\tau,u)\equiv \,
\mathsf{j}\big(\xi_{\pm}(\tau,\mathsf{j}(u))\big)
=
\xi_{\pm}(\tau,u)
\ee
where (\ref{id-xi-j-line}) has been used in the last step.
This result tells us that,
for the modular evolution,
the expression \eqref{me9a} combined with (\ref{xi-map-fund}) holds also for $u\notin A$. 
The above analysis can be adapted to the finite volume case
by replacing $\mathsf{j}(u)$ with $\mathsf{j}_L(u)$ and $\xi(\tau,u)$ with $\xi_{L}(\tau,u)$ 
(see (\ref{j0-map-circle-def}) and (\ref{xi-map-fund-circle})  respectively),
finding the same conclusion in terms of (\ref{xi-map-fund-circle}).

As for the chiral currents $j_\pm$,
from the r.h.s. of (\ref{mod-evo-O-in-B}), \eqref{cp4} and \eqref{J-on-j} we have that
%\be
%\label{J-on-j-a}
%Jj_{\pm}(u)J=-\mathsf{j}'(u)j_{\pm}(\mathsf{j}(u))-\frac{\kappa\mu_{\pm}}{2\pi}\left(1+\mathsf{j}'(u)\right)
%\ee
\begin{eqnarray}
\label{exp-app-Bmod-j}
&  & 
\e^{\ri K \tau} \big( J\, j_{\pm}(u)\,J\, \big) \, \e^{-\ri K \tau} 
=
\mathsf{j}'(u)  \; \e^{\ri K \tau}j_{\pm}(\,\mathsf{j}(u)) \, \e^{-\ri K \tau}
-\frac{\kappa\mu_{\pm}}{2\pi} \big[ 1 -\mathsf{j}'(u) \big]
\nn \\
  \rule{0pt}{.5cm}
 & &  = 
 \,\mathsf{j}'(u) \!
 \left[\,
 \partial_{v}\xi_{\pm}(\tau,v)\big|_{v=\mathsf{j}(u)} \, j_{\pm}\big(\xi_{\pm}(\tau,\mathsf{j}(u))\big)
 -\frac{\kappa\mu_{\pm}}{2\pi}\left(1-\left.\partial_{v}\xi_{\pm}(\tau,v)\right|_{v=\mathsf{j}(u)}\right)\right]
-\frac{\kappa\mu_{\pm}}{2\pi} \big[ 1 - \mathsf{j}'(u) \big]
\nn \\
%  \rule{0pt}{.6cm}
% & & =
% -\,\partial_{u}\xi_{\pm}(\tau,\mathsf{j}(u))\,j_{\pm}\left(\xi_{\pm}(\tau,\mathsf{j}(u))\right)
% +\frac{\kappa\mu_{\pm}}{2\pi}\big( \,\mathsf{j}'(u)-\partial_{u}\xi_{\pm}(\tau,\mathsf{j}(u))\big)
%-\frac{\kappa\mu_{\pm}}{2\pi} \big[ 1+\mathsf{j}'(u)\big]
% \nn \\
   \rule{0pt}{.6cm}
 & & =
 \,\partial_{u}\xi_{\pm}(\tau,\mathsf{j}(u)) 
\,j_{\pm}\big(\xi_{\pm}(\tau,\mathsf{j}(u))\big)
 -\frac{\kappa\mu_{\pm}}{2\pi}\big[1 - \partial_{u}\xi_{\pm}(\tau,\mathsf{j}(u))  \big]  \,.
\end{eqnarray}
By applying $J (\cdot) J$ to (\ref{exp-app-Bmod-j}),
one gets the r.h.s. of (\ref{mod-evo-O-in-B}) specified to this case.
Hence, by employing again \eqref{J-on-j}, we obtain
\begin{eqnarray}
j_{\pm}(\tau,u) 
%& = & 
%J\e^{\ri K \tau}J j_{\pm}(u)J\e^{-\ri K \tau}J\nonumber \\
% & = & 
% - \, \partial_{u}\xi_{\pm}(\tau,\mathsf{j}(u))\,
% J\, j_{\pm}\big(\xi_{\pm}(\tau,\mathsf{j}(u))\big) J
% -\frac{\kappa\mu_{\pm}}{2\pi}\big[1+\partial_{u}\xi_{\pm}(\tau,\mathsf{j}(u))\big]
%\nn \\
 & = & 
 \, \partial_{u}\xi_{\pm}(\tau,\mathsf{j}(u)) 
 \left[\,
 \mathsf{j}'(v)\big|_{v=\xi_{\pm}(\tau,\,\mathsf{j}(u))}  \,
 j_{\pm}\big(\,\mathsf{j}(\xi_{\pm}(\tau,\mathsf{j}(u)))\big)
 -  \frac{\kappa\mu_{\pm}}{2\pi}
 \left(1 - \left.\mathsf{j}'(v)\right|_{v=\xi_{\pm}(\tau,\,\mathsf{j}(u))}  \right)
 \right]
 \nn \\
 &  & 
 - \, \frac{\kappa\mu_{\pm}}{2\pi}\big[1 - \partial_{u}\xi_{\pm}(\tau,\mathsf{j}(u)) \big]
\nn \\
%   \rule{0pt}{.6cm}
% & = & 
% \partial_{u}\tilde{\xi}_{\pm}(\tau,u)j_{\pm}\left(\tilde{\xi}_{\pm}(\tau,u)\right)+\partial_{u}\xi_{\pm}(\tau,\mathsf{j}(u))\frac{\kappa\mu_{\pm}}{2\pi}+\partial_{u}\tilde{\xi}_{\pm}(\tau,u)
%  -\frac{\kappa\mu_{\pm}}{2\pi}\big[1+\partial_{u}\xi_{\pm}(\tau,\mathsf{j}(u))\big]
% \nn \\
    \rule{0pt}{.6cm}
 & = & 
 \partial_{u}\tilde{\xi}_{\pm}(\tau,u) \, j_{\pm} \big(\tilde{\xi}_{\pm}(\tau,u) \big)
 -\frac{\kappa\mu_{\pm}}{2\pi}\big[1-\partial_{u}\tilde{\xi}_{\pm}(\tau,u)\big]
\end{eqnarray}
in terms of (\ref{app-xi-tilde-def}).
Thus,  as for the modular evolution of the chiral currents,
the expression \eqref{cp4} combined with (\ref{xi-map-fund}) holds also for $u\notin A$. 
This conclusion is found also in the finite volume case,
once  $\xi(\tau,u)$ is replaced by  $\xi_{L}(\tau,u)$ given in (\ref{xi-map-fund-circle}),
as one obtains by repeating the above analysis
replacing also $\mathsf{j}(u)$ with $\mathsf{j}_L(u)$ 
given in (\ref{j0-map-circle-def}).

In the finite volume case, for (\ref{cal-T-pm-def}), 
from the r.h.s. of (\ref{mod-evo-O-in-B}), \eqref{J-on-T-circle} and (\ref{met4})
we arrive to
%\begin{equation}
%J\mathcal{T}_{\pm}(u)J=\mathsf{j}_{L}'(u)^{2}\mathcal{T}_{\pm}(\mathsf{j}_{L}(u))+\frac{\kappa\mu_{\pm}^{2}}{4\pi}\left(1-\mathsf{j}_{L}'(u)^{2}\right)-\frac{c}{24\pi}\mathcal{S}_{u}\left[\mathsf{j}_{L}(u)\right]
%\end{equation}
\bea
\label{Delta-j-on-T-circle-app}
& & 
\mathrm{e}^{\mathrm{i}K\tau} \big( J\, \mathcal{T}_{\pm}(u)\,J\, \big)\,\mathrm{e}^{-\mathrm{i}K\tau} 
=\,
%\nn
%\\
% & & =
\mathsf{j}_{L}'(u)^{2} \,\mathrm{e}^{iK\tau} \,\mathcal{T}_{\pm}(\,\mathsf{j}_{L}(u)) \,\mathrm{e}^{-iK\tau}
+\left( \frac{\kappa \mu^2_\pm}{4\pi} - \frac{\pi c}{12 L^2} \right) \left[ \,1-\mathsf{j}'_L(u)^{2} \,\right]
%+\frac{\kappa\mu_{\pm}^{2}}{4\pi}\left(1-\mathsf{j}_{L}'(u)^{2}\right)-\frac{c}{24\pi}\mathcal{S}_{u}\left[\mathsf{j}_{L}(u)\right]
\nn \\
 & & =\,
 \mathsf{j}_{L}'(u)^{2} \; 
 \bigg\{ 
 \left( \partial_{v}\xi_{L,\pm}(\tau,v)\big|_{v=\mathsf{j}_{L}(u)}\right)^{2} \, 
 \mathcal{T}_{\pm}\big(\xi_{\pm}(\tau,\mathsf{j}_L(u))\big)
 +
 \frac{\kappa\mu_{\pm}^{2}}{4\pi}
 \left[\,1-\left(\left.\partial_{v}\xi_{\pm}(\tau,v)\right|_{v=\mathsf{j}_L(u)}\right)^{2} \,\right]
\nn \\
 &  & \hspace{2cm}
   -\, \frac{c}{24\pi} \, \mathcal{S}_{v} [\xi_{\pm}](\tau,v) \big|_{v=\mathsf{j}_{L}(u)}\,
   \bigg\} 
+\left( \frac{\kappa \mu^2_\pm}{4\pi} - \frac{\pi c}{12 L^2} \right) \left[ \,1-\mathsf{j}'_L(u)^{2} \,\right]
\nn \\
\rule{0pt}{.7cm}
 & & =\,
 \big(\partial_{u}\xi_{\pm}(\tau,\mathsf{j}_{L}(u))\big)^{2} \, 
 \mathcal{T}_{\pm}\big(\xi_{\pm}(\tau,\mathsf{j}_L(u))\big)
 +
 \frac{\kappa\mu_{\pm}^{2}}{4\pi}
 \left[ \, 1- \big(\partial_{u}\xi_{\pm}(\tau,\mathsf{j}_{L}(u))\big)^{2} \, \right]
 \nn\\
 \rule{0pt}{.6cm}
 &  & \hspace{.6cm}
% +\frac{\kappa\mu_{\pm}^{2}}{4\pi}\left(1-\mathsf{j}_{L}'(u)^{2}\right)
 - \,\frac{c}{24\pi}\;\mathsf{j}_{L}'(u)^{2}\, \mathcal{S}_{v} [\xi_{\pm}](\tau,v)\big|_{v=\mathsf{j}_{L}(u)}
 - \frac{\pi c}{12 L^2} \left[ \,1-\mathsf{j}'_L(u)^{2} \,\right]
% +\mathcal{S}_{u} [\,\mathsf{j}_{L}](u)
% \right\} 
\nn
 \\  
 \rule{0pt}{.7cm}
 \label{Delta-j-on-T-circle}
 & & 
 =\,
  \big(\partial_{u}\xi_{\pm}(\tau,\mathsf{j}_{L}(u))\big)^{2} \, 
  \mathcal{T}_{\pm}\big(\xi_{\pm}(\tau,\mathsf{j}_L(u))\big)
 +
 \frac{\kappa\mu_{\pm}^{2}}{4\pi}
 \left[ \, 1- \big(\partial_{u}\xi_{\pm}(\tau,\mathsf{j}_{L}(u))\big)^{2} \, \right]
  - 
  \frac{c}{24\pi}\;\mathcal{S}_{u}[\xi_\pm \circ  \mathsf{j}_{L} ](u) 
\nn \\
& &
%  \big(\partial_{u}\xi_{\pm}(\tau,\mathsf{j}_{L}(u))\big)^{2} \, \mathcal{T}_{\pm}(\,\mathsf{j}_{L}(u))
% +
% \left( \frac{\kappa \mu^2_\pm}{4\pi} - \frac{\pi c}{12 L^2} \right) 
% \left[ \, 1- \big(\partial_{u}\xi_{\pm}(\tau,\mathsf{j}_{L}(u))\big)^{2} \, \right]
% \left(\partial_{u}\xi_{L,\pm}(\tau,\mathsf{j}_{L}(u))\right)^{2}\mathcal{T}_{\pm}(\mathsf{j}_{L}(u))
% +\frac{\kappa\mu_{\pm}^{2}}{4\pi}
% \left[1-\big(\partial_{u}\xi_{L,\pm}(\tau,\mathsf{j}_{L}(u))\big)^{2}\right]
% -\frac{c}{24\pi}\mathcal{S}_{u} [\xi_{\pm}] (\tau,\mathsf{j}_{L}(u)) 
\eea
where the following identity for the Schwarzian derivative
\be
\label{schw-prop}
g'(u)^{2} \,\mathcal{S}_{v}[f](v)\big|_{v=g(u)}
=\,
\mathcal{S}_{u}[f\circ g](u) - \mathcal{S}_{u}[g](u)
\ee
and \eqref{id-sch-L} have been employed.
By applying $J(\cdot)J$
to \eqref{Delta-j-on-T-circle-app} and then using \eqref{J-on-T-circle},
we get
\be
\mathcal{T}_{\pm}(\tau,u) 
% & = & J\mathrm{e}^{\mathrm{i}K\tau}J\mathcal{T}_{\pm}J\mathrm{e}^{-\mathrm{i}K\tau}J \\
\,=\,
  \big(\partial_{u} \tilde{\xi}_{\pm}(\tau,\mathsf{j}_{L}(u))\big)^{2} \, \mathcal{T}_{\pm}(\,\mathsf{j}_{L}(u))
 +
 \left( \frac{\kappa \mu^2_\pm}{4\pi} - \frac{\pi c}{12 L^2} \right) 
 \left[ \, 1- \big(\partial_{u} \tilde{\xi}_{\pm}(\tau,\mathsf{j}_{L}(u))\big)^{2} \, \right]
%\left(\partial_{u}\tilde{\xi}_{L,\pm}(\tau,u)\right)^{2}\mathcal{T}_{\pm}\left(\tilde{\xi}_{L,\pm}(\tau,u)\right)+\frac{\kappa\mu_{\pm}^{2}}{4\pi}\left[1-\left(\partial_{u}\tilde{\xi}_{L,\pm}(\tau,u)\right)^{2}\right]-\frac{c}{24\pi}\mathcal{S}_{u}\left[\tilde{\xi}_{L,\pm}(\tau,u)\right] 
\ee
with $\tilde{\xi}_{\pm}(\tau,u)$ being defined in terms of (\ref{xi-map-fund-circle}) as follows
\be
\label{app-xi-tilde-def}
\tilde{\xi}_{\pm}(\tau,u)
\equiv\,
\mathsf{j}_L\big(\xi_{\pm}(\tau,\mathsf{j}_L(u))\big)
=\xi_{\pm}(\tau,u)
\ee
where (\ref{id-xi-j-circle}) has been employed in the last step.
Thus,
for the modular evolution of the chiral operators (\ref{cal-T-pm-def})
we can use \eqref{T-evolved-circle} combined with (\ref{xi-map-fund})
also for $u\notin A$.

%%%%%%%%%%%%%%%%%%%%%%%%%%%%%%%%%%%%%%%%%%%%%%%%%%%%%
%\newpage
\section{Integrals for the quantum noise}
\label{app-noise integrals}
%%%%%%%%%%%%%%%%%%%%%%%%%%%%%%%%%%%%%%%%%%%%%%%%%%%%%

In this appendix we discuss the explicit computation of the integrals occurring in Sec.\,\ref{subsec-noise}.
Let us consider first the following identity 
\be
\label{id-coth-1}
\frac{1}{\sinh^{2}(\pi\tau\pm \ri \epsilon)} 
\,=\,
-\frac{1}{\pi} \, \partial_{\tau} \coth(\pi\tau\pm \ri \epsilon)
\ee
where
\be
\label{id-coth-app}
\coth(\pi\tau\pm \ri \epsilon) = \coth(\pi\tau)\mp \ri\,\delta(\tau)
\ee
where in the r.h.s. the principal value regularization is assumed for the distribution $\coth(\pi\tau)$.
%It is useful to compute its Fourier transform
%\begin{equation}
%\int_{-\infty}^{+\infty}d\tau\,\mathrm{e}^{\mathrm{i}\omega\tau}\,\coth(\pi\tau)= \ri \coth\left(\frac{\omega}{2}\right) .
%\end{equation}

The first integral to consider is the one occurring in (\ref{P-JN-current-ris}), namely
\bea
& &
\label{app-noise-int-current}
\lim_{\varepsilon \to 0} \,
\int_{-\infty}^\infty 
\frac{\e^{\ri \omega t }}{\sinh^2 \! \big[ \pi ( t - \ri \varepsilon)/\beta\big] } 
   \, \rd t
  \,=\,
    - \frac{\beta}{\pi}\,
    \lim_{\varepsilon \to 0} \,\int_{-\infty}^{+\infty}\!
 \mathrm{e}^{\mathrm{i}\beta \omega\tau}\,
 \partial_{\tau} \coth(\pi\tau-  \ri \epsilon) \,\rd \tau
% \frac{\kappa}{2\pi}\int_{-\infty}^{+\infty}\mathrm{e}^{\mathrm{i}\omega\tau}\,\partial_{\tau}\left[\coth(\pi\tau)\right]
\\
\rule{0pt}{.7cm}
& &
=\,
    \ri\;\frac{\beta^2 \omega}{\pi}\,
    \int_{-\infty}^{+\infty}\!
    \big[
    \coth(\pi\tau) + \ri\,\delta(\tau)
    \big]\,
 \mathrm{e}^{\mathrm{i}\beta \omega\tau} \,\rd \tau
 \,=\,
 -\frac{\beta^2}{\pi} \, \big[ \, \omega\, \coth(\beta \omega/2) +\omega \,\big]
 \nn
\eea
%\begin{eqnarray}
%\label{app-noise-int-current}
%& & \hspace{-.45cm}
%\int_{-\infty}^{+\infty}
%\left[ \, \frac{1}{\sinh^{2}(\pi\tau-\ri \epsilon)}+\frac{1}{\sinh^{2}(\pi\tau+\ri \epsilon)} \, \right] 
%\mathrm{e}^{\mathrm{i}\omega\tau} \, \rd \tau
%\\
%\rule{0pt}{.75cm}
% & = & 
% \frac{\kappa}{2\pi}\int_{-\infty}^{+\infty}\!
% \mathrm{e}^{\mathrm{i}\omega\tau}\,
% \partial_{\tau} \coth(\pi\tau) \,\rd \tau
%% \frac{\kappa}{2\pi}\int_{-\infty}^{+\infty}\mathrm{e}^{\mathrm{i}\omega\tau}\,\partial_{\tau}\left[\coth(\pi\tau)\right]
%\,=\,
% \frac{\kappa}{2\pi}(-\mathrm{i}\omega)\int_{-\infty}^{+\infty}\!
% \mathrm{e}^{\mathrm{i}\omega\tau}
% \coth(\pi\tau)  \,\rd \tau
%\,=\,
%% \frac{\kappa}{2\pi}(-\mathrm{i}\omega) \ri \coth\left(\frac{\omega}{2}\right)
%% \,=\,
% \frac{\kappa}{2\pi} \, \omega\coth(\omega/2)
%\nonumber 
%\end{eqnarray}
where we used (\ref{id-coth-1}), (\ref{id-coth-app}) 
and $\int_{-\infty}^{+\infty} \mathrm{e}^{\mathrm{i}\omega\tau} \coth(\pi\tau) \,\rd \tau= \ri \coth (\omega/2)$.
This computation can be adapted to investigate also the other integral we need, 
which occurs in (\ref{P-JN-energy-ris0}).
Indeed, by considering the following identity
\be
\label{sinh4-id}
\frac{1}{\sinh^{4}(\pi\tau\pm \ri \epsilon)} 
\,=\,
\frac{2}{3\pi} \, \partial_{\tau} \coth(\pi\tau\pm \ri \epsilon)
-
\frac{1}{6\pi^{3}} \, \partial_{\tau}^{3} \coth(\pi\tau\pm \ri \epsilon)
\ee
we find that
\bea
& &
\label{app-noise-int-energy}
\lim_{\varepsilon \to 0} \,
\int_{-\infty}^\infty 
\frac{\e^{\ri \omega t }}{\sinh^4 \! \big[ \pi ( t - \ri \varepsilon)/\beta\big] } 
   \, \rd t
   \\
   \rule{0pt}{.7cm}
   & &
  \,=\,
    \frac{\beta}{\pi}\,
    \lim_{\varepsilon \to 0} \,\int_{-\infty}^{+\infty}\!
 \left[\, \frac{2}{3} \, \partial_{\tau} \coth(\pi\tau-\ri\epsilon)
 -
 \frac{1}{6\pi^{2}}\,\partial_{\tau}^{3} \coth(\pi\tau-\ri\epsilon) \right]
 \mathrm{e}^{\mathrm{i}\beta \omega\tau}
 \,\rd \tau
% \frac{\kappa}{2\pi}\int_{-\infty}^{+\infty}\mathrm{e}^{\mathrm{i}\omega\tau}\,\partial_{\tau}\left[\coth(\pi\tau)\right]
\nn
\\
\rule{0pt}{.7cm}
   & &
  \,=\,
   -\,\ri\, \frac{\beta^2 \omega}{\pi}
     \left[\, \frac{2}{3} + \frac{(\beta \omega)^2}{6\pi^{2}} \,\right]
         \lim_{\varepsilon \to 0} \,\int_{-\infty}^{+\infty}\!
 \mathrm{e}^{\mathrm{i}\beta \omega\tau}
 \coth(\pi\tau-  \ri \epsilon) \,\rd \tau
\nn
\\
\rule{0pt}{.6cm}
   & &
     \,=\,
  \frac{\beta^2}{6\pi^3} \, \big[\,4 \pi^2 + (\beta \omega)^2\,\big] \,\big[ \, \omega \coth (\beta\omega / 2) +\omega\, \big]
  \nn
\eea
where (\ref{sinh4-id}) and (\ref{id-coth-app})
have been employed.

\bibliographystyle{nb}

\bibliography{refsModTrans}

\end{document}